\documentclass[preprint,preprintnumbers,aps,superscriptaddress,nofootinbib,tightenlines,floatfix]{revtex4}

\usepackage{amsmath,amssymb,bm}
\usepackage{subfigure}
\usepackage{verbatim}

\newcommand{\beq}{\begin{equation}}
\newcommand{\eeq}{\end{equation}}
\newcommand{\bea}{\begin{eqnarray}}
\newcommand{\eea}{\end{eqnarray}}

\newcommand{\gsim}{\raisebox{-0.7ex}{$\stackrel{\textstyle >}{\sim}$ }}
\newcommand{\lsim}{\raisebox{-0.7ex}{$\stackrel{\textstyle <}{\sim}$ }}

\ifx\pdfoutput\undefined
\usepackage[dvips]{graphicx}
\else
\usepackage[pdftex]{graphicx}
\fi
\usepackage{epstopdf}

\begin{document}
\DeclareGraphicsExtensions{.pdf,.gif,.jpg}
\begin{figure}[!t]

  \vskip -1.5cm
  \leftline{\includegraphics[width=0.25\textwidth]{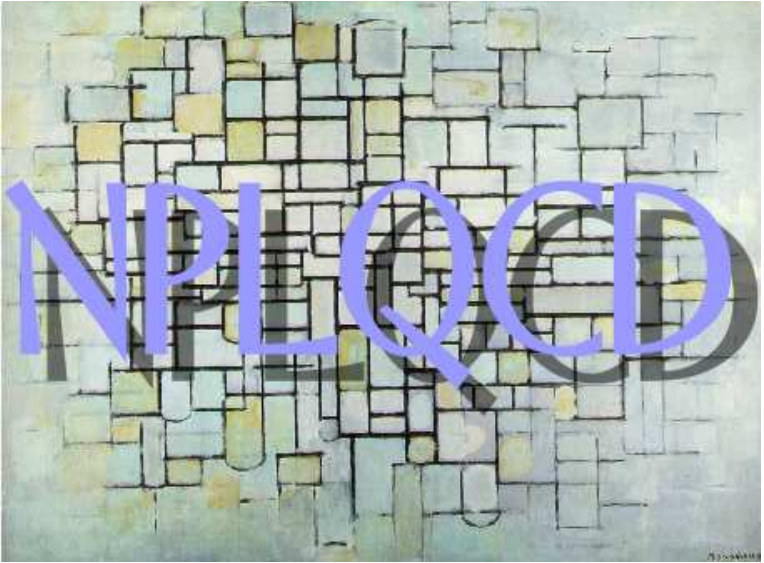}}
\end{figure}

\preprint{\vbox{ 
\hbox{NT@UW-11-01} 
}}

\title{Extracting Scattering Phase-Shifts in Higher Partial-Waves from
  Lattice QCD Calculations
}
\author{Thomas Luu}
\email[]{tluu@llnl.gov}
\affiliation{N
  Section, Lawrence Livermore National Laboratory, Livermore, CA
  94551, USA}
\author{Martin J. Savage}
\email[]{mjs5@u.washington.edu}
\affiliation{Department of Physics, University of Washington, Seattle,
  WA 98195-1560, USA}

\date{\today}

\begin{abstract}
L\"uscher's method is routinely used to determine
meson-meson, meson-baryon and baryon-baryon
s-wave scattering amplitudes below inelastic thresholds
from Lattice QCD calculations - presently at unphysical light-quark masses.  
In this work  we review the formalism and develop the requisite
expressions to extract phase-shifts
describing meson-meson scattering in partial-waves with angular-momentum $l\le
6$ and $l=9$.
The implications of the underlying cubic symmetry, and strategies 
for extracting the phase-shifts from Lattice QCD calculations, are presented,
along with a discussion of the signal-to-noise problem that
afflicts the higher partial-waves.

\end{abstract}


\maketitle

\tableofcontents
\newpage
\section{Introduction\label{sect:intro}  }
\noindent
The s-wave interactions between hadrons are being calculated with Lattice QCD (LQCD) with
increasing precision.  Presently, such calculations are being performed at
unphysical light-quark masses, and in the case of mesonic interactions,
extrapolations to the physical light-quark masses are made possible 
by chiral perturbation theory ($\chi$PT). Unfortunately, such extrapolations
are presently not reliable for baryon-baryon interactions, and 
it is likely that 
LQCD calculations at, or very near, the physical light-quark masses will be
required to make precise predictions for these interactions due to the
fine-tunings that are known to exist in nuclear physics.
In most LQCD calculations, 
periodic boundary-conditions (BC's) are imposed on the quark and gluons
fields in the spatial-directions of
the lattice volume and  L\"uscher's 
method~\cite{Luscher:1986pf,Luscher:1990ux} 
can be  used to extract scattering phase-shifts 
from the energy-eigenvalues of two-hadron states that lie below 
inelastic thresholds.  
As it is the irreducible representations (irreps) of the cubic group
that determine the degeneracies of the eigenstates in the (cubic) 
lattice volume,
it is difficult to determine the phase-shifts, $\delta_l$, beyond the
lowest few partial-waves.
Each of the irreps of the cubic group have a non-zero
overlap with infinitely many irreps of SO(3), and  as a result, the energy-eigenvalues
of two-hadron states 
transforming as a certain irrep of the cubic group receive
contributions from the phase-shifts in an infinite number of partial-waves.  
In contrast, two-particle systems confined in a harmonic oscillator potential 
have a one-to-one relation between phase shifts and the energy-eigenvalues 
since the potential respects SO(3) 
symmetry~\cite{Luu:2010hw,Stetcu:2007ms}~\footnote{The same is true when a 
``spherical-wall'' is imposed on the separation between
hadrons, as has been demonstrated in recent lattice effective field theory 
(LEFT) calculations~\cite{Epelbaum:2010xt}.}.
The mixing of angular momentum in cubic irreps consequently limits the 
precision with which the phase-shift in any
given partial-wave can be extracted in a LQCD calculation.
This was made obvious in the work of Mandula, Zweig and Govaerts,\cite{Mandula:1983ut},
and 
explicitly detailed in L\"uscher's
papers~\cite{Luscher:1986pf,Luscher:1990ux}. 
L\"uscher calculated 
the energies of states in the $A_1^+$ irrep of the cubic group as a
function of $\delta_0$ and $\delta_4$, and gave general expressions for the energies
of states transforming in each of the 
cubic irreps in terms of the $\delta_l$.
The extension of this formalism to systems with non-zero center-of-mass (CM) momentum
was performed by Rummukainen and Gottlieb~\cite{Rummukainen:1995vs}, 
and later by Kim, Sharpe and Sachrajda~\cite{Kim:2005gf}.  Recently finite-volume expressions for three-nucleon systems within cubic volumes have been investigated \cite{Luu:2008fg,Kreuzer:2010ti,Kreuzer:2008bi}.

Fully-dynamical $n_f=2+1$ LQCD calculations of meson-meson interactions in the
isospin-stretch-states ({\it i.e.} no disconnected diagrams) 
are presently enabling predictions of  the s-wave
interactions with percent-level 
precision~\cite{Beane:2005rj,Beane:2007xs,Feng:2009ij,Beane:2006gj,Nagata:2008wk,Beane:2007uh}
(for a recent review, see Ref.~\cite{Beane:2010em}),
and very recently a preliminary calculation of the $\pi^+\pi^+$ d-wave
phase-shift has been performed~\cite{Dudek:2010ew}.
Further, preliminary calculations of $I=0$ $\pi\pi$ scattering, which contain 
disconnected diagrams, have been performed~\cite{Liu:2009uw}.
These calculations were preceded by quenched 
LQCD
calculations~\cite{Sharpe:1992pp,Gupta:1993rn,Kuramashi:1993ka,Kuramashi:1993yu,Fukugita:1994na,Gattringer:2004wr,Fukugita:1994ve,Fiebig:1999hs,Aoki:1999pt,Liu:2001zp,Liu:2001ss,Aoki:2001hc,Aoki:2002in,Aoki:2002sg,Aoki:2002ny,Juge:2003mr,Ishizuka:2003nb,Aoki:2005uf,Aoki:2004wq},
and by early $n_f=2$ LQCD calculations~\cite{Yamazaki:2004qb}.
Meson-baryon systems are starting to be explored in the channels for which
disconnected diagrams are not required in the LQCD calculations~\cite{Torok:2009dg}.
Further, LQCD 
calculations of baryon-baryon interactions are beginning to become
reliable at unphysical pion 
masses~\cite{Beane:2006mx,Beane:2006gf,Aoki:2008hh,Beane:2009py},
and recently the binding energy of the H-dibaryon has been 
calculated~\cite{Beane:2010hg}.
Now that the methodology for extracting s-wave interactions has been
shown to be effective,  it is appropriate to explore the higher partial-waves.  
In the meson sector, a determination of the p-wave phase-shifts
has direct implications for post-dicting the mass and width of
the $\rho$-meson~\cite{Gockeler:2008kc,Frison:2010}, but this requires evaluating
disconnected diagrams in LQCD - calculations that are
computationally expensive.  In nucleon-nuclei scattering, the experimentally
determined p-wave
phase-shifts are thought to be at the heart of the ``$A_y$-puzzle'' in 
nucleon-deuteron scattering.  
Further, it is found phenomenologically that only the phase-shifts in partial waves 
with $l\lsim 4$ are required
to perform relatively  precise calculations of nuclear structure and reactions (at the
physical pion mass).

The formalism required to analyze the $J=1$ coupled-channels, in which
the deuteron is the ground-state, has been put in place by 
Liu, Feng and He~\cite{Liu:2005kr}, 
and exploratory quenched calculations 
of the s-d mixing parameter, $\varepsilon$,
at pion masses of $m_\pi\sim
730, 530$ and $380~{\rm MeV}$ have been performed
in a small number of lattice volumes~\cite{Aoki:2009ji}.  
Further, there has been recent work in developing the phenomenology that goes
beyond L\"uschers formalism in an attempt to explore resonances (and couplings to
multi-hadron final-states) in the single baryon and
meson sectors\cite{Niu:2010rk}.

It is appropriate to point out that there is a substantial amount of
information and technology 
that is directly relevant to this subject, in particular space-groups,
that has been developed for study of condensed matter systems.  
Much of the work in this paper draws directly from various applications found in these fields.
Discussions of space-groups can be found in texts, such as
Ref.~\cite{Dresselhaus} or Ref.~\cite{Tinkham}, 
as are discussions of point-groups, and other
formalisms that impact  the present calculations.

While the papers by  L\"uscher~\cite{Luscher:1986pf,Luscher:1990ux} contain 
the required formalism, 
we take this opportunity to present the explicit relations between the
energy-eigenvalues of two-meson states in a cubic volume 
and the phase-shifts in the partial-waves with $l\le 6$ and $l=9$.
The experimentally measured phase-shifts describing  $\pi\pi$ scattering in the
lowest-lying partial-waves, appropriately parameterized, are used to perform
estimates of the energy-eigenvalues that are expected in LQCD calculations of 
such systems  over a range of lattice volumes. We also discuss the issue of signal-to-noise degradation while performing lattice calculations in higher partial waves.

\section{Formalism\label{sect:formalism} }
\noindent
In the absence of interactions, the states in the cubic volume can be defined
by their behavior under transformations of the cubic group and by their energy. As the
momentum in the volume is quantized in integer multiples of $2\pi/L$, where $L$
is the spatial-extent of the volume, the energy quantum number can be replaced
by the magnitude of the integer triplet defining the momentum, $|{\bf n}|^2$, where 
${\bf n}=(n_x,n_y,n_z)$. Instead of the energy, it is convenient to refer to
the particular
$|{\bf n}|^2$-shell.  For each partial-wave with $l\le 6$ and $l=9$, an irrep
of the cubic group is identified for which $\delta_l$ provides the dominant
contribution to the interaction energy.
Sources and sinks used in LQCD calculations that are
constructed to transform under such  irreps  will
allow for  a determination of the $\delta_l$ at some level of precision.  
The energy of states  with  $|{\bf n}|^2\le 6$
are required to lie below the inelastic threshold in order to obtain 
all of the phase-shifts with  $l\le 6$, thereby requiring
relatively large lattice volumes.  Further, the energy of a state in the 
$|{\bf n}|^2=14$-shell is required to obtain the $l=9$ phase-shift.

A non-zero phase-shift in a given partial-wave will, in general, contribute to
the energy-eigenvalues of two-hadron states in the volume that transform as one
or more
irreducible representations of the full cubic group,
$\Gamma^{(i)}$.~\footnote{
The irreps of the full cubic group are
$\Gamma^{(i)} = A_1^\pm, A_2^\pm, E^\pm, T_1^\pm$, and $T_2^\pm$, and have dimensionality 
$1, 1, 2, 3$ and $3$ respectively. The
superscript denotes the parity of $\Gamma^{(i)}$.}
\begin{table}
\begin{center}
\begin{minipage}[!ht]{16.5 cm}
\caption{
Decomposition of the orbital angular momentum eigenstates,
$|l,m\rangle$, into irreps of the cubic group, $\Gamma^{(i)}$, for $l\le
9$ (see, for instance, Ref.~\protect\cite{Dresselhaus}).
}
\label{tab:Lcontributetab}
\end{minipage}
\begin{tabular}{| c | c | }
\hline 
  \ \ \ {\rm Angular Momentum, l}\ \ \   
& \ \ \ {\rm Irreps of the Cubic Group,} $\Gamma^{(i)}$\ \ \ 
    \\
      \hline
$ 0$ & $A_1^+$ \\
$ 1$ & $T_1^-$  \\
$ 2$ & $E^+\oplus T_2^+$   \\
$ 3$ & $A_2^-\oplus T_1^-\oplus T_2^-$   \\
$ 4$ & $A_1^+\oplus E^+\oplus T_1^+\oplus T_2^+$ \\
$ 5$ & $E^-\oplus T_1^{-(1)}\oplus T_1^{-(2)} \oplus T_2^-$ \\
$ 6$ & $A_1^+\oplus A_2^+\oplus  E^+\oplus T_1^+\oplus T_2^{+(1)}\oplus T_2^{+(2)}$   \\
$ 7$ & $A_2^-\oplus E^-\oplus T_1^{-(1)}\oplus T_1^{-(2)} \oplus T_2^{-(1)} \oplus T_2^{-(2)}$ \\
$ 8$ & $A_1^+ \oplus E^{+(1)} \oplus E^{+(2)} \oplus T_1^{+(1)}  \oplus
T_1^{+(2)} \oplus T_2^{+(1)}\oplus T_2^{+(2)}$   \\
$ 9$ & $A_1^-\oplus A_2^-\oplus E^-\oplus T_1^{-(1)} \oplus T_1^{-(2)}  \oplus
T_1^{-(3)}  \oplus T_2^{-(1)} \oplus T_2^{-(2)}$ \\
\hline
\end{tabular}
\begin{minipage}[t]{16.5 cm}
\vskip 0.5cm
\noindent
\end{minipage}
\end{center}
\end{table}     
Table~\ref{tab:Lcontributetab} shows the decomposition of the 
orbital angular momentum eigenstates, $|l,m\rangle$,
into the $\Gamma^{(i)}$ for $l\le 9$, 
from which it is straightforward to determine 
the $\Gamma^{(i)}$ that have
energy-eigenvalues that 
depend upon a given phase-shift $\delta_l$~\footnote{Each $\Gamma^{(i)}$
  appears at least once in the decomposition of the $|l,m\rangle$ with $l\le 6$
except $A_1^-$ which first appears in the decomposition of the 
$l=9$ irrep~\cite{Dresselhaus}.
It is important to note that the decompositions of the $l=7$ and $l=8$ irreps
contain only $\Gamma^{(i)}$ that also appear in the decomposition of the
$l\le 6$ irreps, and consequently there is no $\Gamma^{(i)}$ for which the
dominant contribution to the interaction energy (in the large volume limit)
is from the $l=7$ and $l=8$ partial-waves.
}.
A cursory study of table~\ref{tab:Lcontributetab} shows 
that $A_1^+$-states  will, in general, receive 
contributions to their energy from interactions with $l=0,4,6,8,...$,
as is well known~\cite{Mandula:1983ut}, and similarly for the other $\Gamma^{(i)}$.
As the dimensionality of an  SO(3) irrep (which is $2l+1$ for $|l,m\rangle$)  
must be equal to the sum of the dimensionalities of
the cubic irreps in its  decomposition, 
cubic irreps will, in general, appear multiple times 
(with multiplicities denoted by $N(\Gamma^{(i)},l)$) 
in the decomposition of an SO(3) irrep.  Multiplicities greater than one occur
for $l\ge 5$.  
The space associated with the $j^{\rm th}$ occurrence of $\Gamma^{(i)}$
in the decomposition of $|l,m\rangle$ is spanned by
the orthonormal basis $\{\  |\Gamma^{(i)},L_z;l;j\rangle\ \}$, where the number of
values of $L_z$ equals 
the dimensionality of $\Gamma^{(i)}$, 
e.g. for $l=5$, the 3-dimensional irrep $T_1^-$ occurs twice,
and the space associated with the second occurrence is spanned by
$\{\  |T_1^-,0;5;2\rangle \ ,\  |T_1^-,1;5;2\rangle \ 
,\   |T_1^-,3;5;2\rangle \}$~\footnote{The $L_z$ quantum number indicates
  that a phase of $e^{iL_z\phi}$ results from a (cubic) rotation of $\phi=n\pi/2$
  about the z-axis, with $n$ an integer.
$L_z=3$ is equivalent to a $L_z=-1$ and  $L_z=2$ is equivalent to $L_z=-2$.
}.
When calculating observables in a cubic volume, 
operators transforming as a component of a spherical tensor of
rank-S,
$\hat O_S^{(\mu)}$,
are most conveniently written as
\begin{eqnarray}
\hat O_S^{(\mu)} & = & 
\sum_{i,j,L_z}\ 
\theta^{(\Gamma^{(i)},j,L_z ; S,\mu)}
\ {|\Gamma^{(i)},L_z ; S; j\rangle }
{\langle\Gamma^{(i)},L_z ; S; j|}
\ \ \ ,
\label{eq:opexp}
\end{eqnarray}
where the values of the $\theta^{(\Gamma^{(i)},j,L_z ; S,\mu)}$ are simply determined by
matrix elements of $\hat O_S^{(\mu)}$ between $|l,m\rangle$, or
$|\Gamma^{(i)}, L_z; S; j\rangle$, or any states forming a basis in which the
projections onto $|\Gamma^{(i)}, L_z; S; j\rangle$
are known.  In determining the energy-eigenvalues of the states in the
volume, it is the scattering amplitude in  a given partial-wave
that is written in the form of eq.~(\ref{eq:opexp}),  with $S=l$.

The relations between the energy-eigenvalues of two-hadrons in a cubic volume 
and their scattering phase-shifts below the inelastic threshold, 
originally derived in the context of
non-relativistic quantum mechanics, were shown to be valid in quantum field
theory (QFT) without modification by
L\"uscher~\cite{Luscher:1986pf,Luscher:1990ux}.
The energy-shifts 
of scattering states
due to the interactions exhibit power-law dependence
upon the volume when the range of the interaction is
negligible compared to the spatial-extent of the volume.  
Corrections arising from the range of the interaction (for the case of
$\pi^+\pi^+$ the range is set by $R\sim 1/(2 m_\pi)$, while for nucleon-nucleon
interactions it is set by $R\sim 1/m_\pi$) are exponentially suppressed for
$L \gg R$, and of the form $\sim e^{-L/R}$ \cite{Bedaque:2006yi}.  
In this work, it is assumed
that these finite-range
corrections are negligible compared to the power-law energy-shifts due to the interactions.
It is straightforward to 
calculate a two-hadron Green function
resulting from an arbitrary source and sink.  The Green-function is generated by
the bubble-diagrams with non-interacting two-hadron states propagating from 
the source through multiple insertions of the T-matrix, and then to the 
sink.
In free-space, the Green function exhibits poles at the location of bound-states
and cuts along the positive real axis.  In the finite-volume, modifications
to the propagation of the two non-interacting hadrons eliminates the cuts on the
positive real axis, replacing them with poles at the location of the
energy-eigenstates. Further, these modifications shift the location of the
poles on the negative real axis (if present in infinite volume).
The energy-eigenvalues, corresponding  to both  bound-states and continuum states
in the infinite-volume limit are determined by solutions to~\cite{Kim:2005gf},
\begin{eqnarray}
{\rm det}\left[\ 
\cos\delta - \sin\delta\ F^{(FV)}\ \right]
& = & 0
\ \ \ ,
\label{eq:evals}
\end{eqnarray}
where $\cot\delta$, $\sin\delta$ and $F^{(FV)}$ are $(l^{\rm max}+1)^2\times
(l^{\rm max}+1)^2$ dimensional matrices when the phase-shifts $\delta_l$ are
non-zero for $l\le l^{\rm max}$ and vanish for $l> l^{\rm max}$.  
Initially, it is convenient to work in the $|l,m\rangle$ basis
in which, for uncoupled channels, $\cot\delta$ and 
$\sin\delta$ are diagonal matrices of the form
\begin{eqnarray}
\cos\delta & = & \cos\delta_{l_1}\ \delta_{l_1,l_2}\ \delta_{m_1,m_2}
\ \ ,\ \ 
\sin\delta \ = \ \sin\delta_{l_1}\ \delta_{l_1,l_2}\ \delta_{m_1,m_2}
\ \ ,
\label{eq:trigs}
\end{eqnarray}
for $l_{1,2}\le l^{\rm max}$,
but in which $F^{(FV)}$ has off-diagonal elements, in general.
$F^{(FV)}$ is a matrix that is a function of the dimensionless quantity 
$\tilde q = { q L\over 2\pi}$, where $q$ is related to the energy of
the interacting two-hadron state,
$E_{H_1 H_2} = \sqrt{q^2 + m_{H_1}^2} + \sqrt{q^2 + m_{H_2}^2}$.
Its matrix elements are of the form
\begin{eqnarray}
\overline F^{(FV)}_{l_1m_1;l_2m_2} & = & 
{(-)^{m_2}\over\tilde q\ \pi^{3/2}}\ \sqrt{(2l_1+1)(2l_2+1)}\ 
\sum_{\overline{l}=|l_1-l_2|}^{|l_1+l_2|}\ 
\nonumber\\
&&
\qquad\qquad
\sum_{\overline{m}=-\overline{l}}^{\overline{l}}\ 
{\sqrt{2\overline{l}+1}\over \tilde q^{\overline{l}}}
\left( 
\begin{array}{ccc}
l_1 & \overline{l} & l_2\\
0&0&0
\end{array}
\right)
\left(
\begin{array}{ccc}
l_1 & \overline{l} & l_2\\
-m_1&-\overline{m}&m_2
\end{array}
\right)\
{\cal Z}_{\overline{l},\overline{m}}(1;\tilde q^2)
\ \ \ ,
\label{eq:FV}
\end{eqnarray}
where the functions ${\cal Z}_{l,m}(1;\tilde q^2)$ are
those defined by L\"uscher~\cite{Luscher:1986pf,Luscher:1990ux},
\begin{eqnarray}
{\cal Z}_{l,m}(s;\tilde q^2) & = & 
\sum_{ {\bf n} } \ 
{ |{\bf n}|^l\ Y_{lm}(\Omega_{\bf n}) \over
\left[\ |{\bf n}|^2 - \tilde q^2\ \right]^s }
\ \ \ ,
\label{eq:Zfun}
\end{eqnarray}
where $Y_{lm}(\Omega)$ are the spherical harmonics.
The function ${\cal Z}_{0,0}(1;\tilde q^2)$ is UV-divergent
and is defined with the same renormalization scheme used to define the
infinite-volume scattering amplitude.
It is useful to first
diagonalize the blocks of $F^{(FV)}$ with $l_1=l_2=l$,  $\overline F^{(FV)}_{l m_1 , l
  m_2}$,
which diagonalizes $F^{(FV)}$ into blocks with dimensions
dictated by the number of occurrences of each $\Gamma^{(i)}$ for $l\le l^{\rm
  max}$,  while leaving the $\sin\delta$ and
$\cos\delta$ matrices diagonal.  
Further diagonalizations that may be  required are confined
within each $\Gamma^{(i)}$.
The determinant in eq.~(\ref{eq:evals}) becomes the product
of determinants resulting from each $\Gamma^{(i)}$, and therefore, 
for $l^{\rm max}=6$,  this procedure requires dealing with  matrices of size 
$4\times 4$ or smaller.

Despite the fact that  eq.~(\ref{eq:evals}) requires forming the determinant of
a finite dimensional matrix, it has infinitely many solutions.
It is derived from a 
Green-function between arbitrary sources and sinks which, in principle,
 can couple to all of the
eigenstates in the volume, manifested in the infinite-sums over
integer-triplets 
that define the ${\cal Z}_{l,m}$-functions.
Therefore, the zero's of the determinant in eq.~(\ref{eq:evals})
define all of the energy-eigenvalues and hence eigenstates.
As discussed previously, 
the energy-spectrum of two non-interacting
hadrons in the cubic volume with periodic BC's, and with
vanishing total momentum can be defined by triplets of integers, ${\bf n}$, 
\begin{eqnarray}
E & = & 
\sqrt{ |{\bf q}_1|^2 + m_1^2}\ +\ \sqrt{ |{\bf q}_2|^2 + m_2^2}
\ \rightarrow\ 
\sqrt{ \left({2\pi\over L}\right)^2 |{\bf n}|^2 + m_1^2}\ +\ 
\sqrt{ \left({2\pi\over L}\right)^2 |{\bf n}|^2 + m_2^2}
\nonumber\\
& =  &
{ |{\bf q}_1|^2\over 2 m_1} \ +\  { |{\bf q}_2|^2\over 2 m_2} 
\ +\ ...
\ \rightarrow \ {2\pi^2\over \mu L^2}\ |{\bf n}|^2\ +\ ...
\ \ \ .
\label{eq:KIN}
\end{eqnarray}
where one hadron carries momentum ${\bf q}_1 = {2\pi\over L} {\bf n}$ and the other
carries momentum ${\bf q}_2 = -{2\pi\over L} {\bf n}$, and the reduced mass of
the system is $\mu^{-1} = m_1^{-1} + m_2^{-1}$.
This (non-interacting) spectrum is recovered in  the above formalism,
in particular eq.~(\ref{eq:evals}),
in the limit that
$\delta_l\rightarrow 0$ in each partial-wave 
from the poles in the 
${\cal  Z}_{l,m}$-functions that exist along the positive real axis.
The degeneracy of any given $|{\bf n}|^2$-shell is
straightforward to determine 
and is recovered from the 
number of states in the $\Gamma^{(i)}$
that span the $|{\bf n}|^2$-shell, as shown in 
table~\ref{tab:multiplicities}.
As the (single hadron) momentum eigenstates in a given  $|{\bf n}|^2$-shell
are degenerate, the corresponding $\Gamma^{(i)}$ are also degenerate.
These degeneracies are lifted by two-particle interactions that induce non-zero $\delta_l$'s.
\begin{table}
\begin{center}
\begin{minipage}[!ht]{16.5 cm}
  \caption{The degeneracies of,
and the number of occurrences of each $\Gamma^{(i)}$ in,
the lowest-lying $|{\bf n}|^2$-shells. 
Note: the $A_1^-$ irrep first appears in the  $|{\bf n}|^2=14$ shell.
\label{tab:multiplicities}}
\end{minipage}
\vskip 0.1in
\begin{tabular}{|c||c||c|c|c|c|c|c|c|c|c|c|}
\hline
$|{\bf n}|^2$ & {\rm degeneracy} &
\quad $A_1^+$\ \ \ \  & \quad $A_2^+$\ \ \ \  & \quad $T_1^+$\ \ \ \
& \quad $T_2^+$\  \ \ \  & \quad $E^+$\ \ \ \  & \quad $A_1^-$\ \ \ \   & \quad
$A_2^-$\ \ \ \  & \quad
$T_1^-$\ \ \ \  & \quad $T_2^-$\ \ \ \ 
& \quad $E^-$\\
\hline
0 & 1 & 1 & - & - & - & - & - & - & - & - & -\\
\hline
1 & 6 & 1 & - & - & - & 1 & - & - & 1 & - & -\\
\hline
2 & 12  & 1 & - & - & 1 & 1 & - & - & 1 & 1 & -\\
\hline
3 & 8 & 1 & - & - & 1 & - & - & 1 & 1 & - & -\\
\hline
4 & 6 & 1 & - & - & - & 1 & - & - & 1 & - & -\\
\hline
5 & 24 & 1 & 1 & 1 & 1 & 2 & - & - & 2 & 2 & -\\
\hline
6 & 24 & 1 & - & 1 & 2 & 1 & - & 1 & 2 & 1 & 1\\
\hline
\vdots &   &   &   &   &   &   &   &   &   &   &   \\
\hline
14 & 48 & 1 & 1 & 3 & 3 & 2 & 1 & 1 & 3 & 3 & 2 \\
\hline
\end{tabular}
\begin{minipage}[t]{16.5 cm}
\vskip 0.5cm
\noindent
\end{minipage}
\end{center}
\end{table}
Table~\ref{tab:multiplicities} shows that all but one of the $\Gamma^{(i)}$ are required to
describe the eigenstates for $|{\bf n}|^2\le 6$, and from
table~\ref{tab:Lcontributetab} it can be concluded that for $\delta_l\ne 0$ 
for $l\le 6$ all of the eigenstates with $|{\bf n}|^2\le 6$ are shifted from
the non-interacting two-hadron energy due to interactions.  However, the $A_1^-$ irrep
first occurs in the $|{\bf n}|^2=14$ shell and its energy is dependent upon
interactions with $l\ge 9$.

\section{Energy-Eigenvalues, Sources and Sinks
\label{sect:formulas}}
\noindent
L\"uschers 
formalism, as detailed in the previous section, is used to construct
explicit relations between the energy-eigenvalues of the $\Gamma^{(i)}$
 and the interaction phase-shifts for $l\le l^{max}=6$, the
results of which are presented in this section.  
Sources and sinks for LQCD calculations that transform as a given 
$\Gamma^{(i)}$ are constructed from the single-hadron momentum-eigenstates, and
Fourier transformed into position-space.
One pair of these sources and sinks would couple only to a single 
energy-eigenstate in the absence of interactions between the hadrons.
As the interactions do not induce mixing between distinct $\Gamma^{(i)}$, these
sources and sinks couple, in principle, to all states that transform in the
same $\Gamma^{(i)}$.
To keep the presentation of results simple, explicit derivations are
deferred to Appendix~\ref{sect:explicit calculation}, where
calculations of the even- and odd-parity systems with $l^{max}=4$ are detailed,
and which 
straightforwardly generalize to any $l^{max}$.

As the hadronic interactions considered in this work result from QCD 
with the strong CP-violating parameter $\theta$ set equal to zero, and
without the electroweak interactions,
parity is a good quantum number. 
Consequently, the contributions to the finite-volume function $F^{(FV)}$ do  not mix
states of opposite parity, and therefore the required calculations 
decompose into the parity-even and parity-odd sectors.
If weak-interactions are included in the analysis, 
as will necessarily be the case
when hadronic parity-violating interactions are calculated 
with LQCD, mixing between the parity-sectors will occur.

\subsection{Positive Parity Systems}
There are five positive parity irreps of the cubic group, $A_1^+$, $A_2^+$, $E^+$,
$T_1^+$, and $T_2^+$ with dimensions $1,1,2,3$,and $3$ respectively.
Table~\ref{tab:Lcontributetab} shows how the interactions in a given
partial-wave contribute to each $\Gamma^{(i)}$.  The energy-eigenvalues,
sources and sinks for the even-parity states are presented in the following
sections:~\ref{irrep:A1p}, \ref{irrep:A2p}, \ref{irrep:Ep}, \ref{irrep:T1p}
and \ref{irrep:T2p}.

\subsubsection{$A_1^+$ Representation\label{irrep:A1p}}

The energy-eigenvalues of $A_1^+$ states depend upon
the phase-shifts in the $l=0, 4,6,8,...$ partial-waves, as can be seen in 
table~\ref{tab:Lcontributetab}.
Diagonalization of the blocks in the finite-volume function of the form
$F^{(FV)}_{l; l}$ for $l=0,4,6$ gives the  states $|A_1^+, L_z; l; j\rangle$,
as defined immediately before eq.~(\ref{eq:opexp}), with
\begin{equation}
\label{eqn:A1 basis states}
\begin{split}
|A_1^+,0; 0; 1\rangle \ =\ &|0,0\rangle \\
|A_1^+,0; 4; 1\rangle \ =\ & \frac{1}{2}\sqrt{\frac{5}{6}}\ |4,4\rangle
\ +\ \frac{1}{2}\sqrt{\frac{7}{3}}\
|4,0\rangle
\ +\ \frac{1}{2}\sqrt{\frac{5}{6}}\ |4,-4\rangle   \\
|A_1^+,0; 6; 1\rangle\ =\ & \frac{\sqrt{7}}{4}\ |6,4\rangle\ -\ \frac{1}{2\sqrt{2}}\ |6,0\rangle\ +\
\frac{\sqrt{7}}{4}\ |6,-4\rangle
\end{split}
\end{equation}
for the $A_1^+$ eigenstate of each $\overline{F}^{(FV)}_{l; l}$ in the
orbital angular momentum (spherical-wave) basis  $|l,m\rangle$.
With these states and the corresponding eigenvalues from $\overline{F}^{(FV)}_{l; l}$, 
the procedures described in Appendix~\ref{sect:explicit calculation} allow for
the contribution to eq.~(\ref{eq:evals}) from $A_1^+$ states to be written as
\begin{equation}
\label{eqn:A1 det}
\text{det}\left[
\begin{pmatrix} 
\text{cot}\delta_0 & 0 & 0\\
0 & \text{cot}\delta_4 & 0\\
0 & 0 & \text{cot}\delta_6
\end{pmatrix}
-
\begin{pmatrix} 
\overline{F}_{0;0}^{(FV, A_1^+)} & \overline{F}_{0;4}^{(FV, A_1^+)}  & \overline{F}_{0;6}^{(FV, A_1^+)}\\
\overline{F}_{4;0}^{(FV, A_1^+)} & \overline{F}_{4;4}^{(FV, A_1^+)} & \overline{F}_{4;6}^{(FV, A_1^+)}\\
\overline{F}_{6;0}^{(FV, A_1^+)} & \overline{F}_{6;4}^{(FV, A_1^+)} & \overline{F}_{6;6}^{(FV, A_1^+)}
\end{pmatrix}
\right]\ =\ 0
\ \ \ ,
\end{equation}
where the finite-volume contributions are
\begin{equation}
\label{eqn:A1 FFV}
\begin{split}
 \overline{F}_{0;0}^{(FV, A_1^+)} &= \frac{\mathcal{Z}_{0,0}\left(1;\tilde{q}^2\right)}{\pi ^{3/2}
  \tilde{q}}
\nonumber\\
 \overline{F}_{0;4}^{(FV, A_1^+)}&=\frac{2 \sqrt{\frac{3}{7}}
  \mathcal{Z}_{4,0}\left(1;\tilde{q}^2\right)}{\pi ^{3/2} \tilde{q}^5}
\nonumber\\
 \overline{F}_{0;6}^{(FV, A_1^+)}&=-\frac{2 \sqrt{2}
  \mathcal{Z}_{6,0}\left(1;\tilde{q}^2\right)}{\pi ^{3/2} \tilde{q}^7}
\nonumber\\
 \overline{F}_{4;4}^{(FV, A_1^+)} &=\frac{\mathcal{Z}_{0,0}\left(1;\tilde{q}^2\right)}{\pi ^{3/2} \tilde{q}}+\frac{108 \mathcal{Z}_{4,0}\left(1;\tilde{q}^2\right)}{143 \pi ^{3/2} \tilde{q}^5}+\frac{80 \mathcal{Z}_{6,0}\left(1;\tilde{q}^2\right)}{11 \sqrt{13} \pi ^{3/2}
   \tilde{q}^7}+\frac{560 \mathcal{Z}_{8,0}\left(1;\tilde{q}^2\right)}{143
   \sqrt{17} \pi ^{3/2} \tilde{q}^9}
\nonumber\\
 \overline{F}_{4;6}^{(FV, A_1^+)} &=-\frac{40 \sqrt{\frac{6}{91}} \mathcal{Z}_{4,0}\left(1;\tilde{q}^2\right)}{11 \pi ^{3/2} \tilde{q}^5}+\frac{42 \sqrt{42} \mathcal{Z}_{6,0}\left(1;\tilde{q}^2\right)}{187 \pi ^{3/2} \tilde{q}^7}-\frac{224 \sqrt{\frac{42}{221}}
   \mathcal{Z}_{8,0}\left(1;\tilde{q}^2\right)}{209 \pi ^{3/2} \tilde{q}^9}
\\
   &\quad-\frac{1008 \sqrt{\frac{2}{13}}
     \mathcal{Z}_{10,0}\left(1;\tilde{q}^2\right)}{323 \pi ^{3/2}
     \tilde{q}^{11}}
\nonumber\\
    \overline{F}_{6;6}^{(FV, A_1^+)}&= \frac{\mathcal{Z}_{0,0}\left(1;\tilde{q}^2\right)}{\pi ^{3/2} \tilde{q}}-\frac{126 \mathcal{Z}_{4,0}\left(1;\tilde{q}^2\right)}{187 \pi ^{3/2} \tilde{q}^5}-\frac{160 \sqrt{13} \mathcal{Z}_{6,0}\left(1;\tilde{q}^2\right)}{3553 \pi ^{3/2}
   \tilde{q}^7}+\frac{840 \mathcal{Z}_{8,0}\left(1;\tilde{q}^2\right)}{209
   \sqrt{17} \pi ^{3/2} \tilde{q}^9}
\\
   &\quad-\frac{2016 \sqrt{21} \mathcal{Z}_{10,0}\left(1;\tilde{q}^2\right)}{7429 \pi ^{3/2} \tilde{q}^{11}}+\frac{30492
   \mathcal{Z}_{12,0}\left(1;\tilde{q}^2\right)}{37145 \pi ^{3/2}
   \tilde{q}^{13}}-\frac{1848 \sqrt{1001}
   \mathcal{Z}_{12,4}\left(1;\tilde{q}^2\right)}{37145 \pi ^{3/2}
   \tilde{q}^{13}} 
\ \ \ ,
\end{split}
\end{equation}
and $ \overline{F}_{i;j}^{(FV, \Gamma^{(i)})}= \overline{F}_{j;i}^{(FV, \Gamma^{(i)})}$.  
Equation~(\ref{eqn:A1 det}) yields an infinite number of energy-eigenvalues and
eigenstates, each of which depend upon the phase-shift in the 
$l=$0, 4, and 6 partial waves.

In the $|{\bf n}|^2$-shells for which there is just one $A_1^+$ state, 
as shown in table~\ref{tab:multiplicities},
its energy-shift due to interactions receives contributions from the 
$l=0,4,6,...$ partial-waves.  
However, in the $|{\bf n}|^2$-shells in which there are multiple $A_1^+$ states 
(first occurring at $|{\bf n}|^2 = 9$), 
the energy-eigenstates are linear combinations of these states.
In the large-volume limit, 
the shift in the energy-eigenvalue of one combination is dominated by the 
interactions in the $l=0$ partial-wave,
the shift in a second combination is dominated by the 
interactions in the $l=4$ partial-wave,
the shift in a third combination is dominated by the 
interactions in the $l=6$ partial-wave,
and so on.
So while the naive argument that $A_1^+$ states receive contributions
from interactions in the $l=0,4,6,...$ partial-waves is generally true, linear
combinations of $A_1^+$ states are formed such that  it is not true in the
infinite-volume limit.
The energy-shift of each occurrence of an $A_1^+$ energy-eigenstate in a given $|{\bf n}|^2$-shell
is dominated by the interaction in a different
partial-wave in the infinite-volume limit.
To demonstrate this point, consider the situation where the phase-shift in the 
$l=6$ partial-wave vanishes, in which case 
eq.~(\ref{eqn:A1 det}) 
becomes a $2\times2$ matrix with the following two solutions:
\begin{multline}\label{eqn:A1 two solutions}
\frac{\text{cot}\delta_0}{2}+\frac{\text{cot}\delta_4}{2} -\frac{\mathcal{Z}_{0,0}\left(1;\tilde q^2\right)}{\pi ^{3/2} \tilde q}-\frac{280 \mathcal{Z}_{8,0}\left(1;\tilde q^2\right)}{143 \sqrt{17}
   \pi ^{3/2} \tilde q^9}-\frac{40 \mathcal{Z}_{6,0}\left(1;\tilde q^2\right)}{11 \sqrt{13} \pi ^{3/2} \tilde q^7}-\frac{54 \mathcal{Z}_{4,0}\left(1;\tilde q^2\right)}{143 \pi ^{3/2} \tilde q^5}\\
{\bf\pm}\frac{1}{2} \sqrt{\left(\frac{560 \mathcal{Z}_{8,0}\left(1;\tilde q^2\right)}{143 \sqrt{17} \pi ^{3/2} \tilde q^9}+\frac{80 \mathcal{Z}_{6,0}\left(1;\tilde q^2\right)}{11 \sqrt{13} \pi ^{3/2} \tilde q^7}+\frac{108 \mathcal{Z}_{4,0}\left(1;\tilde q^2\right)}{143 \pi ^{3/2}
   \tilde q^5}+\text{cot}\delta_0-\text{cot}\delta_4\right)^2+\frac{48 \mathcal{Z}_{4,0}\left(1;\tilde q^2\right)^2}{7 \pi ^3 \tilde q^{10}}}\\
   =0\ .
   \end{multline}
\begin{figure}
\centering
\includegraphics[height=\textwidth,angle=-90]{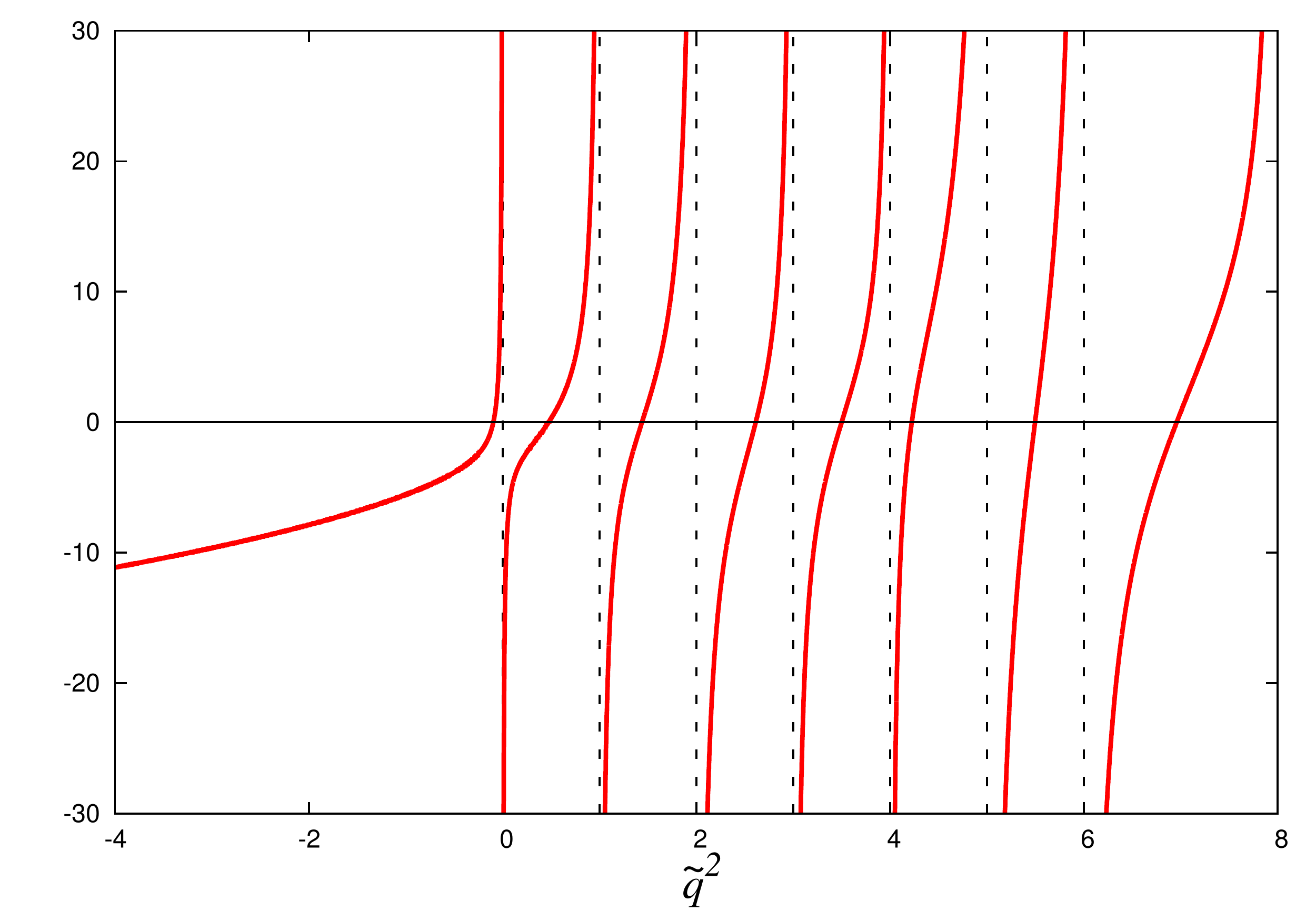}
\caption{The function $\mathcal{Z}_{0,0}\left(1;\tilde q^2\right)$.  
The vertical dashed lines denote the position of the poles of
  the function corresponding to the non-interacting energy-eigenvalues.
\label{fig:A1+ l0}}
\end{figure}
In the case of  $\text{tan}\delta_4\ll\text{tan}\delta_0$, the $l=0$ dominated
solution is
\begin{equation}
q\text{cot}\delta_0\ =\ 
\frac{2}{\sqrt{\pi} L}\mathcal{Z}_{0,0}\left(1;\tilde q^2\right)
\ +\ 
\frac{12288\pi^7}{7L^{10}}\frac{\mathcal{Z}_{4,0}\left(1;\tilde q^2\right)^2}{ [q^9\text{cot}\delta_4]}
\ +\ O\left(\text{tan}^2\delta_4\right)
\ \ ,
\label{eq:A1pl0plusA4}
\end{equation}
and is valid for all $|{\bf n}|^2$-shells.  
If phase shifts in both the $l=4$ and $l=6$ partial-waves vanish,
eq.~(\ref{eqn:A1 det}) 
and eq.~(\ref{eq:A1pl0plusA4})
reduce to the familiar  
result found by L\"uscher~\footnote{The ``S-function'', $S(\tilde q^2)$, used in, 
for example, Ref.~\cite{Beane:2003da}, is related to 
$\mathcal{Z}_{0,0}(1;\tilde q^2)$ by $S(\tilde q^2)
=\sqrt{4\pi}\mathcal{Z}_{0,0}(1;\tilde q^2)$.},
\begin{equation}
q\text{cot}\delta_0\ =\ 
\frac{2}{\sqrt{\pi} L}\ 
\mathcal{Z}_{0,0}\left(1;\tilde
  q^2\right)
\ \ \ ,
\label{eqn:A1+}
\end{equation}
where the function $\mathcal{Z}_{0,0}\left(1;\tilde q^2\right)$
is shown in fig.~\ref{fig:A1+ l0}.
Performing a  large-volume expansion 
of the solution 
(as discussed in
Appendix~\ref{sect:perturbation})
to 
eq.~(\ref{eqn:A1+})
in the $|{\bf n}|^2=9$-shell
gives the energy-eigenvalue 
\begin{eqnarray}
E_{A_1^{+(1)}} & = & 
{1\over 2\mu}\ \left[\ 
{36\pi^2\over L^2}
\ - \
{20\tan\delta_0(|{\bf n}|^2=9)\over L^2}
\ +\ {\cal O}(\tan^2\delta_0)
\ +\ ...
\right]
\ \ \ ,
\label{eq:A1pn9a}
\end{eqnarray}
while the second solution to eq.~(\ref{eqn:A1 two solutions}) has  a perturbative
expansion of the form
\begin{eqnarray}
E_{A_1^{+(2)}} & = & 
{1\over 2\mu}\ \left[\ 
{36\pi^2\over L^2}
\ - \
{8960\tan\delta_4(|{\bf n}|^2=9)\over 243\ L^2}
\ +\ {\cal O}(\tan\delta_6)
\ +\ ...
\right]
\ \ \ ,
\label{eq:A1pn9b}
\end{eqnarray}
where the contribution from the $l=0$ partial-wave is strongly suppressed in the
large-volume limit.  
While the two basis states, $||{\bf n}|^2=9; A_1^+ (1)\rangle$ and 
$||{\bf n}|^2=9; A_1^+ (2)\rangle$, both have a non-vanishing overlap with
$|l,m\rangle = |0,0\rangle$, it is obvious that a linear combination can be formed that has
vanishing overlap.  Inserting the interactions once, as is appropriate for
determining the energy-eigenvalues in large volumes (\emph{i.e.} first order perturbation theory in $1/L$), dictates the form of the 
expansions in eq.~(\ref{eq:A1pn9a}) and  eq.~(\ref{eq:A1pn9b}).

Sources and sinks that have an overlap, and
in general a range of overlaps, with the finite-volume energy-eigenstates 
of hadronic systems are required for LQCD calculations.
While the interactions between hadrons gives rise to energy-eigenstates that are not
products of single-hadron eigenstates of the linear-momentum operator, 
sources and sinks can be
constructed from the single-hadron momentum eigenstates that transform as a
given $\Gamma^{(i)}$, that will have non-zero overlap with the
energy-eigenstates in the same irrep. 
Constructing sources and sinks from single hadrons that 
have equal and opposite momenta ensures that the total momentum
of the combined system vanishes.
The 
relative-momentum-eigenstates of definite parity, ${\cal P}$, 
are denoted by 
\begin{equation}
|\vec{n}\ , \ {\cal P} \rangle\ =
\begin{cases}
\frac{|\vec{n}\rangle
\ +\ {\cal P}\ |-\vec{n}\rangle}{\sqrt{2}}\ &\quad\quad(\vec{n}\ne\vec{0})\\
|\vec{n}\rangle &\quad\quad(\vec{n}=\vec{0}\mbox{ and }\mathcal{P}=+1)\ ,
\end{cases}
\label{eq:statedef}
\end{equation}
where ${\cal P}$ is the parity of the state (${\cal P}=\pm 1$) 
and $\vec{n}=(n_x,n_y,n_z)$ is the  triplet of integers that define the
relative momentum of the two-body system.  The states in eq.~(\ref{eq:statedef}) are eigenstates of the
relative kinetic energy operator $T_{\text{rel}}$, with the eigenvalues
displayed in eq.~(\ref{eq:KIN}).
By  taking appropriate linear combinations of these momentum-eigenstates, 
states in the $A_1^+$ representation (or any other irrep)
can be constructed in each $|{\bf n}|^2$-shell if the shell supports it 
(see table~\ref{tab:multiplicities} and Ref.~\cite{Dresselhaus}).
For example, in the $|{\bf n}|^2=0$ shell the basis-state is
\begin{displaymath}
||{\bf n}|^2=0; \ A_1^+\rangle \ =\ 
|(0,0,0)\ , \ {\cal P}=+1\rangle 
\ ,
\end{displaymath}
while for  $|{\bf n}|^2=1$, the basis-state is 
\begin{displaymath}
||{\bf n}|^2=1; \ A_1^+\rangle \ =\ 
\frac{|(1,0,0)\ , \ {\cal P}=+1\rangle \ +\ 
|(0,1,0)\ ,\ {\cal P}=+1\rangle 
\ +\ |(0,0,1)\ ,\ {\cal P}=+1\rangle }{\sqrt{3}}
\ \ \ .
\end{displaymath}
In general the coefficients of these basis vectors are valid up to an arbitrary phase.  
The complete momentum-space basis for the  $A_1^+$ sources and sinks 
in each $|{\bf  n}|^2$-shell
are presented in table~\ref{tab:A1pSOURCESSINKS} for $|{\bf  n}|^2\le 6$.  
\begin{table}
\begin{center}
\begin{minipage}[!ht]{16.5 cm}
\caption{
The momentum-space structure of $A_1^+$ sources and sinks for $|{\bf n}|^2$=0-3.
These are shown graphically in  figs.~\ref{fig:A1 figures 1} and~\ref{fig:A1 figures 2}.
}
\label{tab:A1pSOURCESSINKS}
\end{minipage}
\begin{small}
\begin{tabular}{|c||c||c||c|  }
\hline
\hline
$|{\bf n}|^2$=0 & $|{\bf n}|^2$=1 & $|{\bf n}|^2$=2 & $|{\bf n}|^2$=3  \\ \hline

$\begin{array}{cc}
 \text{$|$(0,0,0) , +1$\rangle $} & 1
\end{array}$ 
&
$ \begin{array}{cc}
 \text{$|$(1,0,0) , +1$\rangle $} & \frac{1}{\sqrt{3}} \\
 \text{$|$(0,1,0) , +1$\rangle $} & \frac{1}{\sqrt{3}} \\
 \text{$|$(0,0,1) , +1$\rangle $} & \frac{1}{\sqrt{3}}
\end{array} $ 
& 
$\begin{array}{cc}
 \text{$|$(1,1,0) , +1$\rangle $} & \frac{1}{\sqrt{6}} \\
 \text{$|$(1,0,1) , +1$\rangle $} & \frac{1}{\sqrt{6}} \\
 \text{$|$(1,0,-1) , +1$\rangle $} & \frac{1}{\sqrt{6}} \\
 \text{$|$(1,-1,0) , +1$\rangle $} & \frac{1}{\sqrt{6}} \\
 \text{$|$(0,1,1) , +1$\rangle $} & \frac{1}{\sqrt{6}} \\
 \text{$|$(0,1,-1) , +1$\rangle $} & \frac{1}{\sqrt{6}}
\end{array}$
& 
$\begin{array}{cc}
 \text{$|$(1,1,1) , +1$\rangle $} & \frac{1}{2} \\
 \text{$|$(1,1,-1) , +1$\rangle $} & \frac{1}{2} \\
 \text{$|$(1,-1,1) , +1$\rangle $} & \frac{1}{2} \\
 \text{$|$(1,-1,-1) , +1$\rangle $} & \frac{1}{2}
\end{array}$ \\ \hline \hline
\end{tabular}
\end{small}
\begin{minipage}[t]{16.5 cm}
\vskip 0.5cm
\noindent
\end{minipage}
\end{center}
\end{table}     

\begin{table}
\begin{center}
\begin{minipage}[!ht]{16.5 cm}
\caption{
The momentum-space structure of $A_1^+$ sources and sinks for $|{\bf n}|^2$=4-6.
These are shown graphically in  figs.~\ref{fig:A1 figures 2} and~\ref{fig:A1 figures 3}.
}
\label{tab:A1pSOURCESSINKSv2}
\end{minipage}
\begin{small}
\begin{tabular}{|c||c||c|  }
\hline
\hline
$|{\bf n}|^2$=4 & $|{\bf n}|^2$=5 &  $|{\bf n}|^2$=6 \\ \hline

$\begin{array}{cc}
 \text{$|$(2,0,0) , +1$\rangle $} & \frac{1}{\sqrt{3}} \\
 \text{$|$(0,2,0) , +1$\rangle $} & \frac{1}{\sqrt{3}} \\
 \text{$|$(0,0,2) , +1$\rangle $} & \frac{1}{\sqrt{3}}
\end{array}$ 
&
$ \begin{array}{cc|cc}
 \text{$|$(2,1,0) , +1$\rangle $} & \frac{1}{2 \sqrt{3}} &\text{$|$(2,0,1) , +1$\rangle $} & \frac{1}{2 \sqrt{3}} \\
 \text{$|$(2,0,-1) , +1$\rangle $} & \frac{1}{2 \sqrt{3}} &\text{$|$(2,-1,0) , +1$\rangle $} & \frac{1}{2 \sqrt{3}} \\
 \text{$|$(1,2,0) , +1$\rangle $} & \frac{1}{2 \sqrt{3}} & \text{$|$(1,0,2) , +1$\rangle $} & \frac{1}{2 \sqrt{3}} \\
 \text{$|$(1,0,-2) , +1$\rangle $} & \frac{1}{2 \sqrt{3}} & \text{$|$(1,-2,0) , +1$\rangle $} & \frac{1}{2 \sqrt{3}} \\
 \text{$|$(0,2,1) , +1$\rangle $} & \frac{1}{2 \sqrt{3}} & \text{$|$(0,2,-1) , +1$\rangle $} & \frac{1}{2 \sqrt{3}} \\
 \text{$|$(0,1,2) , +1$\rangle $} & \frac{1}{2 \sqrt{3}} & \text{$|$(0,1,-2) , +1$\rangle $} & \frac{1}{2 \sqrt{3}}
\end{array}$ 
& 
$\begin{array}{cc|cc}
 \text{$|$(2,1,1) , +1$\rangle $} & \frac{1}{2 \sqrt{3}} & \text{$|$(2,1,-1) , +1$\rangle $} & \frac{1}{2 \sqrt{3}} \\
 \text{$|$(2,-1,1) , +1$\rangle $} & \frac{1}{2 \sqrt{3}} & \text{$|$(2,-1,-1) , +1$\rangle $} & \frac{1}{2 \sqrt{3}} \\
 \text{$|$(1,2,1) , +1$\rangle $} & \frac{1}{2 \sqrt{3}} & \text{$|$(1,2,-1) , +1$\rangle $} & \frac{1}{2 \sqrt{3}} \\
 \text{$|$(1,1,2) , +1$\rangle $} & \frac{1}{2 \sqrt{3}} & \text{$|$(1,1,-2) , +1$\rangle $} & \frac{1}{2 \sqrt{3}} \\
 \text{$|$(1,-1,2) , +1$\rangle $} & \frac{1}{2 \sqrt{3}} & \text{$|$(1,-1,-2) , +1$\rangle $} & \frac{1}{2 \sqrt{3}} \\
 \text{$|$(1,-2,1) , +1$\rangle $} & \frac{1}{2 \sqrt{3}} & \text{$|$(1,-2,-1) , +1$\rangle $} & \frac{1}{2 \sqrt{3}}
\end{array}$ \\ \hline \hline
\end{tabular}
\end{small}
\begin{minipage}[t]{16.5 cm}
\vskip 0.5cm
\noindent
\end{minipage}
\end{center}
\end{table}     
\begin{figure}
\centering
\mbox{\includegraphics[width=.77\textwidth,angle=0]{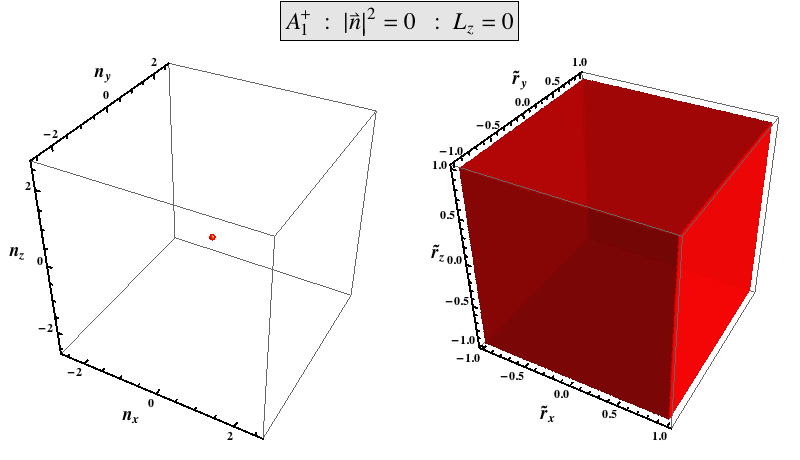}}
\mbox{\includegraphics[width=.77\textwidth,angle=0]{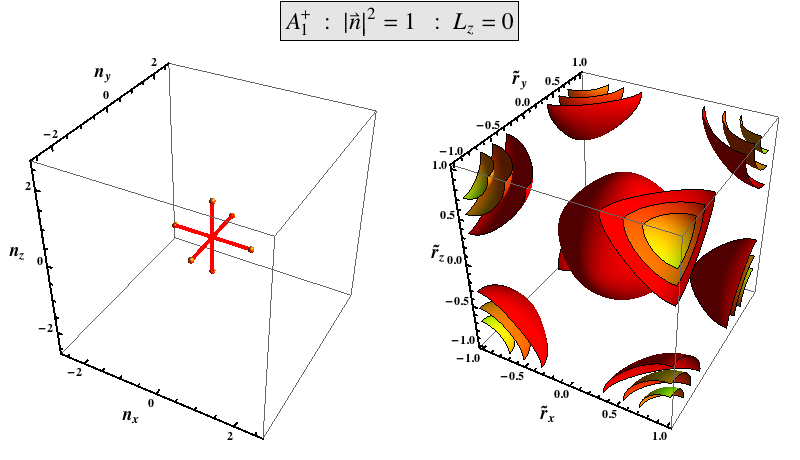}}
\mbox{\includegraphics[width=.77\textwidth,angle=0]{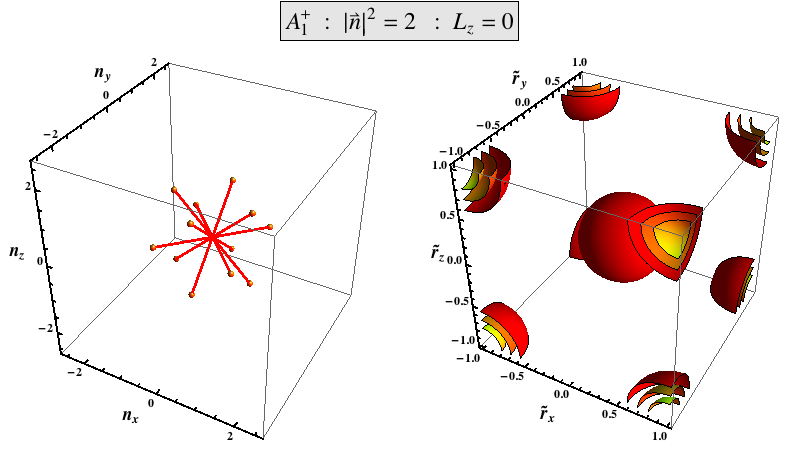}}
\caption{The momentum-space representations (left) and position-space
  representations (right) of two-body relative states in the $A_1^+$
  representation for select $|{\bf n}|^2$ shells.  Here $\tilde r=r/L$.
\label{fig:A1 figures 1}}
\end{figure}
\begin{figure}
\centering
\mbox{\includegraphics[width=.77\textwidth,angle=0]{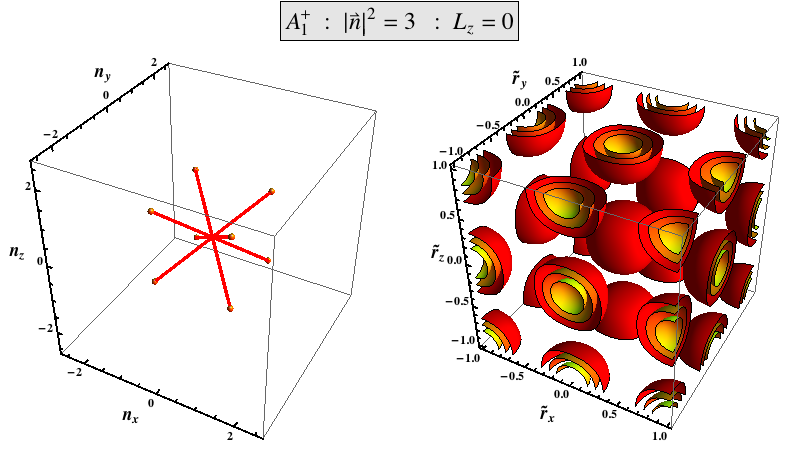}}
\mbox{\includegraphics[width=.77\textwidth,angle=0]{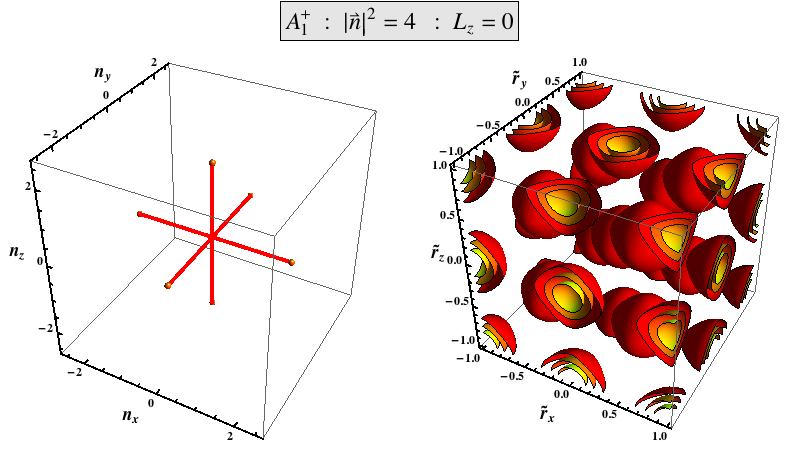}}
\mbox{\includegraphics[width=.77\textwidth,angle=0]{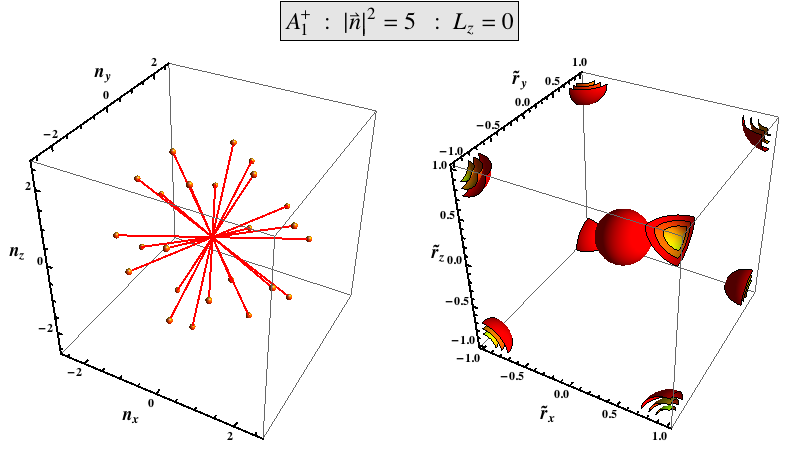}}
\caption{The momentum-space representations (left) and position-space
  representations (right) of two-body relative states in the $A_1^+$
  representation for select $|{\bf n}|^2$-shells. 
\label{fig:A1 figures 2}}
\end{figure}
\begin{figure}
\centering
\mbox{\includegraphics[width=.77\textwidth,angle=0]{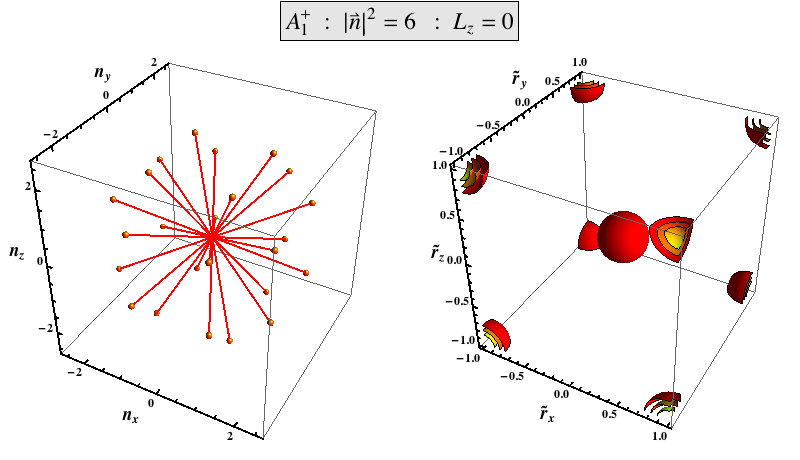}}
\caption{The momentum-space representations (left) and position-space
  representations (right) of two-body relative states in the $A_1^+$
  representation for $|{\bf n}|^2$=6 shell.
\label{fig:A1 figures 3}}
\end{figure}

The momentum-space representations of the sources and sinks
(the left panels in figs.~\ref{fig:A1 figures 1}-\ref{fig:A1 figures 3})
show the ${\bf n}$-vectors that transform as an
$A_1^+$ in the given  $|{\bf n}|^2$-shells.  The widths of the vectors are
proportional to the magnitude of their 
amplitudes and their color denotes the sign (red=positive, blue=negative).   
The position-space representations of the sources and sinks
(the right panels in figs.~\ref{fig:A1 figures 1}-\ref{fig:A1 figures 3})
show the surfaces  of constant
 $\rho_{ {\bf n},{\cal P}}({\bf r})$, defined by
\begin{eqnarray}
\rho_{ {\bf n},{\cal P}}({\bf r})
&  = & 
|\langle\ {\bf r}\ | \ (n_x,n_y,n_z) \ ,\ {\cal P}\ \rangle |^2
\ \ \ ,
\end{eqnarray}  
which are
obtained by Fourier transform. 
In the position-space representations, ${\bf r}$ 
refers to the relative distance between the two particles and in the
figures ${\bf \tilde r}$ is defined to be
${\bf \tilde r}={\bf r}/L$.

\subsubsection{$A_2^+$ Representation}
\label{irrep:A2p}
The other  one-dimensional positive-parity irrep of the cubic group is the $A_2^+$.  
Due to its complexity, the lowest-lying state transforming as a $A_2^+$ is in
the $|{\bf n}|^2=5$ shell, as indicated in table~\ref{tab:multiplicities}.
\begin{figure}[!ht]
\centering
\includegraphics[height=\textwidth,angle=-90]{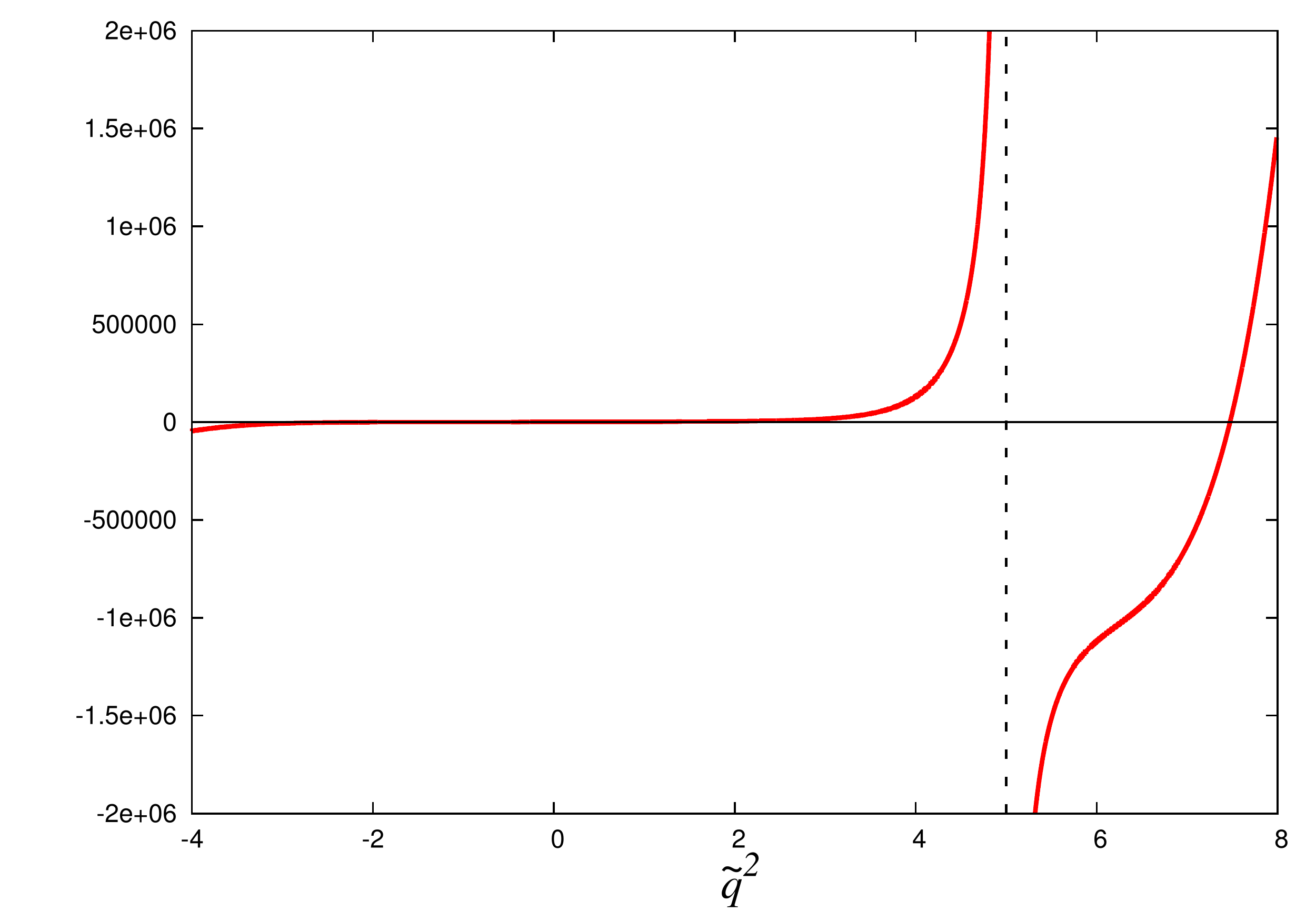}
\caption{The function $\mathcal{X}_{A_2}^+$, as defined in eq.~(\ref{eqn:A2+}),  as a 
function of $\tilde q^2$.  The vertical dashed line denotes the position 
of the pole of the function corresponding to the non-interacting 
energy-eigenvalues.
\label{fig:A2+ l6}}
\end{figure}
Further, the lowest partial-wave contributing to its energy is $l=6$,
and as this analysis is truncated to partial-waves with $l\le 6$, the
contribution to the determinant in eq.~(\ref{eq:evals}) has
the solution 
\begin{eqnarray}
\label{eqn:A2+}
q^{13}\text{cot}\delta_6
&&
\ =\ \left(\frac{2\pi}{L}\right)^{13}\frac{1}{\pi^{3/2}}
\times
\nonumber\\
&&\left(\tilde q^{12}
   \mathcal{Z}_{0,0}\left(1;\tilde q^2\right)+\frac{6 \tilde q^8
     \mathcal{Z}_{4,0}\left(1;\tilde q^2\right)}{17}-\frac{160 \sqrt{13} \tilde
     q^6 \mathcal{Z}_{6,0}\left(1;\tilde q^2\right)}{323}-\frac{40 \tilde q^4
     \mathcal{Z}_{8,0}\left(1;\tilde q^2\right)}{19 \sqrt{17}}\right.
\nonumber\\
&&
\left. 
\ \ \ 
-\frac{2592 \sqrt{21} \tilde q^2 \mathcal{Z}_{10,0}\left(1;\tilde
      q^2\right)}{7429}+\frac{1980 \mathcal{Z}_{12,0}\left(1;\tilde
      q^2\right)}{7429}+\frac{264 \sqrt{1001} \mathcal{Z}_{12,4}\left(1;\tilde
      q^2\right)}{7429}\right)
\nonumber\\
&&\equiv \ 
\left(\frac{2\pi}{L}\right)^{13}\frac{1}{\pi^{3/2}} \mathcal{X}_{A_2}^+\left(\tilde q^2\right)\ .
\end{eqnarray} 
and the associated eigenstate of the $\overline F^{(FV)}_{6;6}$
block is
\begin{equation}
|A_2^+,2: 6; 1\rangle \ =\ 
\frac{1}{4}\sqrt{\frac{11}{2}}|6,2\rangle
\ +\  \frac{1}{4}\sqrt{\frac{11}{2}}|6,-2\rangle 
\ -\  \frac{1}{4}\sqrt{\frac{5}{2}}|6,6\rangle
\ -\  \frac{1}{4}\sqrt{\frac{5}{2}}|6,-6\rangle
 \ \ \ .
\end{equation}
The function $\mathcal{X}_{A_2}^+$ is shown in fig.~\ref{fig:A2+ l6} as a 
function of $\tilde q^2$.  
Its pole at $\tilde q^2$=5, denoted by the vertical dashed line, 
corresponds to the non-interacting ($\delta_6=0$) energy-eigenvalue.  
This is the only $|{\bf n}|^2$-shell with 
$|{\bf n}|^2<6$ which supports the $A_2^+$ irrep, 
as shown  in
table~\ref{tab:multiplicities}. 
In fig.~\ref{fig:A2 figures} we give the graphical representations of the
source and sink that generates this irrep in the $|{\bf n}|^2=5$-shell.
\begin{figure}
\centering
\includegraphics[width=.77\textwidth,angle=0]{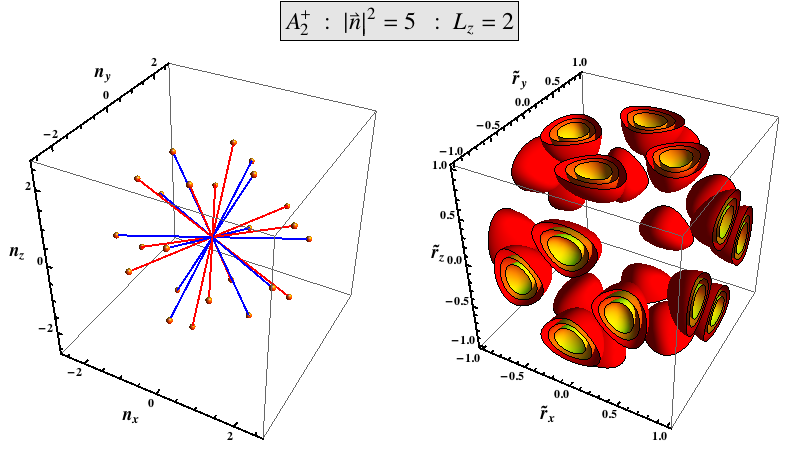}
\caption{The momentum-space representation (left) and position-space 
representation (right) of two-body relative states in the $A_2^+$ 
representation for the $|{\bf n}|^2=5$ shell.
\label{fig:A2 figures}}
\end{figure}
As this is the lowest-lying state whose energy-eigenvalue is insensitive 
to $l<6$ interactions, it is LQCD correlation functions constructed to transform
in the $A_2^+$ irrep that will enable a calculation of $\delta_6$. 
However, as the lowest
energy contributing to an  $A_2^+$ correlation function
occurs in the  $|{\bf n}^2|=5$-shell,
relatively large lattice volumes will be required in order to have this state lie 
below the inelastic threshold.
\begin{table}
\begin{center}
\begin{minipage}[!ht]{16.5 cm}
\caption{
The momentum-space structure of the $A_2^+$ source and sink in the $|{\bf n}|^2=5$-shell.
They are shown graphically in  fig.~\ref{fig:A2 figures}.
}
\label{tab:A2pSOURCESSINKS}
\end{minipage}
\begin{tabular}{|c|  }
\hline
\hline
$|{\bf n}|^2$=5 \\ \hline\hline
$\begin{array}{cc|cc}
 \text{$|$(2,1,0) , +1$\rangle $} & -\frac{1}{2 \sqrt{3}} & \text{$|$(2,0,1) , +1$\rangle $}  &\frac{1}{2 \sqrt{3}} \\
 \text{$|$(2,0,-1) , +1$\rangle $} & \frac{1}{2 \sqrt{3}} & \text{$|$(2,-1,0) , +1$\rangle $}  & -\frac{1}{2 \sqrt{3}} \\
 \text{$|$(1,2,0) , +1$\rangle $} & \frac{1}{2 \sqrt{3}} & \text{$|$(1,0,2) , +1$\rangle $} & -\frac{1}{2 \sqrt{3}} \\
 \text{$|$(1,0,-2) , +1$\rangle $} & -\frac{1}{2 \sqrt{3}} & \text{$|$(1,-2,0) , +1$\rangle $} & \frac{1}{2 \sqrt{3}} \\
 \text{$|$(0,2,1) , +1$\rangle $} & -\frac{1}{2 \sqrt{3}} & \text{$|$(0,2,-1) , +1$\rangle $} & -\frac{1}{2 \sqrt{3}} \\
 \text{$|$(0,1,2) , +1$\rangle $} & \frac{1}{2 \sqrt{3}} & \text{$|$(0,1,-2) , +1$\rangle $} & \frac{1}{2 \sqrt{3}}
\end{array}$\\
\hline
\hline
\end{tabular}
\begin{minipage}[t]{16.5 cm}
\vskip 0.5cm
\noindent
\end{minipage}
\end{center}
\end{table}     
The momentum-space structure of the source and sink that couple to the $A_2^+$
state 
in the $|{\bf n}|^2=5$-shell 
is given in table~\ref{tab:A2pSOURCESSINKS}.

\subsubsection{$E^+$ Representation}
\label{irrep:Ep}
The energy-eigenvalues of states transforming in the $E^+$ irrep receive
contributions from interactions in the $l=2,4,6,... $ partial-waves.
As the $E^+$ irrep is two-dimensional, the contribution to the determinant in 
eq.~(\ref{eq:evals}) results from  
a $6\times 6$ matrix when $l \le 6$.  However, as the two
states in the $E^+$ irrep (with $L_z=0$ and $L_z=2$) 
are degenerate, and orbital-angular momentum is conserved by the interactions
(unlike the situation in the baryon-sector),
the analysis can be reduced to that of a $3\times 3$ matrix.
\begin{figure}[!ht]
\centering
\includegraphics[height=\textwidth,angle=-90]{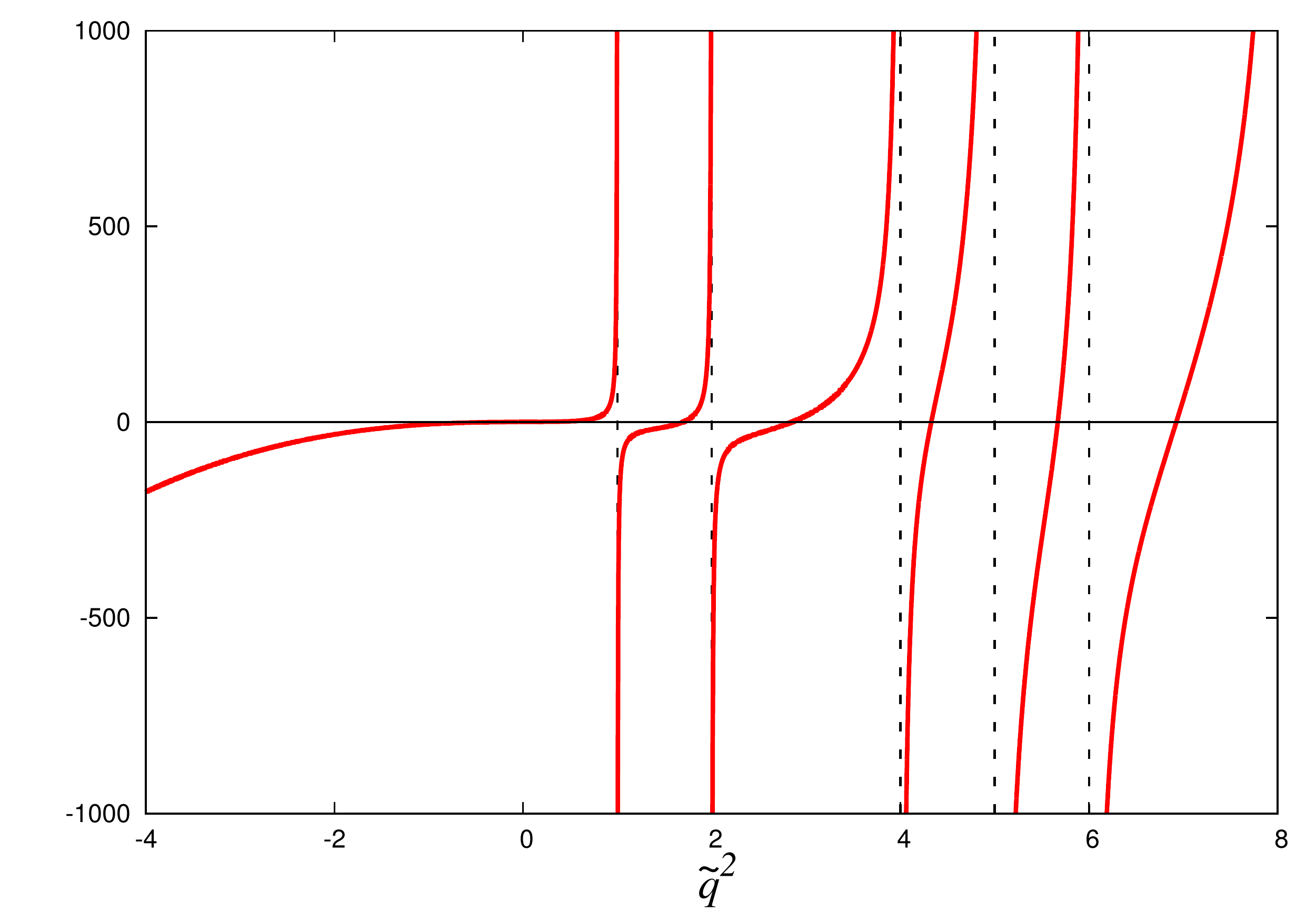}
\caption{The function  $\mathcal{X}_{E}^+$, as defined in  eq.~(\protect\ref{eqn:E+ l=2}),
as a function of $\tilde q^2$.  The vertical dashed lines denote the position 
of the poles in the function corresponding to the non-interacting
energy-eigenvalues.
\label{fig:E+ l=2}}
\end{figure}
The $E^+$ $L_z=0$ states associated with the 
$\overline{F}^{(FV)}_{2;2}$, $\overline{F}^{(FV)}_{4;4}$, and $\overline{F}^{(FV)}_{6;6}$ blocks 
are 
\begin{equation}
\begin{split}
|E^+,0; 2; 1\rangle\ =\ &|2,0 \rangle  \\
|E^+,0; 4; 1\rangle\ =\ & \frac{1}{2}\sqrt{\frac{7}{6}}|4,4\rangle
\ -\ \frac{1}{2}\sqrt{\frac{5}{3}}|4,0\rangle
\ +\ \frac{1}{2}\sqrt{\frac{7}{6}}|4,-4\rangle
\\
|E^+,0; 6; 1\rangle\ =\ & \frac{1}{4}|6,4 \rangle
\ +\ \frac{1}{2}\sqrt{\frac{7}{2}}|6,0\rangle
\ +\ \frac{1}{4}|6,-4 \rangle
\ \ \ \ ,
\end{split}
\end{equation}
and the contribution to 
eq.~(\ref{eq:evals}) becomes
\begin{equation}\label{eqn:E det}
\text{det}\left[
\begin{pmatrix}
\text{cot}\delta_2 & 0 & 0\\
0 & \text{cot}\delta_4 & 0\\
0 & 0 & \text{cot}\delta_6
\end{pmatrix}
-
\begin{pmatrix} 
\overline{F}_{2;2}^{(FV, E^+)} & \overline{F}_{2;4}^{(FV, E^+)}  & \overline{F}_{2;6}^{(FV, E^+)}\\
\overline{F}_{4;2}^{(FV, E^+)} & \overline{F}_{4;4}^{(FV, E^+)} & \overline{F}_{4;6}^{(FV, E^+)}\\
\overline{F}_{6;2}^{(FV, E^+)} & \overline{F}_{6;4}^{(FV, E^+)} & \overline{F}_{6;6}^{(FV, E^+)}
\end{pmatrix}
\right]
\ =\ 0\ ,
\end{equation}
where
\begin{equation}
\begin{split}
\overline{F}_{2;2}^{(FV, E^+)} &=\frac{\mathcal{Z}_{0,0}\left(1;\tilde
    q^2\right)}{\pi ^{3/2} \tilde q}+\frac{6 \mathcal{Z}_{4,0}\left(1;\tilde
    q^2\right)}{7 \pi ^{3/2} \tilde q^5}
\nonumber\\
\overline{F}_{2;4}^{(FV, E^+)}&=-\frac{30 \sqrt{\frac{3}{13}}
  \mathcal{Z}_{6,0}\left(1;\tilde q^2\right)}{11 \pi ^{3/2} \tilde q^7}-\frac{40
  \sqrt{3} \mathcal{Z}_{4,0}\left(1;\tilde q^2\right)}{77 \pi ^{3/2} \tilde q^5}
\nonumber\\
\overline{F}_{2;6}^{(FV, E^+)}&=\frac{8 \sqrt{\frac{14}{1105}} \mathcal{Z}_{8,0}\left(1;\tilde q^2\right)}{\pi ^{3/2} \tilde q^9}+\frac{4 \sqrt{\frac{14}{5}} \mathcal{Z}_{6,0}\left(1;\tilde q^2\right)}{11 \pi ^{3/2} \tilde q^7}+\frac{30 \sqrt{\frac{10}{91}} \mathcal{Z}_{4,0}\left(1;\tilde q^2\right)}{11 \pi
   ^{3/2} \tilde q^5}
\nonumber\\
\overline{F}_{4;4}^{(FV, E^+)} &=\frac{\mathcal{Z}_{0,0}\left(1;\tilde q^2\right)}{\pi ^{3/2} \tilde q}+\frac{392 \mathcal{Z}_{8,0}\left(1;\tilde q^2\right)}{143 \sqrt{17} \pi ^{3/2} \tilde q^9}-\frac{64 \mathcal{Z}_{6,0}\left(1;\tilde q^2\right)}{11 \sqrt{13} \pi ^{3/2} \tilde q^7}+\frac{108
   \mathcal{Z}_{4,0}\left(1;\tilde q^2\right)}{1001 \pi ^{3/2} \tilde q^5}
\nonumber\\
\overline{F}_{4;6}^{(FV, E^+)} &=-\frac{1512 \sqrt{\frac{2}{65}} \mathcal{Z}_{10,0}\left(1;\tilde q^2\right)}{323 \pi ^{3/2} \tilde q^{11}}-\frac{128 \sqrt{\frac{210}{221}} \mathcal{Z}_{8,0}\left(1;\tilde q^2\right)}{209 \pi ^{3/2} \tilde q^9}-\frac{18 \sqrt{210} \mathcal{Z}_{6,0}\left(1;\tilde q^2\right)}{187
   \pi ^{3/2} \tilde q^7}-\frac{8 \sqrt{\frac{30}{91}}
   \mathcal{Z}_{4,0}\left(1;\tilde q^2\right)}{11 \pi ^{3/2} \tilde q^5}
\nonumber\\
   \overline{F}_{6;6}^{(FV, E^+)}&=\frac{\mathcal{Z}_{0,0}\left(1;\tilde q^2\right)}{\pi ^{3/2} \tilde q}+\frac{30492 \mathcal{Z}_{12,0}\left(1;\tilde q^2\right)}{37145 \pi ^{3/2} \tilde q^{13}}+\frac{264 \sqrt{1001} \mathcal{Z}_{12,4}\left(1;\tilde q^2\right)}{37145 \pi ^{3/2} \tilde q^{13}}+\frac{1152 \sqrt{21}
   \mathcal{Z}_{10,0}\left(1;\tilde q^2\right)}{7429 \pi ^{3/2} \tilde q^{11}}\\
   &\quad+\frac{280 \mathcal{Z}_{8,0}\left(1;\tilde q^2\right)}{209 \sqrt{17} \pi ^{3/2} \tilde q^9}+\frac{480 \sqrt{13} \mathcal{Z}_{6,0}\left(1;\tilde q^2\right)}{3553 \pi ^{3/2} \tilde q^7}+\frac{114
   \mathcal{Z}_{4,0}\left(1;\tilde q^2\right)}{187 \pi ^{3/2} \tilde q^5}
\ \ \ \ .
\end{split}
\end{equation}
It is obvious that
the solutions of eq.~(\ref{eqn:E det}) depend upon the $l=2$, 4, and 6
partial-waves  in a non-trivial manner.

In the limit of vanishing interactions in partial-waves with $l>4$,  the 
contribution from the $E^+$ irrep to eq.~(\ref{eqn:E det}) results  from a 
$2\times 2$ matrix, and has solutions 
\begin{multline}
\frac{\text{cot}\delta_2}{2}+\frac{\text{cot}\delta_4}{2}-\frac{\mathcal{Z}_{0,0}\left(1;\tilde q^2\right)}{\pi ^{3/2} q}-\frac{196
   \mathcal{Z}_{8,0}\left(1;\tilde q^2\right)}{143 \sqrt{17} \pi ^{3/2} \tilde q^9}+\frac{32 \mathcal{Z}_{6,0}\left(1;\tilde q^2\right)}{11 \sqrt{13} \pi ^{3/2} \tilde q^7}-\frac{69 \mathcal{Z}_{4,0}\left(1;\tilde q^2\right)}{143 \pi ^{3/2} \tilde q^5}=\\
{\bf\pm}\frac{1}{2} \left[\left(\frac{392 \mathcal{Z}_{8,0}\left(1;\tilde q^2\right)}{143 \sqrt{17} \pi ^{3/2} \tilde q^9}-\frac{64 \mathcal{Z}_{6,0}\left(1;\tilde q^2\right)}{11 \sqrt{13} \pi ^{3/2} \tilde q^7}-\frac{750 \mathcal{Z}_{4,0}\left(1;\tilde q^2\right)}{1001 \pi ^{3/2} \tilde q^5}+\text{cot}\delta_2-\text{cot}\delta_4\right)^2\right. \\
   \left. +4 \left(\frac{30 \sqrt{\frac{3}{13}} \mathcal{Z}_{6,0}\left(1;\tilde
           q^2\right)}{11 \pi ^{3/2} \tilde q^7}+\frac{40 \sqrt{3}
         \mathcal{Z}_{4,0}\left(1;\tilde q^2\right)}{77 \pi ^{3/2} \tilde
         q^5}\right)^2\right]^{1/2}\ .
\label{eq:Epl2l4trunc}
   \end{multline} 
In the limit that $\tan\delta_4 << \tan\delta_2$, the
$l=2$ dominated solutions to eq.~(\ref{eq:Epl2l4trunc})
result from
\begin{equation}
\label{eqn:E+ l=2}
\begin{split}
q^5\text{cot}\delta_2 &=\left(\frac{2\pi}{L}\right)^5\frac{1}{\pi^{3/2}}
\left(\tilde q^4 \mathcal{Z}_{0,0}\left(1;\tilde q^2\right)
+\frac{6}{7} \mathcal{Z}_{4,0}\left(1;\tilde q^2\right)\right)
\ =\ 
\left(\frac{2\pi}{L}\right)^5\frac{1}{\pi^{3/2}}\mathcal{X}_{E}^+\left(\tilde
  q^2\right)
\ ,
\end{split}
\end{equation}
where function  $\mathcal{X}_{E}^+$ is shown in fig.~\ref{fig:E+ l=2}
as a function $\tilde q^2$ \footnote{This expression has been derived previously by R. Briceno \cite{Briceno}.}. 
\begin{figure}
\centering
\mbox{\includegraphics[width=.77\textwidth,angle=0]{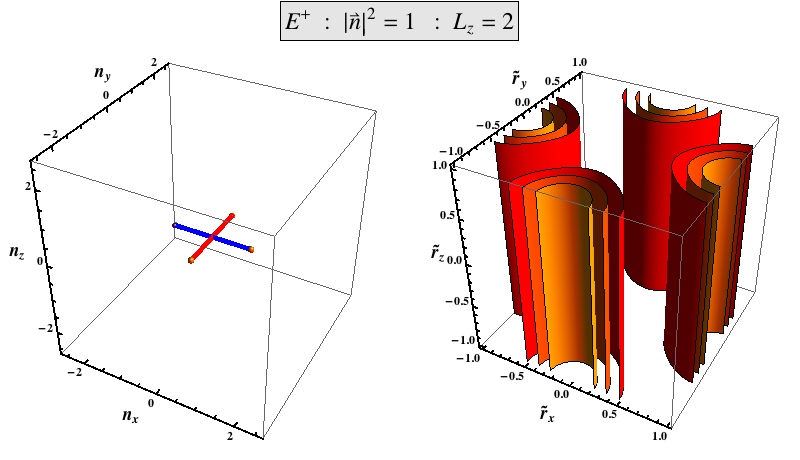}}\\
\mbox{\includegraphics[width=.77\textwidth,angle=0]{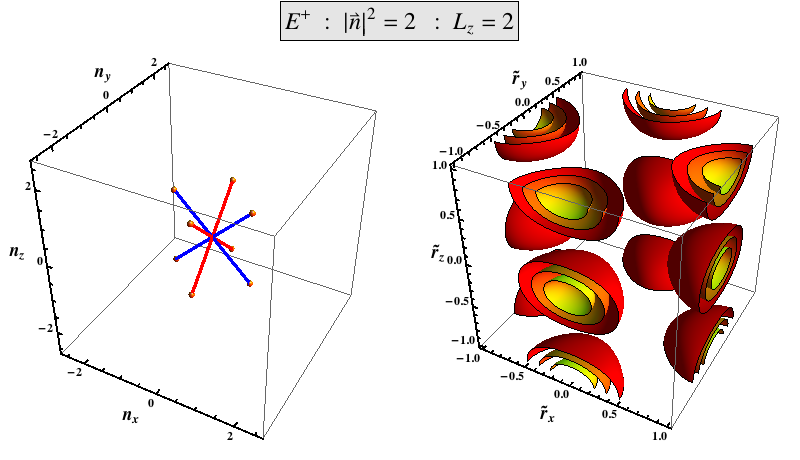}}\\
\mbox{\includegraphics[width=.77\textwidth,angle=0]{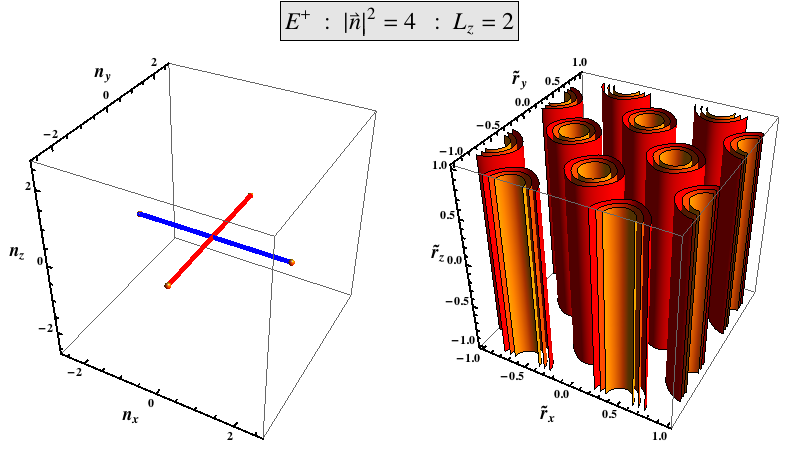}}
\caption{The momentum-space representations (left) and position-space
  representations (right) of two-body relative states in the $E^+$
  representation with $L_z=2$
in the $|{\bf n}|^2$=1, 2, and 4 shells.
\label{fig:E figures}}
\end{figure}
\begin{figure}
\centering
\mbox{\includegraphics[width=.77\textwidth,angle=0]{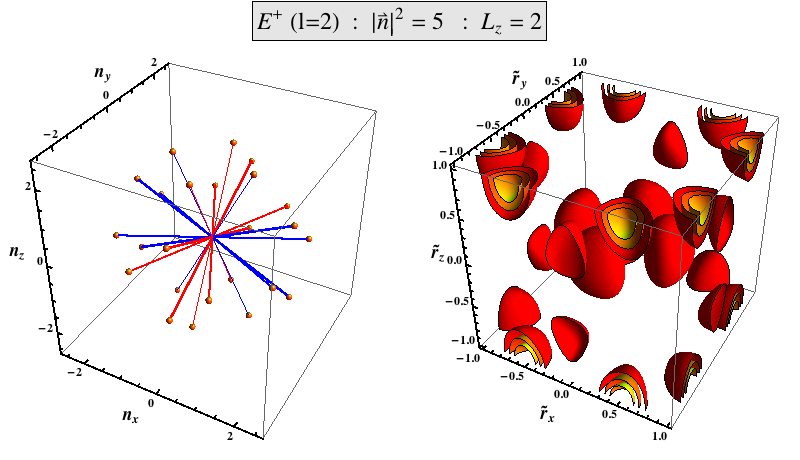}}\\
\mbox{\includegraphics[width=.77\textwidth,angle=0]{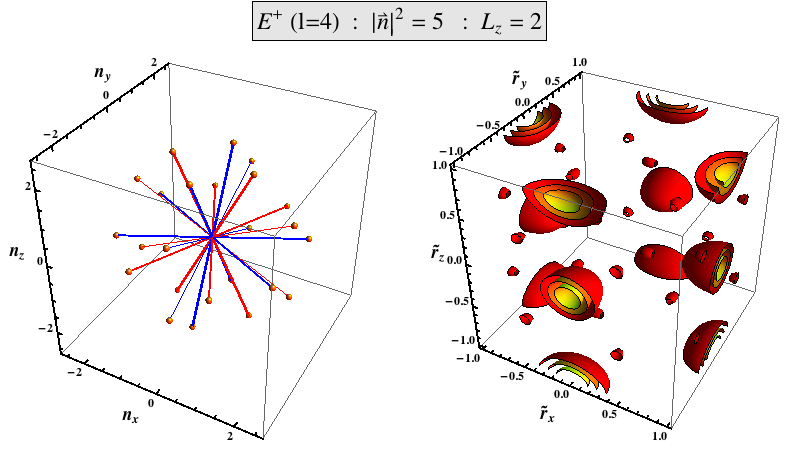}}\\
\mbox{\includegraphics[width=.77\textwidth,angle=0]{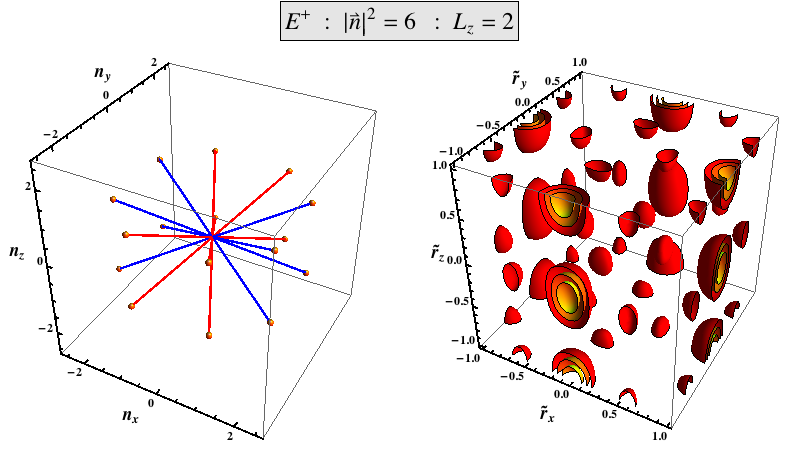}}
\caption{The momentum-space representations (left) and position-space
  representations (right) of two-body relative states in the $E^+$
  representation with $L_z=2$ in the $|{\bf n}|^2$=5 and 6 shells.
\label{fig:E figures 2}}
\end{figure}
The graphical representations  of the
sources and sinks that generate this irrep for 
$|{\bf n}|^2\le6$-shell are shown  in 
figs.~\ref{fig:E figures} and~\ref{fig:E figures 2} in the case of $L_z=0$,
and 
the momentum-space structures are given 
explicitly  in table~\ref{tab:EpSOURCESSINKSLz0} and table~\ref{tab:EpSOURCESSINKSLz2}.

\begin{table}
\begin{center}
\begin{minipage}[!ht]{16.5 cm}
\caption{
The momentum-space structure of $E^+$, $L_z=0$ sources and sinks for $|{\bf n}|^2$=1,2, and 4.
}
\label{tab:EpSOURCESSINKSLz0}
\end{minipage}
\begin{small}
\begin{tabular}{|c||c||c|  }
\hline
\hline
$|{\bf n}|^2$=1 & $|{\bf n}|^2$=2 &  $|{\bf n}|^2$=4 \\ \hline

$\begin{array}{cc}
 \text{$|$(1,0,0) , +1$\rangle $} & \frac{1}{\sqrt{6}} \\
 \text{$|$(0,1,0) , +1$\rangle $} & \frac{1}{\sqrt{6}} \\
 \text{$|$(0,0,1) , +1$\rangle $} & -\sqrt{\frac{2}{3}}
\end{array}$ 
&
$\begin{array}{cc}
 \text{$|$(1,1,0) , +1$\rangle $} & \frac{1}{\sqrt{3}} \\
 \text{$|$(1,0,1) , +1$\rangle $} & -\frac{1}{2 \sqrt{3}} \\
 \text{$|$(1,0,-1) , +1$\rangle $} & -\frac{1}{2 \sqrt{3}} \\
 \text{$|$(1,-1,0) , +1$\rangle $} & \frac{1}{\sqrt{3}} \\
 \text{$|$(0,1,1) , +1$\rangle $} & -\frac{1}{2 \sqrt{3}} \\
 \text{$|$(0,1,-1) , +1$\rangle $} & -\frac{1}{2 \sqrt{3}}
\end{array}$ 
& 
$\begin{array}{cc}
 \text{$|$(2,0,0) , +1$\rangle $} & \frac{1}{\sqrt{6}} \\
 \text{$|$(0,2,0) , +1$\rangle $} & \frac{1}{\sqrt{6}} \\
 \text{$|$(0,0,2) , +1$\rangle $} & -\sqrt{\frac{2}{3}}
\end{array}$ \\ \hline \hline
\end{tabular}
\end{small}
\begin{minipage}[t]{16.5 cm}
\vskip 0.5cm
\noindent
\end{minipage}
\end{center}
\end{table}     
\begin{table}
\begin{center}
\begin{minipage}[!ht]{16.5 cm}
\caption{
The momentum-space structure of $E^+$, $L_z=0$ sources and sinks for $|{\bf n}|^2$=5 and 6.
}
\label{tab:EpSOURCESSINKSLz0v2}
\end{minipage}
\begin{small}
\begin{tabular}{|c||c||c|  }
\hline
\hline
$|{\bf n}|^2$=5$_{(l=2)}$ & $|{\bf n}|^2$=5$_{(l=4)}$ &  $|{\bf n}|^2$=6 \\ \hline

$\begin{array}{cc}
 \text{$|$(2,1,0) , +1$\rangle $} & \frac{5}{2 \sqrt{78}} \\
 \text{$|$(2,0,1) , +1$\rangle $} & \frac{1}{\sqrt{78}} \\
 \text{$|$(2,0,-1) , +1$\rangle $} & \frac{1}{\sqrt{78}} \\
 \text{$|$(2,-1,0) , +1$\rangle $} & \frac{5}{2 \sqrt{78}} \\
 \text{$|$(1,2,0) , +1$\rangle $} & \frac{5}{2 \sqrt{78}} \\
 \text{$|$(1,0,2) , +1$\rangle $} & -\frac{7}{2 \sqrt{78}} \\
 \text{$|$(1,0,-2) , +1$\rangle $} & -\frac{7}{2 \sqrt{78}} \\
 \text{$|$(1,-2,0) , +1$\rangle $} & \frac{5}{2 \sqrt{78}} \\
 \text{$|$(0,2,1) , +1$\rangle $} & \frac{1}{\sqrt{78}} \\
 \text{$|$(0,2,-1) , +1$\rangle $} & \frac{1}{\sqrt{78}} \\
 \text{$|$(0,1,2) , +1$\rangle $} & -\frac{7}{2 \sqrt{78}} \\
 \text{$|$(0,1,-2) , +1$\rangle $} & -\frac{7}{2 \sqrt{78}}
\end{array}$ 
&
$\begin{array}{cc}
 \text{$|$(2,1,0) , +1$\rangle $} & \frac{3}{2 \sqrt{26}} \\
 \text{$|$(2,0,1) , +1$\rangle $} & -\sqrt{\frac{2}{13}} \\
 \text{$|$(2,0,-1) , +1$\rangle $} & -\sqrt{\frac{2}{13}} \\
 \text{$|$(2,-1,0) , +1$\rangle $} & \frac{3}{2 \sqrt{26}} \\
 \text{$|$(1,2,0) , +1$\rangle $} & \frac{3}{2 \sqrt{26}} \\
 \text{$|$(1,0,2) , +1$\rangle $} & \frac{1}{2 \sqrt{26}} \\
 \text{$|$(1,0,-2) , +1$\rangle $} & \frac{1}{2 \sqrt{26}} \\
 \text{$|$(1,-2,0) , +1$\rangle $} & \frac{3}{2 \sqrt{26}} \\
 \text{$|$(0,2,1) , +1$\rangle $} & -\sqrt{\frac{2}{13}} \\
 \text{$|$(0,2,-1) , +1$\rangle $} & -\sqrt{\frac{2}{13}} \\
 \text{$|$(0,1,2) , +1$\rangle $} & \frac{1}{2 \sqrt{26}} \\
 \text{$|$(0,1,-2) , +1$\rangle $} & \frac{1}{2 \sqrt{26}}
\end{array}$ 
& 
$\begin{array}{cc}
 \text{$|$(2,1,1) , +1$\rangle $} & -\frac{1}{2 \sqrt{6}} \\
 \text{$|$(2,1,-1) , +1$\rangle $} & -\frac{1}{2 \sqrt{6}} \\
 \text{$|$(2,-1,1) , +1$\rangle $} & -\frac{1}{2 \sqrt{6}} \\
 \text{$|$(2,-1,-1) , +1$\rangle $} & -\frac{1}{2 \sqrt{6}} \\
 \text{$|$(1,2,1) , +1$\rangle $} & -\frac{1}{2 \sqrt{6}} \\
 \text{$|$(1,2,-1) , +1$\rangle $} & -\frac{1}{2 \sqrt{6}} \\
 \text{$|$(1,1,2) , +1$\rangle $} & \frac{1}{\sqrt{6}} \\
 \text{$|$(1,1,-2) , +1$\rangle $} & \frac{1}{\sqrt{6}} \\
 \text{$|$(1,-1,2) , +1$\rangle $} & \frac{1}{\sqrt{6}} \\
 \text{$|$(1,-1,-2) , +1$\rangle $} & \frac{1}{\sqrt{6}} \\
 \text{$|$(1,-2,1) , +1$\rangle $} & -\frac{1}{2 \sqrt{6}} \\
 \text{$|$(1,-2,-1) , +1$\rangle $} & -\frac{1}{2 \sqrt{6}}
\end{array}$ \\ \hline \hline
\end{tabular}
\end{small}
\begin{minipage}[t]{16.5 cm}
\vskip 0.5cm
\noindent
\end{minipage}
\end{center}
\end{table}     
\begin{table}
\begin{center}
\begin{minipage}[!ht]{16.5 cm}
\caption{
The momentum-space structure of $E^+$, $L_z=2$ sources and sinks for $|{\bf n}|^2$=1, 2, and 4.
These are shown graphically in  fig.~\ref{fig:E figures}.
}
\label{tab:EpSOURCESSINKSLz2}
\end{minipage}
\begin{small}
\begin{tabular}{|c||c||c|  }
\hline
\hline
$|{\bf n}|^2$=1 & $|{\bf n}|^2$=2 &  $|{\bf n}|^2$=4 \\ \hline

$\begin{array}{cc}
 \text{$|$(1,0,0) , +1$\rangle $} & -\frac{1}{\sqrt{2}} \\
 \text{$|$(0,1,0) , +1$\rangle $} & \frac{1}{\sqrt{2}}
\end{array}$ 
&
$\begin{array}{cc}
 \text{$|$(1,0,1) , +1$\rangle $} & -\frac{1}{2} \\
 \text{$|$(1,0,-1) , +1$\rangle $} & -\frac{1}{2} \\
 \text{$|$(0,1,1) , +1$\rangle $} & \frac{1}{2} \\
 \text{$|$(0,1,-1) , +1$\rangle $} & \frac{1}{2}
\end{array}$ 
& 
$\begin{array}{cc}
 \text{$|$(2,0,0) , +1$\rangle $} & -\frac{1}{\sqrt{2}} \\
 \text{$|$(0,2,0) , +1$\rangle $} & \frac{1}{\sqrt{2}}
\end{array}$ \\ \hline \hline
\end{tabular}
\end{small}
\begin{minipage}[t]{16.5 cm}
\vskip 0.5cm
\noindent
\end{minipage}
\end{center}
\end{table}     
\begin{table}
\begin{center}
\begin{minipage}[!ht]{16.5 cm}
\caption{
The momentum-space structure of $E^+$, $L_z=2$ sources and sinks for $|{\bf n}|^2$=5 and 6.
These are shown graphically in  fig.~\ref{fig:E figures 2}.
}
\label{tab:EpSOURCESSINKSLz2v2}
\end{minipage}
\begin{small}
\begin{tabular}{|c||c||c|  }
\hline
\hline
$|{\bf n}|^2$=5$_{(l=2)}$ & $|{\bf n}|^2$=5$_{(l=4)}$ &  $|{\bf n}|^2$=6 \\ \hline

$\begin{array}{cc}
 \text{$|$(2,1,0) , +1$\rangle $} & -\frac{3}{2 \sqrt{26}} \\
 \text{$|$(2,0,1) , +1$\rangle $} & -\sqrt{\frac{2}{13}} \\
 \text{$|$(2,0,-1) , +1$\rangle $} & -\sqrt{\frac{2}{13}} \\
 \text{$|$(2,-1,0) , +1$\rangle $} & -\frac{3}{2 \sqrt{26}} \\
 \text{$|$(1,2,0) , +1$\rangle $} & \frac{3}{2 \sqrt{26}} \\
 \text{$|$(1,0,2) , +1$\rangle $} & -\frac{1}{2 \sqrt{26}} \\
 \text{$|$(1,0,-2) , +1$\rangle $} & -\frac{1}{2 \sqrt{26}} \\
 \text{$|$(1,-2,0) , +1$\rangle $} & \frac{3}{2 \sqrt{26}} \\
 \text{$|$(0,2,1) , +1$\rangle $} & \sqrt{\frac{2}{13}} \\
 \text{$|$(0,2,-1) , +1$\rangle $} & \sqrt{\frac{2}{13}} \\
 \text{$|$(0,1,2) , +1$\rangle $} & \frac{1}{2 \sqrt{26}} \\
 \text{$|$(0,1,-2) , +1$\rangle $} & \frac{1}{2 \sqrt{26}}
\end{array}$ 
&
$\begin{array}{cc}
 \text{$|$(2,1,0) , +1$\rangle $} & -\frac{5}{2 \sqrt{78}} \\
 \text{$|$(2,0,1) , +1$\rangle $} & \frac{1}{\sqrt{78}} \\
 \text{$|$(2,0,-1) , +1$\rangle $} & \frac{1}{\sqrt{78}} \\
 \text{$|$(2,-1,0) , +1$\rangle $} & -\frac{5}{2 \sqrt{78}} \\
 \text{$|$(1,2,0) , +1$\rangle $} & \frac{5}{2 \sqrt{78}} \\
 \text{$|$(1,0,2) , +1$\rangle $} & \frac{7}{2 \sqrt{78}} \\
 \text{$|$(1,0,-2) , +1$\rangle $} & \frac{7}{2 \sqrt{78}} \\
 \text{$|$(1,-2,0) , +1$\rangle $} & \frac{5}{2 \sqrt{78}} \\
 \text{$|$(0,2,1) , +1$\rangle $} & -\frac{1}{\sqrt{78}} \\
 \text{$|$(0,2,-1) , +1$\rangle $} & -\frac{1}{\sqrt{78}} \\
 \text{$|$(0,1,2) , +1$\rangle $} & -\frac{7}{2 \sqrt{78}} \\
 \text{$|$(0,1,-2) , +1$\rangle $} & -\frac{7}{2 \sqrt{78}}
\end{array}$ 
& 
$\begin{array}{cc}
 \text{$|$(2,1,1) , +1$\rangle $} & -\frac{1}{2 \sqrt{2}} \\
 \text{$|$(2,1,-1) , +1$\rangle $} & -\frac{1}{2 \sqrt{2}} \\
 \text{$|$(2,-1,1) , +1$\rangle $} & -\frac{1}{2 \sqrt{2}} \\
 \text{$|$(2,-1,-1) , +1$\rangle $} & -\frac{1}{2 \sqrt{2}} \\
 \text{$|$(1,2,1) , +1$\rangle $} & \frac{1}{2 \sqrt{2}} \\
 \text{$|$(1,2,-1) , +1$\rangle $} & \frac{1}{2 \sqrt{2}} \\
 \text{$|$(1,-2,1) , +1$\rangle $} & \frac{1}{2 \sqrt{2}} \\
 \text{$|$(1,-2,-1) , +1$\rangle $} & \frac{1}{2 \sqrt{2}}
\end{array}$ \\ \hline \hline
\end{tabular}
\end{small}
\begin{minipage}[t]{16.5 cm}
\vskip 0.5cm
\noindent
\end{minipage}
\end{center}
\end{table}     

There are two occurrences of the $E^+$ irrep in the $|{\bf n}|^2=5$-shell.
Linear combinations of the basis states can be formed: one that is
dominated by $\delta_2$,
and one that is 
dominated by $\delta_4$ in the infinite-volume limit, as shown in 
table~\ref{tab:EpSOURCESSINKSLz0} and table~\ref{tab:EpSOURCESSINKSLz2}.
The states 
are defined by  $\langle l,m| |{\bf n}|^2_l; \Gamma^{(i)}, L_z\rangle=
\langle 2,0|5_4; E^+ ,0\rangle = 0$, and the orthogonal combination 
$|5_2; E^+,0\rangle$.  As is the case in the $A_1^+$ sector, these states are
not energy-eigenstates since they have a non-zero projection, in principle, 
onto all $E^+$ states.  
The perturbative expansions of the energy-eigenvalues 
in the large-volume limit 
can be
found in Appendix~\ref{sect:perturbation}.

\subsubsection{$T_1^+$ Representation}
\label{irrep:T1p}
The energy-eigenvalues of states transforming in the $T_1^+$ irrep receive
contributions
from interactions in the $l=4,6,...$ partial-waves.
The $T_1^+$ irrep is three-dimensional, with states identified by $L_z=0,1,3$, 
and provides a contribution to the determinant in 
eq.~(\ref{eq:evals})
that results from a $6\times 6$ matrix for $l\le 6$.  As the three $L_z$-states
are degenerate, the analysis collapses down to that of a $2\times 2$ matrix.
\begin{figure}
\centering
\includegraphics[height=\textwidth,angle=-90]{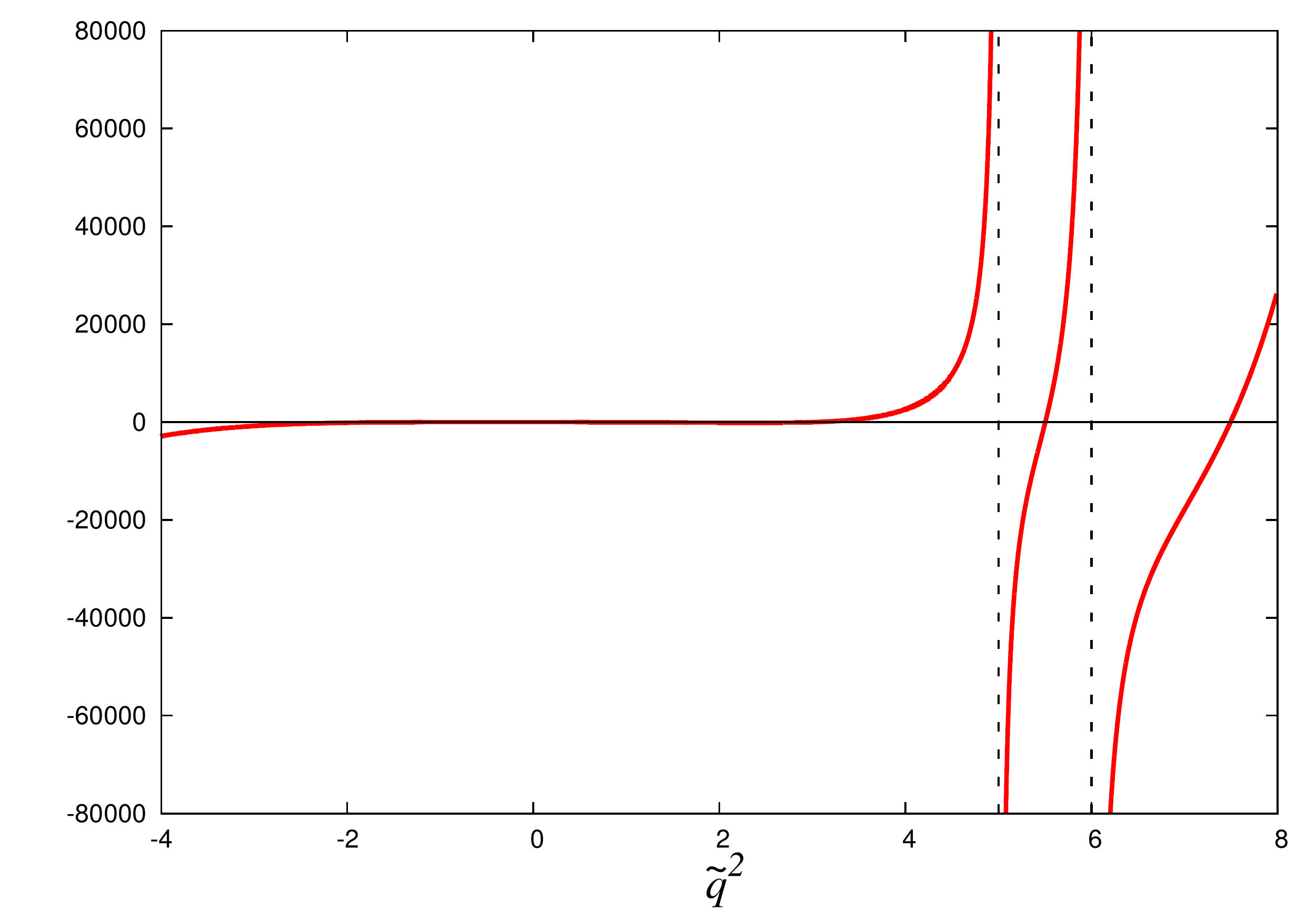}
\caption{The function $\mathcal{X}_{T_1}^{+}$, as defined in eq.~(\protect\ref{eqn:T1+ l=4}), 
as a function of $\tilde q^2$.  The vertical dashed lines denote the 
position of the poles of the function corresponding to the non-interacting
energy-eigenvalues.  
\label{fig:T1+ l=4}}
\end{figure}
The $T_1^+$ $L_z=0$  
states associated with the $\overline{F}^{(FV)}_{4;4}$ and $\overline{F}^{(FV)}_{6;6}$ blocks are 
%
\begin{equation}
\begin{split}
|T_1^+, 0; 4; 1 \rangle\ =\ & \frac{1}{\sqrt{2}}|4,4\rangle
\ -\ \frac{1}{\sqrt{2}}|4,-4\rangle\\
|T_1^+, 0; 6; 1 \rangle\ =\ & \frac{1}{\sqrt{2}}|6,4\rangle
\ -\ \frac{1}{\sqrt{2}}|6,-4\rangle
\ \ \ ,
\end{split}
\end{equation}
and the contribution  to eq.~(\ref{eq:evals}) is
\begin{equation}
\label{eqn:T1+ det}
\text{det}\left[
\begin{pmatrix}
\text{cot}\delta_4 & 0\\
0 & \text{cot}\delta_6
\end{pmatrix}
\ -\ 
\begin{pmatrix} 
\overline{F}_{4;4}^{(FV, T_1^+)} & \overline{F}_{4;6}^{(FV, T_1^+)} \\
\overline{F}_{6;2}^{(FV, T_1^+)} & \overline{F}_{6;6}^{(FV, T_1^+)} 
\end{pmatrix}
\right]=0
\ \ \ \ ,
\end{equation}
where
\begin{equation}
\begin{split}
\overline{F}_{4;4}^{(FV, T_1^+)} &=\frac{\mathcal{Z}_{0,0}\left(1;\tilde q^2\right)}{\pi ^{3/2} \tilde q}-\frac{448 \mathcal{Z}_{8,0}\left(1;\tilde q^2\right)}{143 \sqrt{17} \pi ^{3/2} \tilde q^9}-\frac{4 \mathcal{Z}_{6,0}\left(1;\tilde q^2\right)}{11 \sqrt{13} \pi ^{3/2} \tilde q^7}+\frac{54 \mathcal{Z}_{4,0}\left(1;\tilde q^2\right)}{143
   \pi ^{3/2} \tilde q^5}
\nonumber\\
\overline{F}_{4;6}^{(FV, T_1^+)} &=\frac{576 \sqrt{\frac{21}{65}} \mathcal{Z}_{10,0}\left(1;\tilde q^2\right)}{323 \pi ^{3/2} \tilde q^{11}}+\frac{112 \sqrt{\frac{5}{221}} \mathcal{Z}_{8,0}\left(1;\tilde q^2\right)}{209 \pi ^{3/2} \tilde q^9}+\frac{42 \sqrt{5} \mathcal{Z}_{6,0}\left(1;\tilde q^2\right)}{187 \pi
   ^{3/2} \tilde q^7}-\frac{12 \sqrt{\frac{5}{13}}
   \mathcal{Z}_{4,0}\left(1;\tilde q^2\right)}{11 \pi ^{3/2} \tilde q^5}
\nonumber\\
  \overline{F}_{6;6}^{(FV, T_1^+)}&=\frac{\mathcal{Z}_{0,0}\left(1;\tilde q^2\right)}{\pi ^{3/2} \tilde q}-\frac{26136 \mathcal{Z}_{12,0}\left(1;\tilde q^2\right)}{37145 \pi ^{3/2} \tilde q^{13}}+\frac{1584 \sqrt{1001} \mathcal{Z}_{12,4}\left(1;\tilde q^2\right)}{37145 \pi ^{3/2} \tilde q^{13}}+\frac{624 \sqrt{21}
   \mathcal{Z}_{10,0}\left(1;\tilde q^2\right)}{7429 \pi ^{3/2} \tilde q^{11}}
\\
   &\quad+\frac{120 \mathcal{Z}_{8,0}\left(1;\tilde q^2\right)}{209 \sqrt{17} \pi ^{3/2} \tilde q^9}-\frac{80 \sqrt{13} \mathcal{Z}_{6,0}\left(1;\tilde q^2\right)}{3553 \pi ^{3/2} \tilde q^7}-\frac{96
   \mathcal{Z}_{4,0}\left(1;\tilde q^2\right)}{187 \pi ^{3/2} \tilde q^5}
\ \ \ .
\nonumber
\end{split}
\end{equation}
The solutions to eq.~(\ref{eqn:T1+ det}) are obtained from
\begin{multline}
\frac{\text{cot}\delta_4}{2}+\frac{\text{cot}\delta_6}{2}-\frac{312 \sqrt{21} \mathcal{Z}_{10,0}\left(1;\tilde{q}^2\right)}{7429 \pi ^{3/2} \tilde{q}^{11}}+\frac{13068 \mathcal{Z}_{12,0}\left(1;\tilde{q}^2\right)}{37145 \pi ^{3/2} \tilde{q}^{13}}-\frac{792 \sqrt{1001} \mathcal{Z}_{12,4}\left(1;\tilde{q}^2\right)}{37145 \pi ^{3/2}
   \tilde{q}^{13}}\\
   -\frac{\mathcal{Z}_{0,0}\left(1;\tilde{q}^2\right)}{\pi ^{3/2} \tilde{q}}
   +\frac{15 \mathcal{Z}_{4,0}\left(1;\tilde{q}^2\right)}{221 \pi ^{3/2} \tilde{q}^5}+\frac{106 \mathcal{Z}_{6,0}\left(1;\tilde{q}^2\right)}{323 \sqrt{13} \pi ^{3/2} \tilde{q}^7}+\frac{316
   \mathcal{Z}_{8,0}\left(1;\tilde{q}^2\right)}{247 \sqrt{17} \pi ^{3/2} \tilde{q}^9}=\\
   {\bf \pm}
   \frac{1}{2}\left[
   \left(\frac{624 \sqrt{21} \mathcal{Z}_{10,0}\left(1;\tilde{q}^2\right)}{7429 \pi ^{3/2} \tilde{q}^{11}}-\frac{26136 \mathcal{Z}_{12,0}\left(1;\tilde{q}^2\right)}{37145 \pi ^{3/2} \tilde{q}^{13}}+\frac{1584 \sqrt{1001}
   \mathcal{Z}_{12,4}\left(1;\tilde{q}^2\right)}{37145 \pi ^{3/2} \tilde{q}^{13}}-\frac{2166 \mathcal{Z}_{4,0}\left(1;\tilde{q}^2\right)}{2431 \pi ^{3/2} \tilde{q}^5}\right. \right.\\
   +\left.\left.\frac{252 \mathcal{Z}_{6,0}\left(1;\tilde{q}^2\right)}{3553 \sqrt{13} \pi ^{3/2}
   \tilde{q}^7}+\frac{10072 \mathcal{Z}_{8,0}\left(1;\tilde{q}^2\right)}{2717 \sqrt{17} \pi ^{3/2} \tilde{q}^9}+\text{cot$\delta $}_4-\text{cot$\delta $}_6\right)^2\right.\\
   +\left.\frac{4}{\pi^3} \left(\frac{576 \sqrt{\frac{21}{65}}
   \mathcal{Z}_{10,0}\left(1;\tilde{q}^2\right)}{323 \tilde{q}^{11}}-\frac{12 \sqrt{\frac{5}{13}} \mathcal{Z}_{4,0}\left(1;\tilde{q}^2\right)}{11 \tilde{q}^5}+\frac{42 \sqrt{5} \mathcal{Z}_{6,0}\left(1;\tilde{q}^2\right)}{187 \tilde{q}^7}+\frac{112
   \sqrt{\frac{5}{221}} \mathcal{Z}_{8,0}\left(1;\tilde{q}^2\right)}{209 \tilde{q}^9}\right)^2\right]^{1/2}\ .
   \end{multline}
In the situation where the interaction in the $l=6$ (and higher) partial-wave vanishes, the energy-eigenvalues are sensitive to the $l=4$
interaction alone, and can be found from
\begin{eqnarray}
\label{eqn:T1+ l=4}
  q^9\text{cot}\delta_4
 & = & 
\left(\frac{2\pi}{L}\right)^9\frac{1}{\pi^{3/2}}\ 
  \left(
  \tilde{q}^8\mathcal{Z}_{0,0}\left(1;\tilde q^2\right)-\frac{448 \mathcal{Z}_{8,0}\left(1;\tilde q^2\right)}{143 \sqrt{17}}-\frac{4\tilde{q}^2 \mathcal{Z}_{6,0}\left(1;\tilde q^2\right)}{11 \sqrt{13}}+\frac{54 \tilde{q}^4\mathcal{Z}_{4,0}\left(1;\tilde q^2\right)}{143
   }\right)
\nonumber\\
 & \equiv &  
\left(\frac{2\pi}{L}\right)^9\frac{1}{\pi^{3/2}}\ 
\mathcal{X}_{T_1}^{+}\left(\tilde q^2\right)
\ \ \ ,
\end{eqnarray}
where
the function $\mathcal{X}_{T_1}^{+}\left(\tilde q^2\right)$ is shown in fig.~\ref{fig:T1+ l=4}.

\begin{figure}
\centering
\mbox{\includegraphics[width=.77\textwidth,angle=0]{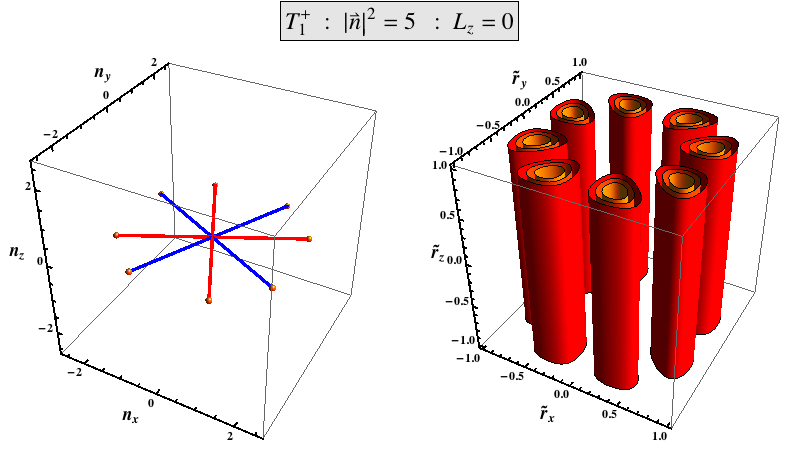}}\\
\mbox{\includegraphics[width=.77\textwidth,angle=0]{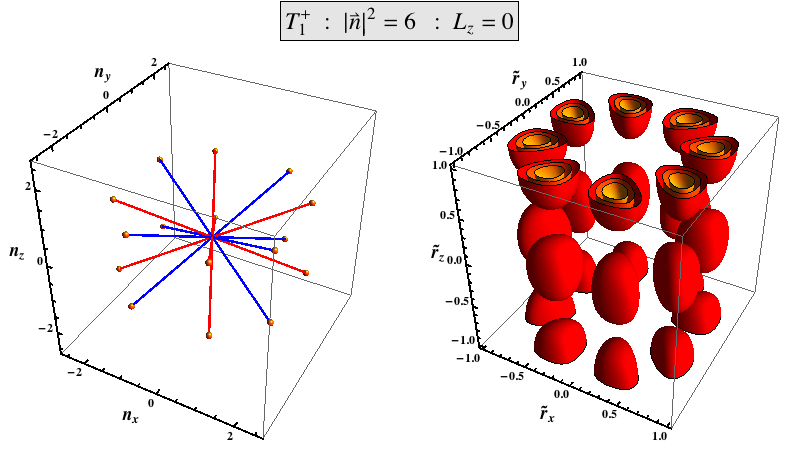}}
\caption{The momentum-space representations (left) and position-space
  representations (right) of two-body relative states in the $T_1^+$
  representation with $L_z=0$ in the $|{\bf n}|^2$=5 and 6 shells.
\label{fig:T1plus figures}}
\end{figure}
The graphical representations  of the
sources and sinks that generate the $T_1^+$ irrep for the low-lying  $|{\bf
  n}|^2$-shells are shown in 
fig.~\ref{fig:T1plus figures},
and  the momentum-space structures for $L_z=0$ are given 
explicitly in table~\ref{tab:T1pSOURCESSINKS}, and for $L_z=1$ in
table~\ref{tab:T1pSOURCESSINKSLz1}.
The structures of the $L_z=3$ sources and sinks are related to those 
with $L_z=1$ by complex conjugation of the coefficients.
\begin{table}
\begin{center}
\begin{minipage}[!ht]{16.5 cm}
\caption{
The momentum-space structure of $T_1^+$, $L_z=0$ sources and sinks.
These are shown graphically in  fig.~\ref{fig:T1plus figures}.
}
\label{tab:T1pSOURCESSINKS}
\end{minipage}
\begin{small}
\begin{tabular}{ |c||c|  }
\hline
\hline
$|{\bf n}|^2$=5 &  $|{\bf n}|^2$=6 \\ \hline

$\begin{array}{cc}
 \text{$|$(2,1,0) , +1$\rangle $} & \frac{1}{2} \\
 \text{$|$(2,-1,0) , +1$\rangle $} & -\frac{1}{2} \\
 \text{$|$(1,2,0) , +1$\rangle $} & -\frac{1}{2} \\
 \text{$|$(1,-2,0) , +1$\rangle $} & \frac{1}{2}
\end{array}$ 
& 
$\begin{array}{cc|cc}
 \text{$|$(2,1,1) , +1$\rangle $} & \frac{1}{2 \sqrt{2}} & \text{$|$(2,1,-1) , +1$\rangle $} & \frac{1}{2 \sqrt{2}} \\
 \text{$|$(2,-1,1) , +1$\rangle $} & -\frac{1}{2 \sqrt{2}} & \text{$|$(2,-1,-1) , +1$\rangle $} & -\frac{1}{2 \sqrt{2}} \\
 \text{$|$(1,2,1) , +1$\rangle $} & -\frac{1}{2 \sqrt{2}} & \text{$|$(1,2,-1) , +1$\rangle $} & -\frac{1}{2 \sqrt{2}} \\
 \text{$|$(1,-2,1) , +1$\rangle $} & \frac{1}{2 \sqrt{2}} & \text{$|$(1,-2,-1) , +1$\rangle $} & \frac{1}{2 \sqrt{2}}
\end{array}$ \\ \hline \hline
\end{tabular}
\end{small}
\begin{minipage}[t]{16.5 cm}
\vskip 0.5cm
\noindent
\end{minipage}
\end{center}
\end{table}     
\begin{table}
\begin{center}
\begin{minipage}[!ht]{16.5 cm}
\caption{
The momentum-space structure of $T_1^+$, $L_z=1$ sources and sinks.
}
\label{tab:T1pSOURCESSINKSLz1}
\end{minipage}
\begin{small}
\begin{tabular}{ |c||c|  }
\hline
\hline
$|{\bf n}|^2$=5 &  $|{\bf n}|^2$=6 \\ \hline

$\begin{array}{cc|cc}
 \text{$|$(2,0,1) , +1$\rangle $} & \frac{1}{2 \sqrt{2}} &\text{$|$(2,0,-1) , +1$\rangle $} & -\frac{1}{2 \sqrt{2}} \\
 \text{$|$(1,0,2) , +1$\rangle $} & -\frac{1}{2 \sqrt{2}} & \text{$|$(1,0,-2) , +1$\rangle $} & \frac{1}{2 \sqrt{2}} \\
 \text{$|$(0,2,1) , +1$\rangle $} & \frac{i}{2 \sqrt{2}} & \text{$|$(0,2,-1) , +1$\rangle $} & -\frac{i}{2 \sqrt{2}} \\
 \text{$|$(0,1,2) , +1$\rangle $} & -\frac{i}{2 \sqrt{2}} & \text{$|$(0,1,-2) , +1$\rangle $} & \frac{i}{2 \sqrt{2}}
\end{array}$ 
& 
$\begin{array}{cc|cc}
 \text{$|$(2,1,1) , +1$\rangle $} & \frac{1}{4} & \text{$|$(2,1,-1) , +1$\rangle $} & -\frac{1}{4} \\
 \text{$|$(2,-1,1) , +1$\rangle $} & \frac{1}{4} & \text{$|$(2,-1,-1) , +1$\rangle $} & -\frac{1}{4} \\
 \text{$|$(1,2,1) , +1$\rangle $} & \frac{i}{4} & \text{$|$(1,2,-1) , +1$\rangle $} & -\frac{i}{4} \\
 \text{$|$(1,1,2) , +1$\rangle $} & -\frac{1}{4}-\frac{i}{4} & \text{$|$(1,1,-2) , +1$\rangle $} & \frac{1}{4}+\frac{i}{4} \\
 \text{$|$(1,-1,2) , +1$\rangle $} & -\frac{1}{4}+\frac{i}{4} & \text{$|$(1,-1,-2) , +1$\rangle $} & \frac{1}{4}-\frac{i}{4} \\
 \text{$|$(1,-2,1) , +1$\rangle $} & -\frac{i}{4} & \text{$|$(1,-2,-1) , +1$\rangle $} & \frac{i}{4}
\end{array}$ \\ \hline \hline
\end{tabular}
\end{small}
\begin{minipage}[t]{16.5 cm}
\vskip 0.5cm
\noindent
\end{minipage}
\end{center}
\end{table}     
%

\subsubsection{$T_2^+$ Representation}
\label{irrep:T2p}
The energy-eigenvalues of states transforming in the $T_2^+$ irrep receive
contributions
from interactions in the $l=2,4,6,...$ partial-waves.
The $T_2^+$ irrep is three-dimensional, with states defined by $L_z=1,2,3$, 
and provides a contribution to the determinant in 
eq.~(\ref{eq:evals})
that results from a $12\times 12$ matrix for $l\le 6$ (it is $12\times 12$ and
not
$9\times 9$ because there are two $T_2^+$'s in the decomposition of $l=6$, see
table~\ref{tab:Lcontributetab}).  
As the three $L_z$-states
are degenerate, the analysis collapses down to that of a $4\times 4$ matrix.
\begin{figure}
\centering
\includegraphics[height=\textwidth,angle=-90]{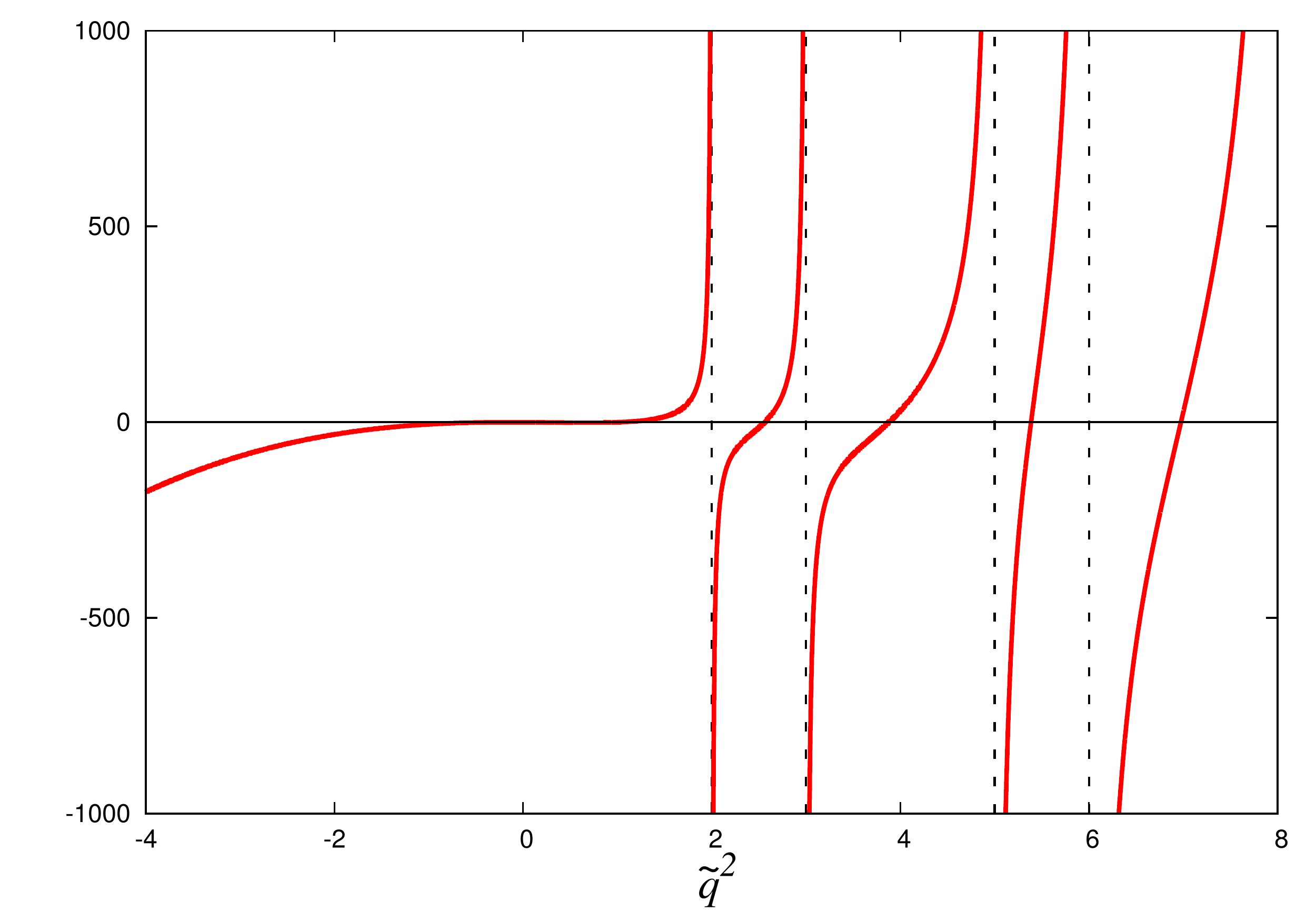}
\caption{The function $\mathcal{X}_{T_2}^+$, as defined in eq.~(\ref{eqn:T2+ l=2}), as a
  function of $\tilde q^2$.  The vertical dashed lines denote the position of
  the poles of the function corresponding to the $T_2^+$ non-interacting energy-eigenvalues.
\label{fig:T2+ l=2}}
\end{figure}
The $T_2^+$ $L_z=2$  
states associated with the $\overline{F}^{(FV)}_{2;2}$, $\overline{F}^{(FV)}_{4;4}$, and $\overline{F}^{(FV)}_{6,6}$ blocks are 
\begin{equation}
\begin{split}
|T_2^+, 2; 2; 1\rangle\ &\ =\ \frac{1}{\sqrt{2}}\left(|2,2\rangle
\ -\ |2,-2\rangle\right)\\
|T_2^+, 2; 4; 1\rangle\ &\ =\ \frac{1}{\sqrt{2}}\left(|4,2\rangle
\ -\ |4,-2\rangle\right)\\
|T_2^+, 2; 6; 1\rangle\ &\ =\ \frac{1}{\sqrt{2}}\left(|6,2\rangle
\ -\ |6,-2\rangle\right)\\
|T_2^+, 2; 6; 2\rangle\ &\ =\ \frac{1}{\sqrt{2}}\left(|6,6\rangle
\ -\ |6,-6\rangle\right)
\end{split}
\label{eq:T2p24}
\end{equation}
With these basis states,
the contribution to the determinant in 
eq.~(\ref{eq:evals}) becomes
\begin{equation}
\label{eqn:T2+ det}
\text{det}\left[
\begin{pmatrix}
\text{cot}\delta_2 & 0 & 0 & 0\\
0 & \text{cot}\delta_4 & 0 & 0\\
0 & 0 & \text{cot}\delta_6 & 0\\
0 & 0 & 0 & \text{cot}\delta_6
\end{pmatrix}
-
\begin{pmatrix} 
\overline{F}_{2;2}^{(FV, T_2^+)}  & \overline{F}_{2;4}^{(FV, T_2^+)}  & \overline{F}_{2;6_1}^{(FV, T_2^+)} & \overline{F}_{2;6_2}^{(FV, T_2^+)}\\
\overline{F}_{4;2}^{(FV, T_2^+)}  & \overline{F}_{4;4}^{(FV, T_2^+)}  & \overline{F}_{4;6_1}^{(FV, T_2^+)} & \overline{F}_{4;6_2}^{(FV, T_2^+)}\\
\overline{F}_{6_1;2}^{(FV, T_2^+)} & \overline{F}_{6_1;4}^{(FV, T_2^+)} & \overline{F}_{6_1;6_1}^{(FV, T_2^+)}& \overline{F}_{6_1,6_2}^{(FV, T_2^+)}\\
\overline{F}_{6_2;2}^{(FV, T_2^+)} & \overline{F}_{6_2;4}^{(FV, T_2^+)} & \overline{F}_{6_2;6_1}^{(FV, T_2^+)}  & \overline{F}_{6_2,6_2}^{(FV, T_2^+)}
\end{pmatrix}
\right]\ =\ 0
\ \ \ ,
\end{equation}
where
\begin{equation}
\begin{split}
\overline{F}_{2,2}^{(FV, T_2^+)}&=\frac{\mathcal{Z}_{0,0}\left(1;\tilde{q}^2\right)}{\pi
  ^{3/2} \tilde{q}}-\frac{4 \mathcal{Z}_{4,0}\left(1;\tilde{q}^2\right)}{7 \pi
  ^{3/2} \tilde{q}^5}
\\
\overline{F}_{2,4}^{(FV, T_2^+)}&=\frac{40 \sqrt{\frac{3}{13}}
  \mathcal{Z}_{6,0}\left(1;\tilde{q}^2\right)}{11 \pi ^{3/2}
  \tilde{q}^7}-\frac{20 \sqrt{3} \mathcal{Z}_{4,0}\left(1;\tilde{q}^2\right)}{77
  \pi ^{3/2} \tilde{q}^5}
\\
\overline{F}_{4,4}^{(FV, T_2^+)}&=\frac{\mathcal{Z}_{0,0}\left(1;\tilde{q}^2\right)}{\pi
  ^{3/2} \tilde{q}}-\frac{54 \mathcal{Z}_{4,0}\left(1;\tilde{q}^2\right)}{77 \pi
  ^{3/2} \tilde{q}^5}+\frac{20 \mathcal{Z}_{6,0}\left(1;\tilde{q}^2\right)}{11
  \sqrt{13} \pi ^{3/2} \tilde{q}^7}
  \\
\overline{F}_{2,6_1}^{(FV, T_2^+)}&\ =\ 
\frac{5 \sqrt{\frac{13}{14}} \mathcal{Z}_{4,0}\left(1;\tilde{q}^2\right)}{11 \pi ^{3/2} \tilde{q}^5}-\frac{5 \sqrt{14} \mathcal{Z}_{6,0}\left(1;\tilde{q}^2\right)}{11 \pi ^{3/2} \tilde{q}^7}
\\
\overline{F}_{4,6_1}^{(FV, T_2^+)}&=-\frac{28 \sqrt{\frac{42}{221}} \text{$\mathcal{Z}$}_{8,0}\left(1;\tilde{q}^2\right)}{19 \pi ^{3/2} \tilde{q}^9}+\frac{10 \sqrt{\frac{6}{91}} \mathcal{Z}_{4,0}\left(1;\tilde{q}^2\right)}{11 \pi ^{3/2} \tilde{q}^5}+\frac{\sqrt{\frac{21}{2}}
   \mathcal{Z}_{6,0}\left(1;\tilde{q}^2\right)}{187 \pi ^{3/2} \tilde{q}^7}+\frac{1008 \sqrt{\frac{2}{13}} \mathcal{Z}_{10,0}\left(1;\tilde{q}^2\right)}{323 \pi ^{3/2} \tilde{q}^{11}}\\
\overline{F}_{6_1,6_1}^{(FV, T_2^+)}&=-\frac{45 \text{$\mathcal{Z}$}_{8,0}\left(1;\tilde{q}^2\right)}{19 \sqrt{17} \pi ^{3/2} \tilde{q}^9}+\frac{\mathcal{Z}_{0,0}\left(1;\tilde{q}^2\right)}{\pi ^{3/2} \tilde{q}}-\frac{59 \mathcal{Z}_{4,0}\left(1;\tilde{q}^2\right)}{187 \pi ^{3/2}
   \tilde{q}^5}+\frac{620 \sqrt{13} \mathcal{Z}_{6,0}\left(1;\tilde{q}^2\right)}{3553 \pi ^{3/2} \tilde{q}^7}\\
   &+\frac{162 \sqrt{21} \mathcal{Z}_{10,0}\left(1;\tilde{q}^2\right)}{7429 \pi ^{3/2} \tilde{q}^{11}}+\frac{3267
   \mathcal{Z}_{12,0}\left(1;\tilde{q}^2\right)}{7429 \pi ^{3/2}
   \tilde{q}^{13}}-\frac{198 \sqrt{1001}
   \mathcal{Z}_{12,4}\left(1;\tilde{q}^2\right)}{7429 \pi ^{3/2}
   \tilde{q}^{13}}
\end{split}
\nonumber
\end{equation}
\begin{equation}
\begin{split}
\overline{F}_{2,6_2}^{(FV, T_2^+)}&\ =\ 
\frac{15 \sqrt{\frac{5}{2002}} \mathcal{Z}_{4,0}\left(1;\tilde{q}^2\right)}{\pi ^{3/2} \tilde{q}^5}+\frac{\sqrt{\frac{14}{55}} \mathcal{Z}_{6,0}\left(1;\tilde{q}^2\right)}{\pi ^{3/2} \tilde{q}^7}-\frac{64 \sqrt{\frac{14}{12155}}
   \mathcal{Z}_{8,0}\left(1;\tilde{q}^2\right)}{3 \pi ^{3/2} \tilde{q}^9}
\\
\overline{F}_{4,6_2}^{(FV, T_2^+)}&=-\frac{2 \sqrt{\frac{30}{1001}} \mathcal{Z}_{4,0}\left(1;\tilde{q}^2\right)}{\pi ^{3/2} \tilde{q}^5}-\frac{9 \sqrt{\frac{105}{22}} \mathcal{Z}_{6,0}\left(1;\tilde{q}^2\right)}{17 \pi ^{3/2} \tilde{q}^7}+\frac{20 \sqrt{\frac{210}{2431}}
   \mathcal{Z}_{8,0}\left(1;\tilde{q}^2\right)}{19 \pi ^{3/2} \tilde{q}^9}+\frac{336 \sqrt{\frac{22}{65}} \mathcal{Z}_{10,0}\left(1;\tilde{q}^2\right)}{323 \pi ^{3/2} \tilde{q}^{11}}
 \\
\overline{F}_{6_1,6_2}^{(FV, T_2^+)}&=\frac{3 \sqrt{\frac{5}{11}} \mathcal{Z}_{4,0}\left(1;\tilde{q}^2\right)}{17 \pi ^{3/2} \tilde{q}^5}+\frac{140 \sqrt{\frac{65}{11}} \mathcal{Z}_{6,0}\left(1;\tilde{q}^2\right)}{323 \pi ^{3/2} \tilde{q}^7}+\frac{5 \sqrt{\frac{5}{187}}
   \mathcal{Z}_{8,0}\left(1;\tilde{q}^2\right)}{57 \pi ^{3/2} \tilde{q}^9}-\frac{666 \sqrt{\frac{231}{5}} \mathcal{Z}_{10,0}\left(1;\tilde{q}^2\right)}{7429 \pi ^{3/2} \tilde{q}^{11}}\\
   &-\frac{1287 \sqrt{\frac{11}{5}}
   \mathcal{Z}_{12,0}\left(1;\tilde{q}^2\right)}{7429 \pi ^{3/2} \tilde{q}^{13}}+\frac{858 \sqrt{\frac{91}{5}} \mathcal{Z}_{12,4}\left(1;\tilde{q}^2\right)}{7429 \pi ^{3/2} \tilde{q}^{13}}
    \\
\overline{F}_{6_2,6_2}^{(FV, T_2^+)}&=\frac{\mathcal{Z}_{0,0}\left(1;\tilde{q}^2\right)}{\pi ^{3/2} \tilde{q}}+\frac{9 \mathcal{Z}_{4,0}\left(1;\tilde{q}^2\right)}{17 \pi ^{3/2} \tilde{q}^5}-\frac{20 \sqrt{13} \mathcal{Z}_{6,0}\left(1;\tilde{q}^2\right)}{323 \pi ^{3/2}
   \tilde{q}^7}+\frac{5 \mathcal{Z}_{8,0}\left(1;\tilde{q}^2\right)}{19 \sqrt{17} \pi ^{3/2} \tilde{q}^9}-\frac{18 \sqrt{21} \mathcal{Z}_{10,0}\left(1;\tilde{q}^2\right)}{7429 \pi ^{3/2} \tilde{q}^{11}}\\
   &-\frac{23991
   \mathcal{Z}_{12,0}\left(1;\tilde{q}^2\right)}{37145 \pi ^{3/2} \tilde{q}^{13}}-\frac{594 \sqrt{1001} \mathcal{Z}_{12,4}\left(1;\tilde{q}^2\right)}{37145 \pi ^{3/2} \tilde{q}^{13}}
\ \ \ .
\end{split}
\end{equation}
The solutions to eq.~(\ref{eqn:T2+ det})
must be determined numerically and will, in general, depend upon the interactions
in the $l=2$, 4, and 6 partial-waves.
In the limit where the interactions in the $l=6$ and higher partial-waves
vanish, leaving contributions only from interactions in the $l=2,4$ partial-waves,
the contribution to the determinant in eq.~(\ref{eqn:T2+ det}) 
collapses down to that of a $2\times2$ matrix, which has solutions
\begin{multline}\label{eqn:T2+ 24 mixing}
\frac{\text{cot$\delta $}_2}{2}+\frac{\text{cot$\delta $}_4}{2}
-\frac{\mathcal{Z}_{0,0}\left(1;\tilde{q}^2\right)}{\pi ^{3/2} \tilde{q}}+\frac{7 \mathcal{Z}_{4,0}\left(1;\tilde{q}^2\right)}{11 \pi ^{3/2} \tilde{q}^5}-\frac{10 \mathcal{Z}_{6,0}\left(1;\tilde{q}^2\right)}{11 \sqrt{13} \pi ^{3/2}
   \tilde{q}^7}  \\
   =\pm\frac{1}{2} \left[\left(-\frac{10 \mathcal{Z}_{4,0}\left(1;\tilde{q}^2\right)}{77 \pi ^{3/2} \tilde{q}^5}+\frac{20 \mathcal{Z}_{6,0}\left(1;\tilde{q}^2\right)}{11 \sqrt{13} \pi ^{3/2} \tilde{q}^7}+\text{cot$\delta
   $}_2-\text{cot$\delta $}_4\right)^2\right.\\
   \quad+\left.4 \left(\frac{20 \sqrt{3} \mathcal{Z}_{4,0}\left(1;\tilde{q}^2\right)}{77 \pi ^{3/2} \tilde{q}^5}-\frac{40 \sqrt{\frac{3}{13}} \mathcal{Z}_{6,0}\left(1;\tilde{q}^2\right)}{11 \pi ^{3/2}
   \tilde{q}^7}\right)^2\right]^{1/2}\ .
\end{multline}
In the limit that $\tan\delta_4 << \tan\delta_2$
the energy-eigenvalues are the solutions to 
\begin{equation}\label{eqn:T2+ l=2}
\begin{split}
q^5\text{cot}\delta_2&=\left(\frac{2\pi}{L}\right)^5\frac{1}{\pi^{3/2}}\left(\tilde{q}^4
  \mathcal{Z}_{0,0}\left(1;\tilde{q}^2\right)-\frac{4
  }{7}\mathcal{Z}_{4,0}\left(1;\tilde{q}^2\right)\right)
\ =\ \left(\frac{2\pi}{L}\right)^5\frac{1}{\pi^{3/2}}\mathcal{X}_{T_2}^+\left(\tilde
  q^2\right)
\ \ \ ,
\end{split}
\end{equation}
where  $\mathcal{X}_{T_2}^+$ is shown as a function of
$\tilde q^2$ in fig.~\ref{fig:T2+ l=2}. 
The $T_2^+$ irrep first appears in the $|{\bf n}|^2=2$-shell, as can be seen in
fig.~\ref{fig:T2+ l=2}.
\begin{figure}
\centering
\mbox{\includegraphics[width=.77\textwidth,angle=0]{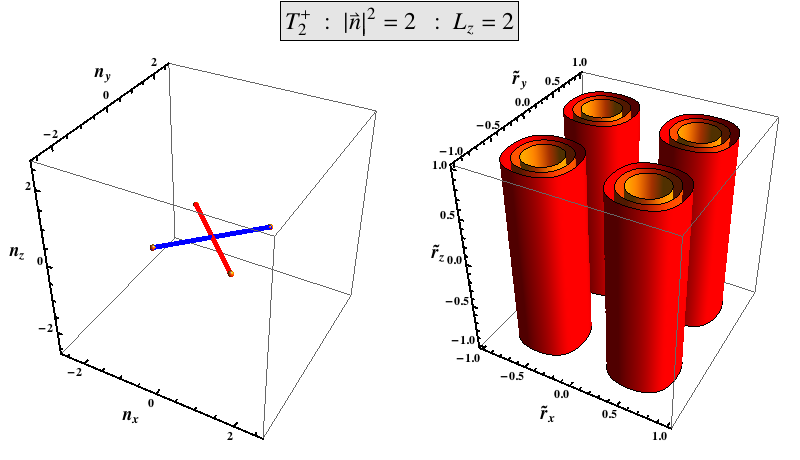}}\\
\mbox{\includegraphics[width=.77\textwidth,angle=0]{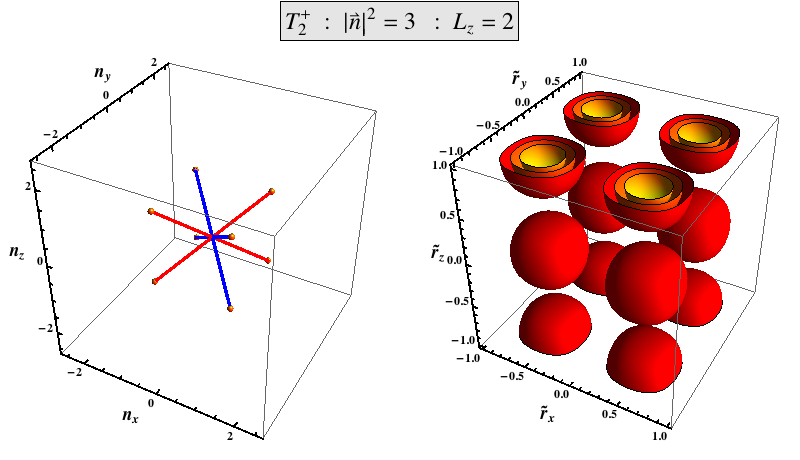}}\\
\mbox{\includegraphics[width=.77\textwidth,angle=0]{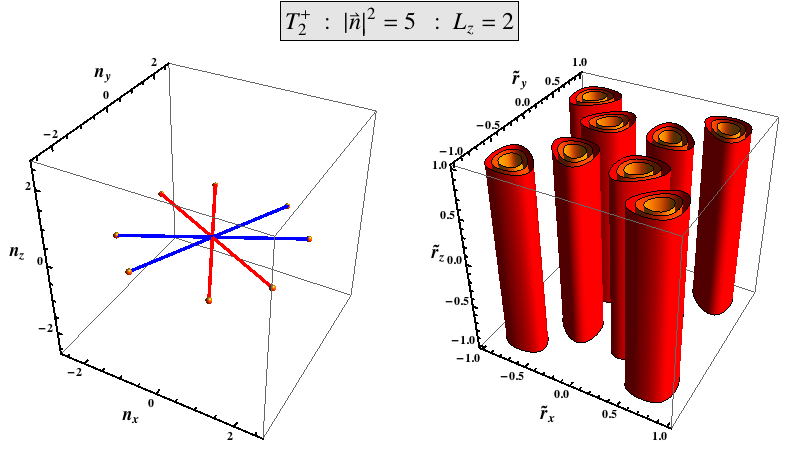}}
\caption{The momentum-space representations (left) and position-space
  representations (right) of two-body relative states in the $T_2^+$
  representation with $L_z=2$ for $|{\bf n}|^2$=2, 3, and 5 shells.
\label{fig:T2+ figures}}
\end{figure}
\begin{figure}
\centering
\mbox{\includegraphics[width=.77\textwidth,angle=0]{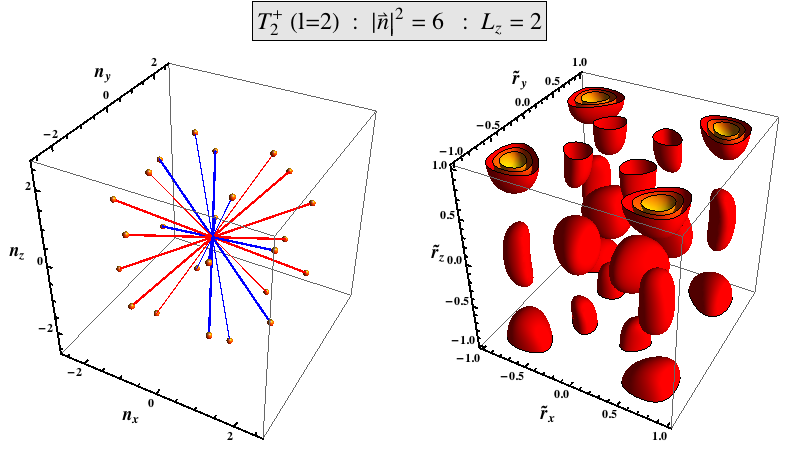}}\\
\mbox{\includegraphics[width=.77\textwidth,angle=0]{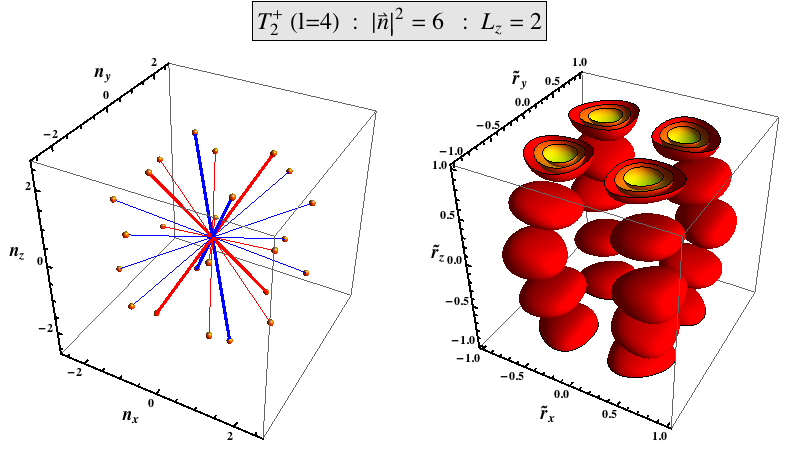}}
\caption{The momentum-space representations (left) and position-space
  representations (right) of two-body relative states in the $T_2^+$
  representation with $L_z=2$ for the $|{\bf n}|^2$=6-shells.
\label{fig:T2+ figures 2}}
\end{figure}
The graphical representations  of the
sources and sinks that generate the $T_2^+$ irrep for the low-lying  $|{\bf
  n}|^2$-shells are shown in 
fig.~\ref{fig:T2+ figures} and fig.~\ref{fig:T2+ figures 2},
and  the momentum-space structures for $L_z=2$ and $L_z=1$ are given 
explicitly in table~\ref{tab:T2pSOURCESSINKSLz2} and 
table~\ref{tab:T2pSOURCESSINKSLz1v2}, respectively.
The structures of the $L_z=3$ sources and sinks are related to those 
with $L_z=1$ by complex conjugation of the coefficients.
\begin{table}
\begin{center}
\begin{minipage}[!ht]{16.5 cm}
\caption{
The momentum-space structure of $T_2^+$, $L_z=2$ sources and sinks for $|{\bf n}|^2$=2, 3, and 5.
These are shown graphically in figs.~\ref{fig:T2+ figures}.
}
\label{tab:T2pSOURCESSINKSLz2}
\end{minipage}
\begin{small}
\begin{tabular}{|c||c||c|  }
\hline
\hline
$|{\bf n}|^2$=2 & $|{\bf n}|^2$=3 &  $|{\bf n}|^2$=5 \\ \hline

$\begin{array}{cc}
 \text{$|$(1,1,0) , +1$\rangle $} & -\frac{1}{\sqrt{2}} \\
 \text{$|$(1,-1,0) , +1$\rangle $} & \frac{1}{\sqrt{2}}
\end{array}$ 
&
$\begin{array}{cc}
 \text{$|$(1,1,1) , +1$\rangle $} & \frac{1}{2} \\
 \text{$|$(1,1,-1) , +1$\rangle $} & \frac{1}{2} \\
 \text{$|$(1,-1,1) , +1$\rangle $} & -\frac{1}{2} \\
 \text{$|$(1,-1,-1) , +1$\rangle $} & -\frac{1}{2}
\end{array}$ 
& 
$\begin{array}{cc}
 \text{$|$(2,1,0) , +1$\rangle $} & -\frac{1}{2} \\
 \text{$|$(2,-1,0) , +1$\rangle $} & \frac{1}{2} \\
 \text{$|$(1,2,0) , +1$\rangle $} & -\frac{1}{2} \\
 \text{$|$(1,-2,0) , +1$\rangle $} & \frac{1}{2}
\end{array}$ \\ \hline \hline
\end{tabular}
\end{small}
\begin{minipage}[t]{16.5 cm}
\vskip 0.5cm
\noindent
\end{minipage}
\end{center}
\end{table}     

\begin{table}
\begin{center}
\begin{minipage}[!ht]{16.5 cm}
\caption{
The momentum-space structure of $T_2^+$, $L_z=2$ sources and sinks for $|{\bf n}|^2$=6.
These are shown graphically in figs.~\ref{fig:T2+ figures 2}.
}
\label{tab:T2pSOURCESSINKSLz0v2}
\end{minipage}
\begin{small}
\begin{tabular}{|c||c|  }
\hline
\hline
$|{\bf n}|^2$=6$_{(l=2)}$ & $|{\bf n}|^2$=6$_{(l=4)}$  \\ \hline

$\begin{array}{cc|cc}
 \text{$|$(2,1,1) , +1$\rangle $} & \frac{1}{3} & \text{$|$(2,1,-1) , +1$\rangle $} & \frac{1}{3} \\
 \text{$|$(2,-1,1) , +1$\rangle $} & -\frac{1}{3} & \text{$|$(2,-1,-1) , +1$\rangle $} & -\frac{1}{3} \\
 \text{$|$(1,2,1) , +1$\rangle $} & \frac{1}{3} & \text{$|$(1,2,-1) , +1$\rangle $} & \frac{1}{3} \\
 \text{$|$(1,1,2) , +1$\rangle $} & \frac{1}{6} & \text{$|$(1,1,-2) , +1$\rangle $} & \frac{1}{6} \\
 \text{$|$(1,-1,2) , +1$\rangle $} & -\frac{1}{6} & \text{$|$(1,-1,-2) , +1$\rangle $} & -\frac{1}{6} \\
 \text{$|$(1,-2,1) , +1$\rangle $} & -\frac{1}{3} & \text{$|$(1,-2,-1) , +1$\rangle $} & -\frac{1}{3}
\end{array}$ 
& 
$\begin{array}{cc|cc}
 \text{$|$(2,1,1) , +1$\rangle $} & -\frac{1}{6 \sqrt{2}} & \text{$|$(2,1,-1) , +1$\rangle $} & -\frac{1}{6 \sqrt{2}} \\
 \text{$|$(2,-1,1) , +1$\rangle $} & \frac{1}{6 \sqrt{2}} & \text{$|$(2,-1,-1) , +1$\rangle $} & \frac{1}{6 \sqrt{2}} \\
 \text{$|$(1,2,1) , +1$\rangle $} & -\frac{1}{6 \sqrt{2}} & \text{$|$(1,2,-1) , +1$\rangle $} & -\frac{1}{6 \sqrt{2}} \\
 \text{$|$(1,1,2) , +1$\rangle $} & \frac{\sqrt{2}}{3} & \text{$|$(1,1,-2) , +1$\rangle $} & \frac{\sqrt{2}}{3} \\
 \text{$|$(1,-1,2) , +1$\rangle $} & -\frac{\sqrt{2}}{3} & \text{$|$(1,-1,-2) , +1$\rangle $} & -\frac{\sqrt{2}}{3} \\
 \text{$|$(1,-2,1) , +1$\rangle $} & \frac{1}{6 \sqrt{2}} & \text{$|$(1,-2,-1) , +1$\rangle $} & \frac{1}{6 \sqrt{2}}
\end{array}$ \\ \hline \hline
\end{tabular}
\end{small}
\begin{minipage}[t]{16.5 cm}
\vskip 0.5cm
\noindent
\end{minipage}
\end{center}
\end{table}     
\begin{table}
\begin{center}
\begin{minipage}[!ht]{16.5 cm}
\caption{
The momentum-space structure of $T_2^+$, $L_z=1$ sources and sinks for $|{\bf n}|^2$=2, 3, and 5.
}
\label{tab:T2pSOURCESSINKSLz1}
\end{minipage}
\begin{small}
\begin{tabular}{|c||c||c|  }
\hline
\hline
$|{\bf n}|^2$=2 & $|{\bf n}|^2$=3 &  $|{\bf n}|^2$=5 \\ \hline

$\begin{array}{cc}
 \text{$|$(1,0,1) , +1$\rangle $} & -\frac{1}{2} \\
 \text{$|$(1,0,-1) , +1$\rangle $} & \frac{1}{2} \\
 \text{$|$(0,1,1) , +1$\rangle $} & -\frac{i}{2} \\
 \text{$|$(0,1,-1) , +1$\rangle $} & \frac{i}{2}
\end{array}$ 
&
$\begin{array}{cc}
 \text{$|$(1,1,1) , +1$\rangle $} & -\frac{1}{2} \\
 \text{$|$(1,1,-1) , +1$\rangle $} & \frac{1}{2} \\
 \text{$|$(1,-1,1) , +1$\rangle $} & \frac{i}{2} \\
 \text{$|$(1,-1,-1) , +1$\rangle $} & -\frac{i}{2}
\end{array}$ 
& 
$\begin{array}{cc|cc}
 \text{$|$(2,0,1) , +1$\rangle $} & -\frac{1}{2 \sqrt{2}} & \text{$|$(2,0,-1) , +1$\rangle $} & \frac{1}{2 \sqrt{2}} \\
 \text{$|$(1,0,2) , +1$\rangle $} & -\frac{1}{2 \sqrt{2}} & \text{$|$(1,0,-2) , +1$\rangle $} & \frac{1}{2 \sqrt{2}} \\
 \text{$|$(0,2,1) , +1$\rangle $} & -\frac{i}{2 \sqrt{2}} & \text{$|$(0,2,-1) , +1$\rangle $} & \frac{i}{2 \sqrt{2}} \\
 \text{$|$(0,1,2) , +1$\rangle $} & -\frac{i}{2 \sqrt{2}} & \text{$|$(0,1,-2) , +1$\rangle $} & \frac{i}{2 \sqrt{2}}
\end{array}$ \\ \hline \hline
\end{tabular}
\end{small}
\begin{minipage}[t]{16.5 cm}
\vskip 0.5cm
\noindent
\end{minipage}
\end{center}
\end{table}     

\begin{table}
\begin{center}
\begin{minipage}[!ht]{16.5 cm}
\caption{
The momentum-space structure of $T_2^+$, $L_z=1$ sources and sinks for $|{\bf n}|^2$=6.
}
\label{tab:T2pSOURCESSINKSLz1v2}
\end{minipage}
\begin{small}
\begin{tabular}{|c||c|  }
\hline
\hline
$|{\bf n}|^2$=6$_{(l=2)}$ & $|{\bf n}|^2$=6$_{(l=4)}$  \\ \hline

$\begin{array}{cc|cc}
 \text{$|$(2,1,1) , +1$\rangle $} & \frac{\sqrt{\frac{5}{2}}}{6} & \text{$|$(2,1,-1) , +1$\rangle $} & -\frac{\sqrt{\frac{5}{2}}}{6} \\
 \text{$|$(2,-1,1) , +1$\rangle $} & \frac{\frac{1}{2}-\frac{2 i}{3}}{\sqrt{10}} & \text{$|$(2,-1,-1) , +1$\rangle $} & -\frac{\frac{1}{2}-\frac{2 i}{3}}{\sqrt{10}} \\
 \text{$|$(1,2,1) , +1$\rangle $} & \frac{\frac{2}{3}+\frac{i}{2}}{\sqrt{10}} & \text{$|$(1,2,-1) , +1$\rangle $} & -\frac{\frac{2}{3}+\frac{i}{2}}{\sqrt{10}} \\
 \text{$|$(1,1,2) , +1$\rangle $} & \frac{1}{3} \sqrt{\frac{4}{5}+\frac{3 i}{5}} & \text{$|$(1,1,-2) , +1$\rangle $} & -\frac{1+\frac{i}{3}}{\sqrt{10}} \\
 \text{$|$(1,-1,2) , +1$\rangle $} & \frac{\frac{1}{3}-i}{\sqrt{10}} & \text{$|$(1,-1,-2) , +1$\rangle $} & -\frac{\frac{1}{3}-i}{\sqrt{10}} \\
 \text{$|$(1,-2,1) , +1$\rangle $} & -\frac{1}{6} i \sqrt{\frac{5}{2}} & \text{$|$(1,-2,-1) , +1$\rangle $} & \frac{1}{6} i \sqrt{\frac{5}{2}}
\end{array}$ 
& 
$\begin{array}{cc|cc}
 \text{$|$(2,1,1) , +1$\rangle $} & \frac{\frac{23}{12}-2 i}{\sqrt{65}} & \text{$|$(2,1,-1) , +1$\rangle $} & -\frac{\frac{23}{12}-2 i}{\sqrt{65}} \\
 \text{$|$(2,-1,1) , +1$\rangle $} & -\frac{\frac{3}{4}-\frac{8 i}{3}}{\sqrt{65}} & \text{$|$(2,-1,-1) , +1$\rangle $} & \frac{\frac{3}{4}-\frac{8 i}{3}}{\sqrt{65}} \\
 \text{$|$(1,2,1) , +1$\rangle $} & -\frac{\frac{8}{3}+\frac{3 i}{4}}{\sqrt{65}} & \text{$|$(1,2,-1) , +1$\rangle $} & \frac{\frac{8}{3}+\frac{3 i}{4}}{\sqrt{65}} \\
 \text{$|$(1,1,2) , +1$\rangle $} & \frac{\frac{1}{4}+\frac{11 i}{12}}{\sqrt{65}} & \text{$|$(1,1,-2) , +1$\rangle $} & -\frac{\frac{1}{4}+\frac{11 i}{12}}{\sqrt{65}} \\
 \text{$|$(1,-1,2) , +1$\rangle $} & \frac{\frac{11}{12}-\frac{i}{4}}{\sqrt{65}} & \text{$|$(1,-1,-2) , +1$\rangle $} & -\frac{\frac{11}{12}-\frac{i}{4}}{\sqrt{65}} \\
 \text{$|$(1,-2,1) , +1$\rangle $} & -\frac{2+\frac{23 i}{12}}{\sqrt{65}} & \text{$|$(1,-2,-1) , +1$\rangle $} & \frac{2+\frac{23 i}{12}}{\sqrt{65}}
\end{array}$ \\ \hline \hline
\end{tabular}
\end{small}
\begin{minipage}[t]{16.5 cm}
\vskip 0.5cm
\noindent
\end{minipage}
\end{center}
\end{table}

The $l=2$ phase-shift was calculated 
from the energies of states 
in both the $E^+$ and $T_2^+$ irreps 
in recent work by Dudek {\it et al}~\cite{Dudek:2010ew}.
Two states in each irrep were calculated below the $2\pi\rightarrow 4\pi$ inelastic threshold
at the pion mass of the calculation.  
The contamination in the extraction of $\delta_2$ from the
higher partial-waves was estimated to be small.

\subsection{Negative Parity Systems}

The analysis of the odd-parity energy-levels, and their associated sources and
sinks,  parallels that of the even-parity states. There are five negative parity irreps of the cubic group, $A_1^-$, $A_2^-$, $E^-$,
$T_1^-$, and $T_2^-$ with dimensions $1,1,2,3$, and $3$ respectively. The energy-eigenvalues,
sources and sinks for the negative-parity states are presented in the following
sections:~\ref{irrep:A2m}, \ref{irrep:Em}, \ref{irrep:T1m}, \ref{irrep:T2m} and \ref{irrep:A1m}.
As discussed previously, the $A_1^-$ irrep first appears 
relatively high in the spectrum, 
in the 
$|{\bf n}|^2=14$ shell, and is sensitive to the $l=9$ and higher partial-waves.

\subsubsection{$A_2^-$ Representation\label{irrep:A2m}}
The energy-eigenvalues of states transforming in the $A_2^-$ irrep 
($L_z=2$)
receive
contributions only from interactions in the $l=3$ partial-wave for $l\le 6$, as
presented in table~\ref{tab:Lcontributetab}.
\begin{figure}
\centering
\includegraphics[height=\textwidth,angle=-90]{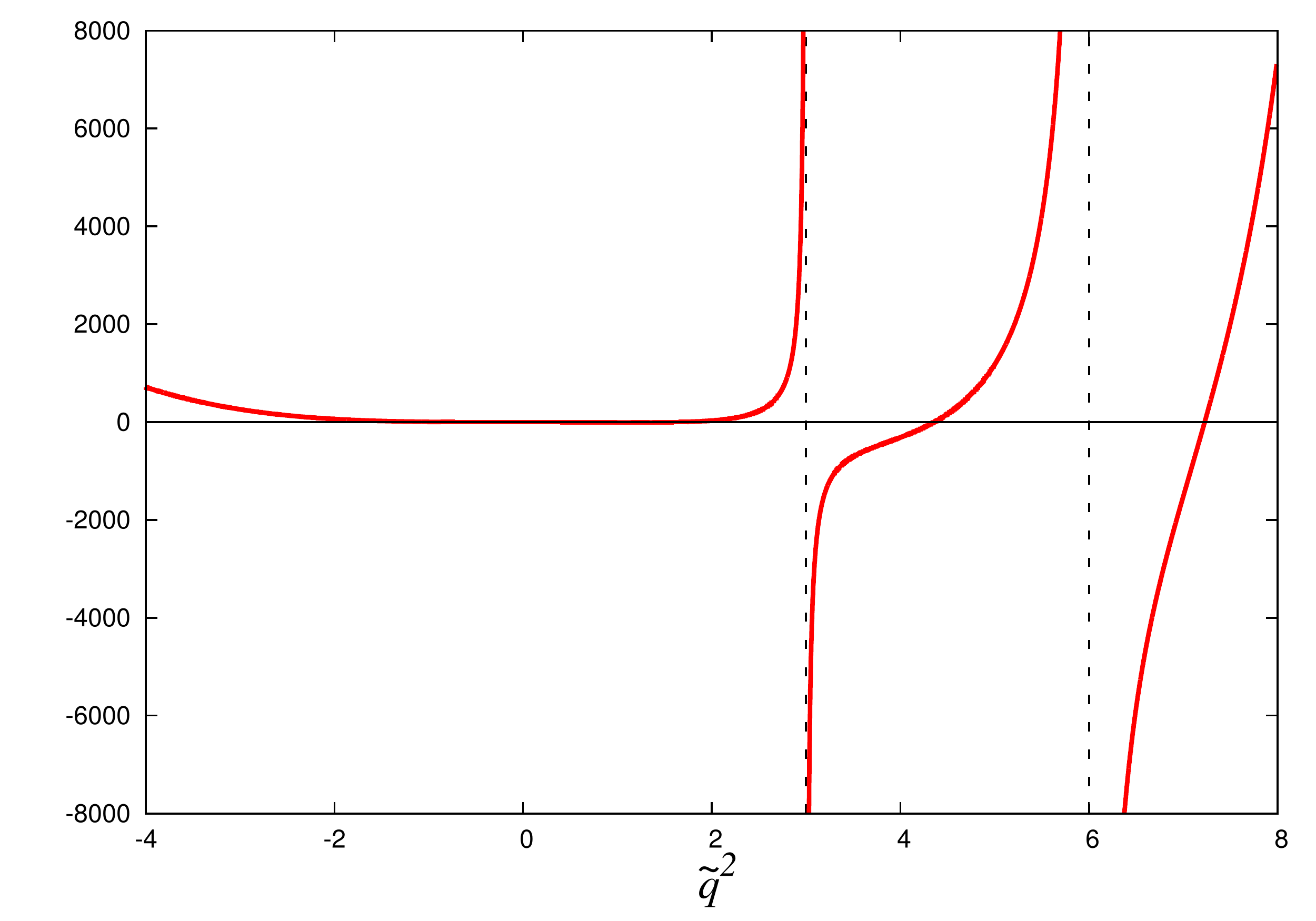}
\caption{The function $\mathcal{X}_{A_2}^-$, as defined in eq.~(\ref{eqn:A2- l=3}), as a
  function of $\tilde q^2$.  The vertical dashed lines denote the position of
  the poles of the function corresponding to the non-interacting 
energy-eigenvalues.
\label{fig:A2- l=3}}
\end{figure}
The $A_2^-$ state associated with the $\overline{F}^{(FV)}_{3;3}$ block is (in the
$|l,m\rangle$ basis) 
\begin{displaymath}
|A_2^-, 2; 3; 1\rangle
\ =\ \frac{1}{\sqrt{2}}|3,2\rangle \ -\ \frac{1}{\sqrt{2}}|3,-2\rangle
\ \ \ \ ,
\end{displaymath}
and the solutions to eq.~(\ref{eq:evals}) from this irrep result from
\begin{equation}
\label{eqn:A2- l=3}
\begin{split}
q^7\text{cot}\delta_3 \ & \ =\ 
\left(\frac{2\pi}{L}\right)^7\frac{1}{\pi^{3/2}}
\left(\tilde{q}^6\mathcal{Z}_{0,0}\left(1;\tilde q^2\right)-\frac{12}{11}\tilde q^2\mathcal{Z}_{4,0}\left(1;\tilde q^2\right)+\frac{80}{11\sqrt{13}}
\mathcal{Z}_{6,0}\left(1;\tilde q^2\right)\right)\\
\ &\ \equiv\ \left(\frac{2\pi}{L}\right)^7\frac{1}{\pi^{3/2}}\  
\mathcal{X}_{A_2}^-\left(\tilde q^2\right) 
\ \ \ ,
\end{split}
\end{equation}
where the function $\mathcal{X}_{A_2}^-(\tilde q^2)$ is shown in fig.~\ref{fig:A2- l=3}.
\begin{figure}
\centering
\mbox{\includegraphics[width=.77\textwidth,angle=0]{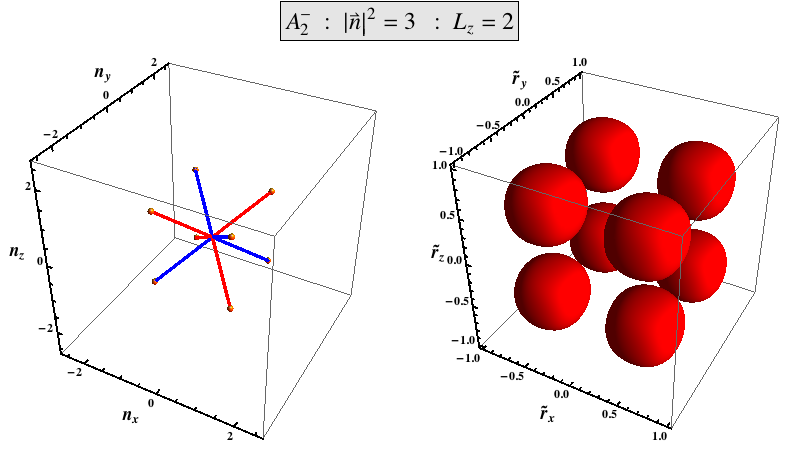}}\\
\mbox{\includegraphics[width=.77\textwidth,angle=0]{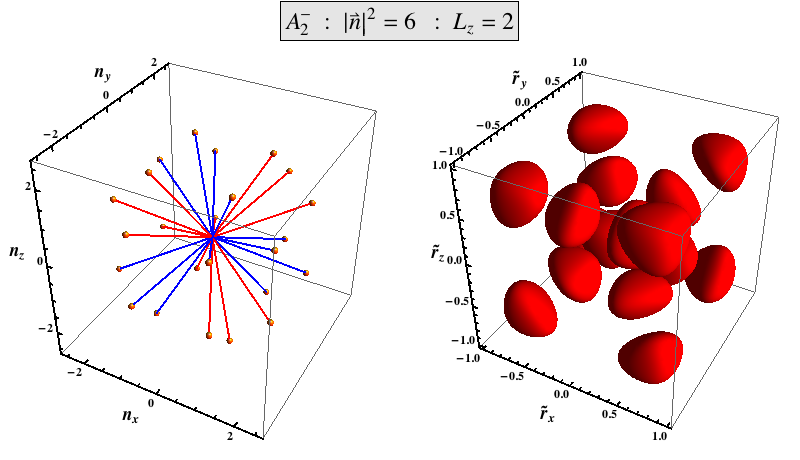}}
\caption{The momentum-space representations (left) and position-space
  representations 
(right) of two-body relative states in the $A_2^-$ representation for the lowest-lying
$|{\bf n}|^2$-shells.\label{fig:A2- figures}
}
\end{figure}
The graphical representations  of the
sources and sinks that generate the $A_2^-$ irrep in the low-lying  
$|{\bf  n}|^2$-shells 
($|{\bf  n}|^2=3$ and $|{\bf  n}|^2=6$)
are shown in  fig.~\ref{fig:A2- figures},
and  the momentum-space structures are given 
explicitly  in table~\ref{tab:A2mSOURCESSINKS}.
\begin{table}
\begin{center}
\begin{minipage}[!ht]{16.5 cm}
\caption{
The momentum-space structure of $A_2^-$ sources and sinks.
These are shown graphically in  fig.~\ref{fig:A2- figures}.
}
\label{tab:A2mSOURCESSINKS}
\end{minipage}
\begin{small}
\begin{tabular}{|c||c|  }
\hline
\hline
$|{\bf n}|^2$=3 & $|{\bf n}|^2$=6  \\ \hline

$\begin{array}{cc}
 \text{$|$(1,1,1) , -1$\rangle $} & \frac{1}{2} \\
 \text{$|$(1,1,-1) , -1$\rangle $} & -\frac{1}{2} \\
 \text{$|$(1,-1,1) , -1$\rangle $} & -\frac{1}{2} \\
 \text{$|$(1,-1,-1) , -1$\rangle $} & \frac{1}{2}
\end{array}$ 
& 
$\begin{array}{cc|cc}
 \text{$|$(2,1,1) , -1$\rangle $} & \frac{1}{2 \sqrt{3}} &\text{$|$(2,1,-1) , -1$\rangle $} & -\frac{1}{2 \sqrt{3}} \\
 \text{$|$(2,-1,1) , -1$\rangle $} & -\frac{1}{2 \sqrt{3}} &\text{$|$(2,-1,-1) , -1$\rangle $} & \frac{1}{2 \sqrt{3}} \\
 \text{$|$(1,2,1) , -1$\rangle $} & \frac{1}{2 \sqrt{3}} & \text{$|$(1,2,-1) , -1$\rangle $} & -\frac{1}{2 \sqrt{3}} \\
 \text{$|$(1,1,2) , -1$\rangle $} & \frac{1}{2 \sqrt{3}} &\text{$|$(1,1,-2) , -1$\rangle $} & -\frac{1}{2 \sqrt{3}} \\
 \text{$|$(1,-1,2) , -1$\rangle $} & -\frac{1}{2 \sqrt{3}} & \text{$|$(1,-1,-2) , -1$\rangle $} & \frac{1}{2 \sqrt{3}} \\
 \text{$|$(1,-2,1) , -1$\rangle $} & -\frac{1}{2 \sqrt{3}} & \text{$|$(1,-2,-1) , -1$\rangle $} & \frac{1}{2 \sqrt{3}}
\end{array}$ \\ \hline \hline
\end{tabular}
\end{small}
\begin{minipage}[t]{16.5 cm}
\vskip 0.5cm
\noindent
\end{minipage}
\end{center}
\end{table}

The $A_2^-$ irrep first appears in the $|{\bf n}|^2=3$-shell and $l=3$
is the lowest contributing partial-wave.  
LQCD calculations of correlation functions from sources and sinks transforming
as $A_2^-$ will provide determinations of $\delta_3$ with contamination from
partial-waves with $l\ge 7$, i.e. the energy of the $A_2^-$ states receive
contributions from $l=3,7,...$.
This is in contrast to states in the $T_2^-$ irrep, which will be considered
subsequently, whose energy-eigenvalues receive contributions from partial-waves with
$l=3,5,...$.
This suggests that the $A_2^-$ irrep is optimal for determining $\delta_3$.

\subsubsection{$E^-$ Representation\label{irrep:Em}}
The energy-eigenvalues of $E^-$ states receive
contributions only from interactions in the $l=5$ partial-wave for $l\le 6$, as
presented in table~\ref{tab:Lcontributetab}.
As the $E^-$ irrep is two-dimensional, the contribution to the determinant in 
eq.~(\ref{eq:evals}) results from a $2\times 2$ matrix for $l \le 6$, which collapses
down to a one-dimensional factor  as the $L_z=0$ and $L_z=2$ states are degenerate.
\begin{figure}
\centering
\includegraphics[height=\textwidth,angle=-90]{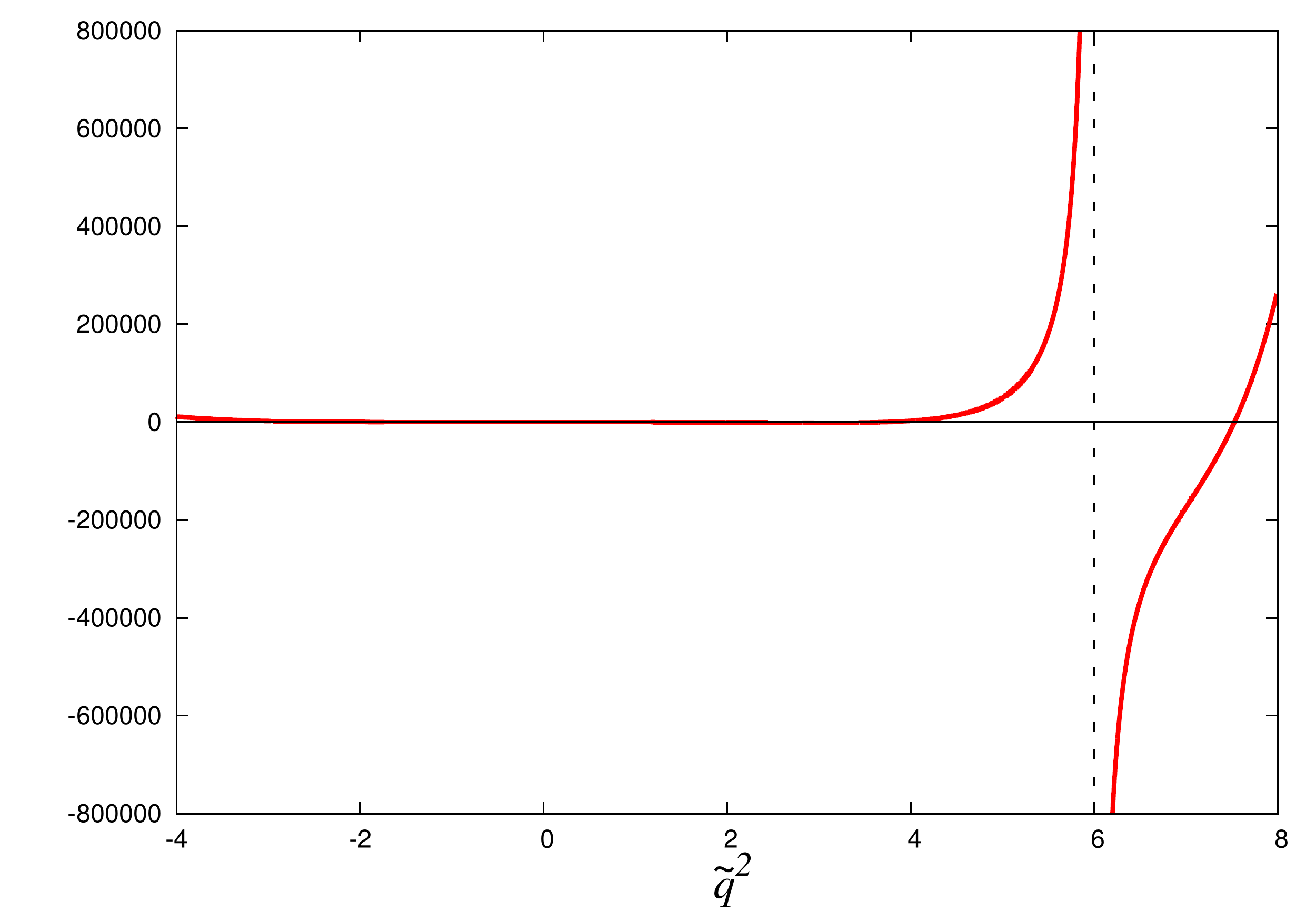}
\caption{The function $\mathcal{X}_{E}^-$, as defined in eq.~(\ref{eqn:E- l=5}), as a function
  of $\tilde q^2$.  The vertical dashed lines denote the position of the poles
  of the function corresponding to the non-interacting energy-eigenvalues. 
\label{fig:E- l=5}}
\end{figure}
The $E^-$  $L_z=0$ state associated with the $\overline{F}^{(FV)}_{5,5}$ block 
is
\begin{displaymath}
|E^-, 0; 5; 1\rangle\ =\ 
\frac{1}{\sqrt{2}}|5,4 \rangle\ -\ \frac{1}{\sqrt{2}}|5,-4 \rangle
\ \ \ \ .
\end{displaymath}
The solution  to eq.~(\ref{eq:evals})
from  the $E^-$ irrep results from
\begin{eqnarray}
\label{eqn:E- l=5}
&& q^{11}\text{cot}\delta_5 \ =\ 
\left(\frac{2\pi}{L}\right)^{11}\frac{1}{\pi^{3/2}}\ \times\ 
\nonumber\\
&&\left(  \tilde q^{10}\mathcal{Z}_{0,0}\left(1;\tilde q^2\right)- \frac{6\tilde q^6\mathcal{Z}_{4,0}\left(1;\tilde q^2\right)}{13}
+\frac{32\tilde q^4\mathcal{Z}_{6,0}\left(1;\tilde q^2\right)}{17\sqrt{13}}
-\frac{672\tilde q^2\mathcal{Z}_{8,0}\left(1;\tilde q^2\right)}{247\sqrt{17}}
+\frac{1152\sqrt{21}\mathcal{Z}_{10,0}\left(1;\tilde q^2\right)}{4199}
\right)
\nonumber\\
& & \qquad  \qquad \equiv \ 
\left(\frac{2\pi}{L}\right)^{11}\frac{1}{\pi^{3/2}}\mathcal{X}_{E}^-\left(\tilde
  q^2\right)
\ \ \ .
\end{eqnarray}
where  the function $\mathcal{X}_{E}^-(\tilde q^2)$ is shown in fig.~\ref{fig:E- l=5}.
\begin{figure}
\centering
\mbox{\includegraphics[width=.77\textwidth,angle=0]{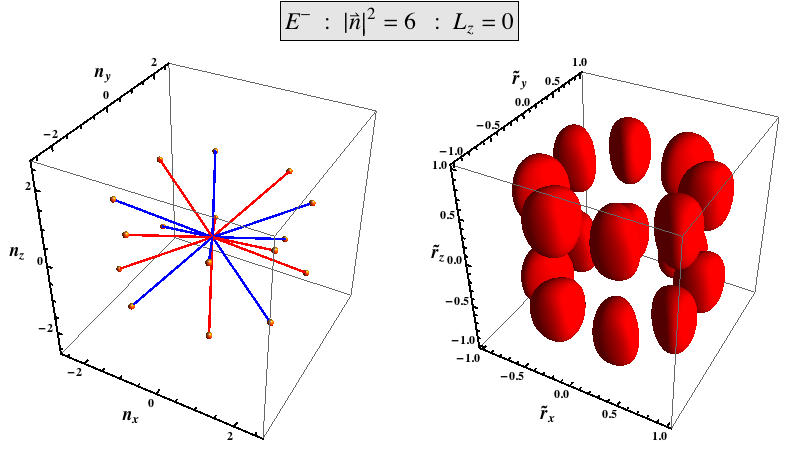}}
\caption{The momentum-space representations (left) and position-space
  representations 
(right) of two-body relative states in the $E^-$ representation with $L_z=0$ for the 
$|{\bf n}|^2=6$ shell.
\label{fig:E- figures}
}
\end{figure}
The graphical representations  of the
source and sink that generate the $E^-$ irrep in the
$|{\bf n}|^2=6$-shell
are shown in fig.~\ref{fig:E- figures},
and  the momentum-space structure is given 
explicitly  in table~\ref{tab:EmSOURCESSINKS}.
\begin{table}
\begin{center}
\begin{minipage}[!ht]{16.5 cm}
\caption{
The momentum-space structure of $E^-$, $L_z=0$, 2 sources and sinks.
The $L_z=0$ case is shown graphically in  fig.~\ref{fig:E- figures}.
}
\label{tab:EmSOURCESSINKS}
\end{minipage}
\begin{tabular}{|c||c|  }
\hline
\hline
$|{\bf n}|^2$=6 $L_z=0$ &$|{\bf n}|^2$=6 $L_z=2$  \\ \hline 
$\begin{array}{cc|cc}
 \text{$|$(2,1,1) , -1$\rangle $} & -\frac{1}{2 \sqrt{2}} & \text{$|$(2,1,-1) , -1$\rangle $} & \frac{1}{2 \sqrt{2}} \\
 \text{$|$(2,-1,1) , -1$\rangle $} & \frac{1}{2 \sqrt{2}} & \text{$|$(2,-1,-1) , -1$\rangle $} & -\frac{1}{2 \sqrt{2}} \\
 \text{$|$(1,2,1) , -1$\rangle $} & \frac{1}{2 \sqrt{2}} & \text{$|$(1,2,-1) , -1$\rangle $} & -\frac{1}{2 \sqrt{2}} \\
 \text{$|$(1,-2,1) , -1$\rangle $} & -\frac{1}{2 \sqrt{2}} & \text{$|$(1,-2,-1) , -1$\rangle $} & \frac{1}{2 \sqrt{2}}
\end{array}$
&
$\begin{array}{cc|cc}
 \text{$|$(2,1,1) , -1$\rangle $} & -\frac{1}{2 \sqrt{6}} & \text{$|$(2,1,-1) , -1$\rangle $} & \frac{1}{2 \sqrt{6}} \\
 \text{$|$(2,-1,1) , -1$\rangle $} & \frac{1}{2 \sqrt{6}} & \text{$|$(2,-1,-1) , -1$\rangle $} & -\frac{1}{2 \sqrt{6}} \\
 \text{$|$(1,2,1) , -1$\rangle $} & -\frac{1}{2 \sqrt{6}} & \text{$|$(1,2,-1) , -1$\rangle $} & \frac{1}{2 \sqrt{6}} \\
 \text{$|$(1,1,2) , -1$\rangle $} & \frac{1}{\sqrt{6}} & \text{$|$(1,1,-2) , -1$\rangle $} & -\frac{1}{\sqrt{6}} \\
 \text{$|$(1,-1,2) , -1$\rangle $} & -\frac{1}{\sqrt{6}} & \text{$|$(1,-1,-2) , -1$\rangle $} & \frac{1}{\sqrt{6}} \\
 \text{$|$(1,-2,1) , -1$\rangle $} & \frac{1}{2 \sqrt{6}} & \text{$|$(1,-2,-1) , -1$\rangle $} & -\frac{1}{2 \sqrt{6}}
\end{array}$\\
\hline
\hline
\end{tabular}
\begin{minipage}[t]{16.5 cm}
\vskip 0.5cm
\noindent
\end{minipage}
\end{center}
\end{table}

The $E^-$ irrep first appears in the $|{\bf n}|^2=6$-shell and $l=5$
is the lowest contributing partial-wave.  
LQCD calculations of correlation functions from sources and sinks transforming
as $E^-$ will provide determinations of $\delta_5$ with contamination from
partial-waves with $l\ge 7$, i.e. the energy of the $E^-$ states receive
contributions from $l=5,7,...$.
The LQCD calculations will need to be performed in relatively large volumes, as
we discuss later, in order for 
the $|{\bf n}|^2=6$ shell to lie below the  inelastic threshold.

\subsubsection{$T_1^-$ Representation\label{irrep:T1m}}
The energy-eigenvalues of states transforming in the $T_1^-$ irrep receive
contributions from interactions in the $l=1,3,5,...$ partial-waves, as
presented in table~\ref{tab:Lcontributetab}.
As the $T_1^-$ irrep is three-dimensional, the contribution to
eq.~(\ref{eq:evals}) is 
the determinant of a 
$12\times 12$ matrix for $l \le 6$ (there are two
$T_1^-$'s in the decomposition of $l=5$), 
which collapses
down to 
the determinant of a 
$4\times 4$ matrix as the $L_z=0$, $L_z=1$ and $L_z=3$ states are degenerate.
The $T_1^-$  $L_z=0$ states associated with the 
$\overline{F}^{(FV)}_{1;1}$, $\overline{F}^{(FV)}_{3;3}$, and $\overline{F}^{(FV)}_{5;5}$ blocks are 
\begin{displaymath}
\begin{split}
|T_1^-, 0; 1; 1\rangle \ &\ =\ |1,0\rangle \\
|T_1^-, 0; 3; 1\rangle \ &\ =\ |3,0\rangle \\
|T_1^-, 0; 5; 1\rangle \ &\ =\ |5,0\rangle \\
|T_1^-, 0; 5; 2\rangle \ &\ =\ \frac{1}{\sqrt{2}}\left(\ |5,4\rangle\ +\ |5,-4\rangle\ \right)
\ \ \ .
\end{split}
\end{displaymath}

\begin{figure}
\centering
\includegraphics[height=\textwidth,angle=-90]{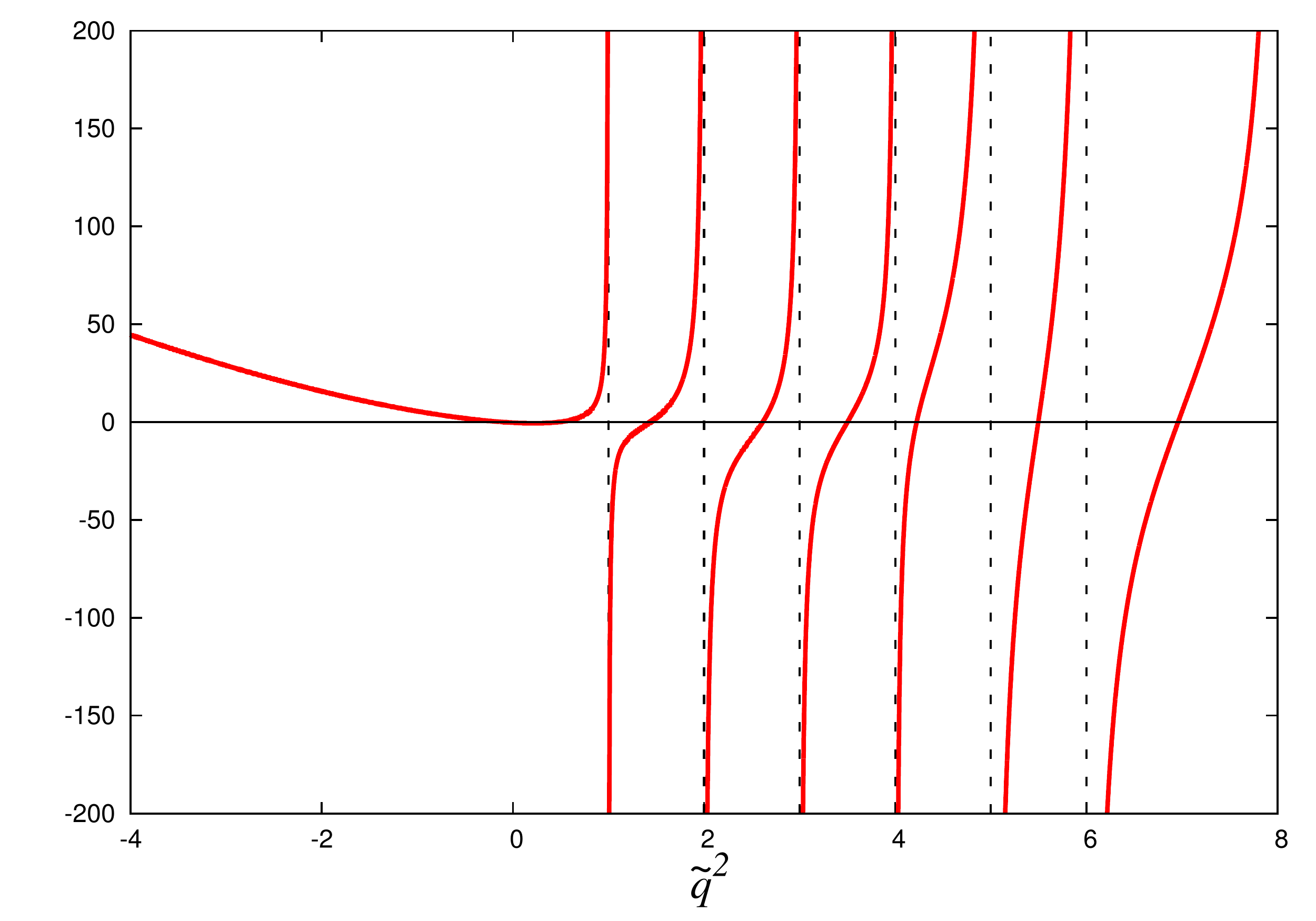}
\caption{The function $\mathcal{X}_{T_1}^-$, as defined in
  eq.~(\protect\ref{eqn:T1- l=1}),
as a function  of $\tilde q^2$.  The vertical dashed lines denote the position of the poles
  of the function corresponding to the non-interacting energy-eigenvalues.
\label{fig:T1- l=1}}
\end{figure}

With these four basis states, the $T_1^-$ contribution to  eq.~(\ref{eq:evals})
becomes
\begin{equation}
\label{eqn:T1- det}
\text{det}\left[
\begin{pmatrix}
\text{cot}\delta_1 & 0 & 0 & 0\\
0 & \text{cot}\delta_3 & 0 & 0\\
0 & 0 & \text{cot}\delta_5 & 0\\
0 & 0 & 0 & \text{cot}\delta_5
\end{pmatrix}
\ -\ 
\begin{pmatrix} 
\overline{F}_{1;1}^{(FV, T_1^-)} & \overline{F}_{1;3}^{(FV, T_1^-)}  &\overline{F}_{1;5_1}^{(FV, T_1^-)} & \overline{F}_{1;5_2}^{(FV, T_1^-)}\\
\overline{F}_{3;1}^{(FV, T_1^-)} & \overline{F}_{3;3}^{(FV, T_1^-)} & \overline{F}_{3;5_1}^{(FV, T_1^-)} & \overline{F}_{3;5_2}^{(FV, T_1^-)}\\
\overline{F}_{5_1;1}^{(FV, T_1^-)} & \overline{F}_{5_1;3}^{(FV, T_1^-)} & \overline{F}_{5_1;5_1}^{(FV, T_1^-)}&  \overline{F}_{5_1,5_2}^{(FV, T_1^-)}\\
\overline{F}_{5_2;1}^{(FV, T_1^-)} & \overline{F}_{5_2;3}^{(FV, T_1^-)} &  \overline{F}_{5_2,5_1}^{(FV, T_1^-)} & \overline{F}_{5_2,5_2}^{(FV, T_1^-)}
\end{pmatrix}
\right]\ =\ 0
\ \ \ \ ,
\end{equation}
where
\begin{equation}
\begin{split}
\overline{F}_{1,1}^{(FV, T_1^-)}&=\frac{\mathcal{Z}_{0,0}\left(1;\tilde q^2\right)}{\pi ^{3/2} \tilde q}
\\
\overline{F}_{1,3}^{(FV, T_1^-)}&=\frac{4 \mathcal{Z}_{4,0}\left(1;\tilde q^2\right)}{\sqrt{21} \pi ^{3/2} \tilde q^5}
\\
\overline{F}_{1,5_1}^{(FV,T_1^-)}&=\frac{5 \mathcal{Z}_{4,0}\left(1;\tilde{q}^2\right)}{\sqrt{33} \pi ^{3/2} \tilde{q}^5}+\frac{6 \sqrt{\frac{3}{143}} \mathcal{Z}_{6,0}\left(1;\tilde{q}^2\right)}{\pi ^{3/2} \tilde{q}^7}
\\
\overline{F}_{1,5_2}^{(FV,T_1^-)}&=\frac{\sqrt{\frac{15}{77}} \mathcal{Z}_{4,0}\left(1;\tilde{q}^2\right)}{\pi ^{3/2} \tilde{q}^5}-\frac{2 \sqrt{\frac{105}{143}} \mathcal{Z}_{6,0}\left(1;\tilde{q}^2\right)}{\pi ^{3/2} \tilde{q}^7}
\\
\overline{F}_{3,3}^{(FV, T_1^-)}&=\frac{\mathcal{Z}_{0,0}\left(1;\tilde q^2\right)}{\pi
  ^{3/2} \tilde q}+\frac{100 \mathcal{Z}_{6,0}\left(1;\tilde q^2\right)}{33
  \sqrt{13} \pi ^{3/2} \tilde q^7}+\frac{6 \mathcal{Z}_{4,0}\left(1;\tilde
    q^2\right)}{11 \pi ^{3/2} \tilde q^5}
\\
\overline{F}_{3,5_1}^{(FV, T_1^-)}&=\frac{60 \mathcal{Z}_{4,0}\left(1;\tilde{q}^2\right)}{13 \sqrt{77} \pi ^{3/2} \tilde{q}^5}+\frac{7 \sqrt{\frac{7}{143}} \mathcal{Z}_{6,0}\left(1;\tilde{q}^2\right)}{3 \pi ^{3/2} \tilde{q}^7}+\frac{56 \sqrt{\frac{7}{187}}
   \mathcal{Z}_{8,0}\left(1;\tilde{q}^2\right)}{13 \pi ^{3/2} \tilde{q}^9}
   \\
\overline{F}_{3,5_2}^{(FV,T_1^-)}&=-\frac{12 \sqrt{\frac{5}{11}} \mathcal{Z}_{4,0}\left(1;\tilde{q}^2\right)}{13 \pi ^{3/2} \tilde{q}^5}+\frac{7 \sqrt{\frac{5}{143}} \mathcal{Z}_{6,0}\left(1;\tilde{q}^2\right)}{\pi ^{3/2} \tilde{q}^7}+\frac{56 \sqrt{\frac{5}{187}}
   \mathcal{Z}_{8,0}\left(1;\tilde{q}^2\right)}{39 \pi ^{3/2} \tilde{q}^9}
   \\
\overline{F}_{5_1,5_1}^{(FV, T_1^-)}&=\frac{\mathcal{Z}_{0,0}\left(1;\tilde{q}^2\right)}{\pi ^{3/2} \tilde{q}}+\frac{6 \mathcal{Z}_{4,0}\left(1;\tilde{q}^2\right)}{13 \pi ^{3/2} \tilde{q}^5}+\frac{80 \mathcal{Z}_{6,0}\left(1;\tilde{q}^2\right)}{51 \sqrt{13} \pi ^{3/2}
   \tilde{q}^7}+\frac{490 \mathcal{Z}_{8,0}\left(1;\tilde{q}^2\right)}{247 \sqrt{17} \pi ^{3/2} \tilde{q}^9}+\frac{756 \sqrt{21} \mathcal{Z}_{10,0}\left(1;\tilde{q}^2\right)}{4199 \pi ^{3/2} \tilde{q}^{11}}
      \\
\overline{F}_{5_1,5_2}^{(FV, T_1^-)}&=\frac{6 \sqrt{\frac{5}{7}} \mathcal{Z}_{4,0}\left(1;\tilde{q}^2\right)}{13 \pi ^{3/2} \tilde{q}^5}+\frac{8 \sqrt{\frac{35}{13}} \mathcal{Z}_{6,0}\left(1;\tilde{q}^2\right)}{17 \pi ^{3/2} \tilde{q}^7}-\frac{154 \sqrt{\frac{35}{17}}
   \mathcal{Z}_{8,0}\left(1;\tilde{q}^2\right)}{741 \pi ^{3/2} \tilde{q}^9}-\frac{2772 \sqrt{\frac{3}{5}} \mathcal{Z}_{10,0}\left(1;\tilde{q}^2\right)}{4199 \pi ^{3/2} \tilde{q}^{11}}
   \\
 \overline{F}_{5_2,5_2}^{(FV, T_1^-)}&  =\frac{\mathcal{Z}_{0,0}\left(1;\tilde{q}^2\right)}{\pi ^{3/2} \tilde{q}}-\frac{6 \mathcal{Z}_{4,0}\left(1;\tilde{q}^2\right)}{13 \pi ^{3/2} \tilde{q}^5}+\frac{32 \mathcal{Z}_{6,0}\left(1;\tilde{q}^2\right)}{17 \sqrt{13} \pi ^{3/2}
   \tilde{q}^7}+\frac{14 \sqrt{17} \mathcal{Z}_{8,0}\left(1;\tilde{q}^2\right)}{247 \pi ^{3/2} \tilde{q}^9}-\frac{84 \sqrt{21} \mathcal{Z}_{10,0}\left(1;\tilde{q}^2\right)}{323 \pi ^{3/2} \tilde{q}^{11}}
\ \ \ .
\end{split}
\end{equation}

In the limit of vanishing interactions in the  $l=5$ partial-wave, 
eq.~(\ref{eqn:T1- det}) collapses down to 
the determinant of 
a $2\times2$ matrix, which has solutions
\begin{multline}
\frac{\text{cot$\delta $}_1}{2}+\frac{\text{cot$\delta $}_3}{2}-\frac{\mathcal{Z}_{0,0}\left(1;\tilde{q}^2\right)}{\pi ^{3/2} \tilde{q}}-\frac{3 \mathcal{Z}_{4,0}\left(1;\tilde{q}^2\right)}{11 \pi ^{3/2} \tilde{q}^5}-\frac{50 \mathcal{Z}_{6,0}\left(1;\tilde{q}^2\right)}{33 \sqrt{13} \pi ^{3/2}
   \tilde{q}^7}
\\
   ={\bf \pm}
{1\over 2}
\sqrt{\left(\frac{6 \mathcal{Z}_{4,0}\left(1;\tilde{q}^2\right)}{11 \pi ^{3/2} \tilde{q}^5}+\frac{100 \mathcal{Z}_{6,0}\left(1;\tilde{q}^2\right)}{33 \sqrt{13} \pi ^{3/2} \tilde{q}^7}+\text{cot$\delta $}_1-\text{cot$\delta
   $}_3\right)^2+\frac{64 \mathcal{Z}_{4,0}\left(1;\tilde{q}^2\right)^2}{21 \pi
 ^3 \tilde{q}^{10}}}\ .
\label{eq:T1ml13}
\end{multline}
In the situation where $\tan\delta_3 << \tan\delta_1$, eq.~(\ref{eq:T1ml13})
can be perturbatively expanded to give the $l=1$ dominant solution
\begin{equation}
\label{eqn:T1- l=1}
\begin{split}
q^3\text{cot}\delta_1&=\left(\frac{2\pi}{L}\right)^3
\frac{1}{\pi^{3/2}}\tilde q^2\mathcal{Z}_{0,0}\left(1;\tilde q^2\right)
\ \ \equiv\ \ \left(\frac{2\pi}{L}\right)^3\frac{1}{\pi^{3/2}}
\mathcal{X}_{T_1}^-\left(\tilde q^2\right)
\ \ \ ,
\end{split}
\end{equation}
where the function $\mathcal{X}_{T_1}^-\left(\tilde q^2\right)$ is shown in
fig.~\ref{fig:T1- l=1}.
\begin{figure}
\centering
\mbox{\includegraphics[width=.77\textwidth,angle=0]{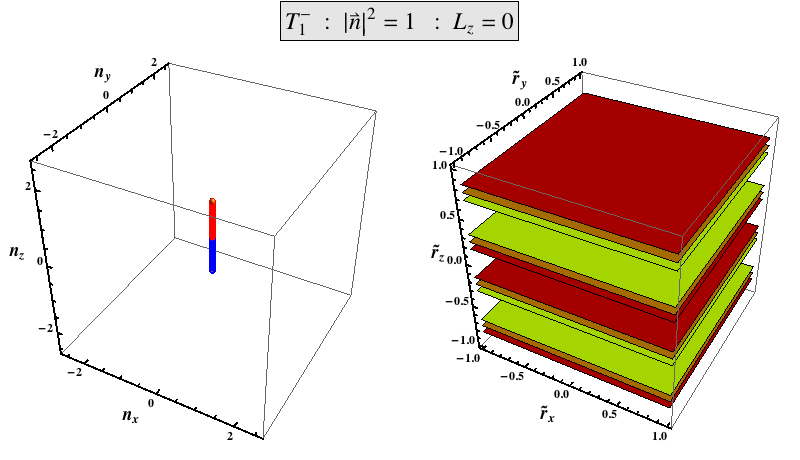}}\\
\mbox{\includegraphics[width=.77\textwidth,angle=0]{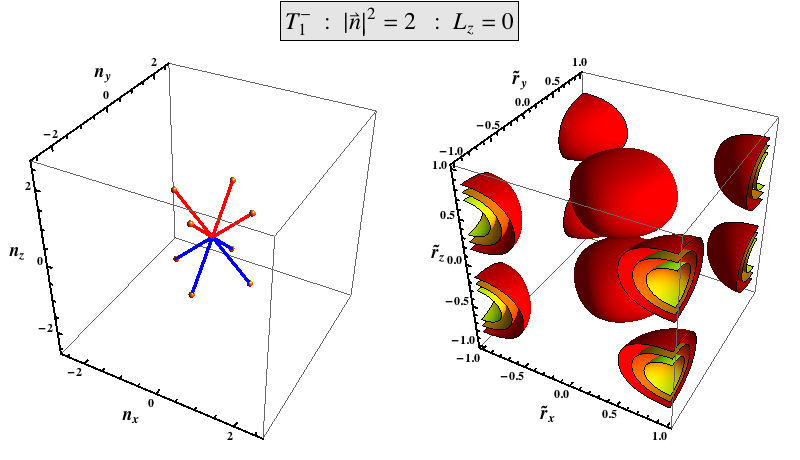}}\\
\mbox{\includegraphics[width=.77\textwidth,angle=0]{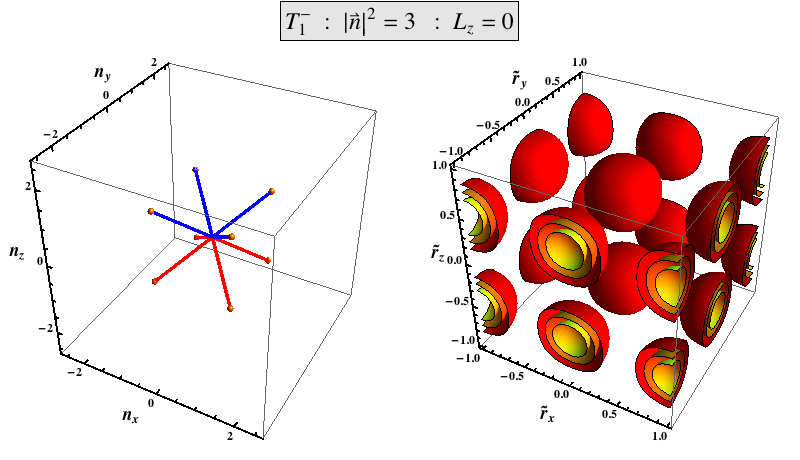}}
\caption{The momentum-space representations (left) and position-space
  representations 
(right) of two-body relative states in the $T_1^-$ representation 
with $L_z=0$
for the
$|{\bf n}|^2=1,2,3$-shells.\label{fig:T1- figures}}
\end{figure}
\begin{figure}
\centering
\mbox{\includegraphics[width=.77\textwidth,angle=0]{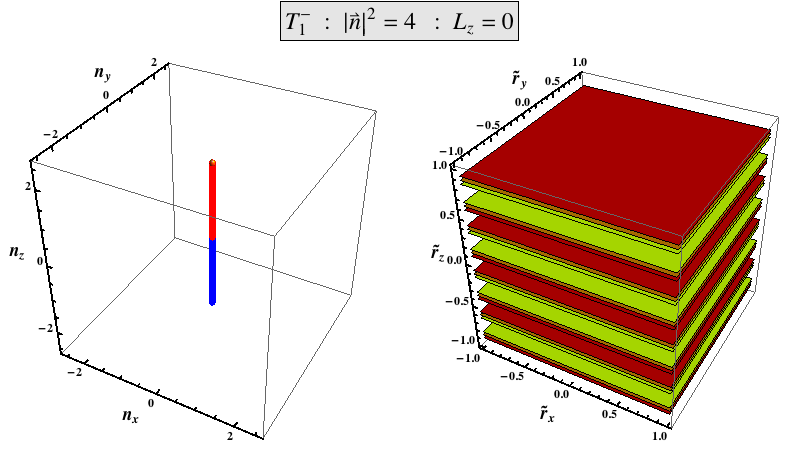}}\\
\mbox{\includegraphics[width=.77\textwidth,angle=0]{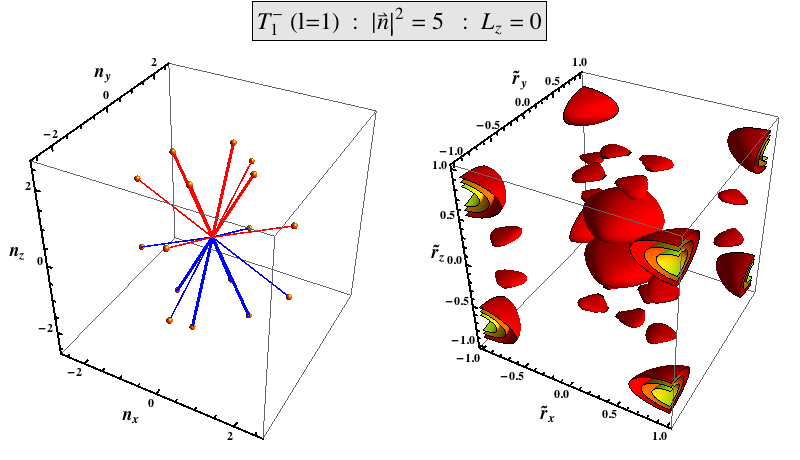}}\\
\mbox{\includegraphics[width=.77\textwidth,angle=0]{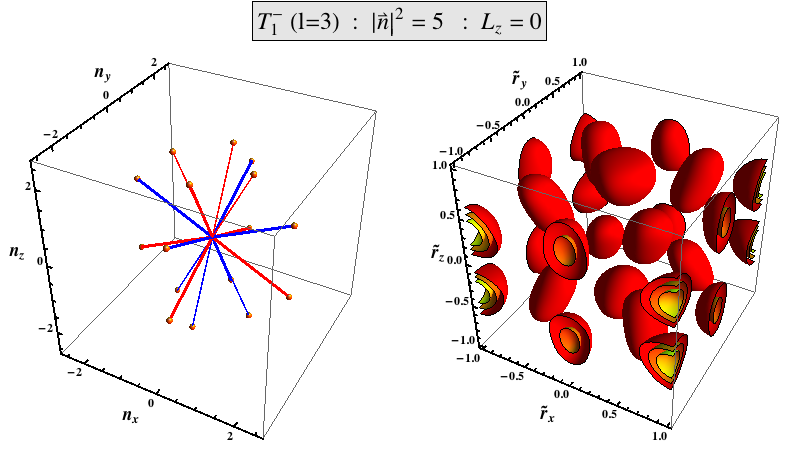}}
\caption{The momentum-space representations (left) and position-space
  representations 
(right) of two-body relative states in the $T_1^-$ representation with $L_z=0$ for the
$|{\bf n}|^2=4,5$-shells.\label{fig:T1- figuresB}
}
\end{figure}
\begin{figure}
\centering
\mbox{\includegraphics[width=.77\textwidth,angle=0]{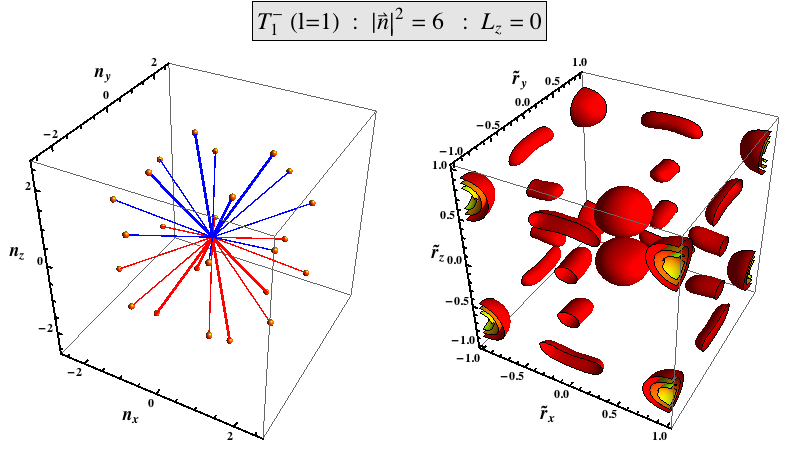}}\\
\mbox{\includegraphics[width=.77\textwidth,angle=0]{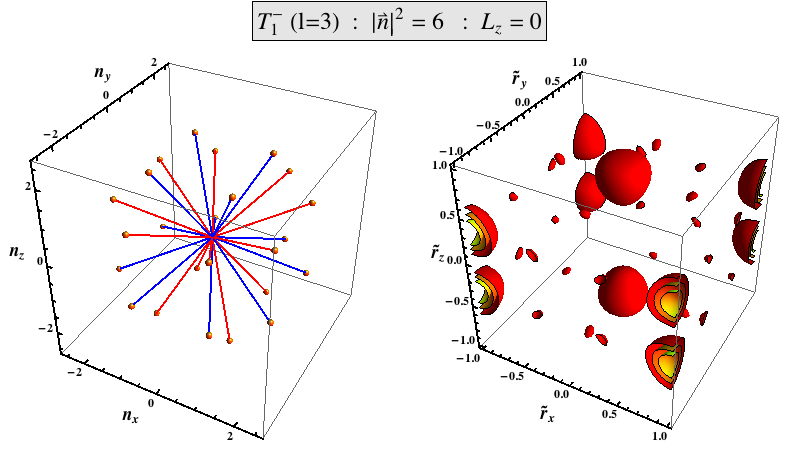}}
\caption{The momentum-space representations (left) and position-space
  representations 
(right) of two-body relative states in the $T_1^-$ representation with $L_z=0$ for the
$|{\bf n}|^2=6$-shells.\label{fig:T1- figuresC}
}
\end{figure}
The graphical representations  of the
source and sink that generate the $T_1^-$ irrep for the lowest-lying  
$|{\bf  n}|^2$-shell
are shown in fig.~\ref{fig:T1- figures}, fig.~\ref{fig:T1- figuresB} 
and fig.~\ref{fig:T1- figuresC},
and  the momentum-space structure is given 
explicitly  in table~\ref{tab:T1mSOURCESSINKS} and 
table~\ref{tab:T1mSOURCESSINKSLz1}.
The structures of the $L_z=3$ sources and sinks are related to those 
with $L_z=1$ by complex conjugation of the coefficients.
\begin{table}
\begin{center}
\begin{minipage}[!ht]{16.5 cm}
\caption{
The momentum-space structure of $T_1^-$, $L_z=0$ sources and sinks for $|{\bf n}|^2$=1-4.
These are shown graphically in  figs.~\ref{fig:T1- figures} and~\ref{fig:T1- figuresB}.
}
\label{tab:T1mSOURCESSINKS}
\end{minipage}
\begin{tabular}{|c||c||c||c|  }
\hline
\hline
$|{\bf n}|^2$=1& $|{\bf n}|^2$=2 & $|{\bf n}|^2$=3 & $|{\bf n}|^2$=4 \\ \hline 
$\begin{array}{cc}
 \text{$|$(0,0,1) , -1$\rangle $} & 1
\end{array}$
&
$\begin{array}{cc}
 \text{$|$(1,0,1) , -1$\rangle $} & \frac{1}{2} \\
 \text{$|$(1,0,-1) , -1$\rangle $} & -\frac{1}{2} \\
 \text{$|$(0,1,1) , -1$\rangle $} & \frac{1}{2} \\
 \text{$|$(0,1,-1) , -1$\rangle $} & -\frac{1}{2}
\end{array}$
&
$\begin{array}{cc}
 \text{$|$(1,1,1) , -1$\rangle $} & -\frac{1}{2} \\
 \text{$|$(1,1,-1) , -1$\rangle $} & \frac{1}{2} \\
 \text{$|$(1,-1,1) , -1$\rangle $} & -\frac{1}{2} \\
 \text{$|$(1,-1,-1) , -1$\rangle $} & \frac{1}{2}
\end{array}$
&
$\begin{array}{cc}
 \text{$|$(0,0,2) , -1$\rangle $} & 1
\end{array}$\\
\hline
\hline
\end{tabular}
\begin{minipage}[t]{16.5 cm}
\vskip 0.5cm
\noindent
\end{minipage}
\end{center}
\end{table}     

\begin{table}
\begin{center}
\begin{minipage}[!ht]{16.5 cm}
\caption{
The momentum-space structure of $T_1^-$, $L_z=0$ sources and sinks for $|{\bf n}|^2$=5 and 6.
These are shown graphically in  figs.~\ref{fig:T1- figuresB} and~\ref{fig:T1- figuresC}.
}
\label{tab:T1mSOURCESSINKSv2}
\end{minipage}
\begin{tabular}{|c||c||c||c|  }
\hline
\hline
$|{\bf n}|^2$=5$_{(l=1)}$ & $|{\bf n}|^2$=5$_{(l=3)}$ & $|{\bf n}|^2$=6$_{(l=1)}$ & $|{\bf n}|^2$=6$_{(l=3)}$ \\ \hline 

$\begin{array}{cc}
 \text{$|$(2,0,1) , -1$\rangle $} & \frac{1}{2 \sqrt{5}} \\
 \text{$|$(2,0,-1) , -1$\rangle $} & -\frac{1}{2 \sqrt{5}} \\
 \text{$|$(1,0,2) , -1$\rangle $} & \frac{1}{\sqrt{5}} \\
 \text{$|$(1,0,-2) , -1$\rangle $} & -\frac{1}{\sqrt{5}} \\
 \text{$|$(0,2,1) , -1$\rangle $} & \frac{1}{2 \sqrt{5}} \\
 \text{$|$(0,2,-1) , -1$\rangle $} & -\frac{1}{2 \sqrt{5}} \\
 \text{$|$(0,1,2) , -1$\rangle $} & \frac{1}{\sqrt{5}} \\
 \text{$|$(0,1,-2) , -1$\rangle $} & -\frac{1}{\sqrt{5}}
\end{array}$
&
$\begin{array}{cc}
 \text{$|$(2,0,1) , -1$\rangle $} & -\frac{1}{\sqrt{5}} \\
 \text{$|$(2,0,-1) , -1$\rangle $} & \frac{1}{\sqrt{5}} \\
 \text{$|$(1,0,2) , -1$\rangle $} & \frac{1}{2 \sqrt{5}} \\
 \text{$|$(1,0,-2) , -1$\rangle $} & -\frac{1}{2 \sqrt{5}} \\
 \text{$|$(0,2,1) , -1$\rangle $} & -\frac{1}{\sqrt{5}} \\
 \text{$|$(0,2,-1) , -1$\rangle $} & \frac{1}{\sqrt{5}} \\
 \text{$|$(0,1,2) , -1$\rangle $} & \frac{1}{2 \sqrt{5}} \\
 \text{$|$(0,1,-2) , -1$\rangle $} & -\frac{1}{2 \sqrt{5}}
\end{array}$
&
$\begin{array}{cc}
 \text{$|$(2,1,1) , -1$\rangle $} & -\frac{1}{2 \sqrt{6}} \\
 \text{$|$(2,1,-1) , -1$\rangle $} & \frac{1}{2 \sqrt{6}} \\
 \text{$|$(2,-1,1) , -1$\rangle $} & -\frac{1}{2 \sqrt{6}} \\
 \text{$|$(2,-1,-1) , -1$\rangle $} & \frac{1}{2 \sqrt{6}} \\
 \text{$|$(1,2,1) , -1$\rangle $} & -\frac{1}{2 \sqrt{6}} \\
 \text{$|$(1,2,-1) , -1$\rangle $} & \frac{1}{2 \sqrt{6}} \\
 \text{$|$(1,1,2) , -1$\rangle $} & -\frac{1}{\sqrt{6}} \\
 \text{$|$(1,1,-2) , -1$\rangle $} & \frac{1}{\sqrt{6}} \\
 \text{$|$(1,-1,2) , -1$\rangle $} & -\frac{1}{\sqrt{6}} \\
 \text{$|$(1,-1,-2) , -1$\rangle $} & \frac{1}{\sqrt{6}} \\
 \text{$|$(1,-2,1) , -1$\rangle $} & -\frac{1}{2 \sqrt{6}} \\
 \text{$|$(1,-2,-1) , -1$\rangle $} & \frac{1}{2 \sqrt{6}}
\end{array}$
&
$\begin{array}{cc}
 \text{$|$(2,1,1) , -1$\rangle $} & \frac{1}{2 \sqrt{3}} \\
 \text{$|$(2,1,-1) , -1$\rangle $} & -\frac{1}{2 \sqrt{3}} \\
 \text{$|$(2,-1,1) , -1$\rangle $} & \frac{1}{2 \sqrt{3}} \\
 \text{$|$(2,-1,-1) , -1$\rangle $} & -\frac{1}{2 \sqrt{3}} \\
 \text{$|$(1,2,1) , -1$\rangle $} & \frac{1}{2 \sqrt{3}} \\
 \text{$|$(1,2,-1) , -1$\rangle $} & -\frac{1}{2 \sqrt{3}} \\
 \text{$|$(1,1,2) , -1$\rangle $} & -\frac{1}{2 \sqrt{3}} \\
 \text{$|$(1,1,-2) , -1$\rangle $} & \frac{1}{2 \sqrt{3}} \\
 \text{$|$(1,-1,2) , -1$\rangle $} & -\frac{1}{2 \sqrt{3}} \\
 \text{$|$(1,-1,-2) , -1$\rangle $} & \frac{1}{2 \sqrt{3}} \\
 \text{$|$(1,-2,1) , -1$\rangle $} & \frac{1}{2 \sqrt{3}} \\
 \text{$|$(1,-2,-1) , -1$\rangle $} & -\frac{1}{2 \sqrt{3}}
\end{array}$\\
\hline
\hline
\end{tabular}
\begin{minipage}[t]{16.5 cm}
\vskip 0.5cm
\noindent
\end{minipage}
\end{center}
\end{table}

\begin{table}
\begin{center}
\begin{minipage}[!ht]{16.5 cm}
\caption{
The momentum-space structure of $T_1^-$, $L_z=1$ sources and sinks for $|{\bf n}|^2$=1-4.
}
\label{tab:T1mSOURCESSINKSLz1}
\end{minipage}
\begin{tabular}{|c||c||c||c|  }
\hline
\hline
$|{\bf n}|^2$=1& $|{\bf n}|^2$=2 & $|{\bf n}|^2$=3 & $|{\bf n}|^2$=4 \\ \hline 
$\begin{array}{cc}
 \text{$|$(1,0,0) , -1$\rangle $} & \frac{1}{\sqrt{2}} \\
 \text{$|$(0,1,0) , -1$\rangle $} & \frac{i}{\sqrt{2}}
\end{array}$
&
$\begin{array}{cc}
 \text{$|$(1,1,0) , -1$\rangle $} & \frac{i}{2} \\
 \text{$|$(1,0,1) , -1$\rangle $} & \frac{1}{4}+\frac{i}{4} \\
 \text{$|$(1,0,-1) , -1$\rangle $} & \frac{1}{4}+\frac{i}{4} \\
 \text{$|$(1,-1,0) , -1$\rangle $} & \frac{1}{2} \\
 \text{$|$(0,1,1) , -1$\rangle $} & -\frac{1}{4}+\frac{i}{4} \\
 \text{$|$(0,1,-1) , -1$\rangle $} & -\frac{1}{4}+\frac{i}{4}
\end{array}$
&
$\begin{array}{cc}
 \text{$|$(1,1,1) , -1$\rangle $} & \frac{1}{2} \\
 \text{$|$(1,1,-1) , -1$\rangle $} & \frac{1}{2} \\
 \text{$|$(1,-1,1) , -1$\rangle $} & -\frac{i}{2} \\
 \text{$|$(1,-1,-1) , -1$\rangle $} & -\frac{i}{2}
\end{array}$
&
$\begin{array}{cc}
 \text{$|$(2,0,0) , -1$\rangle $} & \frac{1}{\sqrt{2}} \\
 \text{$|$(0,2,0) , -1$\rangle $} & \frac{i}{\sqrt{2}}
\end{array}$\\
\hline
\hline
\end{tabular}
\begin{minipage}[t]{16.5 cm}
\vskip 0.5cm
\noindent
\end{minipage}
\end{center}
\end{table}     

\begin{table}
\begin{center}
\begin{minipage}[!ht]{16.5 cm}
\caption{
The momentum-space structure of $T_1^-$, $L_z=1$ sources and sinks for $|{\bf n}|^2$=5 and 6.
}
\label{tab:T1mSOURCESSINKSLz1v2}
\end{minipage}
\begin{tabular}{|c||c||c||c|  }
\hline
\hline
$|{\bf n}|^2$=5$_{(l=1)}$ & $|{\bf n}|^2$=5$_{(l=3)}$ & $|{\bf n}|^2$=6$_{(l=1)}$ & $|{\bf n}|^2$=6$_{(l=3)}$ \\ \hline 

$\begin{array}{cc}
 \text{$|$(2,1,0) , -1$\rangle $} & \frac{1}{2 \sqrt{2}} \\
 \text{$|$(2,0,1) , -1$\rangle $} & \frac{\frac{2}{5}-\frac{i}{5}}{\sqrt{2}} \\
 \text{$|$(2,0,-1) , -1$\rangle $} & \frac{\frac{2}{5}-\frac{i}{5}}{\sqrt{2}} \\
 \text{$|$(2,-1,0) , -1$\rangle $} & \frac{\frac{3}{10}-\frac{2 i}{5}}{\sqrt{2}} \\
 \text{$|$(1,2,0) , -1$\rangle $} & \frac{\frac{2}{5}+\frac{3 i}{10}}{\sqrt{2}} \\
 \text{$|$(1,0,2) , -1$\rangle $} & \frac{\frac{1}{5}-\frac{i}{10}}{\sqrt{2}} \\
 \text{$|$(1,0,-2) , -1$\rangle $} & \frac{\frac{1}{5}-\frac{i}{10}}{\sqrt{2}} \\
 \text{$|$(1,-2,0) , -1$\rangle $} & -\frac{i}{2 \sqrt{2}} \\
 \text{$|$(0,2,1) , -1$\rangle $} & \frac{\frac{1}{5}+\frac{2 i}{5}}{\sqrt{2}} \\
 \text{$|$(0,2,-1) , -1$\rangle $} & \frac{\frac{1}{5}+\frac{2 i}{5}}{\sqrt{2}} \\
 \text{$|$(0,1,2) , -1$\rangle $} & \frac{\frac{1}{10}+\frac{i}{5}}{\sqrt{2}} \\
 \text{$|$(0,1,-2) , -1$\rangle $} & \frac{\frac{1}{10}+\frac{i}{5}}{\sqrt{2}}
\end{array}$
&
$\begin{array}{cc}
 \text{$|$(2,1,0) , -1$\rangle $} & \frac{1}{2 \sqrt{2}} \\
 \text{$|$(2,0,1) , -1$\rangle $} & \frac{\frac{1}{10}+\frac{i}{5}}{\sqrt{2}} \\
 \text{$|$(2,0,-1) , -1$\rangle $} & \frac{\frac{1}{10}+\frac{i}{5}}{\sqrt{2}} \\
 \text{$|$(2,-1,0) , -1$\rangle $} & -\frac{\frac{3}{10}-\frac{2 i}{5}}{\sqrt{2}} \\
 \text{$|$(1,2,0) , -1$\rangle $} & -\frac{\frac{2}{5}+\frac{3 i}{10}}{\sqrt{2}} \\
 \text{$|$(1,0,2) , -1$\rangle $} & -\frac{\frac{1}{5}+\frac{2 i}{5}}{\sqrt{2}} \\
 \text{$|$(1,0,-2) , -1$\rangle $} & -\frac{\frac{1}{5}+\frac{2 i}{5}}{\sqrt{2}} \\
 \text{$|$(1,-2,0) , -1$\rangle $} & -\frac{i}{2 \sqrt{2}} \\
 \text{$|$(0,2,1) , -1$\rangle $} & -\frac{\frac{1}{5}-\frac{i}{10}}{\sqrt{2}} \\
 \text{$|$(0,2,-1) , -1$\rangle $} & -\frac{\frac{1}{5}-\frac{i}{10}}{\sqrt{2}} \\
 \text{$|$(0,1,2) , -1$\rangle $} & \frac{\frac{2}{5}-\frac{i}{5}}{\sqrt{2}} \\
 \text{$|$(0,1,-2) , -1$\rangle $} & \frac{\frac{2}{5}-\frac{i}{5}}{\sqrt{2}}
\end{array}$
&
$\begin{array}{cc}
 \text{$|$(2,1,1) , -1$\rangle $} & \frac{\sqrt{\frac{5}{3}}}{4} \\
 \text{$|$(2,1,-1) , -1$\rangle $} & \frac{\sqrt{\frac{5}{3}}}{4} \\
 \text{$|$(2,-1,1) , -1$\rangle $} & \frac{\frac{3}{4}-i}{\sqrt{15}} \\
 \text{$|$(2,-1,-1) , -1$\rangle $} & \frac{\frac{3}{4}-i}{\sqrt{15}} \\
 \text{$|$(1,2,1) , -1$\rangle $} & \frac{1+\frac{3 i}{4}}{\sqrt{15}} \\
 \text{$|$(1,2,-1) , -1$\rangle $} & \frac{1+\frac{3 i}{4}}{\sqrt{15}} \\
 \text{$|$(1,1,2) , -1$\rangle $} & \frac{1}{2} \sqrt{\frac{2}{15}+\frac{i}{10}} \\
 \text{$|$(1,1,-2) , -1$\rangle $} & \frac{1}{2} \sqrt{\frac{2}{15}+\frac{i}{10}} \\
 \text{$|$(1,-1,2) , -1$\rangle $} & \frac{1}{2} \sqrt{-\frac{2}{15}-\frac{i}{10}} \\
 \text{$|$(1,-1,-2) , -1$\rangle $} & \frac{1}{2} \sqrt{-\frac{2}{15}-\frac{i}{10}} \\
 \text{$|$(1,-2,1) , -1$\rangle $} & -\frac{1}{4} i \sqrt{\frac{5}{3}} \\
 \text{$|$(1,-2,-1) , -1$\rangle $} & -\frac{1}{4} i \sqrt{\frac{5}{3}}
\end{array}$
&
$\begin{array}{cc}
 \text{$|$(2,1,1) , -1$\rangle $} & \frac{i}{2 \sqrt{3}} \\
 \text{$|$(2,1,-1) , -1$\rangle $} & \frac{i}{2 \sqrt{3}} \\
 \text{$|$(2,-1,1) , -1$\rangle $} & -\frac{1}{2 \sqrt{3}} \\
 \text{$|$(2,-1,-1) , -1$\rangle $} & -\frac{1}{2 \sqrt{3}} \\
 \text{$|$(1,2,1) , -1$\rangle $} & -\frac{i}{2 \sqrt{3}} \\
 \text{$|$(1,2,-1) , -1$\rangle $} & -\frac{i}{2 \sqrt{3}} \\
 \text{$|$(1,1,2) , -1$\rangle $} & \frac{1}{2 \sqrt{3}} \\
 \text{$|$(1,1,-2) , -1$\rangle $} & \frac{1}{2 \sqrt{3}} \\
 \text{$|$(1,-1,2) , -1$\rangle $} & -\frac{i}{2 \sqrt{3}} \\
 \text{$|$(1,-1,-2) , -1$\rangle $} & -\frac{i}{2 \sqrt{3}} \\
 \text{$|$(1,-2,1) , -1$\rangle $} & \frac{1}{2 \sqrt{3}} \\
 \text{$|$(1,-2,-1) , -1$\rangle $} & \frac{1}{2 \sqrt{3}}
\end{array}$\\
\hline
\hline
\end{tabular}
\begin{minipage}[t]{16.5 cm}
\vskip 0.5cm
\noindent
\end{minipage}
\end{center}
\end{table}     

The $T_1^-$ irrep first appears in the $|{\bf n}|^2=1$-shell and $l=1$
is the lowest contributing partial-wave.  
LQCD calculations of correlation functions from sources and sinks transforming
as $T_1^-$ will provide determinations of $\delta_1$ with contamination from
partial-waves with $l\ge 3$.
LQCD calculations of the phase-shift in this partial-wave are presently being
performed, and the $\rho$-resonance is beginning to be 
mapped out, e.g. Ref.~\cite{Gockeler:2008kc}.

\subsubsection{$T_2^-$ Representation\label{irrep:T2m}}
\begin{figure}
\centering
\includegraphics[height=\textwidth,angle=-90]{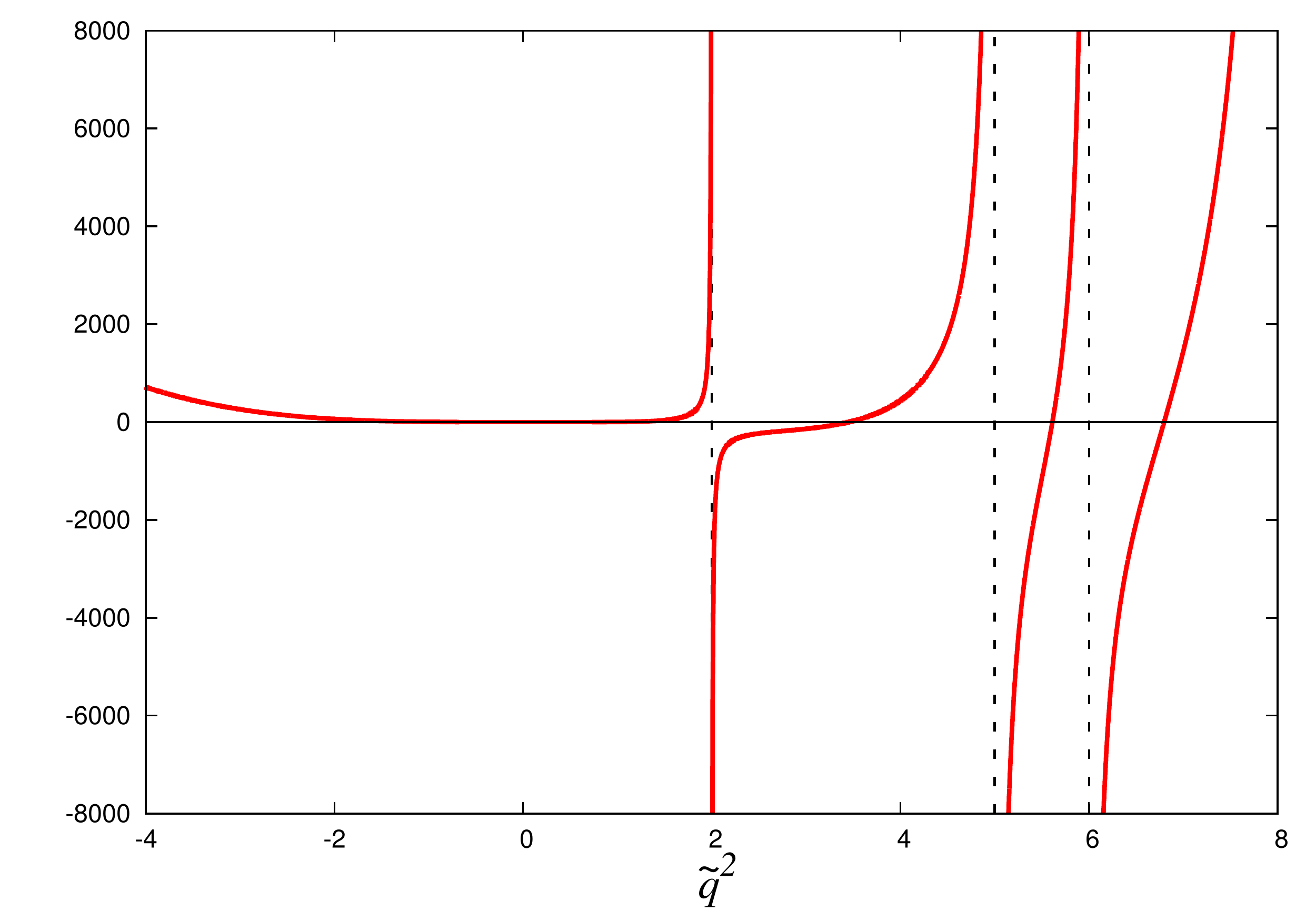}
\caption{The function $\mathcal{X}_{T_2}^-$, as defined in
  eq.~(\protect\ref{eqn:T2- l=3}), 
as a  function of $\tilde q^2$.  The vertical dashed lines denote the position of
  the poles of the function corresponding to the non-interacting energy-eigenvalues. 
\label{fig:T2- l=3}}
\end{figure}

The energy-eigenvalues of states transforming in the $T_2^-$ irrep receive
contributions from interactions in the $l=3,5,...$ partial-waves, as
presented in table~\ref{tab:Lcontributetab}.
As the $T_2^-$ irrep is three-dimensional, the contribution to the determinant in 
eq.~(\ref{eq:evals}) results from a $6\times 6$ matrix for $l \le 6$, 
which collapses
down to the determinant of 
a $2\times 2$ matrix as the $L_z=1$, $L_z=2$ and $L_z=3$ states are degenerate.
The $T_2^-$  $L_z=2$ states associated with the $\overline{F}^{(FV)}_{3;3}$ and 
$F^{(FV)}_{5;5}$ blocks are 
\begin{displaymath}
\begin{split}
|T_2^-, 2; 3; 1 \rangle \ &\ =\ \frac{1}{\sqrt{2}}|3,2\rangle\ +\
\frac{1}{\sqrt{2}}|3,-2\rangle
\\
|T_2^-, 2; 5; 1 \rangle \ &\ =\ \frac{1}{\sqrt{2}}|5,2\rangle\ +\
\frac{1}{\sqrt{2}}|5,-2 \rangle
\ \ \ \ ,
\end{split}
\end{displaymath}
in terms of which, the $T_2^-$ contribution to 
eq.~(\ref{eq:evals}) becomes
\begin{equation}
\label{eqn:T2- det}
\text{det}\left[
\begin{pmatrix}
\text{cot}\delta_3 & 0\\
0 & \text{cot}\delta_5
\end{pmatrix}
\ -\ 
\begin{pmatrix} 
\overline{F}_{3;3}^{(FV, T_2^-)} & \overline{F}_{3;5}^{(FV, T_2^-)} \\
\overline{F}_{5;3}^{(FV, T_2^-)} & \overline{F}_{5;5}^{(FV, T_2^-)} 
\end{pmatrix}
\right]\ =\ 0
\ \ \ ,
\end{equation}
where
\begin{equation}
\begin{split}
\overline{F}_{3;3}^{(FV, T_2^-)}
&=\frac{\mathcal{Z}_{0,0}\left(1;\tilde{q}^2\right)}{\pi ^{3/2}
  \tilde{q}}-\frac{2 \mathcal{Z}_{4,0}\left(1;\tilde{q}^2\right)}{11 \pi ^{3/2}
  \tilde{q}^5}-\frac{60 \mathcal{Z}_{6,0}\left(1;\tilde{q}^2\right)}{11
  \sqrt{13} \pi ^{3/2} \tilde{q}^7}
\nonumber\\
\overline{F}_{3;5}^{(FV, T_2^-)} &=-\frac{20 \mathcal{Z}_{4,0}\left(1;\tilde{q}^2\right)}{13 \sqrt{11} \pi ^{3/2} \tilde{q}^5}-\frac{14 \mathcal{Z}_{6,0}\left(1;\tilde{q}^2\right)}{\sqrt{143} \pi ^{3/2} \tilde{q}^7}+\frac{112 \mathcal{Z}_{8,0}\left(1;\tilde{q}^2\right)}{13
   \sqrt{187} \pi ^{3/2} \tilde{q}^9}
\nonumber\\
\overline{F}_{5;5}^{(FV, T_2^-)}&=\frac{\mathcal{Z}_{0,0}\left(1;\tilde{q}^2\right)}{\pi ^{3/2} \tilde{q}}+\frac{4 \mathcal{Z}_{4,0}\left(1;\tilde{q}^2\right)}{13 \pi ^{3/2} \tilde{q}^5}-\frac{80 \mathcal{Z}_{6,0}\left(1;\tilde{q}^2\right)}{17 \sqrt{13} \pi ^{3/2}
   \tilde{q}^7}-\frac{280 \mathcal{Z}_{8,0}\left(1;\tilde{q}^2\right)}{247
   \sqrt{17} \pi ^{3/2} \tilde{q}^9}-\frac{432 \sqrt{21}
   \mathcal{Z}_{10,0}\left(1;\tilde{q}^2\right)}{4199 \pi ^{3/2}
   \tilde{q}^{11}}
\ \ \ \ .
\end{split}
\label{eq:T2mFs}
\end{equation}
The solutions to this equation result from
\begin{multline}
\frac{\text{cot}\delta_3}{2}+\frac{\text{cot}\delta_5}{2}
   -\frac{\mathcal{Z}_{0,0}\left(1;\tilde{q}^2\right)}{\pi ^{3/2} \tilde{q}}
   -\frac{9 \mathcal{Z}_{4,0}\left(1;\tilde{q}^2\right)}{143 \pi ^{3/2}
     \tilde{q}^5}
+\frac{950 \mathcal{Z}_{6,0}\left(1;\tilde{q}^2\right)}{187 \sqrt{13} \pi ^{3/2} \tilde{q}^7}+\frac{140
   \mathcal{Z}_{8,0}\left(1;\tilde{q}^2\right)}{247 \sqrt{17} \pi ^{3/2} \tilde{q}^9}+\frac{216 \sqrt{21} \mathcal{Z}_{10,0}\left(1;\tilde{q}^2\right)}{4199 \pi ^{3/2} \tilde{q}^{11}} =\\
   {\bf \pm}
   \frac{1}{2}\left[
   \left(-\frac{432 \sqrt{21} \mathcal{Z}_{10,0}\left(1;\tilde{q}^2\right)}{4199 \pi ^{3/2} \tilde{q}^{11}}+\frac{70 \mathcal{Z}_{4,0}\left(1;\tilde{q}^2\right)}{143 \pi ^{3/2} \tilde{q}^5}+\frac{140 \mathcal{Z}_{6,0}\left(1;\tilde{q}^2\right)}{187 \sqrt{13} \pi ^{3/2}
   \tilde{q}^7}-\frac{280 \mathcal{Z}_{8,0}\left(1;\tilde{q}^2\right)}{247 \sqrt{17} \pi ^{3/2} \tilde{q}^9}+\text{cot}\delta_3-\text{cot}\delta_5\right)^2\right.\\
   +\left.\frac{4}{\pi^3} \left(\frac{20\mathcal{Z}_{4,0}\left(1;\tilde{q}^2\right)}{13\sqrt{11} \tilde{q}^5}+\frac{14 \mathcal{Z}_{6,0}\left(1;\tilde{q}^2\right)}{\sqrt{143} \tilde{q}^7}-\frac{112
   \mathcal{Z}_{8,0}\left(1;\tilde{q}^2\right)}{13\sqrt{187}
   \tilde{q}^9}\right)^2\right]^{1/2}
\ \ \ .
\end{multline}
In the limit of vanishing interactions in the  $l=5$ partial-wave, 
eq.~(\ref{eqn:T2- det}) collapses down to 
\begin{equation}\label{eqn:T2- l=3}
   \begin{split}
   q^7\text{cot}\delta_3&=\left(\frac{2\pi}{L}\right)^7\frac{1}{\pi^{3/2}}
   \left(\tilde q^6\mathcal{Z}_{0,0}\left(1;\tilde{q}^2\right) - \frac{2 \tilde q^2\mathcal{Z}_{4,0}\left(1;\tilde{q}^2\right)}{11 }-\frac{60 \mathcal{Z}_{6,0}\left(1;\tilde{q}^2\right)}{11 \sqrt{13}}\right)\\
   &\equiv \left(\frac{2\pi}{L}\right)^7\frac{1}{\pi^{3/2}}
\ \mathcal{X}_{T_2}^-\left(\tilde q^2\right)
 \ \ \ ,
   \end{split}
\end{equation}
where the function $\mathcal{X}_{T_2}^-\left(\tilde q^2\right)$ is shown in
fig.~\ref{fig:T2- l=3}.
\begin{figure}
\centering
\mbox{\includegraphics[width=.77\textwidth,angle=0]{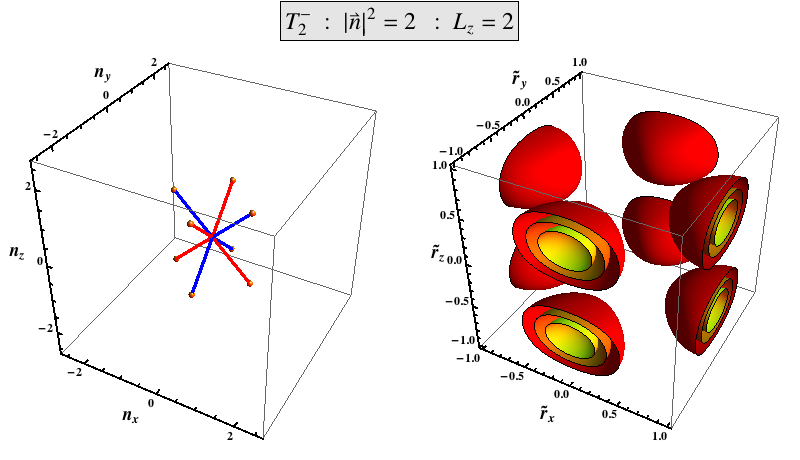}}\\
\mbox{\includegraphics[width=.77\textwidth,angle=0]{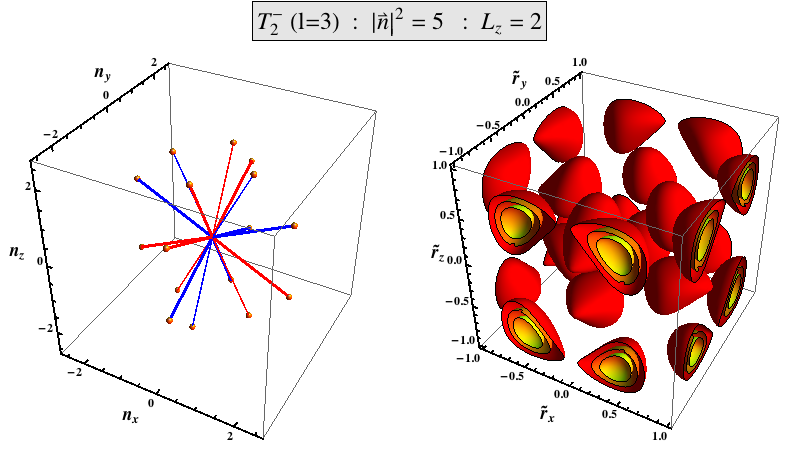}}\\
\mbox{\includegraphics[width=.77\textwidth,angle=0]{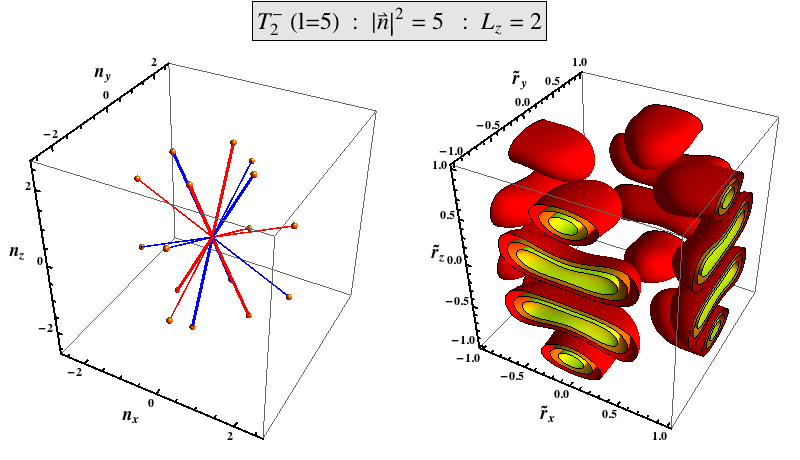}}
\caption{The momentum-space representations (left) and position-space
  representations (right) of two-body relative states in the $T_2^-$
  representation with $L_z=2$ for select $|{\bf n}|^2$-shells.
\label{fig:T2- figures}}
\end{figure}
\begin{figure}
\centering
\mbox{\includegraphics[width=.77\textwidth,angle=0]{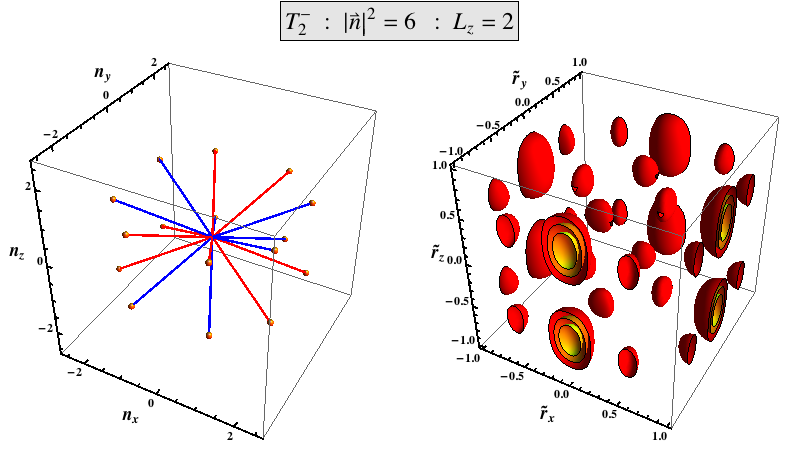}}
\caption{The momentum space representations (left) and position-space
  representations (right) of two-body relative states in the $T_2^-$
  representation with $L_z=2$ for the $|{\bf n}|^2=6$-shell.
\label{fig:T2- figuresB}}
\end{figure}
The graphical representations  of the
source and sink that generate the $T_2^-$ irrep for the lowest-lying  
$|{\bf  n}|^2$-shell
are shown in fig.~\ref{fig:T2- figures} and fig.~\ref{fig:T2- figuresB},
and  the momentum-space structure is given 
explicitly  in tables~\ref{tab:T2mSOURCESSINKS} and 
table~\ref{tab:T2mSOURCESSINKSLz1}.
The structures of the $L_z=3$ sources and sinks are related to those 
with $L_z=1$ by complex conjugation of the coefficients.
\begin{table}
\begin{center}
\begin{minipage}[!ht]{16.5 cm}
\caption{
The momentum-space structure of $T_2^-$, $L_z=2$ sources and sinks.
These are shown graphically in figs.~\ref{fig:T2- figures} and~\ref{fig:T2- figuresB}.
}
\label{tab:T2mSOURCESSINKS}
\end{minipage}
\begin{tabular}{|c||c||c||c|  }
\hline
\hline
$|{\bf n}|^2$=2& $|{\bf n}|^2$=5$_{(l=3)}$ & $|{\bf n}|^2$=5$_{(l=5)}$ & $|{\bf n}|^2$=6 \\ \hline 
$\begin{array}{cc}
 \text{$|$(1,0,1) , -1$\rangle $} & -\frac{1}{2} \\
 \text{$|$(1,0,-1) , -1$\rangle $} & \frac{1}{2} \\
 \text{$|$(0,1,1) , -1$\rangle $} & \frac{1}{2} \\
 \text{$|$(0,1,-1) , -1$\rangle $} & -\frac{1}{2}
\end{array}$
&
$\begin{array}{cc}
 \text{$|$(2,0,1) , -1$\rangle $} & -\frac{1}{\sqrt{5}} \\
 \text{$|$(2,0,-1) , -1$\rangle $} & \frac{1}{\sqrt{5}} \\
 \text{$|$(1,0,2) , -1$\rangle $} & -\frac{1}{2 \sqrt{5}} \\
 \text{$|$(1,0,-2) , -1$\rangle $} & \frac{1}{2 \sqrt{5}} \\
 \text{$|$(0,2,1) , -1$\rangle $} & \frac{1}{\sqrt{5}} \\
 \text{$|$(0,2,-1) , -1$\rangle $} & -\frac{1}{\sqrt{5}} \\
 \text{$|$(0,1,2) , -1$\rangle $} & \frac{1}{2 \sqrt{5}} \\
 \text{$|$(0,1,-2) , -1$\rangle $} & -\frac{1}{2 \sqrt{5}}
\end{array}$
&
$\begin{array}{cc}
 \text{$|$(2,0,1) , -1$\rangle $} & \frac{1}{2 \sqrt{5}} \\
 \text{$|$(2,0,-1) , -1$\rangle $} & -\frac{1}{2 \sqrt{5}} \\
 \text{$|$(1,0,2) , -1$\rangle $} & -\frac{1}{\sqrt{5}} \\
 \text{$|$(1,0,-2) , -1$\rangle $} & \frac{1}{\sqrt{5}} \\
 \text{$|$(0,2,1) , -1$\rangle $} & -\frac{1}{2 \sqrt{5}} \\
 \text{$|$(0,2,-1) , -1$\rangle $} & \frac{1}{2 \sqrt{5}} \\
 \text{$|$(0,1,2) , -1$\rangle $} & \frac{1}{\sqrt{5}} \\
 \text{$|$(0,1,-2) , -1$\rangle $} & -\frac{1}{\sqrt{5}}
\end{array}$
&
$\begin{array}{cc}
 \text{$|$(2,1,1) , -1$\rangle $} & -\frac{1}{2 \sqrt{2}} \\
 \text{$|$(2,1,-1) , -1$\rangle $} & \frac{1}{2 \sqrt{2}} \\
 \text{$|$(2,-1,1) , -1$\rangle $} & -\frac{1}{2 \sqrt{2}} \\
 \text{$|$(2,-1,-1) , -1$\rangle $} & \frac{1}{2 \sqrt{2}} \\
 \text{$|$(1,2,1) , -1$\rangle $} & \frac{1}{2 \sqrt{2}} \\
 \text{$|$(1,2,-1) , -1$\rangle $} & -\frac{1}{2 \sqrt{2}} \\
 \text{$|$(1,-2,1) , -1$\rangle $} & \frac{1}{2 \sqrt{2}} \\
 \text{$|$(1,-2,-1) , -1$\rangle $} & -\frac{1}{2 \sqrt{2}}
\end{array}$\\
\hline
\hline
\end{tabular}
\begin{minipage}[t]{16.5 cm}
\vskip 0.5cm
\noindent
\end{minipage}
\end{center}
\end{table}

\begin{table}
\begin{center}
\begin{minipage}[!ht]{16.5 cm}
\caption{
The momentum-space structure of $T_2^-$, $L_z=1$ sources and sinks.
}
\label{tab:T2mSOURCESSINKSLz1}
\end{minipage}
\begin{tabular}{|c||c||c||c|  }
\hline
\hline
$|{\bf n}|^2$=2& $|{\bf n}|^2$=5$_{(l=3)}$ & $|{\bf n}|^2$=5$_{(l=5)}$ & $|{\bf n}|^2$=6 \\ \hline 
$\begin{array}{cc}
 \text{$|$(1,1,0) , -1$\rangle $} & \frac{1}{2} \\
 \text{$|$(1,0,1) , -1$\rangle $} & -\frac{1}{4}+\frac{i}{4} \\
 \text{$|$(1,0,-1) , -1$\rangle $} & -\frac{1}{4}+\frac{i}{4} \\
 \text{$|$(1,-1,0) , -1$\rangle $} & -\frac{i}{2} \\
 \text{$|$(0,1,1) , -1$\rangle $} & -\frac{1}{4}-\frac{i}{4} \\
 \text{$|$(0,1,-1) , -1$\rangle $} & -\frac{1}{4}-\frac{i}{4}
\end{array}$
&
$\begin{array}{cc}
 \text{$|$(2,1,0) , -1$\rangle $} & \frac{1}{2 \sqrt{2}} \\
 \text{$|$(2,0,1) , -1$\rangle $} & -\frac{\frac{1}{10}-\frac{i}{5}}{\sqrt{2}} \\
 \text{$|$(2,0,-1) , -1$\rangle $} & -\frac{\frac{1}{10}-\frac{i}{5}}{\sqrt{2}} \\
 \text{$|$(2,-1,0) , -1$\rangle $} & -\frac{\frac{3}{10}+\frac{2 i}{5}}{\sqrt{2}} \\
 \text{$|$(1,2,0) , -1$\rangle $} & \frac{\frac{2}{5}-\frac{3 i}{10}}{\sqrt{2}} \\
 \text{$|$(1,0,2) , -1$\rangle $} & -\frac{\frac{1}{5}-\frac{2 i}{5}}{\sqrt{2}} \\
 \text{$|$(1,0,-2) , -1$\rangle $} & -\frac{\frac{1}{5}-\frac{2 i}{5}}{\sqrt{2}} \\
 \text{$|$(1,-2,0) , -1$\rangle $} & -\frac{i}{2 \sqrt{2}} \\
 \text{$|$(0,2,1) , -1$\rangle $} & -\frac{\frac{1}{5}+\frac{i}{10}}{\sqrt{2}} \\
 \text{$|$(0,2,-1) , -1$\rangle $} & -\frac{\frac{1}{5}+\frac{i}{10}}{\sqrt{2}} \\
 \text{$|$(0,1,2) , -1$\rangle $} & -\frac{\frac{2}{5}+\frac{i}{5}}{\sqrt{2}} \\
 \text{$|$(0,1,-2) , -1$\rangle $} & -\frac{\frac{2}{5}+\frac{i}{5}}{\sqrt{2}}
\end{array}$
&
$\begin{array}{cc}
 \text{$|$(2,1,0) , -1$\rangle $} & \frac{1}{2 \sqrt{2}} \\
 \text{$|$(2,0,1) , -1$\rangle $} & -\frac{\frac{2}{5}+\frac{i}{5}}{\sqrt{2}} \\
 \text{$|$(2,0,-1) , -1$\rangle $} & -\frac{\frac{2}{5}+\frac{i}{5}}{\sqrt{2}} \\
 \text{$|$(2,-1,0) , -1$\rangle $} & \frac{\frac{3}{10}+\frac{2 i}{5}}{\sqrt{2}} \\
 \text{$|$(1,2,0) , -1$\rangle $} & -\frac{\frac{2}{5}-\frac{3 i}{10}}{\sqrt{2}} \\
 \text{$|$(1,0,2) , -1$\rangle $} & \frac{\frac{1}{5}+\frac{i}{10}}{\sqrt{2}} \\
 \text{$|$(1,0,-2) , -1$\rangle $} & \frac{\frac{1}{5}+\frac{i}{10}}{\sqrt{2}} \\
 \text{$|$(1,-2,0) , -1$\rangle $} & -\frac{i}{2 \sqrt{2}} \\
 \text{$|$(0,2,1) , -1$\rangle $} & \frac{\frac{1}{5}-\frac{2 i}{5}}{\sqrt{2}} \\
 \text{$|$(0,2,-1) , -1$\rangle $} & \frac{\frac{1}{5}-\frac{2 i}{5}}{\sqrt{2}} \\
 \text{$|$(0,1,2) , -1$\rangle $} & -\frac{\frac{1}{10}-\frac{i}{5}}{\sqrt{2}} \\
 \text{$|$(0,1,-2) , -1$\rangle $} & -\frac{\frac{1}{10}-\frac{i}{5}}{\sqrt{2}}
\end{array}$
&
$\begin{array}{cc}
 \text{$|$(2,1,1) , -1$\rangle $} & \frac{1}{4} \\
 \text{$|$(2,1,-1) , -1$\rangle $} & \frac{1}{4} \\
 \text{$|$(2,-1,1) , -1$\rangle $} & -\frac{1}{4} \\
 \text{$|$(2,-1,-1) , -1$\rangle $} & -\frac{1}{4} \\
 \text{$|$(1,2,1) , -1$\rangle $} & -\frac{i}{4} \\
 \text{$|$(1,2,-1) , -1$\rangle $} & -\frac{i}{4} \\
 \text{$|$(1,1,2) , -1$\rangle $} & -\frac{1}{4}+\frac{i}{4} \\
 \text{$|$(1,1,-2) , -1$\rangle $} & -\frac{1}{4}+\frac{i}{4} \\
 \text{$|$(1,-1,2) , -1$\rangle $} & \frac{1}{4}+\frac{i}{4} \\
 \text{$|$(1,-1,-2) , -1$\rangle $} & \frac{1}{4}+\frac{i}{4} \\
 \text{$|$(1,-2,1) , -1$\rangle $} & -\frac{i}{4} \\
 \text{$|$(1,-2,-1) , -1$\rangle $} & -\frac{i}{4}
\end{array}$\\
\hline
\hline
\end{tabular}
\begin{minipage}[t]{16.5 cm}
\vskip 0.5cm
\noindent
\end{minipage}
\end{center}
\end{table}

The $T_2^-$ irrep first appears in the $|{\bf n}|^2=2$-shell and $l=3$
is the lowest contributing partial-wave.  
LQCD calculations of correlation functions from sources and sinks transforming
as $T_2^-$ will provide determinations of $\delta_3$ with contamination from
partial-waves with $l\ge 5$.

\subsubsection{$A_1^-$ Representation\label{irrep:A1m}}
\begin{figure}
\centering
\includegraphics[height=\textwidth,angle=-90]{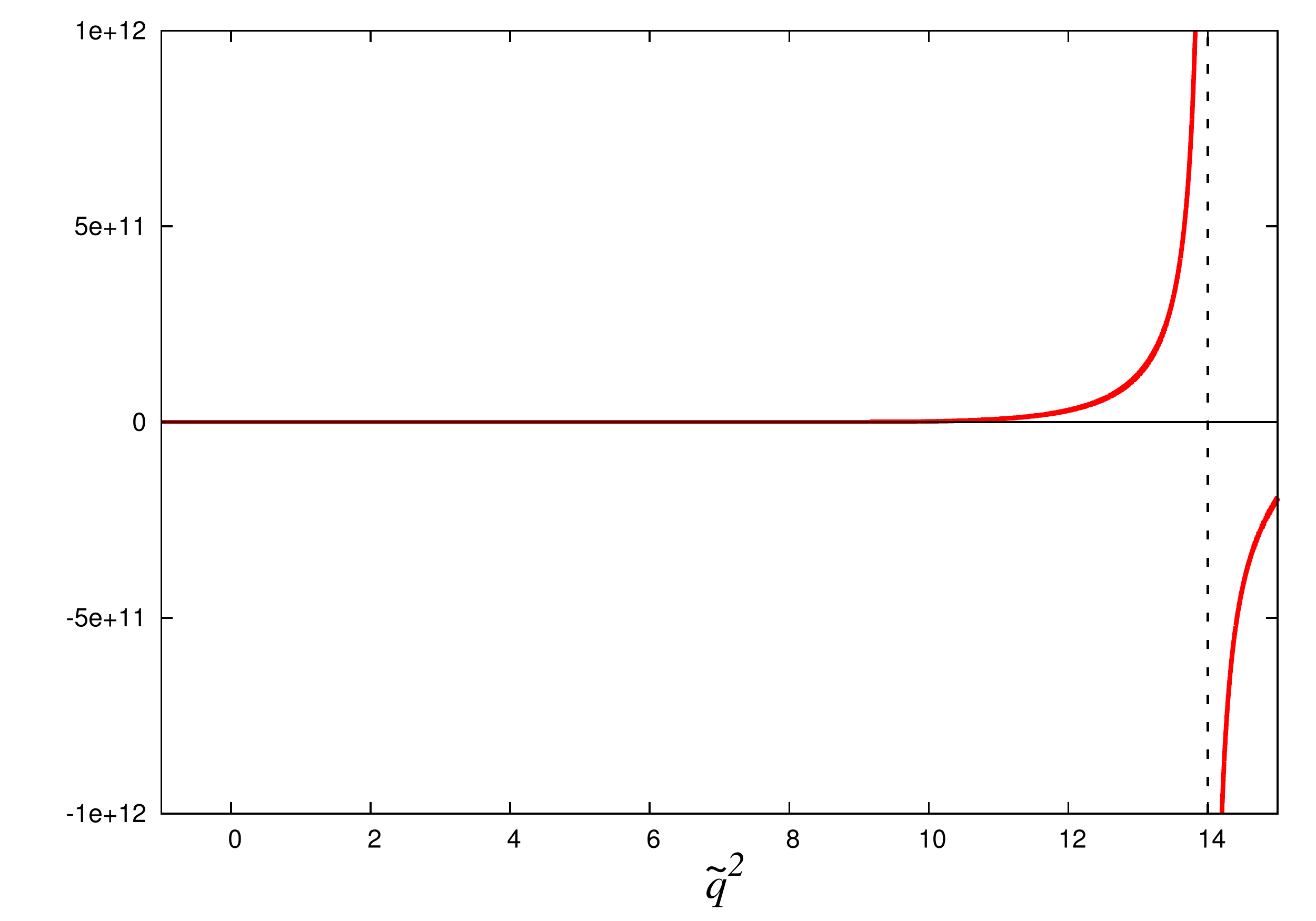}
\caption{The function $\mathcal{X}_{A_1}^-$, as defined in
  eq.~(\protect\ref{eqn:A1- l=9}), 
as a  function of $\tilde q^2$.  The vertical dashed line denotes the position of
  the first pole of the function corresponding to the non-interacting energy-eigenvalue. 
\label{fig:A1- l=9}
}
\end{figure}

The lowest-lying state transforming in the $A_1^-$ irrep is
in the $|{\bf n}|^2=14$-shell.  
The energy-eigenvalues are sensitive to interactions in odd partial-waves with
$l\ge 9$, and the 
energy-splitting in the large volume limit is dominated by the $l=9$
partial-wave. Using the methods of the previous section to isolate the state
and determine the appropriate energy-eigenvalue equation is tedious as
$F^{(FV)}_{9;9}$ is a $19\times 19$ matrix.   Using the following spherical basis state,
\begin{equation}\label{eqn:a1- basis state}
|A_1^-,0;9;1\rangle = \frac{1}{4}\sqrt{\frac{7}{3}}\left(|9,8\rangle-|9,-8\rangle\right)-\frac{1}{4}\sqrt{\frac{17}{3}}\left(|9,4\rangle-|9,-4\rangle\right)\ ,
\end{equation}
the eigenvalue-equation for the interaction in the 
$l=9$ partial-wave is
\begin{eqnarray}
q^{19} \cot\delta_9 & = & 
\left({2\pi\over L}\right)^{19}\ {1\over\pi^{3/2}}\ 
\left(  
\tilde q^{18} {\cal Z}_{0,0}(1;\tilde q^2)
- 
{6
\tilde q^{14} {\cal Z}_{4,0}(1;\tilde q^2) \over 23}
- 
{32\sqrt{13}
\tilde q^{12} {\cal Z}_{6,0}(1;\tilde q^2) \over 115}
\right. \nonumber  \\
&& \left.
- 
{56\sqrt{17}
\tilde q^{10} {\cal Z}_{8,0}(1;\tilde q^2) \over 345}
+ 
{1568\sqrt{7}
\tilde q^{8} {\cal Z}_{10,0}(1;\tilde q^2) \over 3335\sqrt{3}}
+
{308 \tilde q^{6} {\cal Z}_{12,0}(1;\tilde q^2) \over 2139}
\right. \nonumber \\ 
&& \left.
+
{616 \sqrt{1001} \tilde q^{6} {\cal Z}_{12,4}(1;\tilde q^2) \over 20677}
+
{53248 \tilde q^{4} {\cal Z}_{14,0}(1;\tilde q^2) \over 10695\sqrt{29}}
-
{1664\sqrt{11} \tilde q^{2} {\cal Z}_{16,0}(1;\tilde q^2) \over 3565\sqrt{3}}
\right. \nonumber \\
&& \left.
+
{832\sqrt{46189} \tilde q^{2} {\cal Z}_{16,4}(1;\tilde q^2) \over 103385\sqrt{7}}
+
{2206464 \tilde {\cal Z}_{18,0}(1;\tilde q^2) \over 103385\sqrt{37}}
+
{28288\sqrt{3553} {\cal Z}_{18,4}(1;\tilde q^2) \over 20677\sqrt{259}}
\  \right)
\nonumber\\
& = & 
\left({2\pi\over L}\right)^{19}\ {1\over\pi^{3/2}}\ 
\ \mathcal{X}_{A_1}^-\left(\tilde q^2\right)
\ \ \  ,
\label{eqn:A1- l=9}
\end{eqnarray}
where the function $\mathcal{X}_{A_1}^-\left(\tilde q^2\right)$ is shown in
fig.~\ref{fig:A1- l=9}.
\begin{figure}
\centering
\mbox{\includegraphics[width=.77\textwidth,angle=0]{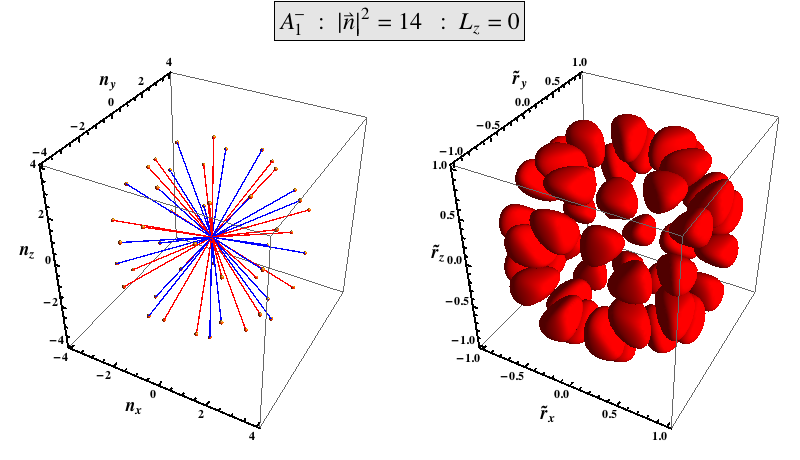}}
\caption{The momentum space representations (left) and position-space
  representations (right) of two-body relative states in the $A_1^-$
  representation with $L_z=0$ for the $|{\bf n}^2|=14$-shell.
\label{fig:A1- figures}}
\end{figure}
The graphical representations  of the
source and sink that generate the $A_1^-$ irrep for the 
$|{\bf  n}|^2=14$-shell
are shown in fig.~\ref{fig:A1- figures},
and  the momentum-space structure is given 
explicitly  in table~\ref{tab:A1mSOURCESSINKS}.
It is interesting to note that these  odd-parity singlet states
require the integers comprising the integer triplet to differ from each other.
The first $|{\bf n}|^2$-shell for which this is possible 
has $|{\bf  n}|^2=14$, and the next has $|{\bf n}|^2=21$.
\begin{table}
\begin{center}
\begin{minipage}[!ht]{16.5 cm}
\caption{
The momentum-space structure of $A_1^-$, $L_z=0$ sources and sinks.
These are shown graphically in fig.~\ref{fig:A1- figures}.
The coefficients of the state vectors are of the form
$c_i\sim \varepsilon^{ |n_x|,|n_y|,|n_z|} {\rm sgn}(n_x) {\rm sgn}(n_y) {\rm sgn}(n_z)$.
}
\label{tab:A1mSOURCESSINKS}
\end{minipage}
\begin{tabular}{|c|  }
\hline
\hline
$|{\bf n}|^2$=14 \\ \hline 
$\begin{array}{cc|cc|cc}
 \text{$|$(1,2,3) , -1$\rangle $} &   \frac{1}{\sqrt{24}} & \text{$|$(1,2,-3) , -1$\rangle $} & -\frac{1}{\sqrt{24}} & \text{$|$(1,-2,3) , -1$\rangle $} & -\frac{1}{\sqrt{24}} \\
 \text{$|$(1,-2,-3) , -1$\rangle $} & \frac{1}{\sqrt{24}} & \text{$|$(1,3,2) , -1$\rangle $} &  -\frac{1}{\sqrt{24}} & \text{$|$(1,3,-2) , -1$\rangle $} &  \frac{1}{\sqrt{24}} \\
 \text{$|$(1,-3,2) , -1$\rangle $} &  \frac{1}{\sqrt{24}} & \text{$|$(1,-3,-2) , -1$\rangle $}& -\frac{1}{\sqrt{24}} & \text{$|$(2,1,3) , -1$\rangle $} &  -\frac{1}{\sqrt{24}} \\
 \text{$|$(2,1,-3) , -1$\rangle $} &  \frac{1}{\sqrt{24}} & \text{$|$(2,-1,3) , -1$\rangle $} &  \frac{1}{\sqrt{24}} & \text{$|$(2,-1,-3) , -1$\rangle $} &-\frac{1}{\sqrt{24}} \\
 \text{$|$(2,3,1) , -1$\rangle $} &   \frac{1}{\sqrt{24}} & \text{$|$(2,3,-1) , -1$\rangle $} & -\frac{1}{\sqrt{24}} & \text{$|$(2,-3,1) , -1$\rangle $} & -\frac{1}{\sqrt{24}} \\
 \text{$|$(2,-3,-1) , -1$\rangle $} & \frac{1}{\sqrt{24}} & \text{$|$(3,2,1) , -1$\rangle $} &  -\frac{1}{\sqrt{24}} & \text{$|$(3,2,-1) , -1$\rangle $} &  \frac{1}{\sqrt{24}} \\
 \text{$|$(3,-2,1) , -1$\rangle $} &  \frac{1}{\sqrt{24}} & \text{$|$(3,-2,-1) , -1$\rangle $} &-\frac{1}{\sqrt{24}} & \text{$|$(3,1,2) , -1$\rangle $} &   \frac{1}{\sqrt{24}} \\
 \text{$|$(3,1,-2) , -1$\rangle $} & -\frac{1}{\sqrt{24}} & \text{$|$(3,-1,2) , -1$\rangle $} & -\frac{1}{\sqrt{24}} & \text{$|$(3,-1,-2) , -1$\rangle $}&  \frac{1}{\sqrt{24}} \\
\end{array}$
\\
\hline
\hline
\end{tabular}
\begin{minipage}[t]{16.5 cm}
\vskip 0.5cm
\noindent
\end{minipage}
\end{center}
\end{table}     
Given the first appearance of this irrep is high in the spectrum, 
a LQCD calculation of the $l=9$ phase-shift will require enormous lattice
volumes in order for the state 
to lie below inelastic thresholds.
Thus, this 
calculation cannot be expected to be performed in the near future.

\section{Discussion\label{sect:discussion}}

\subsection{Strategy for Extracting Phase-Shifts from Lattice QCD}
\noindent
In Lattice QCD calculations, sources and sinks with the quantum numbers of the
hadronic states of interest generate correlation functions, which in general
are sums of exponentials with arguments that depend upon the energy of the
eigenstates in the lattice volume.
One path for LQCD calculations to follow is to form
sources and/or sinks
that transform as irreps of the cubic group  from the eigenstates of
linear-momentum (generated by the Fourier-transform of single-hadron objects).
Clearly, such sources are not the energy-eigenstates in the lattice-volume due
to the interactions between the particles, and as such these sources and sinks
will couple, in principle, to all states in the lattice-volume with the appropriate quantum
numbers.  At large times, the correlation function will depend exponentially
upon the energy of the lowest eigenstate. 
For instance, the correlation function resulting from 
an  $A_1^+$ source constructed from eigenstates with 
$|{\bf n}|^2 \gg 0$ will depend upon the energy of the
lowest-lying $A_1^+$ irrep at large times.
Of course, the overlap onto the lowest-lying
$A_1^+$ irrep may be small, in which case the ground-state dominates only after a
large number of time-slices.  
It is then clear that the optimal way to extract the $\delta_l$ for $l\le 6$ is
to determine the lowest energy-eigenvalue 
of the cubic irrep that is dominated by each $\delta_l$.
These irreps are shown in table~\ref{tab:Lextracttab}, along with the 
$|{\bf n}|^2$-shell of the lowest-lying energy-eigenstate that contributes to
the corresponding correlation functions.
\begin{table}
\begin{center}
\begin{minipage}[!ht]{16.5 cm}
\caption{
The $|{\bf n}|^2$-shell of the lowest-lying energy-eigenstate  
transforming as $\Gamma^{(i)}$, and the
angular-momentum of the dominant 
interaction in the large-volume limit.
}
\label{tab:Lextracttab}
\end{minipage}
\begin{tabular}{| c | c | c |}
\hline 
  \ \ \ $|{\bf n}|^2$ \ \ \   
& \ \ \ $\Gamma$\ \ \  
& \ \ \ $l$\ \ \ 
    \\
      \hline
$ 0$ & $A_1^+$ & $0$  \\
$ 1$ & $T_1^-$ & $1$  \\
$ 1$ & $E^+$   & $2$  \\
$ 2$ & $T_2^+$ & $2$  \\
$ 2$ & $T_2^-$ & $3$  \\
$ 3$ & $A_2^-$ & $3$  \\
$ 5$ & $T_1^+$ & $4$  \\
$ 5$ & $A_2^+$ & $6$  \\
$ 6$ & $E^-$   & $5$  \\
$ 14$ & $A_1^-$   & $9$  \\
\hline
\end{tabular}
\begin{minipage}[t]{16.5 cm}
\vskip 0.5cm
\noindent
\end{minipage}
\end{center}
\end{table}     
Table~\ref{tab:Lextracttab} shows that with just the lowest two 
$|{\bf n}|^2$-shells, $|{\bf n}|^2=0,1$, the phase-shifts in the lowest three
partial-waves, $\delta_{0,1,2}$, can be determined.  In order to determine the
phase-shifts in all partial-waves with $l\le 6$, correlation functions must be
formed for states that have ground-states in the shells up to
$|{\bf n}|^2=6$.
Determining the phase-shift for $l\ge 7$ can be seen to be decidedly more
difficult than for $l\le 6$ as there is only one further irrep of the cubic
group, the $A_1^-$ which first occurs in the $|{\bf n}|^2=14$ and is dominated
by the interactions in the $l=9$ partial-wave in the large volume limit.

For shells in which there are 
multiple occurrences of a given $\Gamma^{(i)}$, 
the partial-diagonalization of the states in the infinite-volume limit
in terms of the angular-momentum of the interactions is possible. 
However, sources cannot be constructed to isolate these states due to
interactions, and in general, closely spaced-states will be encountered in the spectrum.

\subsection{Expectations for the $\pi\pi$ Energy-Eigenvalues}
\noindent
In order to estimate the computational resources required to extract the
$\pi\pi$ phase-shifts in higher partial-waves, the experimentally determined
(and parameterized) phase-shifts can be used to determine the
energy-eigenvalues for a range of lattice volumes.
\begin{figure}[h]
  \centering
     \includegraphics[width=0.49\textwidth]{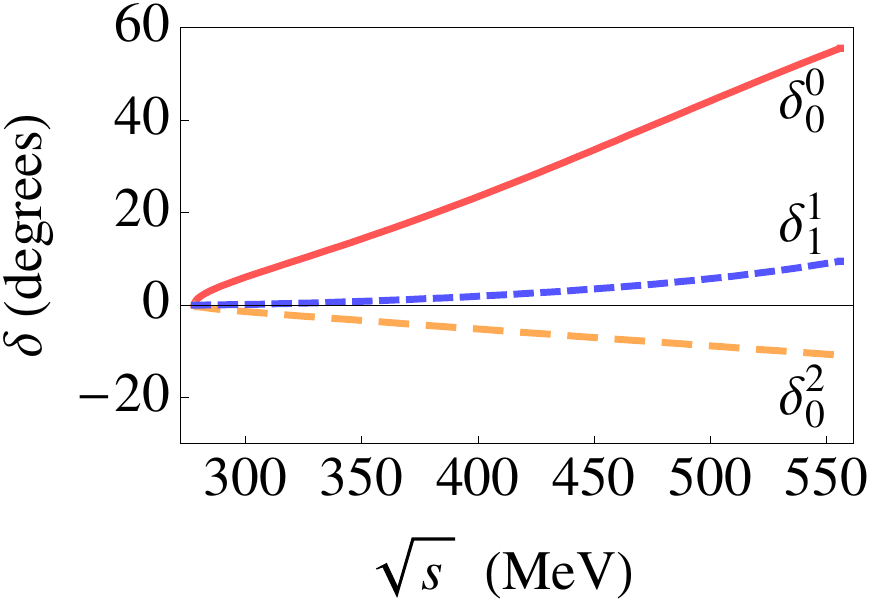}
     \includegraphics[width=0.49\textwidth]{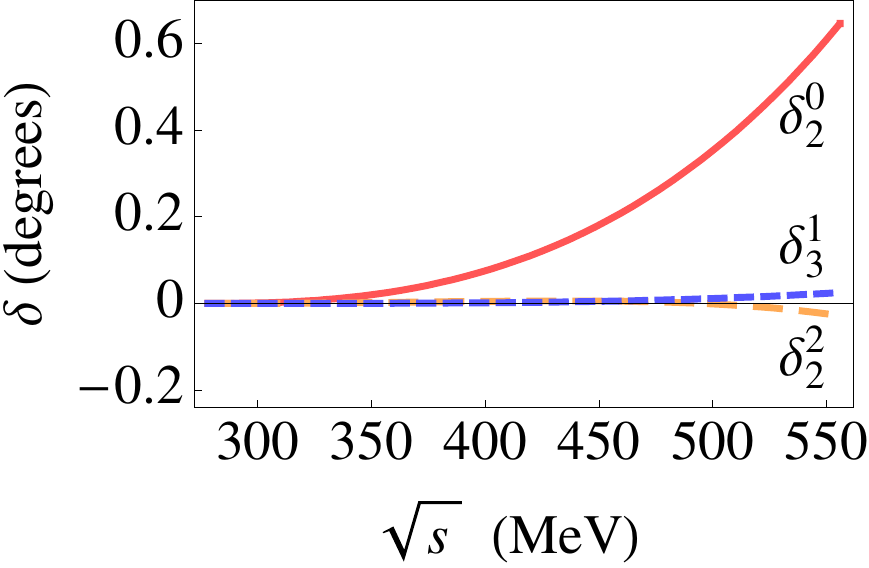}
     \caption{ The $\pi\pi$ phase-shifts, $\delta_l^I$, as a function of
       $\sqrt{s} < 4 m_\pi$ for $l=0,1,2,3$, 
as parameterized in Ref.~\protect\cite{Kaminski:2006qe}.
}
  \label{fig:pipiexpt}
\end{figure}
The $\pi\pi$ phase-shifts for $l=0, 1, 2$, and 3 partial-waves 
extracted from experimental data and parameterized with functions that satisfy
unitarity and analyticity,  and specifically incorporate any lowest-lying
resonances in the channel~\cite{Kaminski:2006qe}~\footnote{
The real-part of the inverse scattering
  amplitude, $\cot\delta_l^I$, is expanded as a power-series in the function
$w(s) = {\sqrt{s}-\sqrt{s_i-s}\over \sqrt{s}+\sqrt{s_i-s}}$,
where $s_i$ is the energy above which inelastic processes cannot be neglected.
}
are shown in fig.~\ref{fig:pipiexpt}.  
The central values of the parameters describing each partial-wave
provided in Ref.~\cite{Kaminski:2006qe} are used in the analysis,
but, as only
estimates of the energy-eigenvalues are being explored,  a systematic
propagation of the uncertainties has not been performed.  Further, we assume isospin symmetry in our analysis.
As L\"uscher's formalism is valid below inelastic thresholds,
only the phase-shifts in the kinematic regime $\sqrt{s}<4 m_\pi$ 
(at the physical pion mass) are considered.

\begin{figure}[h]
  \centering
     \includegraphics[width=0.75\textwidth]{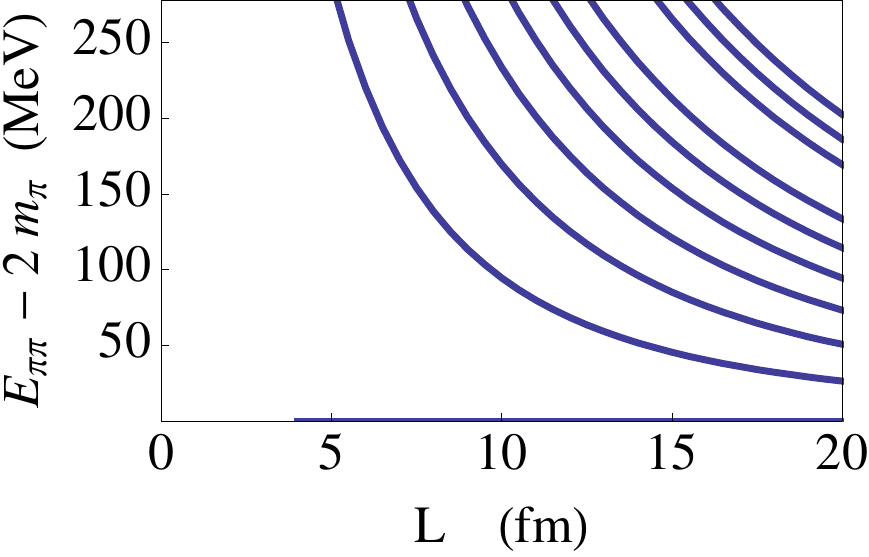}
     \caption{ The energy-eigenvalues associated with two non-interacting pions as a
       function of the spatial-extent of the lattice volume, $L$.  
The spectrum results from the momentum of each pion being restricted to 
${\bf q} =
       {2\pi\over L} {\bf n}$ for all possible triplets of integers, ${\bf n}$,
       due to the periodic boundary conditions imposed on the quark and gluon fields.
The maximum value shown on the vertical-axis corresponds to  
the inelastic threshold $\sqrt{s}=4 m_\pi$. 
For reference, a spatial-extent
of $L=6~{\rm fm}$ corresponds to $m_\pi L\sim 4.2$.
}
  \label{fig:pipifree}
\end{figure}
\begin{figure}[h]
  \centering
     \includegraphics[width=0.48\textwidth]{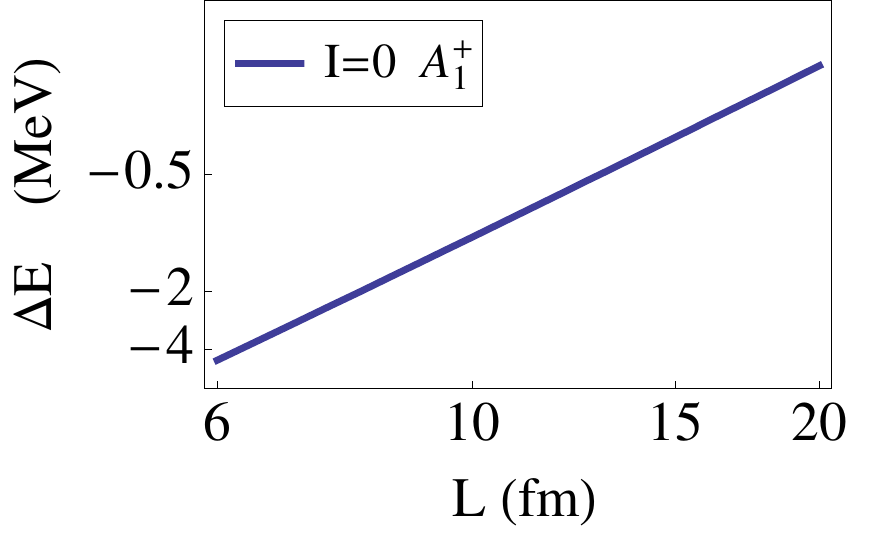}
     \includegraphics[width=0.48\textwidth]{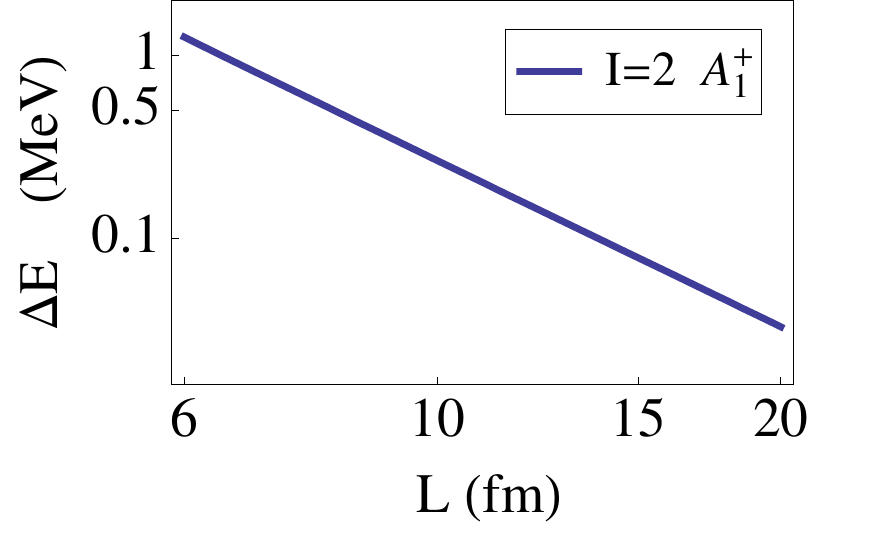}
     \caption{ The expected $\pi\pi$ energy-shifts in the $|{\bf n}|^2=0$ shell due to 
strong interactions.  The left panel shows the shift in the $I=0$ $A_1^+$ irrep 
(dominated by $\delta_0^0$), while the right panel 
 shows the shift in the $I=2$ $A_1^+$ irrep 
(dominated by $\delta_0^2$).
Both the L-axis and the $\Delta E$-axis are scaled logarithmically ($\log_{10}$).
}
  \label{fig:N2eq0Ieq0AND2PeqEven}
\end{figure}
\begin{figure}[h]
  \centering
     \includegraphics[width=0.32\textwidth]{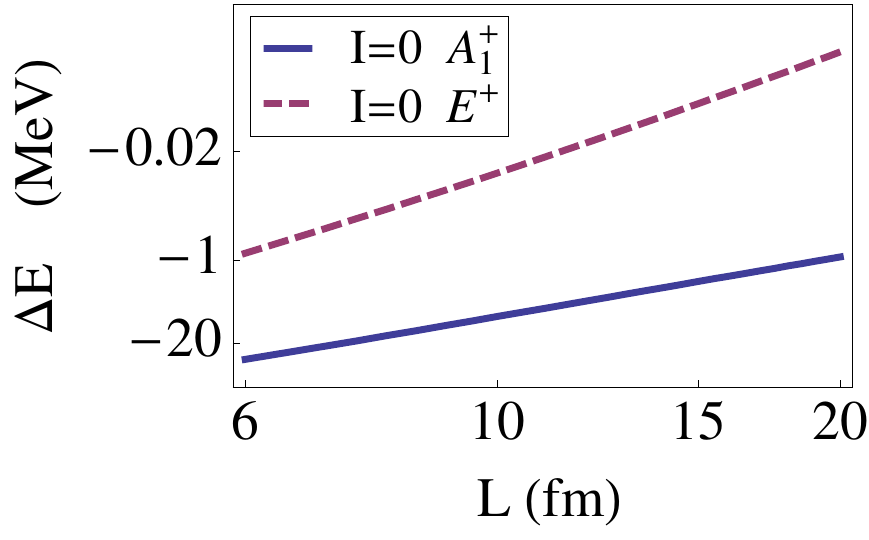}
     \includegraphics[width=0.32\textwidth]{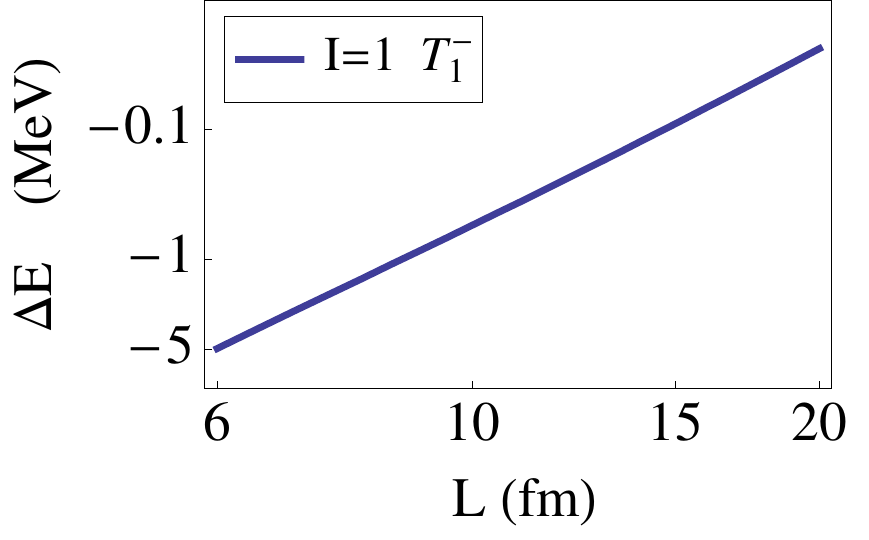}
     \includegraphics[width=0.32\textwidth]{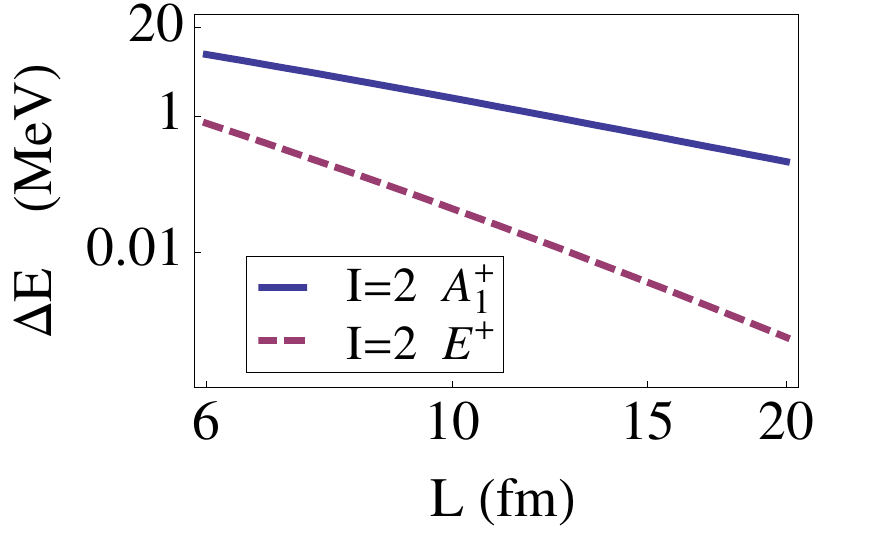}
     \caption{ The  expected $\pi\pi$ energy-shifts
in the $|{\bf n}|^2=1$ shell due to 
strong interactions.  The left panel shows the shift in the $I=0$ $A_1^+$, $E^+$ irreps
(dominated by $\delta_0^0$ and $\delta_2^0$, respectively), 
the center panel shows the shift in the $I=1$ $T_1^-$ irrep 
(dominated by $\delta_1^1$) 
and the right panel 
 shows the shift in the $I=2$ $A_1^+$, $E^+$ irreps 
(dominated by $\delta_0^2$ and $\delta_2^2$, respectively).
Both the L-axis and the $\Delta E$-axis are scaled logarithmically ($\log_{10}$).
}
  \label{fig:N2eq1Ieq0AND1AND2PeqEven}
\end{figure}
\begin{figure}[h]
  \centering
     \includegraphics[width=0.32\textwidth]{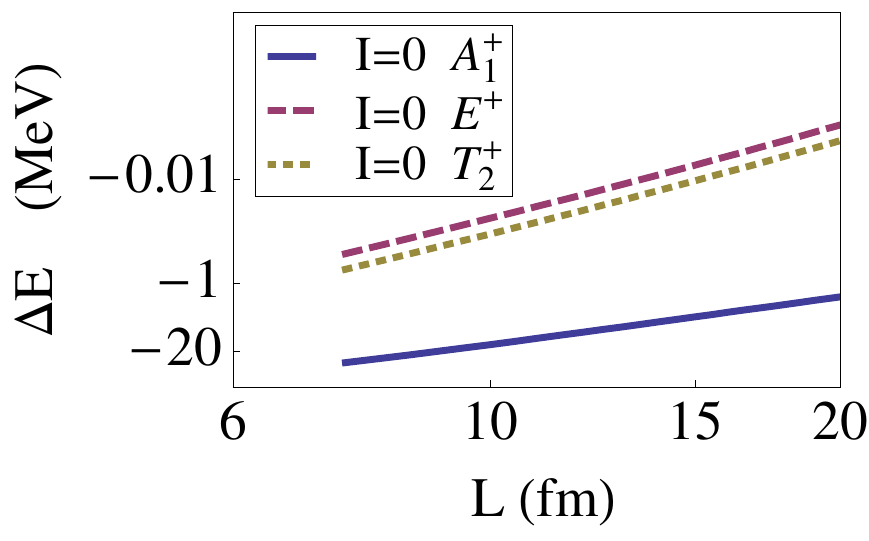}
     \includegraphics[width=0.32\textwidth]{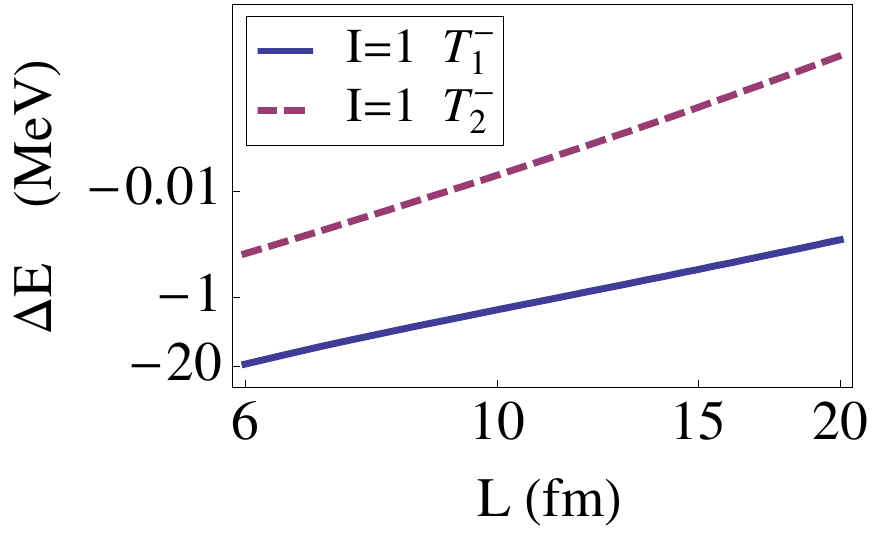}
     \includegraphics[width=0.32\textwidth]{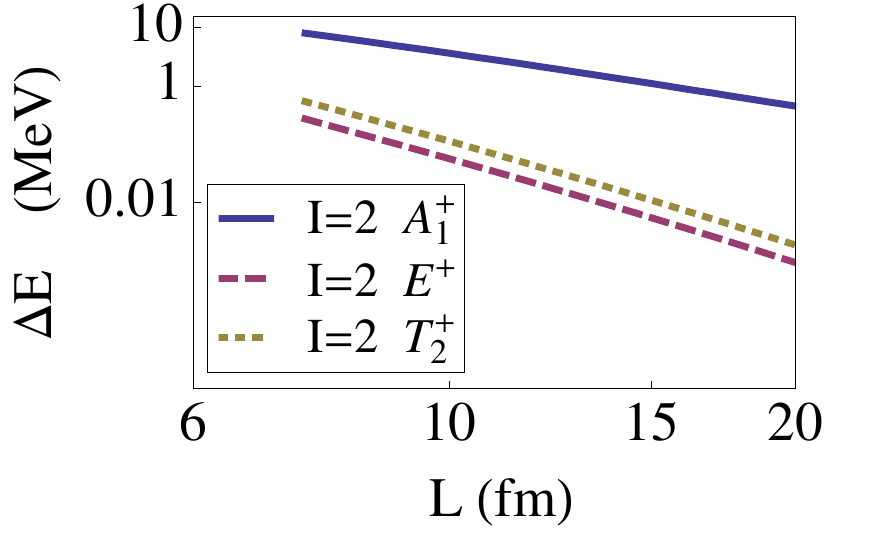}
     \caption{ The  expected $\pi\pi$ energy-shifts
in the $|{\bf n}|^2=2$ shell due to 
strong interactions.  The left panel shows the shift in the $I=0$ $A_1^+$,
$E^+$, $T_2^+$  irreps 
(dominated by $\delta_0^0$, $\delta_2^0$ and $\delta_2^0$, respectively), 
the center panel shows the shift in the $I=1$ $T_1^-$, $T_2^-$ irrep 
(dominated by $\delta_1^1$ and  $\delta_3^1$) 
and the right panel 
 shows the shift in the $I=2$ $A_1^+$, $E^+$, $T_2^+$ irreps 
(dominated by $\delta_0^2$,  $\delta_2^2$ and $\delta_2^2$, respectively).
Both the L-axis and the $\Delta E$-axis are scaled logarithmically ($\log_{10}$).
}
  \label{fig:N2eq2Ieq0AND1AND2}
\end{figure}
\begin{figure}[h]
  \centering
     \includegraphics[width=0.32\textwidth]{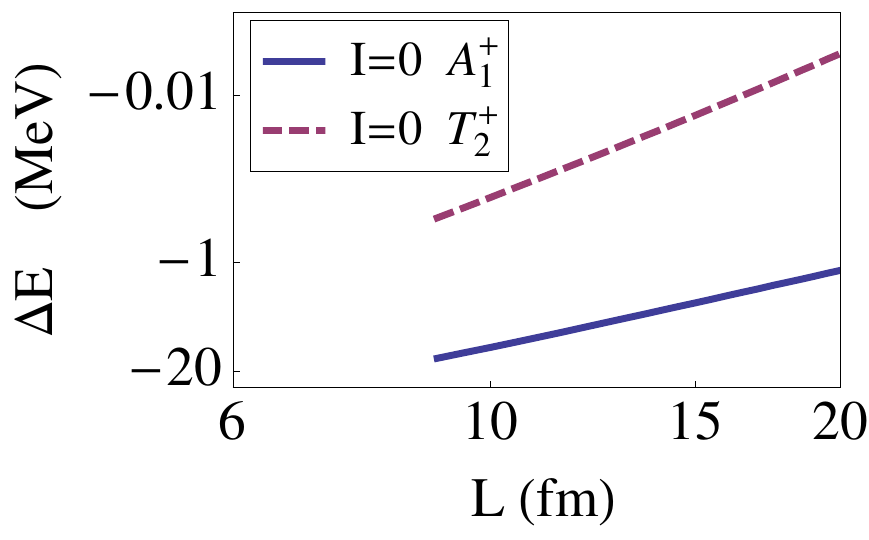}
     \includegraphics[width=0.32\textwidth]{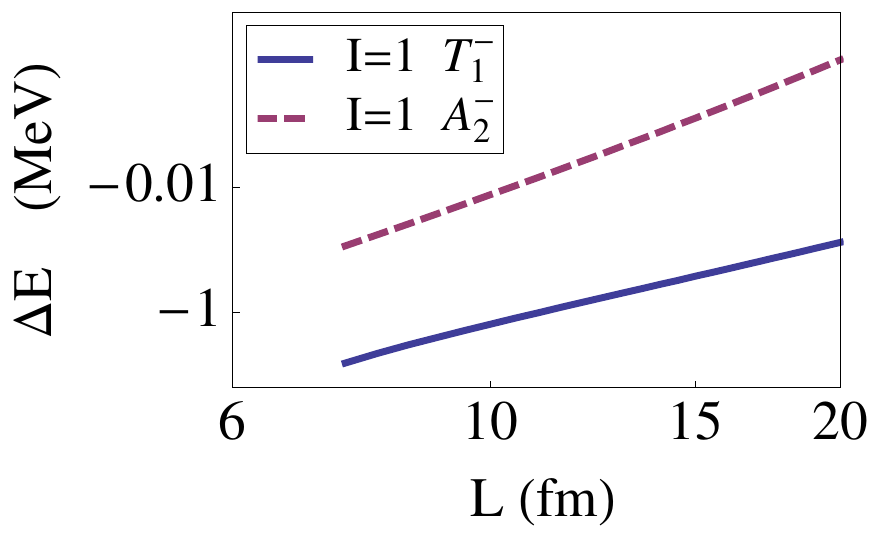}
     \includegraphics[width=0.32\textwidth]{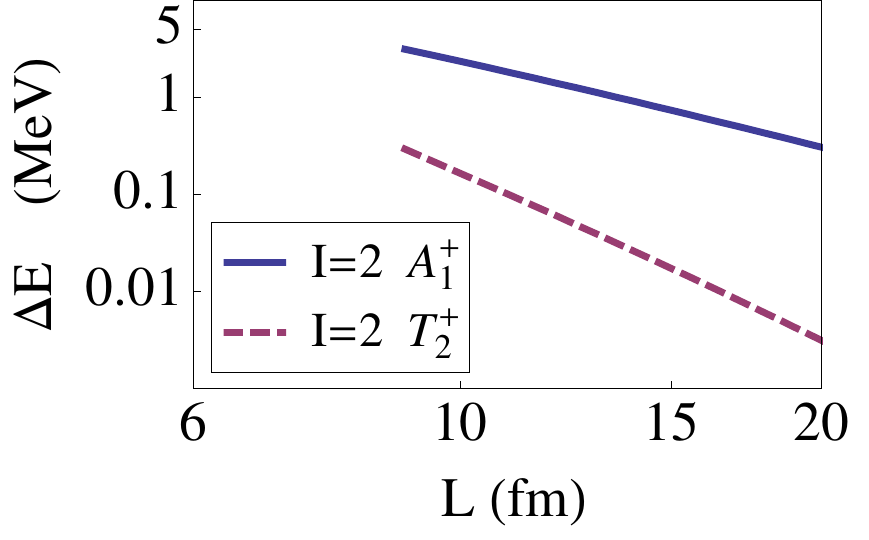}
     \caption{ The  expected $\pi\pi$ energy-shifts
in the $|{\bf n}|^2=3$ shell due to 
strong interactions.  The left panel shows the shift in the $I=0$ $A_1^+$, $T_2^+$  irreps 
(dominated by $\delta_0^0$ and $\delta_2^0$, respectively), 
the center panel shows the shift in the $I=1$ $T_1^-$, $A_2^-$ irreps 
(dominated by $\delta_1^1$ and  $\delta_3^1$, respectively) 
and the right panel 
 shows the shift in the $I=2$ $A_1^+$, $T_2^+$ irreps 
(dominated by $\delta_0^2$ and $\delta_2^2$, respectively).
Both the L-axis and the $\Delta E$-axis are scaled logarithmically ($\log_{10}$).
}
  \label{fig:N2eq3Ieq0AND1AND2}
\end{figure}
\begin{figure}[h]
  \centering
     \includegraphics[width=0.32\textwidth]{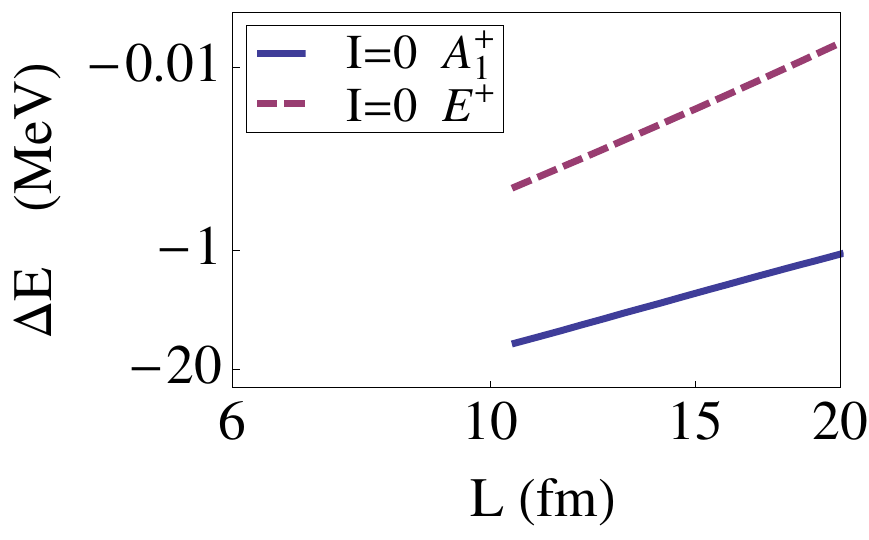}
     \includegraphics[width=0.32\textwidth]{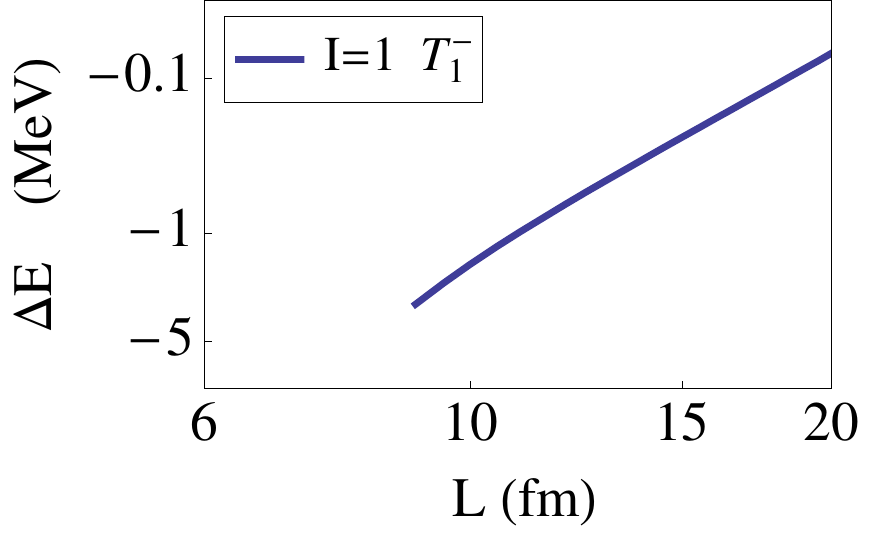}
     \includegraphics[width=0.32\textwidth]{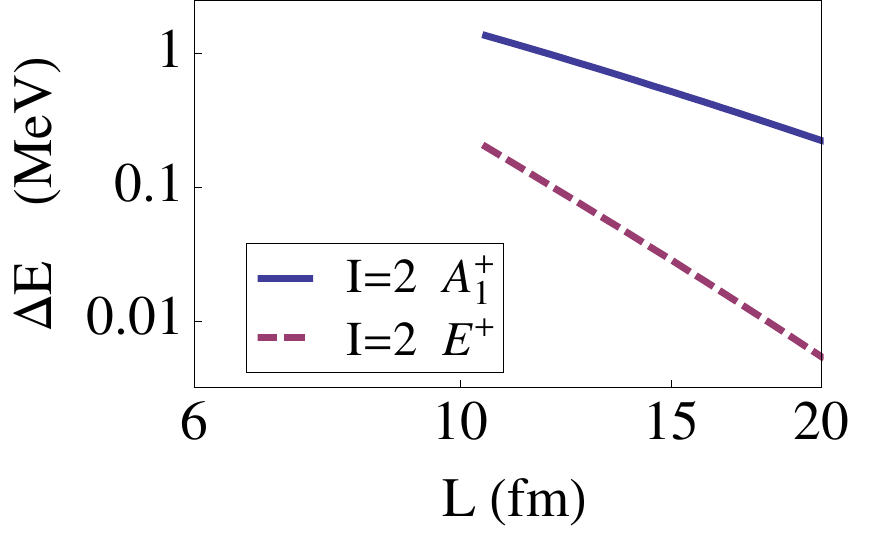}
     \caption{ The  expected $\pi\pi$ energy-shifts
in the $|{\bf n}|^2=4$ shell due to 
strong interactions.  The left panel shows the shift in the $I=0$ $A_1^+$,
$E^+$ irreps
(dominated by $\delta_0^0$ and $\delta_2^0$, respectively), 
the center panel shows the shift in the $I=1$ $T_1^-$ irrep 
(dominated by $\delta_1^1$) 
and the right panel 
 shows the shift in the $I=2$ $A_1^+$, $E^+$ irreps 
(dominated by $\delta_0^2$ and $\delta_2^2$, respectively).
Both the L-axis and the $\Delta E$-axis are scaled logarithmically ($\log_{10}$).
}
  \label{fig:N2eq4Ieq0AND1AND2}
\end{figure}

Figure~\ref{fig:pipifree} shows the energy-eigenvalues associated with two 
non-interacting pions 
with vanishing total momentum
in the lattice volume
(also shown  in fig.~1 of Ref.~\cite{Luscher:1986pf}).
At the physical pion mass, 
LQCD calculations in
volumes with $L\gsim 6~{\rm fm}$ are highly desirable
in order to suppress the exponential corrections that are not included in the
formalism of L\"uscher \cite{Bedaque:2006yi}.
The energy-shift between the
non-interacting state and the interacting 
$I=0$ $A_1^+$ and $I=2$ $A_1^+$ 
states in  the  $|{\bf n}|^2=0$ shell are shown in
fig.~\ref{fig:N2eq0Ieq0AND2PeqEven}.
The energy-shifts for the 
eigenstates in the
$|{\bf n}|^2=1,2,3,4$ shells are shown in 
figs.~\ref{fig:N2eq1Ieq0AND1AND2PeqEven}-\ref{fig:N2eq4Ieq0AND1AND2}, respectively.
The energy-shifts for the $|{\bf n}|^2=0$ $A_1^+$ states and the $|{\bf n}|^2=1$
$T_1^-$ state can also be found in fig.~7 of Ref.~\cite{Luscher:1986pf}.
The energy-shifts of the states due to the s-wave and p-wave interactions
are of comparable size.
As the s-wave interactions are currently
being calculated in volumes
with $L\sim 3.5~{\rm fm}$, we do not anticipate
significant difficulty in performing these calculations at the physical pion mass
in lattices with $L\gsim 6~{\rm fm}$.
In contrast, the energy-shifts of states due to the d-wave ($l=2$) and f-wave
($l=3$)
interactions are
more than an order of magnitude smaller than those of the $A_1^+$ irrep.
Significantly more computational resources will be
required to extract the phase-shifts beyond the s-wave and p-wave.
It is difficult to make estimates for the energy-shifts due to 
interactions beyond the 
f-wave as the experimental measurements of these
phase-shifts  have large uncertainties
or are absent.  Given the results obtained for $l\le 3$, it is 
not difficult to speculate as to the size of the energy-shifts of
partial-waves beyond $l=3$, and the associated difficulty in their extraction from
LQCD calculations.

In order to estimate the amount of mixing of higher partial waves to a given phase-shift
from the energy-eigenvalues,
it is important to understand the expected contributions 
from (all of) the partial-waves. 
\begin{figure}[h]
  \centering
     \includegraphics[width=0.5\textwidth]{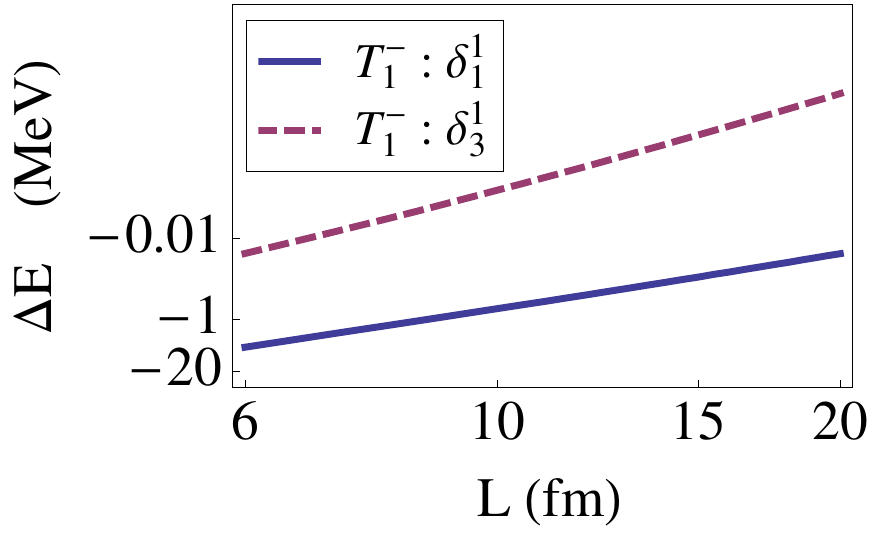}
     \caption{The contributions to the energy-splitting of the $T_1^-$ irrep in
       the $|{\bf n}|^2=1$-shell due to $\delta_1^1$ and $\delta_3^1$.
Both the L-axis and the $\Delta E$-axis are scaled logarithmically ($\log_{10}$).
}
  \label{fig:N2eq1L1L3}
\end{figure}
The energy-splitting of the $T_1^-$ irrep in
the $|{\bf n}|^2=1$-shell from the $l=1$ phase-shift, $\delta_1^1$, 
and the $l=3$ phase-shift, $\delta_3^1$, are shown in fig.~\ref{fig:N2eq1L1L3}.
As expected, the contribution from $\delta_3^1$ is approximately
two-orders of magnitude smaller than that from $\delta_1^1$ over the range of
lattice volumes for which the analysis is applicable.
Therefore, to high precision it is sufficient to use the perturbative expansion
of the energy-splitting in terms of $\tan\delta_1^1$.

\subsection{Signal-to-Noise Issues} 
\noindent
There is also a signal-to-noise ``problem'' in 
the extraction of the $\delta_l$ for $l\ge 1$ as 
the signal-to-noise ratio degrades exponentially at large times.
To demonstrate this behavior we return to the argument given by
Lepage~\cite{Lepage:1989hd}. 
Consider the correlation function resulting from a
source that creates a $\pi^+\pi^+$-state that
transforms in the $E^+$ irrep of the cubic group, 
\begin{eqnarray}
\overline{x}(t)\ =\   \langle \hat \theta_{E^+}(t)\rangle &=&
  \ \langle 0 | \ S_{E^+}(t) S_{E^+}^\dagger (0)
  \ |0\rangle
  \ \rightarrow\ Z_{E^+} \ e^{- E_{0,E^+}^{(\pi^+\pi^+)} t}
  \ \ ,
  \label{eq:GfunEplus}
\end{eqnarray}
where $S_{E^+}(t)$ annihilates a $\pi^+\pi^+$ in the $E^+$ irrep at the time
$t$.
At large times this correlation function depends exponentially upon the
ground-state energy which, in the absence of interactions, is 
$E_{0,E^+}^{(\pi^+\pi^+)} \ =\ 2 \sqrt{ \left({2\pi\over L}\right)^2 + m_\pi^2}$.
The variance of this correlation function is given by 
\begin{eqnarray}
\sigma^2(t)\ =\  \langle \left( \hat \theta_{E^+} (t)\right)^2\rangle 
\ -\ \langle \hat \theta_{E^+}(t)\rangle^2
&=&
\ \langle 0 | \ S_{E^+}(t) S^\dagger_{E^+}(t) S_{E^+}^\dagger (0) S_{E^+} (0)
\ |0\rangle
\ -\ 
\ \langle \hat \theta_{E^+}(t)\rangle^2
\nonumber\\
&=&
\sum_{\Gamma\in E^+\otimes E^+} 
\ C_\Gamma\ 
\ \langle 0 | \ \tilde {\cal S}_{\Gamma}(t) \tilde {\cal S}_\Gamma(0)  \ |0\rangle
\ -\ 
\ \langle \hat \theta_{E^+}(t)\rangle^2
\nonumber\\
& \rightarrow & \tilde Z_{A_1^+} \ e^{- 2 E_{0,A_1^+}^{(\pi^+\pi^+)} t}
\ -\ 
 Z_{E^+}^2 \ e^{- 2 E_{0,E^+}^{(\pi^+\pi^+)} t}
  \ \ ,
  \label{eq:Gfunvar}
\end{eqnarray}
where $C_\Gamma$ are the Clebsch-Gordan coefficients in the expansion 
$E^+\otimes E^+ = A_1^+\oplus A_2^+\oplus
E^+$~\cite{Dresselhaus,Moore:2006ng,Basak:2005ir}. The energy, $ \Delta_{\rm NS}$, that dictates the
long-time behavior of the variance correlation function is that of the
lowest-lying irrep composed of four pions, which is the lowest-lying
$A_1^+$-irrep that has an energy of 
$2 E_{0,A_1^+}^{(\pi^+\pi^+)} \ =\ 4 m_\pi$ in the absence of interactions.
Therefore, at large times, the noise-to-signal ratio behaves as 
\begin{eqnarray}
{\sigma(t)\over \overline{x}(t)} & \rightarrow & 
{ \sqrt{ \tilde Z_{A_1^+}}\over Z_{E^+}}\ 
e^{ \Delta_{\rm NS} t}
\ \ \  ,\ \ \ 
 \Delta_{\rm NS} \ = \ 2\ \left[\ \sqrt{ \left({2\pi\over L}\right)^2 + m_\pi^2}
- m_\pi\ \right]
\ \ \ ,
\end{eqnarray}
which grows exponentially at large times. 

This argument generalizes to all of the cubic
irreps, and the extraction of the $\delta_l$ for each $l\ge 1$ suffers from
a signal-to-noise problem, with an energy-scale that is approximately 
\begin{eqnarray}
\Delta_{\rm NS}^{(|{\bf n}|^2)} \ = \ 2\ \left[\ \sqrt{ \left({2\pi\over
        L}\right)^2 |{\bf n}|^2 + m_\pi^2}
- m_\pi\ \right]
\ \ \ .
\end{eqnarray}
Obviously, in large volumes, 
the degradation of the signal 
obtained for low-lying states
will  not
dramatically impact the determination of the energy-eigenvalues as the energy-scale
behaves as $\Delta_{\rm NS}^{(|{\bf n}|^2)} \rightarrow |{\bf n}|^2/(m_\pi
L^2)$.  However, for present-day calculations in modest lattice
volumes, the degradation of the signal may impact the extraction of the
phase-shifts in higher partial-waves,
and numerical exploration is required to determine its impact.

\section{Conclusion
\label{sect:conclusion}}
\noindent
We have explored the phenomenology of L\"uscher's method in the extraction of the
phase-shifts in higher partial-waves describing meson-meson scattering below
inelastic thresholds using Lattice QCD,
the formalism for which is contained in the works of 
L\"uscher. The lowest-lying s-wave and p-wave interactions
were explored in those works, and at the time, Lattice QCD calculations of
scattering beyond the s-wave and p-wave were in the distant future.  
However, the rapidly increasing computational resources that are being directed
towards Lattice QCD calculations will allow for the calculation of the
phase-shifts in higher partial-waves in both mesonic and baryonic processes in
the near future.
We have considered the low-lying spectrum of two-meson states in a finite
cubic lattice-volume, and determined contributions from the partial-waves with
$l\le 6$ and $l=9$.  
There are a sufficient number of irreps of the cubic group that
will allow for the calculation of the phase-shifts, $\delta_l$
for $l\le 6$, and possibly $l=9$.
There are no irreps of the cubic group with ground-state energy-splittings
that are dominated by
interactions in the $l=7$ and $l=8$ partial-waves.  As such, there appears to
be no clean way to calculate these phase-shifts from Lattice QCD calculations
performed in cubic volumes.  
High precision calculations of the
energy of states in other irreps may allow for their extraction by forming
differences of energies, but this will require significantly more computational
resources than the extraction of the phase-shifts in lower partial-waves.
We have provided the structure of  sources that will produce
the irreps of the cubic group, in both momentum and position space, that will
generate the relevant states in  LQCD calculations.  
Further, we have given
the 
explicit formula, and their perturbative solutions, that are required to
analyze the results of such LQCD calculations.
We recapitulate the leading contributions to the 
energy-eigenvalue equations and their solutions in the large-volume
limit in  table~\ref{tab:recap} and table~\ref{tab:LowestPerturb}, respectively.
\begin{table}
\begin{center}
\begin{minipage}[!ht]{16.5 cm}
\caption{
A summary of 
the energy-eigenvalue equations for the lowest-lying state in each
$\Gamma^{(i)}$
arising from the interaction in the dominant partial-wave.
}
\label{tab:recap}
\end{minipage}
\begin{tabular}{| c | l | c | }
\hline\hline
 & & \\
\ \ $l$ [$\Gamma^{(i)}$]\ \  & \qquad Leading Eigenvalue Equation \qquad
&  \ \ Section  \qquad \\
 & & \\
\hline 
 & & \\
0 [$A_1^+$] \qquad
& \ 
$q\text{cot}\delta_0\ =\ \frac{2}{\sqrt{\pi} L}\
\mathcal{Z}_{0,0}\left(1;\tilde q^2\right)$ &  
\ref{irrep:A1p}\\
 & & \\
\hline
 & & \\ 
1 [$T_1^-$] \qquad 
& \ 
$q^3\text{cot}\delta_1\ =\left(\frac{2\pi}{L}\right)^3
\frac{1}{\pi^{3/2}}\tilde q^2\mathcal{Z}_{0,0}\left(1;\tilde q^2\right)$ &  \ref{irrep:T1m}\\
 & & \\
\hline
 & & \\
2 [$E^+$]  \qquad
& \ 
$q^5\text{cot}\delta_2\ =\left(\frac{2\pi}{L}\right)^5\frac{1}{\pi^{3/2}}
\left(\tilde q^4 \mathcal{Z}_{0,0}\left(1;\tilde q^2\right)
+\frac{6}{7} \mathcal{Z}_{4,0}\left(1;\tilde q^2\right)\right)$ & \ref{irrep:Ep}\\
2 [$T_2^+$]  \qquad 
& \ $q^5\text{cot}\delta_2\ =
\left(\frac{2\pi}{L}\right)^5\frac{1}{\pi^{3/2}}
\left(\tilde{q}^4 \mathcal{Z}_{0,0}\left(1;\tilde{q}^2\right)
-\frac{4 }{7}\mathcal{Z}_{4,0}\left(1;\tilde{q}^2\right)\right)$ & \ref{irrep:T2p}\\
 & & \\
\hline
 & & \\
3 [$T_2^-$]  \qquad
& \ 
$q^7\text{cot}\delta_3\ =\left(\frac{2\pi}{L}\right)^7\frac{1}{\pi^{3/2}}
   \left(\tilde q^6\mathcal{Z}_{0,0}\left(1;\tilde{q}^2\right) 
- \frac{2 \tilde q^2\mathcal{Z}_{4,0}\left(1;\tilde{q}^2\right)}{11 }
-\frac{60 \mathcal{Z}_{6,0}\left(1;\tilde{q}^2\right)}{11 \sqrt{13}}\right)
   $ &  \ref{irrep:T2m}\\
3 [$A_2^-$]  \qquad 
&\  
$q^7\text{cot}\delta_3\ =\left(\frac{2\pi}{L}\right)^7\frac{1}{\pi^{3/2}}
\left(\tilde{q}^6\mathcal{Z}_{0,0}\left(1;\tilde q^2\right)
-\frac{12}{11}\tilde q^2\mathcal{Z}_{4,0}\left(1;\tilde q^2\right)+\frac{80}{11\sqrt{13}}
\mathcal{Z}_{6,0}\left(1;\tilde q^2\right)\right)
   $ & \ref{irrep:A2m}\\
 & & \\
   \hline
 & & \\ 
4 [$T_1^+$]  \qquad
& \ $q^9\text{cot}\delta_4\ =\left(\frac{2\pi}{L}\right)^9\frac{1}{\pi^{3/2}}\ 
  \left(
  \tilde{q}^8\mathcal{Z}_{0,0}\left(1;\tilde q^2\right)
-\frac{448 \mathcal{Z}_{8,0}\left(1;\tilde q^2\right)}{143 \sqrt{17}}
-\frac{4\tilde{q}^2 \mathcal{Z}_{6,0}\left(1;\tilde q^2\right)}{11 \sqrt{13}}
+\frac{54 \tilde{q}^4\mathcal{Z}_{4,0}\left(1;\tilde q^2\right)}{143
   }\right)$ &  \ref{irrep:T1p}\\
 & & \\
\hline
 & & \\
5 [$E^-$]  \qquad
& \ $\begin{array}{l} 
q^{11}\text{cot}\delta_5\ = \left(\frac{2\pi}{L}\right)^{11}\frac{1}{\pi^{3/2}}\times\\
\left(  \tilde q^{10}\mathcal{Z}_{0,0}\left(1;\tilde q^2\right)
- \frac{6\tilde q^6\mathcal{Z}_{4,0}\left(1;\tilde q^2\right)}{13}
+\frac{32\tilde q^4\mathcal{Z}_{6,0}\left(1;\tilde q^2\right)}{17\sqrt{13}}
-\frac{672\tilde q^2\mathcal{Z}_{8,0}\left(1;\tilde q^2\right)}{247\sqrt{17}}
+\frac{1152\sqrt{21}\mathcal{Z}_{10,0}\left(1;\tilde q^2\right)}{4199} \right)
 \end{array}
 $  & \ref{irrep:Em}\\
 & & \\
\hline
 & & \\ 
6 [$A_2^+$]  \qquad
& \ $\begin{array}l
q^{13}\text{cot}\delta_6\ =\left(\frac{2\pi}{L}\right)^{13}\frac{1}{\pi^{3/2}}\times
\\
\left(\tilde q^{12}
  \mathcal{Z}_{0,0}\left(1;\tilde q^2\right)+\frac{6 \tilde q^8
   \mathcal{Z}_{4,0}\left(1;\tilde q^2\right)}{17}-\frac{160 \sqrt{13} \tilde
        q^6 \mathcal{Z}_{6,0}\left(1;\tilde q^2\right)}{323}-\frac{40 \tilde q^4
             \mathcal{Z}_{8,0}\left(1;\tilde q^2\right)}{19 \sqrt{17}}\right.
\\
\quad\quad\quad\quad\quad\quad\quad\quad\quad\quad
\left. -\frac{2592 \sqrt{21} \tilde q^2 \mathcal{Z}_{10,0}\left(1;\tilde
      q^2\right)}{7429}+\frac{1980 \mathcal{Z}_{12,0}\left(1;\tilde
     q^2\right)}{7429}+\frac{264 \sqrt{1001} \mathcal{Z}_{12,4}\left(1;\tilde
     q^2\right)}{7429}\right)
     
\end{array}
$
 &  \ref{irrep:A2p}\\
 & & \\
\hline
 & & \\ 
9 [$A_1^-$]  \qquad
& \ $\begin{array}l
q^{19} \cot\delta_9 \ = \ 
\left({2\pi\over L}\right)^{19}\ {1\over\pi^{3/2}}\times
\\
\left(  
\tilde q^{18} {\cal Z}_{0,0}(1;\tilde q^2)
- 
{6
\tilde q^{14} {\cal Z}_{4,0}(1;\tilde q^2) \over 23}
- 
{32\sqrt{13}
\tilde q^{12} {\cal Z}_{6,0}(1;\tilde q^2) \over 115}
- 
{56\sqrt{17}
\tilde q^{10} {\cal Z}_{8,0}(1;\tilde q^2) \over 345}
\right. \\
\left.
+ 
{1568\sqrt{7}
\tilde q^{8} {\cal Z}_{10,0}(1;\tilde q^2) \over 3335\sqrt{3}}
+
{308 \tilde q^{6} {\cal Z}_{12,0}(1;\tilde q^2) \over 2139}
+
{616 \sqrt{1001} \tilde q^{6} {\cal Z}_{12,4}(1;\tilde q^2) \over 20677}
+
{53248 \tilde q^{4} {\cal Z}_{14,0}(1;\tilde q^2) \over 10695\sqrt{29}}
\right. \\
\left.
-
{1664\sqrt{11} \tilde q^{2} {\cal Z}_{16,0}(1;\tilde q^2) \over 3565\sqrt{3}}
+
{832\sqrt{46189} \tilde q^{2} {\cal Z}_{16,4}(1;\tilde q^2) \over 103385\sqrt{7}}
+
{2206464 \tilde {\cal Z}_{18,0}(1;\tilde q^2) \over 103385\sqrt{37}}
+
{28288\sqrt{3553} {\cal Z}_{18,4}(1;\tilde q^2) \over 20677\sqrt{259}}
\  \right)
\end{array}
$
&  \ref{irrep:A1m} \\
& & \\
\hline
\hline
\end{tabular}
\begin{minipage}[t]{16.5 cm}
\vskip 0.5cm
\noindent
\end{minipage}
\end{center}
\end{table}  
\begin{table}
\begin{center}
\begin{minipage}[!ht]{16.5 cm}
\caption{The perturbative expansions of the energy-eigenvalues of the 
lowest-lying state in each $\Gamma^{(i)}$.
}
\label{tab:LowestPerturb}
\end{minipage}
\begin{tabular}{| c | l |}
\hline 
\hline 
& \\
  \ \ \ $|{\bf n}|^2$ \ $[\Gamma^{(i)}]$ \ \   
& \ \ \ \qquad $|{\rm q}|^2$\ \ \ 
\\
& \\
\hline
& \\
$0$\  [$A_1^+$] & \  
$-{4\pi a_0\over L^3}\ \left[\ 1\ - 2.8373 \left({a_0\over
      L}\right) + 6.3752 \left({a_0\over L}\right)^2\ \right] \ + \ ...$
\ \ \ 
\\
& \\
\hline
& \\
$1$\  [$T_1^-$] & \  
${4\pi^2\over L^2}\ 
\left[\ 
1 - {3\over \pi^2}\tan\delta_1 \left( 1 - 0.3653 \tan\delta_1 \right)
\ +\ ... \right]$
\\
$1$\  [$E^+$] & \  
${4\pi^2\over L^2}\ 
\left[\ 
1 - {15\over 2 \pi^2}\tan\delta_2 \left( 1 - 1.5672 \tan\delta_2 \right)
\ +\ ... \right]$
\\
& \\
\hline
& \\
$2$\  [$T_2^+$] & \  
${4\pi^2\over L^2}\ 
\left[\ 
2 - {15\over 2\sqrt{2} \pi^2}\tan\delta_2 \left( 1 - 0.4830 \tan\delta_2 \right)
\ +\ ... \right]$
\\
$2$\  [$T_2^-$] & \  
${4\pi^2\over L^2}\ 
\left[\ 
2 - {105\over 8\sqrt{2} \pi^2}\tan\delta_3 
\ +\ ... \right]$
\\
& \\
\hline
& \\
$3$\  [$A_2^-$] &  \ 
${4\pi^2\over L^2}\ 
\left[\ 
3 - {140\over 9\sqrt{3} \pi^2}\tan\delta_3 
\ +\ ... \right]$
\\
& \\
\hline
& \\
$5$\  [$T_1^+$] &  \ 
${4\pi^2\over L^2}\ 
\left[\ 
5 - {2268\over 125\sqrt{5} \pi^2}\tan\delta_4
\ +\ ... \right]$
\\
$5$\  [$A_2^+$] & \ 
${4\pi^2\over L^2}\ 
\left[\ 
5 - {162162\over 3125\sqrt{5} \pi^2}\tan\delta_6
\ +\ ... \right]$
\\
& \\
\hline
& \\
$6$\  [$E^-$] & \ 
${4\pi^2\over L^2}\ 
\left[\ 
6 - {385\over 12\sqrt{6} \pi^2}\tan\delta_5
\ +\ ... \right]$
\\
& \\
\hline
& \\
$14$\  [$A_1^-$] & \ 
${4\pi^2\over L^2}\ 
\left[\ 
14 - {4208972625\over 46118408\sqrt{14} \pi^2}\tan\delta_9
\ +\ ... \right]$
\\
& \\
\hline 
\hline
\end{tabular}
\begin{minipage}[t]{16.5 cm}
\vskip 0.5cm
\noindent
\end{minipage}
\end{center}
\end{table}     

Experimental measurements of the $\pi\pi$ phase-shifts are difficult,
with precision currently at the few-percent level in the s-wave and p-wave, but much
larger uncertainties are associated with the phase-shifts 
in the higher partial-waves.   
It would appear that Lattice QCD calculations will be able to provide
low-energy meson-meson phase-shifts in the low partial-waves
with significantly more precision than the corresponding 
experimental measurements.
While the contributions to energy-splittings rapidly become smaller
with higher partial-waves, we conclude that it is presently possible to extract
phase-shifts for partial-waves with $l\le 3$.  
The implications of the recent preliminary calculations~\cite{Dudek:2010ew}
of the $l=2$ phase-shift, $\delta_2^2$, at unphysical pion masses
are very encouraging for future calculations.

\appendix

\section{Block-Diagonalization of $F^{(FV)}$ 
\label{sect:explicit calculation}}
\noindent
As the number of partial-waves with non-zero phase-shifts
increases, so does the complexity of the
calculation of the energy-eigenvalues in a finite cubic volume.
To illustrate the method for determining the
energy-eigenvalues, we provide the details of the calculation when
$\delta_l\ne 0$ for $l\le 4$.  As is true in all cases involving
parity-conserving interactions, 
the analysis in the
even-parity sector ($l=0,2,4$) decouples from that in the odd-parity sector
($l=1,3$).
The calculations that are required for $\delta_l\ne 0$ for $l> 4$ become more
complicated due to the dimensionality of the matrices involved, the
contributions from the ${\cal Z}_{l,m}$ for higher values of $l$, and 
to multiple occurrences of the same cubic irreps.  This last feature means
that the diagonalization of the finite-volume functions, $F_{l;l}^{(FV)}$, are not
dictated entirely by geometry due to mixing between the multiple occurrences of
a given $\Gamma^{(i)}$.
However, such calculations are a straightforward extension of what follows.

\subsection{Odd-Parity Sector with $\delta_{1,3}\ne 0$}
\noindent
In the odd-parity sector with only $\delta_{1,3}\ne 0$, the finite-volume
  corrections are encapsulated in $ F^{(FV)}_-$ which is a $10\times
  10$ matrix.  It has block form
\begin{eqnarray}
F^{(FV)}_-
& = & 
\left(
\begin{array}{c|c}
\\
 F^{(FV)}_{1;1} & 
 F^{(FV)}_{1;3}\\ \\
\hline\\
F^{(FV)}_{3;1} &  F^{(FV)}_{3;3}\\ \\
\end{array}
\right)
\ \ \ ,
\label{eq:F3}
\end{eqnarray}
where the component matrices in the $|l,m\rangle$-basis are~\footnote{
Explicitly, the basis is
$\{ |1,1\rangle, |1,0\rangle , |1,-1\rangle , |3,3\rangle,  |3,2\rangle,
|3,1\rangle,  |3,0\rangle,  |3,-1\rangle,  |3,-2\rangle,  |3,-3\rangle \}$.
}
\begin{eqnarray}
F^{(FV)}_{1;1}
& = & 
{1\over\pi^{3/2}}\ {1\over \tilde q}\ {\cal Z}_{0,0}(1;\tilde q^2)\ 
\ {\rm diag}\left(1,1,1\right)
\nonumber\\
 F^{(FV)}_{1;3}
& = & \left( F^{(FV)}_{3;1}\right)^T\ =\
{1\over\pi^{3/2}}\ {1\over \tilde q^5}\ {\cal Z}_{4,0}(1;\tilde q^2)\ 
\ {2\over\sqrt{21}}\ 
\left(
\begin{array}{ccccccc}
0&0& -\sqrt{3\over 2}&0&0&0& -\sqrt{5\over 2} \\
0&0&0&2&0&0&0\\
 -\sqrt{5\over 2}&0&0&0& -\sqrt{3\over 2}&0&0
\end{array}
\right)
\nonumber\\
\overline F^{(FV)}_{3;3}
& = &  
{1\over\pi^{3/2}}\ {1\over \tilde q}\ {\cal Z}_{0,0}(1;\tilde q^2)\ 
\ {\rm diag}\left(1,1,1,1,1,1,1\right)
\nonumber\\
&& 
\ +\ 
{1\over\pi^{3/2}}\ {1\over \tilde q^5}\ {\cal Z}_{4,0}(1;\tilde q^2)\ 
{1\over 11}\ 
\left(
\begin{array}{ccccccc}
3&0&0&0&\sqrt{15}&0&0\\
0&-7&0&0&0&5&0\\
0&0&1&0&0&0&\sqrt{15}\\
0&0&0&6&0&0&0\\
\sqrt{15}&0&0&0&1&0&0\\
0&5&0&0&0&-7&0\\
0&0&\sqrt{15}&0&0&0&3\\
\end{array}
\right)\ 
\\
&& 
\ +\ 
{1\over\pi^{3/2}}\ {1\over \tilde q^7}\ {\cal Z}_{6,0}(1;\tilde q^2)\ 
{5\over 33\sqrt{13}}\
\left(
\begin{array}{ccccccc}
-1&0&0&0&7\sqrt{15}&0&0\\
0&6&0&0&0&-42&0\\
0&0&-15&0&0&0&7\sqrt{15}\\
0&0&0&20&0&0&0\\
7\sqrt{15}&0&0&0&-15&0&0\\
0&-42&0&0&0&6&0\\
0&0&7\sqrt{15}&0&0&0&-1\\
\end{array}
\right)\ 
\ \ \ ,
\nonumber
\label{eq:F13}
\end{eqnarray}
where the relevant relations between the ${\cal
  Z}_{l,m}$, that can be found in eq.~(\ref{eq:Z4m6m8m10m12m}),
have been used,
and where ${\rm diag}(a,b,...)$ denotes a diagonal matrix.
It is convenient to first diagonalize the $F^{(FV)}_{l;l}$ blocks
($F^{(FV)}_{1;1}$ is already diagonal in this basis).
The block-diagonal matrix, $S_-$, is defined to have the form
\begin{eqnarray}
S_- & = & 
\left(
\begin{array}{c|c}
S_{11} & 0 \\
\hline
0 & S_{33}
\end{array}
\right)
\ \ \ ,
\label{eq:S11S33block}
\end{eqnarray}
and when acting on $F^{(FV)}_-$ produces a matrix, $\overline{F}^{(FV)}_- \ = \ 
S_-.F^{(FV)}_-.S^\dagger_-$,
which can be re-arranged into
block-diagonal form where each block is associated with 
a $\Gamma^{(i)}$.
The matrices $\cos\delta$ and
$\sin\delta$ in eq.~(\ref{eq:trigs})
are invariant under this transformation,
\begin{eqnarray}
\overline{\cos}\delta & = & 
{\rm diag}\left(c_1,c_1,c_1,c_3,c_3,c_3,c_3,c_3,c_3,c_3\right)
\ \ \ ,
\end{eqnarray}
where $c_1$ and $c_3$ denote $\cos\delta_1$ and $\cos\delta_3$, respectively,
and similarly for $\overline{\sin}\delta$.
The components of $S_-$ in eq.~(\ref{eq:S11S33block}) are 
\begin{eqnarray}
S_{11} & = & 
\left(
\begin{array}{ccc}
1&0&0\\ 0&1&0 \\ 0&0&1
\end{array}
\right)
\ \ ,\ \ 
S_{33}\ =\ 
\left(
\begin{array}{ccccccc}
0&0&-\sqrt{5\over 8}&0&0&0&\sqrt{3\over 8}\\
0&{1\over\sqrt{2}}&0&0&0&{1\over\sqrt{2}}&0\\
-\sqrt{3\over 8}&0&0&0&\sqrt{5\over 8}&0&0\\
0&-{1\over\sqrt{2}}&0&0&0&{1\over\sqrt{2}}&0\\
0&0&\sqrt{3\over 8}&0&0&0&\sqrt{5\over 8}\\
\sqrt{5\over 8}&0&0&0&\sqrt{3\over 8}&0&0\\
0&0&0&1&0&0&0
\end{array}
\right)
\ \ \ ,
\label{eq:S11S33}
\end{eqnarray}
and the matrix $\overline{F}^{(FV)}_-$ is of the form 
\begin{eqnarray}
\overline{F}^{(FV)}_-
& = & 
\left(
\begin{array}{c|c}
\\
\overline{F}^{(FV)}_{1;1} & 
\overline{F}^{(FV)}_{1;3}\\ \\
\hline\\
\overline{F}^{(FV)}_{3;1} & 
\overline{F}^{(FV)}_{3;3}\\ \\
\end{array}
\right)
\ \ \ ,
\label{eq:F3bar}
\end{eqnarray}
where
\begin{eqnarray}
\overline{F}^{(FV)}_{1;1}
& = & 
{1\over\pi^{3/2}}\ {1\over \tilde q}\ {\cal Z}_{0,0}(1;\tilde q^2)\ 
\ {\rm diag}\left(1,1,1\right)
\nonumber\\
\overline{F}^{(FV)}_{1;3}
& = & 
\left(\overline{F}^{(FV)}_{3;1}\right)^T\ =\
{1\over\pi^{3/2}}\ {1\over \tilde q^5}\ {\cal Z}_{4,0}(1;\tilde q^2)\ 
\ {4\over\sqrt{21}}\ 
\left(
\begin{array}{ccccccc}
 0 & 0 & 0 & 0 & -1 & 0 & 0 \\
 0 & 0 & 0 & 0 & 0 & 0 & 1 \\
 0 & 0 & 0 & 0 & 0 & -1 & 0
\end{array}
\right)
\nonumber\\
\overline{F}^{(FV)}_{3;3}
& = & 
{1\over\pi^{3/2}}\ {1\over \tilde q}\ {\cal Z}_{0,0}(1;\tilde q^2)\ 
\ {\rm diag}\left(1,1,1,1,1,1,1\right)
\nonumber\\
&& 
\ +\ 
{1\over\pi^{3/2}}\ {1\over \tilde q^5}\ {\cal Z}_{4,0}(1;\tilde q^2)\ 
{2\over 11}\ 
\ {\rm diag}\left(-1,-1,-1,-6,3,3,3\right)
 \nonumber\\
&&
\ +\ 
{1\over\pi^{3/2}}\ {1\over \tilde q^7}\ {\cal Z}_{6,0}(1;\tilde q^2)\ 
{20\over 33\sqrt{13}}\
\ {\rm diag}\left(-9,-9,-9,12,5,5,5\right)
\ \ \ .
\label{eq:F13bar}
\end{eqnarray}
From the structure of $\overline{F}^{(FV)}_-$ it is clear that the ordering
of the $\Gamma^{(i)}$
along the diagonal of $\overline{F}^{(FV)}_{3;3}$
is $T_2^-$, $A_2^-$ and $T_1^-$, respectively, and 
the equations that dictate the energy-eigenvalues of each of the $\Gamma^{(i)}$, given
in eqs.~(\ref{eqn:A2- l=3}), (\ref{eq:T1ml13}), and~(\ref{eqn:T2- l=3}), 
follow directly from eq.~(\ref{eq:F13bar}). 
The matrix $S_-$ that diagonalizes $F^{(FV)}_-$ is independent of the ${\cal
  Z}_{l,m}$ functions because, with each relevant $\Gamma^{(i)}$ 
occurring only once, the
decomposition depends upon geometry only.

\subsection{Even-Parity Sector with $\delta_{0,2,4}\ne 0$}
\noindent
In the even-parity sector with only $\delta_{0,2,4}\ne 0$, the finite-volume
  corrections are encapsulated in $F^{(FV)}_+$ which is a $15\times
  15$ matrix.  It has block  form
\begin{eqnarray}
F^{(FV)}_+ & = & 
\left(
\begin{array}{c | c | c}
& & \\
F^{(FV)}_{0;0} & F^{(FV)}_{0;2} & F^{(FV)}_{0;4} \\ 
& & \\
\hline
& & \\
 F^{(FV)}_{2;0} &F^{(FV)}_{2;2} &F^{(FV)}_{2;4} \\ 
& & \\
\hline\
 & & \\
 F^{(FV)}_{4;0} & F^{(FV)}_{4;2} &F^{(FV)}_{4;4} \\ 
& & \\
\end{array}
\right)
\ \ \ ,
\label{eq:F024}
\end{eqnarray}
where
\begin{eqnarray}
 F^{(FV)}_{0;0} & = & 
{1\over\pi^{3/2}}\ {1\over \tilde q}\ {\cal Z}_{0,0}(1;\tilde q^2)
\ \ ,\ \ 
F^{(FV)}_{0;2} \ = \
\left(
\begin{array}{ccccc}
0&0&0&0&0
\end{array}
\right)
\nonumber\\
 F^{(FV)}_{0;4} & = & 
{1\over\pi^{3/2}}\ {1\over \tilde q^5}\ {\cal Z}_{4,0}(1;\tilde q^2)\ 
\left(
\begin{array}{ccccccccc}
 \sqrt{\frac{5}{14}} & 0 & 0 & 0 & 1 & 0 & 0 & 0 & \sqrt{\frac{5}{14}}
\end{array}
\right)
\nonumber\\
 F^{(FV)}_{2;2} & = & 
{1\over\pi^{3/2}}\ {1\over \tilde q}\ {\cal Z}_{0,0}(1;\tilde q^2)\ 
\ {\rm diag}\left(1,1,1,1,1\right)
 \nonumber\\
&&
\ +\ 
{1\over\pi^{3/2}}\ {1\over \tilde q^5}\ {\cal Z}_{4,0}(1;\tilde q^2)\ 
{1\over 7}
\left(
\begin{array}{ccccc}
 1 & 0 & 0 & 0 & 5 \\
 0 & -4 & 0 & 0 & 0 \\
 0 & 0 & 6 & 0 & 0 \\
 0 & 0 & 0 & -4 & 0 \\
 5 & 0 & 0 & 0 & 1
\end{array}
\right)
\nonumber
\end{eqnarray}
\begin{eqnarray}
 F^{(FV)}_{2;4} & = & 
{1\over\pi^{3/2}}\ {1\over \tilde q^5}\ {\cal Z}_{4,0}(1;\tilde q^2)\ 
{10\sqrt{3}\over 77\sqrt{2}}
\left(
\begin{array}{ccccccccc}
 0 & 0 & -3 \sqrt{2} & 0 & 0 & 0 & -\sqrt{2} & 0 & 0 \\
 0 & 0 & 0 & 1 & 0 & 0 & 0 & -\sqrt{7} & 0 \\
 -2 \sqrt{\frac{7}{3}} & 0 & 0 & 0 & 2 \sqrt{\frac{10}{3}} & 0 & 0 & 0 & -2 \sqrt{\frac{7}{3}} \\
 0 & -\sqrt{7} & 0 & 0 & 0 & 1 & 0 & 0 & 0 \\
 0 & 0 & -\sqrt{2} & 0 & 0 & 0 & -3 \sqrt{2} & 0 & 0
\end{array}
\right)
\nonumber\\
& + & 
{1\over\pi^{3/2}}\ {1\over \tilde q^7}\ {\cal Z}_{6,0}(1;\tilde q^2)\ 
{5\sqrt{3}\over 11\sqrt{13}}
\left(
\begin{array}{ccccccccc}
 0 & 0 & 1 & 0 & 0 & 0 & -7 & 0 & 0 \\
 0 & 0 & 0 & -2 \sqrt{2} & 0 & 0 & 0 & 2 \sqrt{14} & 0 \\
 -\sqrt{\frac{21}{2}} & 0 & 0 & 0 & \sqrt{15} & 0 & 0 & 0 & -\sqrt{\frac{21}{2}} \\
 0 & 2 \sqrt{14} & 0 & 0 & 0 & -2 \sqrt{2} & 0 & 0 & 0 \\
 0 & 0 & -7 & 0 & 0 & 0 & 1 & 0 & 0
\end{array}
\right)
\nonumber
\end{eqnarray}
\begin{eqnarray}
F^{(FV)}_{4;4} & = & 
{1\over\pi^{3/2}}\ {1\over \tilde q}\ {\cal Z}_{0,0}(1;\tilde q^2)\ 
\ {\rm diag}\left(1,1,1,1,1,1,1,1,1\right)
\nonumber\\
& + & 
{1\over\pi^{3/2}}\ {1\over \tilde q^5}\ {\cal Z}_{4,0}(1;\tilde q^2)\ 
{27\over 1001}
\left(
\begin{array}{ccccccccc}
 14 & 0 & 0 & 0 & \sqrt{70} & 0 & 0 & 0 & 0 \\
 0 & -21 & 0 & 0 & 0 & 5 \sqrt{7} & 0 & 0 & 0 \\
 0 & 0 & -11 & 0 & 0 & 0 & 15 & 0 & 0 \\
 0 & 0 & 0 & 9 & 0 & 0 & 0 & 5 \sqrt{7} & 0 \\
 \sqrt{70} & 0 & 0 & 0 & 18 & 0 & 0 & 0 & \sqrt{70} \\
 0 & 5 \sqrt{7} & 0 & 0 & 0 & 9 & 0 & 0 & 0 \\
 0 & 0 & 15 & 0 & 0 & 0 & -11 & 0 & 0 \\
 0 & 0 & 0 & 5 \sqrt{7} & 0 & 0 & 0 & -21 & 0 \\
 0 & 0 & 0 & 0 & \sqrt{70} & 0 & 0 & 0 & 14
\end{array}
\right)
\nonumber\\
& + & 
{1\over\pi^{3/2}}\ {1\over \tilde q^7}\ {\cal Z}_{6,0}(1;\tilde q^2)\ 
{1\over 11\sqrt{13}}
\left(
\begin{array}{ccccccccc}
 -4 & 0 & 0 & 0 & 6 \sqrt{70} & 0 & 0 & 0 & 0 \\
 0 & 17 & 0 & 0 & 0 & -3 \sqrt{7} & 0 & 0 & 0 \\
 0 & 0 & -22 & 0 & 0 & 0 & -42 & 0 & 0 \\
 0 & 0 & 0 & -1 & 0 & 0 & 0 & -3 \sqrt{7} & 0 \\
 6 \sqrt{70} & 0 & 0 & 0 & 20 & 0 & 0 & 0 & 6 \sqrt{70} \\
 0 & -3 \sqrt{7} & 0 & 0 & 0 & -1 & 0 & 0 & 0 \\
 0 & 0 & -42 & 0 & 0 & 0 & -22 & 0 & 0 \\
 0 & 0 & 0 & -3 \sqrt{7} & 0 & 0 & 0 & 17 & 0 \\
 0 & 0 & 0 & 0 & 6 \sqrt{70} & 0 & 0 & 0 & -4
\end{array}
\right)
\nonumber\\
& + & 
{1\over\pi^{3/2}}\ {1\over \tilde q^9}\ {\cal Z}_{8,0}(1;\tilde q^2)\ 
{7\over 143\sqrt{17}}
\left(
\begin{array}{ccccccccc}
 1 & 0 & 0 & 0 & \sqrt{70} & 0 & 0 & 0 & 65 \\
 0 & -8 & 0 & 0 & 0 & -8 \sqrt{7} & 0 & 0 & 0 \\
 0 & 0 & 28 & 0 & 0 & 0 & 28 & 0 & 0 \\
 0 & 0 & 0 & -56 & 0 & 0 & 0 & -8 \sqrt{7} & 0 \\
 \sqrt{70} & 0 & 0 & 0 & 70 & 0 & 0 & 0 & \sqrt{70} \\
 0 & -8 \sqrt{7} & 0 & 0 & 0 & -56 & 0 & 0 & 0 \\
 0 & 0 & 28 & 0 & 0 & 0 & 28 & 0 & 0 \\
 0 & 0 & 0 & -8 \sqrt{7} & 0 & 0 & 0 & -8 & 0 \\
 65 & 0 & 0 & 0 & \sqrt{70} & 0 & 0 & 0 & 1
\end{array}
\right)
\ \ \ ,
\end{eqnarray}
which can be made partially block-diagonalized by
\begin{eqnarray}
S_+ & = & 
\left(
\begin{array}{c|c|c}
S_{0,0} & 0 & 0 \\
\hline
0 & S_{22} & 0 \\
\hline
0 & 0 & S_{44}
\end{array}
\right)
\ \ \ ,
\label{eq:S00S22S44block}
\end{eqnarray}
with
\begin{eqnarray}
S_{0,0} & = & 1
\ \ ,\ \ 
S_{22} \ = \ 
\left(
\begin{array}{ccccc}
-{1\over\sqrt{2}}&0&0&0&{1\over\sqrt{2}}\\
0&0&0&1&0\\
0&1&0&0&0\\
{1\over\sqrt{2}}&0&0&0&{1\over\sqrt{2}}\\
0&0&1&0&0
\end{array}
\right)
\nonumber\\
S_{44}& = & 
\left(
\begin{array}{ccccccccc}
{\sqrt{7}\over 2\sqrt{6}}&0&0&0&-{\sqrt{5}\over 2\sqrt{3}}&0&0&0&{\sqrt{7}\over
  2\sqrt{6}}\\
0&0&{1\over\sqrt{2}}&0&0&0&{1\over\sqrt{2}}&0&0\\
-{1\over\sqrt{2}}&0&0&0&0&0&0&0&{1\over\sqrt{2}}\\
0&0&0&{\sqrt{7}\over 2\sqrt{2}}&0&0&0&{1\over 2\sqrt{2}}&0\\
0&{1\over 2\sqrt{2}}&0&0&0&{\sqrt{7}\over 2\sqrt{2}}&0&0&0\\
{\sqrt{5}\over 2\sqrt{6}}&0&0&0&{\sqrt{7}\over 2\sqrt{3}}&0&0&0&{\sqrt{5}\over
  2\sqrt{6}}\\
0&0&0&-{1\over 2\sqrt{2}}&0&0&0&{\sqrt{7}\over 2\sqrt{2}}&0\\
0&0&-{1\over\sqrt{2}}&0&0&0&{1\over\sqrt{2}}&0&0\\
0&-{\sqrt{7}\over 2\sqrt{2}}&0&0&0&{1\over 2\sqrt{2}}&0&0&0
\end{array}
\right)
\ \ \ .
\label{eq:S00S22S44}
\end{eqnarray}
After this partial-diagonalization, finite-volume function becomes
$\overline{F}^{(FV)}_+ \ = \ 
S_+.F^{(FV)}_+.S^\dagger_+$ where
\begin{eqnarray}
\overline{F}^{(FV)}_{0,0} & = & 
{1\over\pi^{3/2}}\ {1\over \tilde q}\ {\cal Z}_{0,0}(1;\tilde q^2)
\ \ ,\ \ 
\overline{F}^{(FV)}_{0;2} \ = \ 
\left(
\begin{array}{ccccc}
0&0&0&0&0
\end{array}
\right)
\nonumber\\
\overline{F}^{(FV)}_{0;4} & = & 
{1\over\pi^{3/2}}\ {1\over \tilde q^5}\ {\cal Z}_{4,0}(1;\tilde q^2)\ 
{2\sqrt{3}\over\sqrt{7}}
\left(
\begin{array}{ccccccccc}
 0 & 0 & 0 & 0 & 0 & 1 & 0 & 0 & 0
\end{array}
\right)
\nonumber\\
\overline{F}^{(FV)}_{2;2} & = & 
{1\over\pi^{3/2}}\ {1\over \tilde q}\ {\cal Z}_{0,0}(1;\tilde q^2)\ 
\ {\rm diag}\left(1,1,1,1,1\right)
\nonumber\\
& + & 
{1\over\pi^{3/2}}\ {1\over \tilde q^5}\ {\cal Z}_{4,0}(1;\tilde q^2)\ 
{2\over 7}
\ {\rm diag}\left(-2,-2,-2,3,3\right)
\nonumber
\end{eqnarray}
\begin{eqnarray}
\overline{F}^{(FV)}_{2;4} & = & 
{1\over\pi^{3/2}}\ {1\over \tilde q^5}\ {\cal Z}_{4,0}(1;\tilde q^2)\ 
{20\sqrt{3} \over 77}\ 
\left(
\begin{array}{ccccccccc}
 0 & 0 & 0 & 0 & 0 & 0 & 0 & -1 & 0 \\
 0 & 0 & 0 & 0 & 0 & 0 & 0 & 0 & 1 \\
 0 & 0 & 0 & 0 & 0 & 0 & -1 & 0 & 0 \\
 0 & -2 & 0 & 0 & 0 & 0 & 0 & 0 & 0 \\
 -2 & 0 & 0 & 0 & 0 & 0 & 0 & 0 & 0
\end{array}
\right)
\nonumber\\
& + & 
{1\over\pi^{3/2}}\ {1\over \tilde q^7}\ {\cal Z}_{6,0}(1;\tilde q^2)\ 
{40\sqrt{3} \over 11 \sqrt{13}}
\left(
\begin{array}{ccccccccc}
 0 & 0 & 0 & 0 & 0 & 0 & 0 & 1 & 0 \\
 0 & 0 & 0 & 0 & 0 & 0 & 0 & 0 & -1 \\
 0 & 0 & 0 & 0 & 0 & 0 & 1 & 0 & 0 \\
 0 & -\frac{3}{4} & 0 & 0 & 0 & 0 & 0 & 0 & 0 \\
 -\frac{3}{4} & 0 & 0 & 0 & 0 & 0 & 0 & 0 & 0
\end{array}
\right)
\ \ \ .
\end{eqnarray}
\begin{eqnarray}
\overline{F}^{(FV)}_{4;4} & = & 
{1\over\pi^{3/2}}\ {1\over \tilde q}\ {\cal Z}_{0,0}(1;\tilde q^2)\ 
\ {\rm diag}\left(1,1,1,1,1,1,1,1,1\right)
\nonumber\\
& + & 
{1\over\pi^{3/2}}\ {1\over \tilde q^5}\ {\cal Z}_{4,0}(1;\tilde q^2)\ 
{54\over 1001}
\ {\rm diag}\left(2,2,7,7,7,14,-13,-13,-13\right)
\nonumber\\
& + & 
{1\over\pi^{3/2}}\ {1\over \tilde q^7}\ {\cal Z}_{6,0}(1;\tilde q^2)\ 
{4\over 11\sqrt{13}}
\ {\rm diag}\left(-16,-16,-1,-1,-1,20,5,5,5\right)
\nonumber\\
& + & 
{1\over\pi^{3/2}}\ {1\over \tilde q^9}\ {\cal Z}_{8,0}(1;\tilde q^2)\ 
{392\over 1001\sqrt{17}}
\ {\rm diag}\left(7,7,-8,-8,-8,10,0,0,0\right)
\ \ \ .
\end{eqnarray}
It is clear from the form of the matrix $\overline{F}^{(FV)}_+$ that the ordering
of the $\Gamma^{(i)}$ in the  $\overline{F}^{(FV)}_{4;4}$-block
is $E^+$, $T_1^+$, $A_1^+$ and $T_2^+$, respectively,
and 
the equations that dictate the energy-eigenvalues of each of the $\Gamma^{(i)}$, given
in eqs.~(\ref{eqn:A1 two solutions}), (\ref{eq:Epl2l4trunc}), 
(\ref{eqn:T1+ l=4}), and~(\ref{eqn:T2+ 24 mixing}), follow directly from these expressions.

\section{${\cal Z}_{l,m}(1;\tilde q^2)$  Functions
\label{sect:zeta functions}}
\noindent
The two-hadron Green functions in the finite lattice volume depend upon
summations over plane-wave states subject to periodic boundary conditions
and with amplitudes that depend upon the strength of the interactions in 
each of the partial-waves that generate the two-hadron T-matrix.
The summations that define the energy-eigenvalues in the volume 
are~\cite{Luscher:1986pf,Luscher:1990ux}
\begin{eqnarray}
{\cal Z}_{l,m}(1;\tilde q^2) & = & 
\sum_{ {\bf n} } \ 
{ |{\bf n}|^l\ Y_{lm}(\Omega_{\bf n}) \over
\left[\ |{\bf n}|^2 - \tilde q^2\ \right] }
\ \ \ ,
\label{eq:Zfunseq1}
\end{eqnarray}
a special case of the sums defined in eq.~(\ref{eq:Zfun}).
The $l=0$ summation is special as it requires UV regulation in order to be
defined, while sums with $l\ge 1$ are finite due to contribution from the
solitary $Y_{lm}$. 
However, brute-force evaluation of the sums is quite inefficient and L\"uscher
presented a method to evaluate the sums that exponentially accelerates their 
evaluation~\cite{Luscher:1986pf,Luscher:1990ux}, making use of the Poisson
resummation
formula.
In this appendix, we reproduce L\"uscher's results, and then present each of the
${\cal Z}_{l,m}(1;\tilde q^2)$ that contribute to the energy-eigenvalues
considered in the body of this paper.

Numerical evaluation of the ${\cal Z}_{0,0}(1;\tilde q^2)$ can be evaluated by
brute force through the definition
\begin{eqnarray}
{\cal Z}_{0,0}(1;\tilde q^2) & = & 
{1\over\sqrt{4\pi}}\ 
\lim_{\Lambda_{\bf n} \rightarrow\infty}\ 
\left[\ 
\sum_{ {\bf n} }^{\Lambda_{\bf n}} \ {1\over |{\bf n}|^2-\tilde q^2}
\ -\ 4\pi \Lambda_{\bf n}
\ \right]
\ \ \ ,
\label{eq:Zfun00}
\end{eqnarray}
or through the exponentially-accelerated 
relation~\cite{Luscher:1986pf,Luscher:1990ux,Detmold:2004qn}~\footnote{The Poisson resummation formula
\begin{eqnarray}
\sum_{\bf n}\ \delta^3({\bf y}-{\bf n})
& = & 
\sum_{\bf m}\ e^{i 2\pi {\bf m}\cdot {\bf y}}
\ \ \ \ ,
\label{eq:PResum}
\end{eqnarray}
has been used in obtaining eq.~(\ref{eq:Zfun00b}).
}
\begin{eqnarray}
{\cal Z}_{0,0}(1;\tilde q^2) & = & 
\pi e^{\tilde q^2} (2 \tilde q^2 -1)\ +\ 
{e^{\tilde q^2}\over 2\sqrt{\pi}}\ \sum_{\bf n}\ {e^{-|{\bf n}|^2}\over |{\bf
    n}|^2-\tilde q^2}
\nonumber\\
&&
\ -\ {\pi\over 2}\ \int_0^1\ dt\ {e^{t \tilde q^2}\over t^{3/2}}\ 
\left(\ 4 t^2 \tilde q^4\ -\ \sum_{ {\bf m}\ne {\bf 0}} e^{ -\pi^2
    |\overline{\bf m}|^2\over t}
\ \right)
\ \ \ .
\label{eq:Zfun00b}
\end{eqnarray}
For $l\ne 0$, the exponentially accelerated evaluation can be accomplished 
with~\footnote{
We have used a relation that is similar to that used by L\"uscher \cite{Luscher:1990ux},
\begin{eqnarray}
\int\ d^3{\bf x}\ g({\bf x})\ e^{-\lambda |{\bf x}|^2}\ e^{i2\pi {\bf p}\cdot
  {\bf x}}
& = & 
 g({-i\over 2\pi} \nabla_{\bf p})\ 
\int\ d^3{\bf x}\ e^{-\lambda |{\bf x}|^2}\ e^{i2\pi {\bf p}\cdot
  {\bf x}}
\ =\ 
\left({\pi\over\lambda}\right)^{3/2}\ g({i\pi\over\lambda} {\bf p})\ 
e^{-{\pi^2 |{\bf p}|^2\over\lambda}}
\ \ \ .
\end{eqnarray}
}
\begin{eqnarray}
{\cal Z}_{l,m}(1;\tilde q^2) & = & 
\sum_{ {\bf n} } \ 
{ |{\bf n}|^l\ Y_{lm}(\Omega_{\bf n}) \ e^{-\Lambda (  |{\bf n}|^2 - \tilde q^2)}
\over
\left[\ |{\bf n}|^2 - \tilde q^2\ \right] }
\nonumber\\
&&\ +\ 
\sum_{\bf p}
\int_0^\Lambda\ d\lambda\ 
\left({\pi\over \lambda}\right)^{l+3/2}\ e^{\lambda \tilde q^2}\ 
|{\bf p}|^l\  Y_{lm}(\Omega_{\bf p})\ e^{-{\pi^2 |{\bf p}|^2\over\lambda}}
\ \ \ .
\label{eq:Zfunseq1b}
\end{eqnarray}

There are exact relations that exist between the 
${\cal Z}_{l,m}(1;\tilde q^2)$ for fixed $l$:
\begin{eqnarray}
{\cal Z}_{4,\pm4}(1;\tilde q^2) & = & \sqrt{5\over 14} {\cal Z}_{4,0}(1;\tilde q^2)
\nonumber\\
{\cal Z}_{6,\pm 4}(1;\tilde q^2) & = & -\sqrt{7\over 2}\  {\cal
  Z}_{6,0}(1;\tilde q^2)
\nonumber\\
{\cal Z}_{8,\pm 4}(1;\tilde q^2) & = & {\sqrt{154}\over 33} \  {\cal
  Z}_{8,0}(1;\tilde q^2)
\ \ ,\ \ 
{\cal Z}_{8,\pm 8}(1;\tilde q^2) \ = \ {\sqrt{1430}\over 66} \  {\cal
  Z}_{8,0}(1;\tilde q^2)
\nonumber\\
{\cal Z}_{10,\pm 4}(1;\tilde q^2) & = &  -\sqrt{66\over 65} \  {\cal
  Z}_{10,0}(1;\tilde q^2)
\ \ ,\ \ 
{\cal Z}_{10,\pm 8}(1;\tilde q^2) \ = \   -\sqrt{187\over 130} \  {\cal
  Z}_{10,0}(1;\tilde q^2)
\nonumber\\
{\cal Z}_{12,\pm 8}(1;\tilde q^2) & = & 
 \sqrt{429\over 646} \  {\cal  Z}_{12,0}(1;\tilde q^2)
\ -\ 4\sqrt{42\over 323}\  {\cal  Z}_{12,\pm 4}(1;\tilde q^2)
\nonumber\\
{\cal Z}_{12,\pm 12}(1;\tilde q^2) & = & 
4\sqrt{91\over 7429}  \  {\cal  Z}_{12,0}(1;\tilde q^2)
\ +\ 
9\sqrt{11\over 7429}\  {\cal  Z}_{12,\pm 4}(1;\tilde q^2)
\nonumber
\label{eq:Z4m6m8m10m12m}
\end{eqnarray}
\begin{eqnarray}
{\cal Z}_{14,\pm 4}(1;\tilde q^2) & = & 
-{3\over 2} \sqrt{143\over 595}\ 
{\cal Z}_{14,0}(1;\tilde q^2) 
\ \ ,\ \ 
{\cal Z}_{14,\pm 8}(1;\tilde q^2) \ = \ 
-\sqrt{741\over 1190}\ 
{\cal Z}_{14,0}(1;\tilde q^2) 
\nonumber\\
{\cal Z}_{14,\pm 12}(1;\tilde q^2) & = & 
-{1\over 2} \sqrt{437\over 119}\ 
{\cal Z}_{14,0}(1;\tilde q^2) 
\nonumber\\
{\cal Z}_{16,\pm 8}(1;\tilde q^2) & = & 
-6\sqrt{6\over 805} {\cal Z}_{16,4}(1;\tilde q^2) 
\ +\ \sqrt{442\over 2185} {\cal Z}_{16,0}(1;\tilde q^2) 
\nonumber\\
{\cal Z}_{16,\pm 12}(1;\tilde q^2) & = & 
-{31\over 5}\sqrt{13\over 483} {\cal Z}_{16,4}(1;\tilde q^2) 
\ +\ {16\over 5}\sqrt{17\over 437} {\cal Z}_{16,0}(1;\tilde q^2) 
\nonumber\\
{\cal Z}_{16,\pm 16}(1;\tilde q^2) & = & 
4\sqrt{754\over 74865} {\cal Z}_{16,4}(1;\tilde q^2) 
\ +\ 7\sqrt{493\over 135470} {\cal Z}_{16,0}(1;\tilde q^2) 
\nonumber\\
{\cal Z}_{18,\pm 8}(1;\tilde q^2) & = & 
-{58\over 5}\sqrt{22\over 161} {\cal Z}_{18,4}(1;\tilde q^2) 
\ -\ {3\over 5}\sqrt{646\over 23} {\cal Z}_{18,0}(1;\tilde q^2) 
\nonumber\\
{\cal Z}_{18,\pm 12}(1;\tilde q^2) & = & 
{501\over 5}\sqrt{11\over 4669} {\cal Z}_{18,4}(1;\tilde q^2) 
\ +\ {16\over 5}\sqrt{323\over 667} {\cal Z}_{18,0}(1;\tilde q^2) 
\nonumber\\
{\cal Z}_{18,\pm 16}(1;\tilde q^2) & = & 
-4\sqrt{3162\over 23345} {\cal Z}_{18,4}(1;\tilde q^2) 
\ -\ \sqrt{19437\over 6670} {\cal Z}_{18,0}(1;\tilde q^2) 
\ \ \ .
\label{eq:Z14to18}
\end{eqnarray}

Unlike the cases of $l=0,4,6,8,10,14$ which have only one
occurrence of the 
$A_1^+$ irrep in their decomposition, $l=12,16,18$ have two, and as such 
the ${\cal Z}_{12,\pm 4 k}$ ($k$ is an integer) are not simply proportional to
${\cal Z}_{12,0}$, as demonstrated  in eq.~(\ref{eq:Z4m6m8m10m12m}), and a
similar statement can be made about 
${\cal Z}_{16,\pm 4 k}$ and  ${\cal Z}_{18,\pm 4 k}$.

\begin{figure}[h]
  \centering
\input{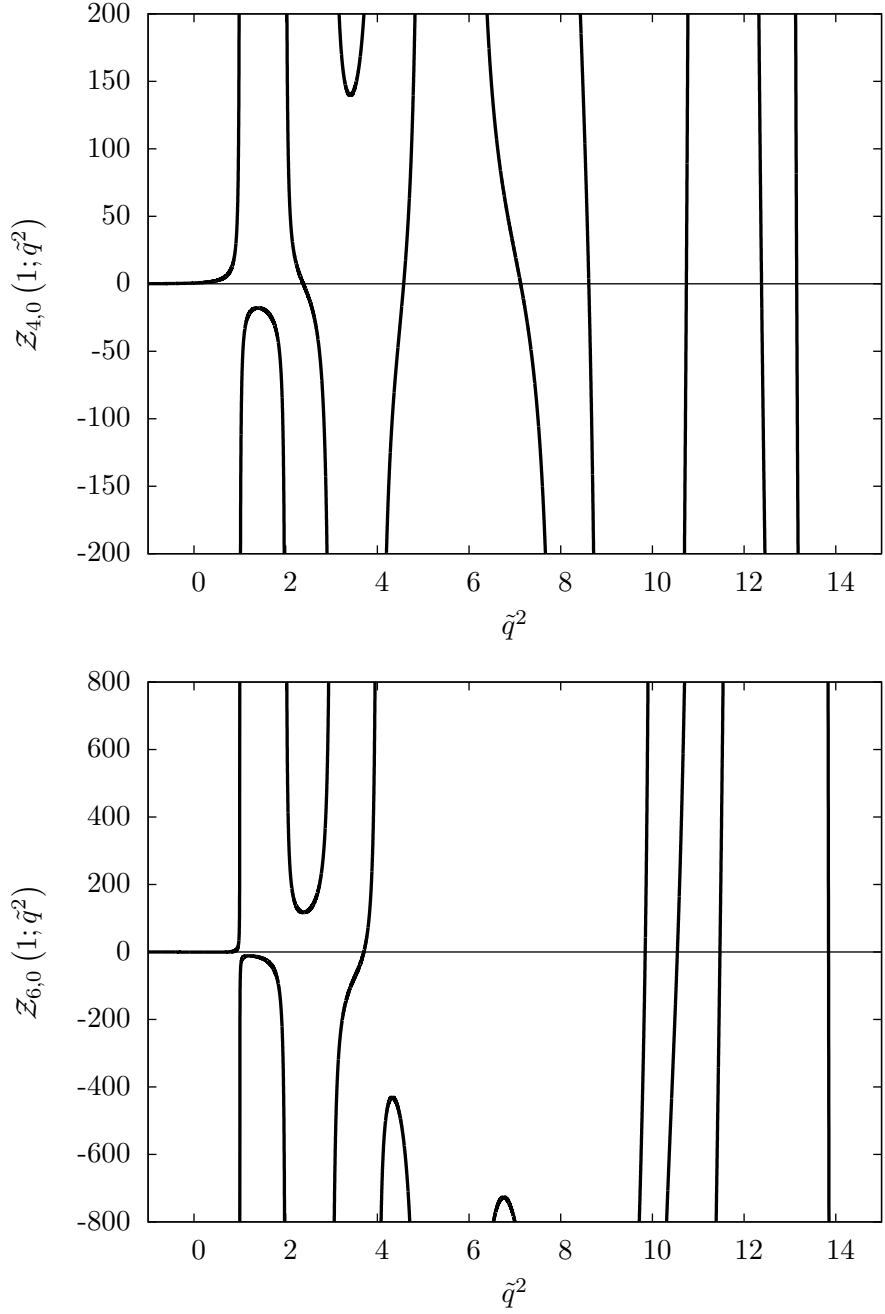}
\caption{ The function ${\cal Z}_{4,0}(1;\tilde q^2)$ (top-panel) and 
${\cal Z}_{6,0}(1;\tilde q^2)$ (bottom-panel).}
  \label{fig:Z40Z60}
\end{figure}

In an effort to better understand the origins of the structure of the functions
determining the energy-eigenvalues of each of the $\Gamma^{(i)}$,
it is useful to explicitly display the functions  ${\cal Z}_{l,m}(1;\tilde
q^2)$.
The function ${\cal Z}_{0,0}(1;\tilde q^2)$ is shown in the body of this paper
in fig.~\ref{fig:A1+ l0}.  
As discussed by L\"uscher, the functions ${\cal Z}_{l,m}(1;\tilde q^2)$ vanish
for all odd-$l$, and also vanishes for $l=2$.  
The function ${\cal Z}_{4,0}(1;\tilde q^2)$ is shown in fig.~\ref{fig:Z40Z60}, and
exhibits some structure that is not present in  ${\cal Z}_{0,0}(1;\tilde q^2)$.
There are branches of ${\cal Z}_{4,0}(1;\tilde q^2)$ that are non-monotonic,
for instance, between $\tilde q^2=1$ and  $\tilde q^2=2$. 
This behavior is found in all of the ${\cal Z}_{l,m}$'s with $l>0$.
The functions ${\cal Z}_{8,0}(1;\tilde q^2)$ and
${\cal Z}_{10,0}(1;\tilde q^2)$ are shown in fig.~\ref{fig:Z80Z100},
\begin{figure}[h]
  \centering
\input{Z80100.tex}
     \caption{ The function ${\cal Z}_{8,0}(1;\tilde q^2)$ (top-panel)
and ${\cal Z}_{10,0}(1;\tilde q^2)$ (bottom-panel).  
}
  \label{fig:Z80Z100}
\end{figure}
${\cal Z}_{12,0}(1;\tilde q^2)$ and
${\cal Z}_{12,4}(1;\tilde q^2)$  in fig.~\ref{fig:Z120Z124},
\begin{figure}[h]
  \centering
  \input{Z120124.tex}
     \caption{ The function ${\cal Z}_{12,0}(1;\tilde q^2)$ (top-panel)
and ${\cal Z}_{12,4}(1;\tilde q^2)$ (bottom-panel).  
}
  \label{fig:Z120Z124}
\end{figure}
${\cal Z}_{14,0}(1;\tilde q^2)$ in fig.~\ref{fig:Z140},
\begin{figure}[h]
  \centering
  \input{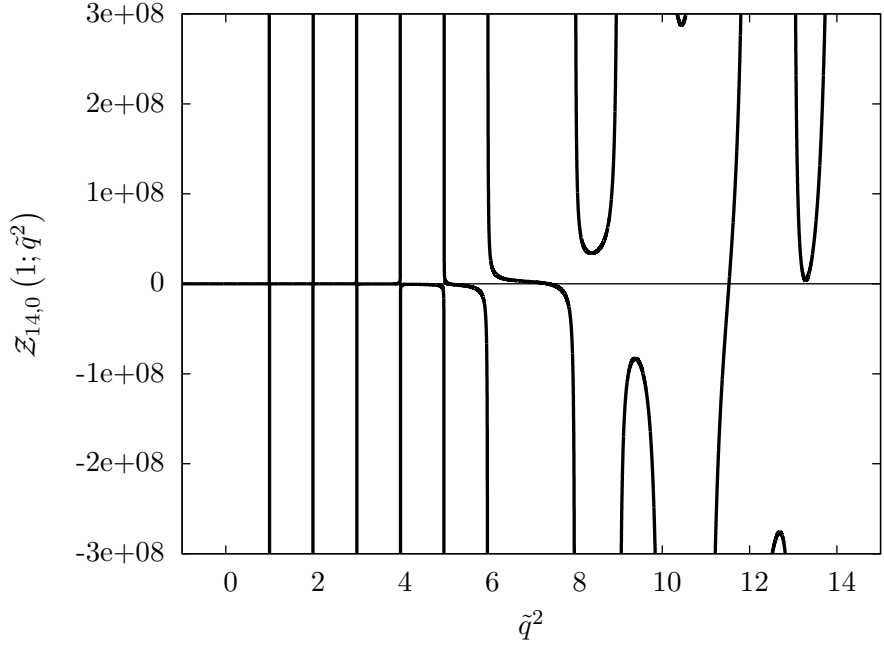}
     \caption{ The function ${\cal Z}_{14,0}(1;\tilde q^2)$.}
 \label{fig:Z140}
\end{figure}
${\cal Z}_{16,0}(1;\tilde q^2)$ and
${\cal Z}_{16,4}(1;\tilde q^2)$  in fig.~\ref{fig:Z160Z164},
\begin{figure}[h]
  \centering
  \input{Z160164.tex}
     \caption{ The functions ${\cal Z}_{16,0}(1;\tilde q^2)$ (top-panel)
and ${\cal Z}_{16,4}(1;\tilde q^2)$ (bottom-panel).  
}
  \label{fig:Z160Z164}
\end{figure}
and finally ${\cal Z}_{18,0}(1;\tilde q^2)$ and
${\cal Z}_{18,4}(1;\tilde q^2)$  in fig.~\ref{fig:Z180Z184}.
\begin{figure}[h]
  \centering
  \input{Z180184.tex}
     \caption{ The functions ${\cal Z}_{18,0}(1;\tilde q^2)$ (top-panel)
and ${\cal Z}_{18,4}(1;\tilde q^2)$ (bottom-panel).  
}
  \label{fig:Z180Z184}
\end{figure}

In constructing the perturbative expressions for the energy-eigenvalues in
terms of the $\delta_l$, the leading contributions result from the residue of
the pole of the leading function.  We present a few of these residues 
of the ${\cal Z}_{l,m}(1;\tilde q^2)$ functions
 in 
table~\ref{tab:resZLM} and table~\ref{tab:resZLMb}.
\begin{table}
\begin{center}
\begin{minipage}[t]{16.5 cm}
\caption{Residues of the functions 
$ \sqrt{4\pi}\  {\cal Z}_{l,m}(1;\tilde q^2)$ for $l\le 12$, 
i.e. the coefficient of
  $-{1\over\tilde Q^2}$ where $\tilde q^2 = |{\bf n}|^2 + \tilde Q^2$.
}
\label{tab:resZLM}
\end{minipage}
\begin{tabular}{| c || c | c | c | c | c | c | c |}
\hline \hline
  $|{\bf n}|^2$   
   &    $R\left[  \sqrt{4\pi} {\cal Z}_{0,0}  \right]$
   &    $R\left[  \sqrt{4\pi} {\cal Z}_{4,0}  \right]$
   &    $R\left[  \sqrt{4\pi} {\cal Z}_{6,0}  \right]$
   &    $R\left[  \sqrt{4\pi} {\cal Z}_{8,0} \right]$
   &    $R\left[  \sqrt{4\pi} {\cal Z}_{10,0}  \right]$
   &    $R\left[  \sqrt{4\pi} {\cal Z}_{12,0}  \right]$
   &    $R\left[  \sqrt{4\pi} {\cal Z}_{12,4}  \right]$
    \\
      \hline\hline
$0$ & 1 & 0 & 0  & 0 & 0 & 0 & 0
\\
$1$ & 6 & ${21\over 2}$ & ${3\sqrt{13} \over 4}$  & ${99\sqrt{17}\over 32}$ &
${65\sqrt{21}\over 64}$ & ${3715\over 256}$ & ${75\sqrt{1001}\over 512}$ 
\\
$2$ & 12 & -21 & $-{39\sqrt{13}\over 2}$   & ${891\sqrt{17}\over 16}$ &
$-{65\sqrt{21}\over 32}$ & -${43885\over 128}$ & $-{8085\sqrt{1001}\over 256}$
\\
$3$ & 8 & -84 &   ${48\sqrt{13}}$   & ${99\sqrt{17}}$ & $-520\sqrt{21}$
& ${2225\over 2}$ & $-{555\sqrt{1001}\over 4}$
\\
$4$ & 6 & 168 &   ${48\sqrt{13}}$ & ${792\sqrt{17}}$ & $1040\sqrt{21}$ 
& $59440$ & $ 600\sqrt{1001}$
\\
$5$ & 24 & 210 &  $-{255\sqrt{13}}$ & $-{13365\sqrt{17}\over 8}$ &
$-{204425\sqrt{21}\over 16}$ & ${21172895\over 64}$ & ${1262535\sqrt{1001}\over 128}$
\\
$6$ & 24 & -378 &  ${333\sqrt{13}}$ & $-{51381\sqrt{17}\over 8}$ &
${408915\sqrt{21}\over 16}$ & ${58441035\over 64}$  & $-{324765\sqrt{1001}\over 128}$ 
\\
$14$ & $48$ & $-4116$ & $-1806\sqrt{13}$ & $-{455301\sqrt{17}\over 4}$ &
${19718335\sqrt{21}\over 8}$ & ${2472346835\over 32}$ &
${309589035\sqrt{1001}\over 64}$
\\
\hline
\end{tabular}
\begin{minipage}[t]{16.5 cm}
\vskip 0.5cm
\noindent
\end{minipage}
\end{center}
\end{table}     
\begin{table}
\begin{center}
\begin{minipage}[t]{16.5 cm}
\caption{Residues of the functions 
$ \sqrt{4\pi}\  {\cal Z}_{l,m}(1;\tilde q^2)$ for $14\le l \le 18$, 
i.e. the coefficient of
  $-{1\over\tilde Q^2}$ where $\tilde q^2 = |{\bf n}|^2 + \tilde Q^2$.
}
\label{tab:resZLMb}
\end{minipage}
\begin{tabular}{| c || c | c | c | c | c |}
\hline \hline
  $|{\bf n}|^2$   
   &    $R\left[  \sqrt{4\pi} {\cal Z}_{14,0}  \right]$
   &    $R\left[  \sqrt{4\pi} {\cal Z}_{16,0}  \right]$
   &    $R\left[  \sqrt{4\pi} {\cal Z}_{16,4}  \right]$
   &    $R\left[  \sqrt{4\pi} {\cal Z}_{18,0} \right]$
   &    $R\left[  \sqrt{4\pi} {\cal Z}_{18,4}  \right]$
    \\
      \hline\hline
$1$  &  $\frac{595 \sqrt{29}}{512}$ & $\frac{22819 \sqrt{33}}{8192}$ &
$\frac{33 \sqrt{323323}}{4096}$ & $
\frac{20613 \sqrt{37}}{16384}$ &
  $ -\frac{39 \sqrt{920227}}{8192}$ \\
$2$  &$ -\frac{52955 \sqrt{29}}{256}$ & $\frac{2627491 \sqrt{33}}{4096} $&
$\frac{5937 \sqrt{323323}}{2048}$ 
& $-\frac{234021\sqrt{37}}{8192} $& $\frac{9063 \sqrt{920227}}{4096}$ \\
$3$  & $3570 \sqrt{29}$ & $-\frac{135261 \sqrt{33}}{16}$ & $\frac{873
  \sqrt{323323}}{8}$ & $-12453 \sqrt{37}$ &
$ 18 \sqrt{920227}$ \\
$4$  & $19040 \sqrt{29}$ & $182552 \sqrt{33}$ & $528 \sqrt{323323} $& $329808 \sqrt{37}$ & $-1248 \sqrt{920227}$ \\
$5$  & $-\frac{9356375 \sqrt{29}}{128}$ & $\frac{3105175795 \sqrt{33}}{2048}$ &
$\frac{1267665 \sqrt{323323}}{1024}$
 & $-\frac{2195267925 \sqrt{37}}{4096}$ & $-\frac{12933225 \sqrt{920227}}{2048}$ \\
$6$  & $-\frac{144911655 \sqrt{29}}{128}$ & $\frac{1849263939 \sqrt{33}}{2048}$
& $\frac{15619473 \sqrt{323323}}{1024}$ & $-\frac{40787860977
   \sqrt{37}}{4096}$ & $-\frac{26504469 \sqrt{920227}}{2048}$
\\
$14$ & $ \frac{6869655205 \sqrt{29}}{64}$ &
$-\frac{12080581399901\sqrt{33}}{1024}$ 
& $\frac{4554268593\sqrt{323323}}{512}$ & 
$\frac{44894607368667 \sqrt{37}}{2048} $ & 
$\frac{51258039399 \sqrt{920227}}{1024}$
\\
\hline
\end{tabular}
\begin{minipage}[t]{16.5 cm}
\vskip 0.5cm
\noindent
\end{minipage}
\end{center}
\end{table}     
%

\section{Perturbative Expressions
\label{sect:perturbation}}
\noindent
In many instances the energy-shifts due to the interactions are small because
the phase-shift is small and/or the lattice volume is large.  
In such
instances, a perturbative expression can be used to extract the phase-shift
from an energy-eigenvalue instead of solving the full expression, as
discussed by L\"uscher~\cite{Luscher:1986pf,Luscher:1990ux}.
The energy-eigenvalues for a given
$\Gamma^{(i)}$ in a given $|{\bf n}|^2$-shell, and more specifically
$\tilde q^2$, can be expanded  
in terms of  
the dimensionless quantity
$\left( L^{(2l+1}q^{2l+1}\cot\delta_l^I \right)^{-1}$~\footnote{The
  corresponding expansion that is appropriate for systems near unitarity, an
  expansion in terms of 
 $\left( L q \cot\delta \right)$ 
for s-wave interactions
that is small for large scattering lengths,
can be found in Ref.~\cite{Beane:2003da}.
}.
In the case of a single partial-wave, 
the general form for the energy of an irrep, $\Gamma^{(i)}$, in the  $|{\bf n}|^2$-shell is 
\begin{eqnarray}
q^{2l+1}\cot\delta_l^I & = & 
{1\over\pi L} \ \left({2\pi\over L}\right)^{2 l}
\left[ 
{\alpha_{-1}^{(|{\bf n}|^2,\Gamma^{(i)})}\over \delta \tilde q^2}
\ +\ \alpha_{0}^{(|{\bf n}|^2,\Gamma^{(i)})}
\ +\ \alpha_{1}^{(|{\bf n}|^2,\Gamma^{(i)})}\delta  \tilde q^2
\ +\  ... \ \right]
\ \ \ ,
\label{eq:pertparam}
\end{eqnarray}
where the solutions to eq.~(\ref{eq:pertparam}) can be written as
\begin{eqnarray}
\tilde q^2_{|{\bf n}|^2,\Gamma} & = & |{\bf n}|^2 + \delta  \tilde q^2 = \left({qL\over
    2\pi}\right)^2
\nonumber\\
 & = & 
|{\bf n}|^2\ +\ 
g^{(|{\bf n}|^2,\Gamma^{(i)})}_0\ \tan\delta_l^I\ 
\left(\ 1\ +\ g^{(|{\bf n}|^2,\Gamma^{(i)})}_1 \tan\delta_l^I
\ +\ g^{(|{\bf n}|^2,\Gamma^{(i)})}_2 \tan^2\delta_l^I\ +\
  ...\right)
\nonumber\\
 & + & 
h^{(|{\bf n}|^2,\Gamma^{(i)})}_0 \ 
{ {d\over d(Lq)^2}\left( (Lq)^{2l+1}\cot\delta_l^I \right)\over 
\left[ (Lq)^{2l+1}\cot\delta_l^I\right]^3 }
\ +\ ...
\ \ \ ,
\label{eq:pertexp}
\end{eqnarray}
where the phase-shift is evaluated at the unperturbed energy of the state,
and the coefficients are
\begin{eqnarray}
g^{(|{\bf n}|^2,\Gamma^{(i)})}_0
& = & 
{ \alpha_{-1}^{(|{\bf n}|^2,\Gamma^{(i)})}
\over 2\pi^2 |{\bf n}|^{2l+1} }
\ \ ,\ \ 
g^{(|{\bf n}|^2,\Gamma^{(i)})}_1
\ = \ 
{ \alpha_{0}^{(|{\bf n}|^2,\Gamma^{(i)})}
\over 2\pi^2 |{\bf n}|^{2l+1} }
\nonumber\\
g^{(|{\bf n}|^2,\Gamma^{(i)})}_2
& = & 
{ (\alpha_{0}^{(|{\bf n}|^2,\Gamma^{(i)})})^2 
+ \alpha_{-1}^{(|{\bf n}|^2,\Gamma^{(i)})} \alpha_{1}^{(|{\bf n}|^2,\Gamma^{(i)})}
\over 2\pi^2 |{\bf n}|^{2l+1} }
\ \ \ .
\label{eq:galp}
\end{eqnarray}

For the terms in eq.~(\ref{eq:pertexp}), the contributions scale as 
\begin{eqnarray}
\tilde q^2_{|{\bf n}|^2,\Gamma} & \sim & 
{\cal O}\left(1\right)
+ 
{\cal O}\left({1\over L^{2l+1}}\right)
+ 
{\cal O}\left({1\over L^{4l+2}}\right)
+ 
{\cal O}\left({1\over L^{6l+3}}\right)
+ 
{\cal O}\left({1\over L^{4l+4}}\right)
+...
\ ,
\label{eq:pertexpparam}
\end{eqnarray}
from which it can be determined when the contributions from 
higher partial-waves become important.  
For instance, the energy-shifts in the 
$T_1^-$ irrep are  dominated by $\delta_1$, and the expansion
is of the form 
\begin{eqnarray}
\tilde q^2_{|{\bf n}|^2,T_1^-} & \sim & 
{\cal O}\left(1\right)
\ +\ 
{\cal O}\left({1\over L^{3}}\right)
\ +\ 
{\cal O}\left({1\over L^{6}}\right)
\ +\ 
{\cal O}\left({1\over L^{9}}\right)
\ +\ 
{\cal O}\left({1\over L^{8}}\right)
\ +\ ...
\ \ \ ,
\label{eq:pertexpparamT1m}
\end{eqnarray}
respectively, and the $l=3$ partial-wave first contributes at 
$ {\cal O}\left({1\over L^{7}}\right)$.
In the case of the $T_1^+$ irrep, which is dominated by $\delta_4$,
the  expansion is of the form 
\begin{eqnarray}
\tilde q^2_{|{\bf n}|^2,T_1^+} & \sim & 
{\cal O}\left(1\right)
\ +\ 
{\cal O}\left({1\over L^{9}}\right)
\ +\ 
{\cal O}\left({1\over L^{18}}\right)
\ +\ 
{\cal O}\left({1\over L^{27}}\right)
\ +\ 
{\cal O}\left({1\over L^{20}}\right)
\ +\ ...
\ \ \ ,
\label{eq:pertexpparamT1p}
\end{eqnarray}
and $\delta_6$ contributions are of the form
$ {\cal O}\left(L^{-13}\right)$.
Therefore, the order at which the higher partial-waves contribute in the
large-volume limit depends upon the $\Gamma^{(i)}$.

The perturbative expansions of the lowest few $A_1^+$ energy-eigenvalues 
in terms of the $l=0$ phase-shift $\delta_0$
were given by
L\"uscher~\cite{Luscher:1986pf}, and here we simply extend those results to
levels with $|{\bf n}|^2\le 6$ with the coefficients given in table~\ref{tab:A1pCOEFFS}.
\begin{table}
\begin{center}
\begin{minipage}[!ht]{16.5 cm}
\caption{
The coefficients, $g_i^{(|{\bf n}|^2,A_1^+)}$ that contribute to the perturbative expansion
of the energy-eigenvalues of states in the $A_1^+$ irrep of the cubic group, as
given in eq.~(\ref{eq:pertexp}), in terms of s-wave phase-shift $\delta_0$.
}
\label{tab:A1pCOEFFS}
\end{minipage}
\begin{tabular}{| c | c | c | c| }
\hline
$|{\bf n}|^2$ & $g^{(|{\bf n}|^2,A_1^+)}_0$  & $g^{(|{\bf n}|^2,A_1^+)}_1$  
& $g^{(|{\bf n}|^2,A_1^+)}_2$ \\
\hline 
 1 & $-\frac{3}{\pi ^2}$ & -0.06137 & -0.3542 \\
 2 & $-\frac{3 \sqrt{2}}{\pi ^2}$ & -0.1826 & -0.3618 \\
 3 & $-\frac{4}{\sqrt{3} \pi ^2}$ & -0.1981 & -0.1996 \\
 4 & $-\frac{3}{2 \pi ^2}$ & 0.2415 & -0.1328 \\
 5 & $-\frac{12}{\sqrt{5} \pi ^2}$ & 0.1590 & -0.5155 \\
 6 & $-\frac{2 \sqrt{6}}{\pi ^2}$ & -0.4798 & -0.2025\\
\hline
\end{tabular}
\begin{minipage}[t]{16.5 cm}
\vskip 0.5cm
\noindent
\end{minipage}
\end{center}
\end{table}     
The energy of the 
$A_1^+$ state in the $|{\bf n}|^2=0$ level can be expressed in terms of the
s-wave scattering
parameters defining the low-energy behavior of the phase-shift, and it is
well-known that
\begin{eqnarray}
\tilde q^2_{|{\bf 0}|^2,A_1^+} & = & 
-{a_0\over \pi L}\ 
\left(\ 1\ +\ c_1 \left({a_0\over L}\right) 
\ +\ c_2 \left({a_0\over L}\right)^2\ +\ ...
\right) 
 +\ ...
\ \ \ .
\label{eq:pertexpneq0}
\end{eqnarray}
where the particle-physics convention for defining the scattering length has
been used, and the coefficients are $c_1 = -2.8373$ and $c_2=6.3752$.

An important point to note is that the perturbative 
energy-shifts that are presented 
in table~\ref{tab:A1pCOEFFS}-table~\ref{tab:T2pCOEFFS}
are for one of the occurrences of
the $\Gamma^{(i)}$ that form a given $|{\bf n}|^2$-shell.
Other occurrences are unperturbed at leading order.
When multiple occurrences of a given irrep
appear in a given $|{\bf n}|^2$-shell, the leading interactions will perturb
the energy of one combination, while leaving the other states unperturbed,  but
the interactions in 
higher partial-waves will perturb these remaining states.
The expansion coefficients for the lowest-lying $T_1^-$ (dominated by
$\delta_1^1$), 
the $E^+$ and $T_2^+$ (both dominated by $\delta_2^I$)
are shown in table~\ref{tab:T1mCOEFFS}, table~\ref{tab:EpCOEFFS}, and 
table~\ref{tab:T2pCOEFFS}, respectively. 
We note that the coefficients in the perturbative expansion of the
energy-eigenstates in the $T_1^-$ irrep given in table~\ref{tab:T1mCOEFFS}
differ from those given by
L\"uscher~\cite{Luscher:1990ux}.
This can be attributed to the fact that 
$q^{(2l+1)}\text{cot}\delta_l$ 
can be expanded in a power-series in energy about 
threshold~\footnote{This is the effective
range expansion which is valid below the threshold of the t-channel cut, 
$|{\bf q}| = m_\pi$ for $\pi\pi\rightarrow\pi\pi$.},
as performed in this work, 
while $q\text{cot}\delta_l$ does not have such an expansion for $l>0$.
\begin{table}
\begin{center}
\begin{minipage}[!ht]{16.5 cm}
\caption{
The coefficients, $g_i^{(|{\bf n}|^2,T_1^{-(1)})}$ that contribute to the perturbative expansion
of the energy-eigenvalues of states in the (first occurrence of the) 
$T_1^-$ irrep of the cubic group, as
given in eq.~(\ref{eq:pertexp}), in terms of $l=1$ phase-shift $\delta_1$.
}
\label{tab:T1mCOEFFS}
\end{minipage}
\begin{tabular}{| c | c | c | c| }
\hline
$|{\bf n}|^2$ & $g^{(|{\bf n}|^2,T_1^{-(1)})}_0$  & $g^{(|{\bf n}|^2,T_1^{-(1)}}_1$  
& $g^{(|{\bf n}|^2,T_1^{-(1)})}_2$ \\
\hline 
 1 & $-\frac{3}{\pi ^2} $& -0.3653 & -0.2058 \\
 2 & $-\frac{3 \sqrt{2}}{\pi ^2}$ & -0.3975 & -0.1979 \\
 3 & $-\frac{4}{\sqrt{3} \pi ^2}$ & -0.2761 & -0.1471 \\
 4 & $-\frac{3}{2 \pi ^2}$ & 0.2035 & -0.1589 \\
 5 & $-\frac{12}{\sqrt{5} \pi ^2}$ & 0.05024 & -0.5555 \\
 6 & $-\frac{2 \sqrt{6}}{\pi ^2} $ & -0.5625 & -0.07659\\
\hline
\end{tabular}
\begin{minipage}[t]{16.5 cm}
\vskip 0.5cm
\noindent
\end{minipage}
\end{center}
\end{table}     
\begin{table}
\begin{center}
\begin{minipage}[!ht]{16.5 cm}
\caption{
The coefficients, $g_i^{(|{\bf n}|^2,E^{+(1)})}$ that contribute to the perturbative expansion
of the energy-eigenvalues of states in the (first occurrence of the) 
$E^+$ irrep of the cubic group, as
given in eq.~(\ref{eq:pertexp}), in terms of $l=2$ phase-shift $\delta_2$.
}
\label{tab:EpCOEFFS}
\end{minipage}
\begin{tabular}{| c | c | c | c| }
\hline
$|{\bf n}|^2$ & $g^{(|{\bf n}|^2,E^{+(1)})}_0$  & $g^{(|{\bf n}|^2,E^{+(1)})}_1$  
& $g^{(|{\bf n}|^2,E^{+(1)})}_2$ \\
\hline 
 1 & $-\frac{15}{2 \pi ^2}$ & -1.5672 & 2.5842 \\
 2 & $-\frac{15}{4 \sqrt{2} \pi ^2}$ & -0.8065 & 0.54 \\
 4 & $-\frac{15}{4 \pi ^2}$ & 0.3272 & -0.421 \\
 5 & $-\frac{78}{5 \sqrt{5} \pi ^2}$ & -0.447 & -0.4331 \\
 6 & $-\frac{5 }{2 \pi ^2}\sqrt{\frac{3}{2}} $& -0.7884 & 0.3746\\
\hline
\end{tabular}
\begin{minipage}[t]{16.5 cm}
\vskip 0.5cm
\noindent
\end{minipage}
\end{center}
\end{table}     
\begin{table}
\begin{center}
\begin{minipage}[!ht]{16.5 cm}
\caption{
The coefficients, $g_i^{(|{\bf n}|^2,T_2^{+(1)})}$ that contribute to the perturbative expansion
of the energy-eigenvalues of states in the  (first occurrence of the) 
$T_2^+$ irrep of the cubic group, as
given in eq.~(\ref{eq:pertexp}), in terms of $l=2$ phase-shift $\delta_2$.
}
\label{tab:T2pCOEFFS}
\end{minipage}
\begin{tabular}{| c | c | c | c| }
\hline
$|{\bf n}|^2$ & $g^{(|{\bf n}|^2,T_2^{+(1)})}_0$  & $g^{(|{\bf n}|^2,T_2^{+(1)})}_1$  
& $g^{(|{\bf n}|^2,T_2^{+(1)})}_2$ \\
\hline 
 2 & $-\frac{15}{2 \sqrt{2} \pi ^2}$ & -0.4830 & -0.1828 \\
 3 & $-\frac{20}{3 \sqrt{3} \pi ^2}$ & -0.6737 & 0.2128 \\
 5 & $-\frac{48}{5 \sqrt{5} \pi ^2}$ & 0.2004 & -0.4515 \\
 6 & $-\frac{5 \sqrt{\frac{3}{2}}}{\pi ^2}$ & -0.5497 & -0.0902 \\
\hline
\end{tabular}
\begin{minipage}[t]{16.5 cm}
\vskip 0.5cm
\noindent
\end{minipage}
\end{center}
\end{table}     
For the remaining irreps, the $T_2^-$, $A_2^-$, $T_1^+$, $A_2^+$ and $E^-$, the
expansion converges rapidly with just one non-trivial term, ${\cal O}\left(\tan\delta_l^I\right)$.
The leading coefficients for these expansion of the energy-eigenvalues for each
of these irreps are given in table~\ref{tab:higherCOEFFS},
along with the coefficients in the expansions for the second occurrences of the
$E^+$, $T_2^+$, $T_1^-$ and $T_2^-$.
The perturbative expansion of the lowest-lying $A_1^-$ state is given in 
table~\ref{tab:LowestPerturb}.
\begin{table}
\begin{center}
\begin{minipage}[!ht]{16.5 cm}
\caption{
The coefficients, $g_0^{(|{\bf n}|^2,\Gamma)}$ that contribute to the perturbative expansion
of the energy-eigenvalues of states in the 
$A_2^-$, $T_2^-$, $T_1^+$, $E^-$ and $A_2^+$, as
given in eq.~(\ref{eq:pertexp}), in terms of dominant phase-shifts $\delta_3$, 
$\delta_3$, $\delta_4$, $\delta_5$, and $\delta_6$, respectively.
Also given are the coefficients
in the  perturbative expansion of the second occurrence of $E^+$, $T_2^+$, $T_1^-$ and $T_2^-$,
in terms of dominant phase-shifts $\delta_4$, 
$\delta_4$, $\delta_3$, and $\delta_5$, respectively.
}
\label{tab:higherCOEFFS}
\end{minipage}
\begin{tabular}{| c | c | c | c | c | c| c | c | c | c | }
\hline
$|{\bf n}|^2$ 
& $g^{(|{\bf n}|^2,A_2^-)}_0$  
& $g^{(|{\bf n}|^2,T_2^-)}_0$  
& $g^{(|{\bf n}|^2,T_1^+)}_0$  
& $g^{(|{\bf n}|^2,E^-  )}_0$  
& $g^{(|{\bf n}|^2,A_2^+)}_0$ 
& $g^{(|{\bf n}|^2,E^{+(2)})}_0$  
& $g^{(|{\bf n}|^2,T_2^{+(2)})}_0$  
& $g^{(|{\bf n}|^2,T_1^{-(2)})}_0$  
& $g^{(|{\bf n}|^2,T_2^{-(2)})}_0$ 
\\ 
\hline 
 2 & 0 & $-\frac{105}{8 \sqrt{2} \pi ^2}$ & 0 & 0 & 0 &0&0&0&0\\
 3 & $-\frac{140}{9 \sqrt{3} \pi ^2}$ & 0 & 0 & 0 & 0 &0&0&0&0\\
 5 & 0 & $-\frac{84}{5 \sqrt{5} \pi ^2}$ & $-{2268\over 125\sqrt{5}\pi^2}$ & 0
 & $-{162162\over 3125\sqrt{5}\pi^2}$ & $-{23814\over 1625\sqrt{5}\pi^2}$ &
0 & $-{252\over 25\sqrt{5}\pi^2}$ & $-{74844\over 3125\sqrt{5}\pi^2}$
\\
 6 & $-\frac{35 \sqrt{\frac{2}{3}}}{3 \pi ^2}$ & $-\frac{35}{4 \sqrt{6} \pi
   ^2}$ & $-{35\sqrt{2}\over 4\sqrt{3} \pi ^2}$ & $-{385\over 12\sqrt{6} \pi
   ^2}$ & 0
& 0 & $-{245\sqrt{2}\over 36\sqrt{3}\pi^2}$ &  $-{175\sqrt{2}\over
  36\sqrt{3}\pi^2}$
& 0 \\
\hline
\end{tabular}
\begin{minipage}[t]{16.5 cm}
\vskip 0.5cm
\noindent
\end{minipage}
\end{center}
\end{table}     

\begin{acknowledgments}
\noindent
We would like to thank David Kaplan for inspiring discussions.
The work of
TL was performed under the auspices of the U.S.~Department of Energy
by Lawrence Livermore National Laboratory under Contract
DE-AC52-07NA27344 and the UNEDF SciDAC grant DE-FC02-07ER41457. 
The work of MJS was supported in part by the
U.S.~Dept.~of Energy under Grant No.~DE-FG02-97ER41014. 
\end{acknowledgments}

\newpage

%
%

\end{document}